\documentclass{amsbook}
\usepackage{graphicx}
\usepackage{amscd}
\usepackage{amsmath}
\usepackage{amsfonts}
\usepackage{amssymb}

\let \frak = \mathfrak

\theoremstyle{plain}
\newtheorem{theorem}{Theorem}

\newtheorem{corollary}[theorem]{Corollary}

\newtheorem{definition}[theorem]{Definition}

\newtheorem{lemma}[theorem]{Lemma}
\newtheorem{notation}[theorem]{Notation}

\newtheorem{proposition}[theorem]{Proposition}

\numberwithin{equation}{chapter}
\numberwithin{theorem}{chapter}

\begin{document}
\frontmatter
\title[Groups and Representations]{An Elementary Introduction to Groups and Representations}
\author{Brian C. Hall}
\address{University of Notre Dame\\
Department of Mathematics\\
Notre Dame IN 46556 USA}
\email{bhall@nd.edu}
\maketitle
\tableofcontents

\section{Preface}

These notes are the outgrowth of a graduate course on Lie groups I taught at
the University of Virginia in 1994. In trying to find a text for the course I
discovered that books on Lie groups either presuppose a knowledge of
differentiable manifolds or provide a mini-course on them at the beginning.
Since my students did not have the necessary background on manifolds, I faced
a dilemma: either use manifold techniques that my students were not familiar
with, or else spend much of the course teaching those techniques instead of
teaching Lie theory. To resolve this dilemma I chose to write my own notes
using the notion of a \textit{matrix Lie group}. A matrix Lie group is simply
a closed subgroup of $\mathsf{GL}(n;\mathbb{C}).$ Although these are often
called simply ``matrix groups,'' my terminology emphasizes that every matrix
group \textit{is} a Lie group.

This approach to the subject allows me to get started quickly on Lie group
theory proper, with a minimum of prerequisites. Since most of the interesting
examples of Lie groups are matrix Lie groups, there is not too much loss of
generality. Furthermore, the proofs of the main results are ultimately similar
to standard proofs in the general setting, but with less preparation.

Of course, there is a price to be paid and certain constructions (e.g.
covering groups) that are easy in the Lie group setting are problematic in the
matrix group setting. (Indeed the universal cover of a matrix Lie group need
not be a matrix Lie group.) On the other hand, the matrix approach suffices
for a first course. Anyone planning to do research in Lie group theory
certainly needs to learn the manifold approach, but even for such a person it
might be helpful to start with a more concrete approach. And for those in
other fields who simply want to learn the basics of Lie group theory, this
approach allows them to do so quickly.

These notes also use an atypical approach to the theory of semisimple Lie
algebras, namely one that starts with a detailed calculation of the
representations of $\mathsf{sl}(3;\mathbb{C})$. My own experience was that the
theory of Cartan subalgebras, roots, Weyl group, etc., was pretty difficult to
absorb all at once. I have tried, then, to motivate these constructions by
showing how they are used in the representation theory of the simplest
representative Lie algebra. (I also work out the case of $\mathsf{sl}%
(2;\mathbb{C}),$ but this case does not adequately illustrate the general theory.)

In the interests of making the notes accessible to as wide an audience as
possible, I have included a very brief introduction to abstract groups, given
in Chapter 1. In fact, not much of abstract group theory is needed, so the
quick treatment I give should be sufficient for those who have not seen this
material before.

I am grateful to many who have made corrections, large and small, to the
notes, including especially Tom Goebeler, Ruth Gornet, and Erdinch Tatar.

\mainmatter

\chapter{Groups}

\section{Definition of a Group, and Basic Properties}

\begin{definition}
A \textbf{group} is a set $G$, together with a map of $G\times G$ into $G$
(denoted $g_{1}\ast g_{2}$) with the following properties:

First, associativity: for all $g_{1},g_{2}\in G$,
\begin{equation}
g_{1}\ast(g_{2}\ast g_{3})=(g_{1}\ast g_{2})\ast g_{3}\text{.}%
\label{Associativity}%
\end{equation}
Second, there exists an element $e$ in $G$ such that for all $g\in G$,
\begin{equation}
g\ast e=e\ast g=g\text{.}\label{Identity}%
\end{equation}
and such that for all $g\in G$, there exists $h\in G$ with
\begin{equation}
g\ast h=h\ast g=e\text{.}\label{Inverses}%
\end{equation}

If $g\ast h=h\ast g$ for all $g,h\in G$, then the group is said to be
\textbf{commutative}\emph{\ }(or \textbf{abelian}).
\end{definition}

The element $e$ is (as we shall see momentarily) unique, and is called the
\textbf{identity element} of the group, or simply the \textbf{identity}. Part
of the definition of a group is that multiplying a group element $g$ by the
identity on \textit{either the right or the left} must give back $g$.

The map of $G\times G$ into $G$ is called the \textbf{product operation} for
the group. Part of the definition of a group $G$ is that the product operation
map $G\times G$ into $G$, i.e., that the product of two elements of $G$ be
again an element of $G$. This property is referred to as \textbf{closure}.

Given a group element $g$, a group element $h$ such that $g*h=h*g=e$ is called
an \textbf{inverse} of $g$. We shall see momentarily that each group element
has a \textit{unique} inverse.

Given a set and an operation, there are four things that must be checked to
show that this is a group: \textit{closure}, \textit{associativity}, existence
of an \textit{identity}, and existence of \textit{inverses}.

\begin{proposition}
[Uniqueness of the Identity]Let $G$ be a group, and let $e,f\in G$ be such
that for all $g\in G$%
\begin{align*}
e\ast g  & =g\ast e=g\\
f\ast g  & =g\ast f=g\text{.}%
\end{align*}
Then $e=f$.
\end{proposition}

\begin{proof}
Since $e$ is an identity, we have
\[
e\ast f=f\text{.}%
\]
On the other hand, since $f$ is an identity, we have
\[
e\ast f=e\text{.}%
\]
Thus $e=e\ast f=f$.
\end{proof}

\begin{proposition}
[Uniqueness of Inverses]Let $G$ be a group, $e$ the (unique) identity of $G$,
and $g,h,k$ arbitrary elements of $G$. Suppose that
\begin{align*}
g\ast h  & =h\ast g=e\\
g\ast k  & =k\ast g=e.
\end{align*}
Then $h=k$.
\end{proposition}

\begin{proof}
We know that $g\ast h=g\ast k$ $(=e)$. Multiplying on the left by $h$ gives
\[
h\ast(g\ast h)=h\ast(g\ast k)\text{.}%
\]
By associativity, this gives
\[
(h\ast g)\ast h=(h\ast g)\ast k\text{,}%
\]
and so
\begin{align*}
e\ast h=e\ast k\\
h=k\text{.}%
\end{align*}
\end{proof}

This is what we wanted to prove.

\begin{proposition}
\label{one-sided}Let $G$ be a group, $e$ the identity element of $G$, and $g$
an arbitrary element of $G$. Suppose $h\in G$ satisfies \textit{either} $h\ast
g=e$ or $g\ast h=e$. Then $h$ is the (unique) inverse of $g$.
\end{proposition}

\begin{proof}
To show that $h$ is the inverse of $g$, we must show \textit{both} that $h\ast
g=e$ and $g\ast h=e$. Suppose we know, say, that $h\ast g=e$. Then our goal is
to show that this implies that $g\ast h=e$.

Since $h\ast g=e$,
\[
g\ast(h\ast g)=g\ast e=g\text{.}%
\]
By associativity, we have
\[
(g\ast h)\ast g=g\text{.}%
\]
Now, by the definition of a group, $g$ has an inverse. Let $k$ be that
inverse. (Of course, in the end, we will conclude that $k=h$, but we cannot
assume that now.) Multiplying on the right by $k$ and using associativity
again gives
\begin{align*}
((g\ast h)\ast g)\ast k=g\ast k=e\\
(g\ast h)\ast(g\ast k)=e\\
(g\ast h)\ast e=e\\
g\ast h=e\text{.}%
\end{align*}

A similar argument shows that if $g\ast h=e$, then $h\ast g=e$.
\end{proof}

Note that in order to show that $h\ast g=e$ implies $g\ast h=e$, we used the
fact that $g$ has an inverse, since it is an element of a group. In more
general contexts (that is, in some system which is not a group), one may have
$h\ast g=e$ but not $g\ast h=e$. (See Exercise \ref{counter-example}.)

\begin{notation}
For any group element $g$, its unique inverse will be denoted $g^{-1}$.
\end{notation}

\begin{proposition}
[Properties of Inverses]\label{inverse.properties}Let $G$ be a group, $e$ its
identity, and $g,h$ arbitrary elements of $G$. Then
\begin{align*}
\left(  g^{-1}\right)  ^{-1}=g\\
\left(  gh\right)  ^{-1}=h^{-1}g^{-1}\\
e^{-1}=e\text{.}%
\end{align*}
\end{proposition}

\begin{proof}
Exercise.
\end{proof}

\section{Some Examples of Groups\label{ex}}

From now on, we will denote the product of two group elements $g_{1}$ and
$g_{2}$ simply by $g_{1}g_{2}$, instead of the more cumbersome $g_{1}\ast
g_{2}$. Moreover, since we have associativity, we will write simply
$g_{1}g_{2}g_{3}$ in place of $(g_{1}g_{2})g_{3}$ or $g_{1}(g_{2}g_{3})$.

\subsection{The trivial group}

The set with one element, $e$, is a group, with the group operation being
defined as $ee=e$. This group is commutative.

Associativity is automatic, since both sides of (\ref{Associativity}) must be
equal to $e$. Of course, $e$ itself is the identity, and is its own inverse.
Commutativity is also automatic.

\subsection{The integers}

The set $\mathbb{Z}$ of integers forms a group with the product operation
being addition. This group is commutative.

First, we check \textit{closure}, namely, that addition maps $\mathbb{Z}%
\times\mathbb{Z}$ into $\mathbb{Z}$, i.e., that the sum of two integers is an
integer. Since this is obvious, it remains only to check
\textit{associativity}, \textit{identity}, and \textit{inverses}. Addition is
associative; zero is the additive identity (i.e., $0+n=n+0=n$, for all
$n\in\mathbb{Z}$); each integer $n$ has an additive inverse, namely, $-n$.
Since addition is commutative, $\mathbb{Z}$ is a commutative group.

\subsection{The reals and $\mathbb{R}^{n}$}

The set $\mathbb{R}$ of real numbers also forms a group under the operation of
addition. This group is commutative. Similarly, the $n$-dimensional Euclidean
space $\mathbb{R}^{n}$ forms a group under the operation of vector addition.
This group is also commutative.

The verification is the same as for the integers.

\subsection{Non-zero real numbers under multiplication}

The set of non-zero real numbers forms a group with respect to the operation
of multiplication. This group is commutative.

Again we check closure: the product of two non-zero real numbers is a non-zero
real number. Multiplication is associative; one is the multiplicative
identity; each non-zero real number $x$ has a multiplicative inverse, namely,
$\frac{1}{x}$. Since multiplication of real numbers is commutative, this is a
commutative group.

This group is denoted $\mathbb{R}^{*}$.

\subsection{Non-zero complex numbers under multiplication}

The set of non-zero complex numbers forms a group with respect to the
operation of complex multiplication. This group is commutative.

This group in denoted $\mathbb{C}^{*}$.

\subsection{Complex numbers of absolute value one under multiplication}

The set of complex numbers with absolute value one (i.e., of the form
$e^{i\theta}$) forms a group under complex multiplication. This group is commutative.

This group is the unit circle, denoted $S^{1}$.

\subsection{Invertible matrices}

For each positive integer $n$, the set of all $n\times n$ invertible matrices
with real entries forms a group with respect to the operation of matrix
multiplication. This group in non-commutative, for $n\geq2$.

We check closure: the product of two invertible matrices is invertible, since
$\left(  AB\right)  ^{-1}=B^{-1}A^{-1}$. Matrix multiplication is associative;
the identity matrix (with ones down the diagonal, and zeros elsewhere) is the
identity element; by definition, an invertible matrix has an inverse. Simple
examples show that the group is non-commutative, except in the trivial case
$n=1$. (See Exercise \ref{gl2}.)

This group is called the \textbf{general linear group} (over the reals), and
is denoted $\mathsf{GL}(n;\mathbb{R}\mathbf{)}$.

\subsection{Symmetric group (permutation group)}

The set of one-to-one, onto maps of the set $\left\{  1,2,\cdots n\right\}  $
to itself forms a group under the operation of composition. This group is
non-commutative for $n\geq3$.

We check closure: the composition of two one-to-one, onto maps is again
one-to-one and onto. Composition of functions is associative; the identity map
(which sends 1 to 1, 2 to 2, etc.) is the identity element; a one-to-one, onto
map has an inverse. Simple examples show that the group is non-commutative, as
long as $n$ is at least 3. (See Exercise \ref{sn}.)

This group is called the \textbf{symmetric group}, and is denoted $S_{n}$. A
one-to-one, onto map of $\left\{  1,2,\cdots n\right\}  $ is a permutation,
and so $S_{n}$ is also called the \textbf{permutation group}. The group
$S_{n}$ has $n!$ elements.

\subsection{Integers \textbf{mod} $n$}

The set $\left\{  0,1,\cdots n-1\right\}  $ forms a group under the operation
of addition \textbf{mod }$n$. This group is commutative.

Explicitly, the group operation is the following. Consider $a,b\in\left\{
0,1\cdots n-1\right\}  $. If $a+b<n$, then $a+b$ $\mathbf{mod}$ $n=a+b$, if
$a+b\geq n$, then $a+b$ $\mathbf{mod}$ $n=a+b-n$. (Since $a$ and $b$ are less
than $n$, $a+b-n$ is less than $n$; thus we have closure.) To show
associativity, note that both $(a+b\ \mathbf{mod}\ n)+c\ \mathbf{mod}\ n$ and
$a+(b+c\ \mathbf{mod}\ n)\ \mathbf{mod}\ n$ are equal to $a+b+c$, minus some
multiple of $n$, and hence differ by a multiple of $n$. But since both are in
the set $\left\{  0,1,\cdots n-1\right\}  $, the only possible multiple on $n$
is zero. Zero is still the identity for addition \textbf{mod} $n$. The inverse
of an element $a\in\left\{  0,1,\cdots n-1\right\}  $ is $n-a$. (Exercise:
check that $n-a$ is in $\left\{  0,1,\cdots n-1\right\}  $, and that
$a+(n-a)\ \mathbf{mod}\ n=0$.) The group is commutative because ordinary
addition is commutative.

This group is referred to as ``$\mathbb{Z}\mathbf{\ mod\ }n$,'' and is denoted
$\mathbb{Z}_{n}$.

\section{Subgroups, the Center, and Direct Products\label{sub}}

\begin{definition}
A \textbf{subgroup} of a group $G$ is a subset $H$ of $G$ with the following properties:

\begin{enumerate}
\item \label{id}The identity is an element of $H$.

\item \label{inv}If $h\in H$, then $h^{-1}\in H$.

\item \label{closure}If $h_{1},h_{2}\in H$, then $h_{1}h_{2}\in H$ .
\end{enumerate}
\end{definition}

The conditions on $H$ guarantee that $H$ is a group, with the same product
operation as $G$ (but restricted to $H$). Closure is assured by (\ref{closure}%
), associativity follows from associativity in $G$, and the existence of an
identity and of inverses is assured by (\ref{id}) and (\ref{inv}).

\subsection{Examples}

Every group $G$ has at least two subgroups: $G$ itself, and the one-element
subgroup $\left\{  e\right\}  $. (If $G$ itself is the trivial group, then
these two subgroups coincide.) These are called the \textbf{trivial subgroups}
of $G$.

The set of even integers is a subgroup of $\mathbb{Z}$: zero is even, the
negative of an even integer is even, and the sum of two even integers is even.

The set $H$ of $n\times n$ real matrices with determinant one is a subgroup of
$\mathsf{GL}(n;\mathbb{R})$. The set $H$ is a sub\textit{set} of
$\mathsf{GL}(n;\mathbb{R})$ because any matrix with determinant one is
invertible. The identity matrix has determinant one, so \ref{id} is satisfied.
The determinant of the inverse is the reciprocal of the determinant, so
\ref{inv} is satisfied; and the determinant of a product is the product of the
determinants, so \ref{closure} is satisfied. This group is called the
\textbf{special linear group} (over the reals), and is denoted $\mathsf{SL}%
(n;\mathbb{R})$.

Additional examples, as well as some non-examples, are given in Exercise
\ref{subgroups}.

\begin{definition}
The \textbf{center} of a group $G$ is the set of all $g\in G$ such that
$gh=hg$ for all $h\in G$.
\end{definition}

It is not hard to see that the center of any group $G$ is a subgroup $G$.

\begin{definition}
Let $G$ and $H$ be groups, and consider the Cartesian product of $G$ and $H$,
i.e., the set of ordered pairs $(g,h)$ with $g\in G,h\in H$. Define a product
operation on this set as follows:
\[
(g_{1},h_{1})(g_{2},h_{2})=(g_{1}g_{2},h_{1}h_{2})\text{.}%
\]
This operation makes the Cartesian product of $G$ and $H$ into a group, called
the \textbf{direct product} of $G$ and $H$ and denoted $G\times H$.
\end{definition}

It is a simple matter to check that this operation truly makes $G\times H$
into a group. For example, the identity element of $G\times H$ is the pair
$(e_{1},e_{2})$, where $e_{1}$ is the identity for $G$, and $e_{2}$ is the
identity for $H$.

\section{Homomorphisms and Isomorphisms}

\begin{definition}
Let $G$ and $H$ be groups. A map $\phi:G\rightarrow H$ is called a
\textbf{homomorphism} if $\phi(g_{1}g_{2})=\phi(g_{1})\phi(g_{2})$ for all
$g_{1},g_{2}\in G$. If in addition, $\phi$ is one-to-one and onto, then $\phi$
is called an \textbf{isomorphism}. An isomorphism of a group with itself is
called an \textbf{automorphism}.
\end{definition}

\begin{proposition}
Let $G$ and $H$ be groups, $e_{1}$ the identity element of $G$, and $e_{2}$
the identity element of $H$. If $\phi:G\rightarrow H$ is a homomorphism, then
$\phi(e_{1})=e_{2}$, and $\phi(g^{-1})=\phi(g)^{-1}$ for all $g\in G$.
\end{proposition}

\begin{proof}
Let $g$ be any element of $G$. Then $\phi(g)=\phi(ge_{1})=\phi(g)\phi(e_{1})$.
Multiplying on the left by $\phi(g)^{-1}$ gives $e_{2}=\phi(e_{1})$. Now
consider $\phi(g^{-1})$. Since $\phi(e_{1})=e_{2}$, we have $e_{2}=\phi
(e_{1})=\phi(gg^{-1})=\phi(g)\phi(g^{-1})$. In light of Prop. \ref{one-sided},
we conclude that $\phi(g^{-1})$ is the inverse of $\phi(g)$.
\end{proof}

\begin{definition}
Let $G$ and $H$ be groups, $\phi:G\rightarrow H$ a homomorphism, and $e_{2}$
the identity element of $H$. The \textbf{kernel} of $\phi$ is the set of all
$g\in G$ for which $\phi(g)=e_{2}$.
\end{definition}

\begin{proposition}
Let $G$ and $H$ be groups, and $\phi:G\rightarrow H$ a homomorphism. Then the
kernel of $\phi$ is a subgroup of $G$.
\end{proposition}

\begin{proof}
Easy.
\end{proof}

\subsection{Examples}

Given any two groups $G$ and $H$, we have the trivial homomorphism from $G$ to
$H$: $\phi(g)=e$ for all $g\in G$. The kernel of this homomorphism is all of
$G$.

In any group $G$, the identity map ($id(g)=g$) is an automorphism of $G$,
whose kernel is just $\left\{  e\right\}  $.

Let $G=H=\mathbb{Z}$, and define $\phi(n)=2n$. This is a homomorphism of
$\mathbb{Z}$ to itself, but not an automorphism. The kernel of this
homomorphism is just $\left\{  0\right\}  $.

The determinant is a homomorphism of $\mathsf{GL}(n,\mathbb{R}\mathbf{)}$ to
$\mathbb{R}^{*}$. The kernel of this map is $\mathsf{SL}\left(  n,\mathbb{R}%
\right)  $.

Additional examples are given in Exercises \ref{zmodn} and \ref{inner}.

If there exists an isomorphism from $G$ to $H$, then $G$ and $H$ are said to
be \textbf{isomorphic}, and this relationship is denoted $G\cong H$. (See
Exercise \ref{isomorphisms}.) Two groups which are isomorphic should be
thought of as being (for all practical purposes) the same group.

\section{Exercises}

Recall the definitions of the groups $\mathsf{GL}(n;\mathbb{R})$, $S_{n}$,
$\mathbb{R}^{*}$, and $\mathbb{Z}_{n}$ from Sect. \ref{ex}, and the definition
of the group $\mathsf{SL}(n;\mathbb{R})$ from Sect. \ref{sub}.

\begin{enumerate}
\item  Show that the center of any group $G$ is a subgroup $G$.

\item \label{subgroups}In (a)-(f), you are given a group $G$ and a subset $H$
of $G$. In each case, determine whether $H$ is a subgroup of $G$.

(a) $G=\mathbb{Z},\ H=\left\{  \text{odd integers}\right\}  $

(b) $G=\mathbb{Z},\ H=\left\{  \text{multiples of 3}\right\}  $

(c) $G=\mathsf{GL}(n;\mathbb{R}),\ H=\left\{  A\in\mathsf{GL}(n;\mathbb{R}%
)\left|  \det A\text{ is an integer}\right.  \right\}  $

(d) $G=\mathsf{SL}(n;\mathbb{R}),\ H=\left\{  A\in\mathsf{SL}(n;\mathbb{R}%
)\left|  \text{all the entries of }A\text{ are integers}\right.  \right\}  $

\textit{Hint}: recall Kramer's rule for finding the inverse of a matrix.

(e) $G=\mathsf{GL}(n;\mathbb{R}),\ H=\left\{  A\in\mathsf{GL}(n;\mathbb{R}%
)\left|  \text{all of the entries of }A\text{ are rational}\right.  \right\}  $

(f) $G=\mathbb{Z}_{9},\ H=\left\{  0,2,4,6,8\right\}  $

\item \label{inverses}Verify the properties of inverses in Prop.
\ref{inverse.properties}.

\item \label{isomorphisms}Let $G$ and $H$ be groups. Suppose there exists an
isomorphism $\phi$ from $G$ to $H$. Show that there exists an isomorphism from
$H$ to $G$.

\item  Show that the set of positive real numbers is a subgroup of
$\mathbb{R}^{\ast}$. Show that this group is isomorphic to the group
$\mathbb{R}$.

\item  Show that the set of automorphisms of any group $G$ is itself a group,
under the operation of composition. This group is the \textbf{automorphism
group} of $G$, $Aut(G)$.

\item \label{inner}Given any group $G$, and any element $g$ in $G$, define
$\phi_{g}:G\rightarrow G$ by $\phi_{g}(h)=ghg^{-1}$. Show that $\phi_{g}$ is
an automorphism of $G$. Show that the map $g\rightarrow\phi_{g}$ is a
homomorphism of $G$ into $Aut(G)$, and that the kernel of this map is the
center of $G$.

\textit{Note}: An automorphism which can be expressed as $\phi_{g}$ for some
$g\in G$ is called an \textbf{inner automorphism}; any automorphism of $G$
which is not equal to any $\phi_{g}$ is called an \textbf{outer automorphism}.

\item \label{gl2}Give an example of two $2\times2$ invertible real matrices
which do not commute. (This shows that $\mathsf{GL}(2,\mathbf{R})$ is not commutative.)

\item  Show that in any group $G$, the center of $G$ is a subgroup.

\item \label{sn}An element $\sigma$ of the permutation group $S_{n}$ can be
written in two-row form,
\[
\sigma=\left(
\begin{array}
[c]{cccc}%
1 & 2 & \cdots &  n\\
\sigma_{1}^{{}} & \sigma_{2} & \cdots & \sigma_{n}%
\end{array}
\right)
\]
where $\sigma_{i}$ denotes $\sigma(i)$. Thus
\[
\sigma=\left(
\begin{array}
[c]{ccc}%
1 & 2 & 3\\
2 & 3 & 1
\end{array}
\right)
\]
is the element of $S_{3}$ which sends 1 to 2, 2 to 3, and 3 to 1. When
multiplying (i.e., composing) two permutations, one performs the one on the
right first, and then the one on the left. (This is the usual convention for
composing functions.)

Compute
\[
\left(
\begin{array}
[c]{ccc}%
1 & 2 & 3\\
2 & 1 & 3
\end{array}
\right)  \left(
\begin{array}
[c]{ccc}%
1 & 2 & 3\\
1 & 3 & 2
\end{array}
\right)
\]
and
\[
\left(
\begin{array}
[c]{ccc}%
1 & 2 & 3\\
1 & 3 & 2
\end{array}
\right)  \left(
\begin{array}
[c]{ccc}%
1 & 2 & 3\\
2 & 1 & 3
\end{array}
\right)
\]
Conclude that $S_{3}$ is not commutative.

\item \label{counter-example}Consider the set $\mathbb{N}\mathbf{=}\left\{
0,1,2,\cdots\right\}  $ of natural numbers, and the set $\mathcal{F}$ of
\textit{all} functions of $\mathbb{N}$ to itself. Composition of functions
defines a map of $\mathcal{F}\times\mathcal{F}$ into $\mathcal{F}$, which is
associative. The identity ($id(n)=n$) has the property that $id\circ f=f\circ
id=f$, for all $f$ in $\mathcal{F}$. However, since we do not restrict to
functions which are one-to-one and onto, not every element of $\mathcal{F}$
has an inverse. Thus $\mathcal{F}$ is not a group.

Give an example of two functions $f,g$ in $\mathcal{F}$ such that $f\circ
g=id$, but $g\circ f\neq id$. (Compare with Prop. \ref{one-sided}.)

\item \label{zmodn}Consider the groups $\mathbb{Z}$ and $\mathbb{Z}_{n}$. For
each $a$ in $\mathbb{Z}$, define $a\ \mathbf{mod\ }n$ to be the unique element
$b$ of $\left\{  0,1,\cdots n-1\right\}  $ such that $a$ can be written as
$a=kn+b$, with $k$ an integer. Show that the map $a\rightarrow
a\ \mathbf{mod\ }n$ is a homomorphism of $\mathbb{Z}$ into $\mathbb{Z}_{n}$.

\item  Let $G$ be a group, and $H$ a subgroup of $G$. $H$ is called a
\textbf{normal subgroup} of $G$ if given any $g\in G$, and $h\in H$,
$ghg^{-1}$ is in $H$.

Show that any subgroup of a commutative group is normal. Show that in any
group $G$, the trivial subgroups $G$ and $\{e\}$ are normal. Show that the
center of any group is a normal subgroup. Show that if $\phi$ is a
homomorphism from $G$ to $H$, then the kernel of $\phi$ is a normal subgroup
of $G$.

Show that $\mathsf{SL}(n;\mathbb{R})$ is a normal subgroup of $\mathsf{GL}%
(n;\mathbb{R})$.

\textit{Note}: a group $G$ with no normal subgroups other than $G$ and
$\left\{  e\right\}  $ is called \textbf{simple}.
\end{enumerate}

\chapter{Matrix Lie Groups}

\section{Definition of a Matrix Lie Group}

Recall that the \textbf{general linear group} over the reals, denoted
$\mathsf{GL}(n;\mathbb{R})$, is the group of all $n\times n$ invertible
matrices with real entries. We may similarly define $\mathsf{GL}%
(n;\mathbb{C})$ to be the group of all $n\times n$ invertible matrices with
complex entries. Of course, $\mathsf{GL}(n;\mathbb{R})$ is contained in
$\mathsf{GL}(n;\mathbb{C})$.

\begin{definition}
\label{matrix.converge}Let $A_{n}$ be a sequence of complex matrices. We say
that $A_{n}$ \textbf{converges} to a matrix $A$ if each entry of $A_{n}$
converges to the corresponding entry of $A$, i.e., if $\left(  A_{n}\right)
_{ij}$ converges to $A_{ij}$ for all $1\leq i,j\leq n$.
\end{definition}

\begin{definition}
\label{matrix.group}A \textbf{matrix Lie group} is any subgroup $H$ of
$\mathsf{GL}(n;\mathbb{C})$ with the following property: if $A_{n}$ is any
sequence of matrices in $H$, and $A_{n}$ converges to some matrix $A$, then
either $A\in H$, or $A$ is not invertible.
\end{definition}

The condition on $H$ amounts to saying that $H$ is a closed subset of
$\mathsf{GL}(n;\mathbb{C})$. (This is not the same as saying that $H$ is
closed in the space of all matrices.) Thus Definition \ref{matrix.group} is
equivalent to saying that a matrix Lie group is a \textbf{closed subgroup} of
$\mathsf{GL}(n;\mathbb{C})$.

The condition that $H$ be a \textit{closed} subgroup, as opposed to merely a
subgroup, should be regarded as a technicality, in that most of the
\textit{interesting} subgroups of $\mathsf{GL}(n;\mathbb{C})$ have this
property. (Almost all of the matrix Lie groups $H$ we will consider have the
stronger property that if $A_{n}$ is any sequence of matrices in $H$, and
$A_{n}$ converges to some matrix $A$, then $A\in H$.)

There is a topological structure on the set of $n\times n$ complex matrices
which goes with the above notion of convergence. This topological structure is
defined by identifying the space of $n\times n$ matrices with $\mathbb{C}%
^{n^{2}}$ in the obvious way and using the usual topological structure on
$\mathbb{C}^{n^{2}}$.

\subsection{Counterexamples}

An example of a subgroup of $\mathsf{GL}(n;\mathbb{C})$ which is not closed
(and hence is not a matrix Lie group) is the set of all $n\times n$ invertible
matrices all of whose entries are real and rational. This is in fact a
subgroup of $\mathsf{GL}(n;\mathbb{C})$, but not a closed subgroup. That is,
one can (easily) have a sequence of invertible matrices with rational entries
converging to an invertible matrix with some irrational entries. (In fact,
\textit{every} real invertible matrix is the limit of some sequence of
invertible matrices with rational entries.)

Another example of a group of matrices which is not a matrix Lie group is the
following subgroup of $\mathsf{GL}(2,\mathbb{C})$. Let $a$ be an irrational
real number, and let
\[
H=\left\{  \left(
\begin{array}
[c]{cc}%
e^{it} & 0\\
0 & e^{ita}%
\end{array}
\right)  \left|  t\in\mathbb{R}\right.  \right\}
\]
Clearly, $H$ is a subgroup of $\mathsf{GL}(2,\mathbb{C})$. Because $a$ is
irrational, the matrix $-I$ is not in $H$, since to make $e^{it}$ equal to
$-1$, we must take $t$ to be an odd integer multiple of $\pi$, in which case
$ta$ cannot be an odd integer multiple of $\pi$. On the other hand, by taking
$t=(2n+1)\pi$ for a suitably chosen integer $n$, we can make $ta$ arbitrarily
\textit{close }to an odd integer multiple of $\pi$. (It is left to the reader
to verify this.) Hence we can find a sequence of matrices in $H $ which
converges to $-I$, and so $H$ is not a matrix Lie group. See Exercise
\ref{not.closed}.

\section{Examples of Matrix Lie Groups}

Mastering the subject of Lie groups involves not only learning the general
theory, but also familiarizing oneself with examples. In this section, we
introduce some of the most important examples of (matrix) Lie groups.

\subsection{The general linear groups $\mathsf{GL}(n;\mathbb{R})$ and
$\mathsf{GL}(n;\mathbb{C})$}

The general linear groups (over $\mathbb{R}$ or $\mathbb{C}$) are themselves
matrix Lie groups. Of course, $\mathsf{GL}(n;\mathbb{C})$ is a subgroup of
itself. Furthermore, if $A_{n}$ is a sequence of matrices in $\mathsf{GL}%
(n;\mathbb{C})$ and $A_{n}$ converges to $A$, then by the definition of
$\mathsf{GL}(n;\mathbb{C})$, either $A$ is in $\mathsf{GL}(n;\mathbb{C})$, or
$A$ is not invertible.

Moreover, $\mathsf{GL}(n;\mathbb{R})$ is a subgroup of $\mathsf{GL}%
(n;\mathbb{C})$, and if $A_{n}\in\mathsf{GL}(n;\mathbb{R})$, and $A_{n}$
converges to $A$, then the entries of $A$ are real. Thus either $A$ is not
invertible, or $A\in\mathsf{GL}(n;\mathbb{R})$.

\subsection{The special linear groups \textsf{SL}$\left(  n;\mathbb{R}\right)
$ and \textsf{SL}$\left(  n;\mathbb{C}\right)  $}

The \textbf{special linear group} (over $\mathbb{R}$ or $\mathbb{C}$) is the
group of $n\times n$ invertible matrices (with real or complex entries) having
determinant one. Both of these are subgroups of $\mathsf{GL}(n;\mathbb{C})$,
as noted in Chapter 1. Furthermore, if $A_{n}$ is a sequence of matrices with
determinant one, and $A_{n}$ converges to $A$, then $A$ also has determinant
one, because the determinant is a continuous function. Thus \textsf{SL}%
$\left(  n;\mathbb{R}\right)  $ and \textsf{SL}$\left(  n;\mathbb{C}\right)  $
are matrix Lie groups.

\subsection{The orthogonal and special orthogonal groups, $\mathsf{O}(n)$ and
\textsf{SO}$(n)$}

An $n\times n$ real matrix $A$ is said to be \textbf{orthogonal} if the column
vectors that make up $A$ are orthonormal, that is, if
\[
\sum_{i=1}^{n}A_{ij}A_{ik}=\delta_{jk}%
\]
Equivalently, $A$ is orthogonal if it preserves the inner product, namely, if
$\left\langle x,y\right\rangle =\left\langle Ax,Ay\right\rangle $ for all
vectors $x,y$ in $\mathbb{R}^{n}$. ($\,$Angled brackets denote the usual inner
product on $\mathbb{R}^{n}$, $\left\langle x,y\right\rangle =\sum_{i}%
x_{i}y_{i}$.) Still another equivalent definition is that $A$ is orthogonal if
$A^{tr}A=I$, i.e., if $A^{tr}=A^{-1}$. ($A^{tr}$ is the transpose of $A$,
$\left(  A^{tr}\right)  _{ij}=A_{ji}$.) See Exercise \ref{orthogonal}.

Since $\det A^{tr}=\det A$, we see that if $A$ is orthogonal, then
$\det(A^{tr}A)=\left(  \det A\right)  ^{2}=\det I=1$. Hence $\det A=\pm1$, for
all orthogonal matrices $A$.

This formula tells us, in particular, that every orthogonal matrix must be
invertible. But if $A$ is an orthogonal matrix, then
\[
\left\langle A^{-1}x,A^{-1}y\right\rangle =\left\langle A\left(
A^{-1}x\right)  ,A\left(  A^{-1}x\right)  \right\rangle =\left\langle
x,y\right\rangle
\]
Thus the inverse of an orthogonal matrix is orthogonal. Furthermore, the
product of two orthogonal matrices is orthogonal, since if $A$ and $B$ both
preserve inner products, then so does $AB$. Thus the set of orthogonal
matrices forms a group.

The set of all $n\times n$ real orthogonal matrices is the \textbf{orthogonal
group} $\mathsf{O}(n)$, and is a subgroup of $\mathsf{GL}(n;\mathbb{C})$. The
limit of a sequence of orthogonal matrices is orthogonal, because the relation
$A^{tr}A=I$ is preserved under limits. Thus $\mathsf{O}(n)$ is a matrix Lie group.

The set of $n\times n$ orthogonal matrices with determinant one is the
\textbf{special orthogonal group} $\mathsf{SO}(n)$. Clearly this is a subgroup
of $\mathsf{O}(n)$, and hence of $\mathsf{GL}(n;\mathbb{C})$. Moreover, both
orthogonality and the property of having determinant one are preserved under
limits, and so $\mathsf{SO}(n)$ is a matrix Lie group. Since elements of
$\mathsf{O}(n)$ already have determinant $\pm1$, $\mathsf{SO}(n) $ is ``half''
of $\mathsf{O}(n)$.

Geometrically, elements of $\mathsf{O}(n)$ are either rotations, or
combinations of rotations and reflections. The elements of $\mathsf{SO}(n)$
are just the rotations.

See also Exercise \ref{so2}.

\subsection{The unitary and special unitary groups, $\mathsf{U}(n)$ and
$\mathsf{SU}(n)$}

An $n\times n$ complex matrix $A$ is said to be \textbf{unitary} if the column
vectors of $A$ are orthonormal, that is, if
\[
\sum_{i=1}^{n}\overline{A_{ij}}A_{ik}=\delta_{jk}%
\]
Equivalently, $A$ is unitary if it preserves the inner product, namely, if
$\left\langle x,y\right\rangle =\left\langle Ax,Ay\right\rangle $ for all
vectors $x,y$ in $\mathbb{C}^{n}$. (Angled brackets here denote the inner
product on $\mathbb{C}^{n}$, $\left\langle x,y\right\rangle =\sum_{i}%
\overline{x_{i}}y_{i}$. We will adopt the convention of putting the complex
conjugate on the left.) Still another equivalent definition is that $A$ is
unitary if $A^{*}A=I$, i.e., if $A^{*}=A^{-1}$. ($A^{*}$ is the adjoint of
$A$, $\left(  A^{*}\right)  _{ij}=\overline{A_{ji}}$.) See Exercise
\ref{unitary}.

Since $\det A^{*}=\overline{\det A}$, we see that if $A$ is unitary, then
$\det\left(  A^{*}A\right)  =\left|  \det A\right|  ^{2}=\det I=1$. Hence
$\left|  \det A\right|  =1$, for all unitary matrices $A$.

This in particular shows that every unitary matrix is invertible. The same
argument as for the orthogonal group shows that the set of unitary matrices
forms a group.

The set of all $n\times n$ unitary matrices is the \textbf{unitary group}
$\mathsf{U}(n)$, and is a subgroup of $\mathsf{GL}(n;\mathbb{C})$. The limit
of unitary matrices is unitary, so $\mathsf{U}(n)$ is a matrix Lie group. The
set of unitary matrices with determinant one is the \textbf{special unitary
group} $\mathsf{SU}(n)$. It is easy to check that $\mathsf{SU}(n)$ is a matrix
Lie group. Note that a unitary matrix can have determinant $e^{i\theta}$ for
any $\theta$, and so $\mathsf{SU}(n)$ is a smaller subset of $\mathsf{U}(n)$
than $\mathsf{SO}(n)$ is of $\mathsf{O}(n)$. (Specifically, $\mathsf{SO}(n)$
has the same dimension as $\mathsf{O}(n)$, whereas $\mathsf{SU}(n)$ has
dimension one less than that of $\mathsf{U}(n)$.)

See also Exercise \ref{su2}.

\subsection{The complex orthogonal groups, $\mathsf{O}(n;\mathbb{C})$ and
$\mathsf{SO}(n;\mathbb{C})$}

Consider the bilinear form $\left(  \ \right)  $ on $\mathbf{C}^{n}$ defined
by $(x,y)=\sum x_{i}y_{i}$. This form is not an inner product, because of the
lack of a complex conjugate in the definition. The set of all $n\times n$
complex matrices $A$ which preserve this form, (i.e., such that
$(Ax,Ay)=(x,y)$ for all $x,y\in\mathbf{C}^{n}$) is the \textbf{complex
orthogonal group} $\mathsf{O}(n;\mathbb{C})$, and is a subgroup of
$\mathsf{GL}(n;\mathbb{C})$. (The proof is the same as for $\mathsf{O}(n)$.)
An $n\times n$ complex matrix $A$ is in $\mathsf{O}(n;\mathbb{C})$ if and only
if $A^{tr}A=I$. It is easy to show that $\mathsf{O}(n;\mathbb{C})$ is a matrix
Lie group, and that $\det A=\pm1$, for all $A$ in $\mathsf{O}(n;\mathbb{C})$.
Note that $\mathsf{O}(n;\mathbb{C})$ is \textit{not} the same as the unitary
group $\mathsf{U}(n)$. The group $\mathsf{SO}(n;\mathbb{C})$ is defined to be
the set of all $A$ in $\mathsf{O}(n;\mathbb{C})$ with $\det A=1$. Then
$\mathsf{SO}(n;\mathbb{C})$ is also a matrix Lie group.

\subsection{The generalized orthogonal and Lorentz groups}

Let $n$ and $k$ be positive integers, and consider $\mathbb{R}^{n+k}$. Define
a symmetric bilinear form $\left[  \ \right]  _{n+k}$ on $\mathbb{R}^{n+k}$ by
the formula
\begin{equation}
\left[  x,y\right]  _{n,k}=x_{1}y_{1}+\cdots+x_{n}y_{n}-x_{n+1}y_{n+1}%
\cdots-y_{n+k}x_{n+k}\label{bilinear}%
\end{equation}
The set of $(n+k)\times(n+k)$ real matrices $A$ which preserve this form
(i.e., such that $\left[  Ax,Ay\right]  _{n,k}=\left[  x,y\right]  _{n,k}$ for
all $x,y\in\mathbb{R}^{n+k}$) is the \textbf{generalized orthogonal group}
$\mathsf{O}(n;k)$, and it is a subgroup of $\mathsf{GL}(n+k;\mathbb{R})$ (Ex.
\ref{generalized}). Since $\mathsf{O}(n;k)$ and $\mathsf{O}(k;n)$ are
essentially the same group, we restrict our attention to the case $n\geq k$.
It is not hard to check that $\mathsf{O}(n;k)$ is a matrix Lie group.

If $A$ is an $(n+k)\times(n+k)$ real matrix, let $A^{(i)}$ denote the
$i^{\text{th}}$ column vector of $A$, that is
\[
A^{(i)}=\left(
\begin{array}
[c]{c}%
A_{1,i}^{}\\
\vdots\\
A_{n+k,i}%
\end{array}
\right)
\]
Then $A$ is in $\mathsf{O}(n;k)$ if and only if the following conditions are
satisfied:
\begin{equation}%
\begin{array}
[c]{cccc}%
\left[  A^{(i)},A^{(j)}\right]  _{n,k} & = & 0 & i\neq j\\
\left[  A^{(i)},A^{(i)}\right]  _{n,k} & = & 1 & 1\leq i\leq n\\
\left[  A^{(i)},A^{(i)}\right]  _{n,k} & = & -1 & n+1\leq i\leq n+k
\end{array}
\end{equation}

Let $g$ denote the $(n+k)\times(n+k)$ diagonal matrix with ones in the first
$n$ diagonal entries, and minus ones in the last $k$ diagonal entries. Then
$A$ is in $\mathsf{O}(n;k)$ if and only if $A^{tr}gA=g$ (Ex. \ref{generalized}%
). Taking the determinant of this equation gives $(\det A)^{2}\det g=\det g$,
or ($\det A)^{2}=1$. Thus for any $A$ in $\mathsf{O}(n;k)$, $\det A=\pm1$.

The group $\mathsf{SO}(n;k)$ is defined to be the set of matrices in
$\mathsf{O}(n;k)$ with $\det A=1$. This is a subgroup of $\mathsf{GL}%
(n+k;\mathbb{R})$, and is a matrix Lie group.

Of particular interest in physics is the \textbf{Lorentz group} $\mathsf{O}%
(3;1)$. (Sometimes the phrase Lorentz group is used more generally to refer to
the group $\mathsf{O}(n;1)$ for any $n\geq1$.) See also Exercise \ref{so11}.

\subsection{The symplectic groups $\mathsf{Sp}(n;\mathbb{R})$, $\mathsf{Sp}%
(n;\mathbb{C})$, and $\mathsf{Sp}(n)$}

The special and general linear groups, the orthogonal and unitary groups, and
the symplectic groups (which will be defined momentarily) make up the
\textbf{classical groups}. Of the classical groups, the symplectic groups have
the most confusing definition, partly because there are three sets of them
($\mathsf{Sp}(n;\mathbb{R})$, $\mathsf{Sp}(n;\mathbb{C})$, and $\mathsf{Sp}(n)
$), and partly because they involve skew-symmetric bilinear forms rather than
the more familiar symmetric bilinear forms. To further confuse matters, the
notation for referring to these groups is not consistent from author to author.

Consider the skew-symmetric bilinear form $B$ on $\mathbb{R}^{2n}$ defined as
follows:
\begin{equation}
B\left[  x,y\right]  =\sum_{i=1}^{n}x_{i}y_{n+i}-x_{n+i}y_{i}\label{skew.form}%
\end{equation}
The set of all $2n\times2n$ matrices $A$ which preserve $B$ (i.e., such that
$B\left[  Ax,Ay\right]  =B\left[  x,y\right]  $ for all $x,y\in\mathbb{R}^{2n}
$) is the \textbf{real symplectic group} $\mathsf{Sp}(n;\mathbb{R})$, and it
is a subgroup of $\mathsf{GL}(2n;\mathbb{R})$. It is not difficult to check
that this is a matrix Lie group (Exercise \ref{symplectic}). This group arises
naturally in the study of classical mechanics. If $J$ is the $2n\times2n$
matrix
\[
J=\left(
\begin{array}
[c]{cc}%
0 & I\\
-I & 0
\end{array}
\right)
\]
then $B\left[  x,y\right]  =\left\langle x,Jy\right\rangle $, and it is
possible to check that a $2n\times2n$ real matrix $A$ is in $\mathsf{Sp}%
(n;\mathbb{R})$ if and only if $A^{tr}JA=J$. (See Exercise \ref{symplectic}.)
Taking the determinant of this identity gives $\left(  \det A\right)  ^{2}\det
J=\det J$, or $\left(  \det A\right)  ^{2}=1$. This shows that $\det A=\pm1$,
for all $A\in\mathsf{Sp}(n;\mathbb{R})$. In fact, $\det A=1$ for all
$A\in\mathsf{Sp}(n;\mathbb{R})$, although this is not obvious.

One can define a bilinear form on $\mathbb{C}^{n}$ by the same formula
(\ref{skew.form}). (This form is bilinear, not Hermitian, and involves no
complex conjugates.) The set of $2n\times2n$ complex matrices which preserve
this form is the \textbf{complex symplectic group}\emph{\ }$\mathsf{Sp}%
(n;\mathbb{C})$. A $2n\times2n$ complex matrix $A$ is in $\mathsf{Sp}%
(n;\mathbb{C})$ if and only if $A^{tr}JA=J$. (Note: this condition involves
$A^{tr}$, \textit{not} $A^{\ast}$.) This relation shows that $\det A=\pm1$,
for all $A\in\mathsf{Sp}(n;\mathbb{C})$. In fact $\det A=1$, for all
$A\in\mathsf{Sp}(n;\mathbb{C})$.

Finally, we have the \textbf{compact symplectic group} $\mathsf{Sp}(n)$
defined as
\[
\mathsf{Sp}(n)=\mathsf{Sp}\left(  n;\mathbb{C}\right)  \cap\mathsf{U}(2n).
\]
See also Exercise \ref{sp1}. For more information and a proof of the fact that
$\det A=1$, for all $A\in\mathsf{Sp}(n;\mathbb{C})$, see Miller, Sect. 9.4.
What we call $\mathsf{Sp}\left(  n;\mathbb{C}\right)  $ Miller calls
$\mathsf{Sp}(n)$, and what we call $\mathsf{Sp}(n)$, Miller calls
$\mathsf{USp}(n)$.

\subsection{The Heisenberg group $H$}

The set of all $3\times3$ real matrices $A$ of the form
\begin{equation}
A=\left(
\begin{array}
[c]{ccc}%
1 & a & b\\
0 & 1 & c\\
0 & 0 & 1
\end{array}
\right) \label{heisenberg}%
\end{equation}
where $a$, $b$, and $c$ are arbitrary real numbers, is the \textbf{Heisenberg
group}. It is easy to check that the product of two matrices of the form
(\ref{heisenberg}) is again of that form, and clearly the identity matrix is
of the form (\ref{heisenberg}). Furthermore, direct computation shows that if
$A$ is as in (\ref{heisenberg}), then
\[
A^{-1}=\left(
\begin{array}
[c]{ccc}%
1 & -a & ac-b\\
0 & 1 & -c\\
0 & 0 & 1
\end{array}
\right)
\]
Thus $H$ is a subgroup of $\mathsf{GL}(3;\mathbb{R})$. Clearly the limit of
matrices of the form (\ref{heisenberg}) is again of that form, and so $H$ is a
matrix Lie group.

It is not evident at the moment why this group should be called the Heisenberg
group. We shall see later that the Lie algebra of $H$ gives a realization of
the Heisenberg commutation relations of quantum mechanics. (See especially
Chapter 5, Exercise \ref{ccr}.)

See also Exercise \ref{heisenberg.center}.

\subsection{The groups $\mathbb{R}^{\ast}$, $\mathbb{C}^{\ast}$, $S^{1}$,
$\mathbb{R}$, and $\mathbb{R}^{n}$}

Several important groups which are not naturally groups of matrices can (and
will in these notes) be thought of as such.

The group $\mathbb{R}^{*}$ of non-zero real numbers under multiplication is
isomorphic to $\mathsf{GL}(1,\mathbb{R})$. Thus we will regard $\mathbb{R}%
^{*}$ as a matrix Lie group. Similarly, the group $\mathbb{C}^{*}$ of non-zero
complex numbers under multiplication is isomorphic to $\mathsf{GL}%
(1;\mathbb{C})$, and the group $S^{1}$ of complex numbers with absolute value
one is isomorphic to $\mathsf{U}(1)$.

The group $\mathbb{R}$ under addition is isomorphic to $\mathsf{GL}%
(1;\mathbb{R})^{+}$ ($1\times1$ real matrices with positive determinant) via
the map $x\rightarrow\left[  e^{x}\right]  $. The group $\mathbb{R}^{n}$ (with
vector addition) is isomorphic to the group of diagonal real matrices with
positive diagonal entries, via the map
\[
(x_{1},\cdots,x_{n})\rightarrow\left(
\begin{array}
[c]{ccc}%
e^{x_{1}} &  & 0\\
& \ddots & \\
0 &  & e^{x_{n}}%
\end{array}
\right)  \text{.}%
\]

\subsection{The Euclidean and Poincar\'{e} groups}

The Euclidean group $\mathsf{E}(n)$ is by definition the group of all
one-to-one, onto, distance-preserving maps of $\mathbb{R}^{n}$ to itself, that
is, maps $f:\mathbb{R}^{n}\rightarrow\mathbb{R}^{n}$ such that $d\left(
f\left(  x\right)  ,f\left(  y\right)  \right)  =d\left(  x,y\right)  $ for
all $x,y\in\mathbb{R}^{n}.$ Here $d$ is the usual distance on $\mathbb{R}%
^{n},$ $d\left(  x,y\right)  =\left|  x-y\right|  .$ Note that we don't assume
\textit{anything} about the structure of $f$ besides the above properties. In
particular, $f$ need not be linear. The orthogonal group $\mathsf{O}(n)$ is a
subgroup of $\mathsf{E}(n)$, and is the group of all \textit{linear}
distance-preserving maps of $\mathbb{R}^{n}$ to itself. The set of
translations of $\mathbb{R}^{n}$ (i.e., the set of maps of the form
$T_{x}(y)=x+y$) is also a subgroup of $\mathsf{E}(n)$.

\begin{proposition}
Every element $T$ of $\mathsf{E}(n)$ can be written uniquely as an orthogonal
linear transformation followed by a translation, that is, in the form
\[
T=T_{x}R
\]
with $x\in\mathbb{R}^{n}$, and $R\in\mathsf{O}(n)$.
\end{proposition}

We will not prove this here. The key step is to prove that every one-to-one,
onto, distance-preserving map of $\mathbb{R}^{n}$ to itself which fixes the
origin must be linear.

Following Miller, we will write an element $T=T_{x}R$ of $\mathsf{E}(n)$ as a
pair $\{x,R\}$. Note that for $y\in\mathbb{R}^{n}$,
\[
\left\{  x,R\right\}  y=Ry+x
\]
and that
\[
\{x_{1},R_{1}\}\{x_{2},R_{2}\}y=R_{1}(R_{2}y+x_{2})+x_{1}=R_{1}R_{2}%
y+(x_{1}+R_{1}x_{2})
\]
Thus the product operation for $\mathsf{E}(n)$ is the following:
\begin{equation}
\{x_{1},R_{1}\}\{x_{2},R_{2}\}=\{x_{1}+R_{1}x_{2},R_{1}R_{2}\}\label{product}%
\end{equation}
The inverse of an element of $\mathsf{E}(n)$ is given by
\[
\{x,R\}^{-1}=\{-R^{-1}x,R^{-1}\}
\]

Now, as already noted, $\mathsf{E}(n)$ is not a subgroup of $\mathsf{GL}%
(n;\mathbb{R})$, since translations are not linear maps. However,
$\mathsf{E}(n) $ is isomorphic to a subgroup of $\mathsf{GL}(n+1;\mathbb{R})$,
via the map which associates to $\{x,R\}\in\mathsf{E}(n)$ the following
matrix
\begin{equation}
\left(
\begin{array}
[c]{cccc}%
&  &  & x_{1}\\
& R &  & \vdots\\
&  &  &  x_{n}\\
0 & \cdots & 0 & 1
\end{array}
\right) \label{euclid}%
\end{equation}
This map is clearly one-to-one, and it is a simple computation to show that it
is a homomorphism. Thus $\mathsf{E}(n)$ is isomorphic to the group of all
matrices of the form (\ref{euclid}) (with $R\in\mathsf{O}(n)$). The limit of
things of the form (\ref{euclid}) is again of that form, and so we have
expressed the Euclidean group $\mathsf{E}(n)$ as a matrix Lie group.

We similarly define the Poincar\'{e} group $\mathsf{P}(n;1)$ to be the group
of all transformations of $\mathbb{R}^{n+1}$ of the form
\[
T=T_{x}A
\]
with $x\in\mathbb{R}^{n+1}$, $A\in\mathsf{O}(n;1)$. This is the group of
affine transformations of $\mathbb{R}^{n+1}$ which preserve the Lorentz
``distance'' $d_{L}(x,y)=(x_{1}-y_{1})^{2}+\cdots+(x_{n}-y_{n})^{2}%
-(x_{n+1}-y_{n+1})^{2}$. (An affine transformation is one of the form
$x\rightarrow Ax+b,$ where $A$ is a linear transformation and $b$ is
constant.) The group product is the obvious analog of the product
(\ref{product}) for the Euclidean group.

The Poincar\'{e} group $\mathsf{P}(n;1)$ is isomorphic to the group of
$(n+2)\times(n+2)$ matrices of the form
\begin{equation}
\left(
\begin{array}
[c]{cccc}%
&  &  & x_{1}\\
& A &  & \vdots\\
&  &  &  x_{n+1}\\
0 & \cdots & 0 & 1
\end{array}
\right) \label{poin}%
\end{equation}
with $A\in\mathsf{O}(n;1)$. The set of matrices of the form (\ref{poin}) is a
matrix Lie group.

\section{Compactness}

\begin{definition}
A matrix Lie group $G$ is said to be \textbf{compact} if the following two
conditions are satisfied:

\begin{enumerate}
\item \label{compact.closed}If $A_{n}$ is any sequence of matrices in $G$, and
$A_{n}$ converges to a matrix $A$, then $A$ is in $G$.

\item \label{compact.bounded}There exists a constant $C$ such that for all
$A\in G$, $\left|  A_{ij}\right|  \leq C$ for all $1\leq i,j\leq n$.
\end{enumerate}
\end{definition}

This is not the usual topological definition of compactness. However, the set
of all $n\times n$ complex matrices can be thought of as $\mathbb{C}^{n^{2}}$.
The above definition says that $G$ is compact if it is a \textit{closed,
bounded} subset of $\mathbb{C}^{n^{2}}$. It is a standard theorem from
elementary analysis that a subset of $\mathbb{C}^{m}$ is compact (in the usual
sense that every open cover has a finite subcover) if and only if it is closed
and bounded.

All of our examples of matrix Lie groups except $\mathsf{GL}(n;\mathbb{R})$
and $\mathsf{GL}(n;\mathbb{C})$ have property (\ref{compact.closed}). Thus it
is the boundedness condition (\ref{compact.bounded}) that is most important.

The property of compactness has very important implications. For example, if
$G$ is compact, then every irreducible unitary representation of $G$ is finite-dimensional.

\subsection{Examples of compact groups}

The groups $\mathsf{O}(n)$ and $\mathsf{SO}(n)$ are compact. Property
(\ref{compact.closed}) is satisfied because the limit of orthogonal matrices
is orthogonal and the limit of matrices with determinant one has determinant
one. Property (\ref{compact.bounded}) is satisfied because if $A$ is
orthogonal, then the column vectors of $A$ have norm one, and hence $\left|
A_{ij}\right|  \leq1$, for all $1\leq i,j\leq n$. A similar argument shows
that $\mathsf{U}(n)$, $\mathsf{SU}(n)$, and $\mathsf{Sp}(n)$ are compact.
(This includes the unit circle, $S^{1}\cong\mathsf{U}(1)$.)

\subsection{Examples of non-compact groups}

All of the other examples given of matrix Lie groups are non-compact.
$\mathsf{GL}(n;\mathbb{R})$ and $\mathsf{GL}(n;\mathbb{C})$ violate property
(\ref{compact.closed}), since a limit of invertible matrices may be
non-invertible. $\mathsf{SL}\left(  n;\mathbb{R}\right)  $ and $\mathsf{SL}%
\left(  n;\mathbb{C}\right)  $ violate (\ref{compact.bounded}), except in the
trivial case $n=1$, since
\[
A_{n}=\left(
\begin{array}
[c]{ccccc}%
n &  &  &  & \\
& \frac1n &  &  & \\
&  & 1 &  & \\
&  &  & \ddots & \\
&  &  &  & 1
\end{array}
\right)
\]
has determinant one, no matter how big $n$ is.

The following groups also violate (\ref{compact.bounded}), and hence are
non-compact: $\mathsf{O}(n;\mathbb{C})$ and $\mathsf{SO}(n;\mathbb{C})$;
$\mathsf{O}(n;k)$ and $\mathsf{SO}(n;k)$ ($n\geq1$, $k\geq1$); the Heisenberg
group $H$; $\mathsf{Sp}\left(  n;\mathbb{R}\right)  $ and $\mathsf{Sp}\left(
n;\mathbb{C}\right)  $; $\mathsf{E}(n)$ and $\mathsf{P}(n;1)$; $\mathbb{R}$
and $\mathbb{R}^{n}$; $\mathbb{R}^{*}$ and $\mathbb{C}^{*}$. It is left to the
reader to provide examples to show that this is the case.

\section{Connectedness}

\begin{definition}
\label{connectedness}A matrix Lie group $G$ is said to be \textbf{connected}
if given any two matrices $A$ and $B$ in $G$, there exists a continuous path
$A(t)$, $a\leq t\leq b$, lying in $G$ with $A(a)=A$, and $A(b)=B$.
\end{definition}

This property is what is called \textbf{path-connected} in topology, which is
not (in general) the same as connected. However, it is a fact (not
particularly obvious at the moment) that a matrix Lie group is connected if
and only if it is path-connected. So in a slight abuse of terminology we shall
continue to refer to the above property as connectedness. (See Section
\ref{lie.groups}.)

A matrix Lie group $G$ which is not connected can be decomposed (uniquely) as
a union of several pieces, called \textbf{components}, such that two elements
of the same component can be joined by a continuous path, but two elements of
different components cannot.

\begin{proposition}
If $G$ is a matrix Lie group, then the component of $G$ containing the
identity is a subgroup of $G$.
\end{proposition}

\begin{proof}
Saying that $A$ and $B$ are both in the component containing the identity
means that there exist continuous paths $A(t)$ and $B(t)$ with $A(0)=B(0)=I$,
$A(1)=A$, and $B(1)=B$. But then $A(t)B(t)$ is a continuous path starting at
$I$ and ending at $AB$. Thus the product of two elements of the identity
component is again in the identity component. Furthermore, $A(t)^{-1}$ is a
continuous path starting at $I$ and ending at $A^{-1}$, and so the inverse of
any element of the identity component is again in the identity component. Thus
the identity component is a subgroup.
\end{proof}

\begin{proposition}
The group $\mathsf{GL}(n;\mathbb{C})$ is connected for all $n\geq1$.
\end{proposition}

\begin{proof}
Consider first the case $n=1$. A $1\times1$ invertible complex matrix $A$ is
of the form $A=\left[  \lambda\right]  $ with $\lambda\in\mathbb{C}^{\ast}$,
the set of non-zero complex numbers. But given any two non-zero complex
numbers, we can easily find a continuous path which connects them and does not
pass through zero.

For the case $n\geq1$, we use the Jordan canonical form. Every $n\times n$
complex matrix $A$ can be written as
\[
A=CBC^{-1}%
\]
where $B$ is the Jordan canonical form. The only property of $B$ we will need
is that $B$ is upper-triangular:
\[
B=\left(
\begin{array}
[c]{ccc}%
\lambda_{1} &  & \ast\\
& \ddots & \\
0 &  & \lambda_{n}%
\end{array}
\right)
\]
If $A$ is invertible, then all the $\lambda_{i}$'s must be non-zero, since
$\det A=\det B=\lambda_{1}\cdots\lambda_{n}$.

Let $B(t)$ be obtained by multiplying the part of $B$ above the diagonal by
$(1-t)$, for $0\leq t\leq1$, and let $A(t)=CB(t)C^{-1}$. Then $A(t)$ is a
continuous path which starts at $A$ and ends at $CDC^{-1}$, where $D$ is the
diagonal matrix
\[
D=\left(
\begin{array}
[c]{ccc}%
\lambda_{1} &  & 0\\
& \ddots & \\
0 &  & \lambda_{n}%
\end{array}
\right)
\]
This path lies in $\mathsf{GL}(n;\mathbb{C})$ since $\det A(t)=\lambda
_{1}\cdots\lambda_{n}$ for all $t$.

But now, as in the case $n=1$, we can define $\lambda_{i}(t)$ which connects
each $\lambda_{i}$ to 1 in $\mathbf{C}^{\ast}$, as $t$ goes from 1 to 2. Then
we can define
\[
A(t)=C\left(
\begin{array}
[c]{ccc}%
\lambda_{1}(t) &  & 0\\
& \ddots & \\
0 &  & \lambda_{n}(t)
\end{array}
\right)  C^{-1}%
\]
This is a continuous path which starts at $CDC^{-1}$ when $t=1$, and ends at
$I$ ($=CIC^{-1}$) when $t=2$. Since the $\lambda_{i}(t)$'s are always
non-zero, $A(t)$ lies in $\mathsf{GL}(n;\mathbb{C})$.

We see, then, that every matrix $A$ in $\mathsf{GL}(n;\mathbb{C})$ can be
connected to the identity by a continuous path lying in $\mathsf{GL}%
(n;\mathbb{C})$. Thus if $A$ and $B$ are two matrices in $\mathsf{GL}%
(n;\mathbb{C})$, they can be connected by connecting each of them to the identity.
\end{proof}

\begin{proposition}
\label{slnc.connect}The group $\mathsf{SL}\left(  n;\mathbb{C}\right)  $ is
connected for all $n\geq1$.
\end{proposition}

\begin{proof}
The proof is almost the same as for $\mathsf{GL}(n;\mathbb{C})$, except that
we must be careful to preserve the condition $\det A=1$. Let $A$ be an
arbitrary element of $\mathsf{SL}\left(  n;\mathbb{C}\right)  $. The case
$n=1$ is trivial, so we assume $n\geq2$. We can define $A(t)$ as above for
$0\leq t\leq1$, with $A(0)=A$, and $A(1)=CDC^{-1}$, since $\det A(t)=\det
A=1$. Now define $\lambda_{i}(t)$ as before for $1\leq i\leq n-1$, and define
$\lambda_{n}(t)$ to be $\left[  \lambda_{1}(t)\cdots\lambda_{n-1}(t)\right]
^{-1}$. (Note that since $\lambda_{1}\cdots\lambda_{n}=1$, $\lambda
_{n}(0)=\lambda_{n}$.) This allows us to connect $A$ to the identity while
staying within $\mathsf{SL}\left(  n;\mathbb{C}\right)  $.
\end{proof}

\begin{proposition}
The groups $\mathsf{U}(n)$ and $\mathsf{SU}(n)$ are connected, for all
$n\geq1$.
\end{proposition}

\begin{proof}
By a standard result of linear algebra, every unitary matrix has an
orthonormal basis of eigenvectors, with eigenvalues of the form $e^{i\theta}$.
It follows that every unitary matrix $U$ can be written as
\begin{equation}
U=U_{1}\left(
\begin{array}
[c]{ccc}%
e^{i\theta_{1}} &  & 0\\
& \ddots & \\
0 &  & e^{i\theta_{n}}%
\end{array}
\right)  U_{1}^{-1}\label{un.diagonal}%
\end{equation}
with $U_{1}$ unitary and $\theta_{i}\in\mathbb{R}$. Conversely, as is easily
checked, every matrix of the form (\ref{un.diagonal}) is unitary. Now define
\[
U(t)=U_{1}\left(
\begin{array}
[c]{ccc}%
e^{i(1-t)\theta_{1}} &  & 0\\
& \ddots & \\
0 &  & e^{i(1-t)\theta_{n}}%
\end{array}
\right)  U_{1}^{-1}%
\]
As $t$ ranges from 0 to 1, this defines a continuous path in $\mathsf{U}(n)$
joining $U$ to $I$. This shows that $\mathsf{U}(n)$ is connected.

A slight modification of this argument, as in the proof of Proposition
\ref{slnc.connect}, shows that $\mathsf{SU}(n)$ is connected.
\end{proof}

\begin{proposition}
The group $\mathsf{GL}(n;\mathbb{R})$ is not connected, but has two
components. These are $\mathsf{GL}(n;\mathbb{R})^{+}$, the set of $n\times n$
real matrices with positive determinant, and $\mathsf{GL}(n;\mathbb{R})^{-}$,
the set of $n\times n$ real matrices with negative determinant.
\end{proposition}

\begin{proof}
$\mathsf{GL}(n;\mathbb{R})$ cannot be connected, for if $\det A>0$ and $\det
B<0$, then any continuous path connecting $A$ to $B$ would have to include a
matrix with determinant zero, and hence pass outside of $\mathsf{GL}%
(n;\mathbb{R})$.

The proof that $\mathsf{GL}(n;\mathbb{R})^{+}$ is connected is given in
Exercise \ref{connect.gln}. Once $\mathsf{GL}(n;\mathbb{R})^{+}$ is known to
be connected, it is not difficult to see that $\mathsf{GL}(n;\mathbb{R})^{-}$
is also connected. For let $C$ be any matrix with negative determinant, and
take $A,B$ in $\mathsf{GL}(n;\mathbb{R})^{-}$. Then $C^{-1}A$ and $C^{-1}B$
are in $\mathsf{GL}(n;\mathbb{R})^{+}$, and can be joined by a continuous path
$D(t)$ in $\mathsf{GL}(n;\mathbb{R})^{+}$. But then $CD(t)$ is a continuous
path joining $A$ and $B$ in $\mathsf{GL}(n;\mathbb{R})^{-}$.
\end{proof}

The following table lists some matrix Lie groups, indicates whether or not the
group is connected, and gives the number of components.
\[%
\begin{array}
[c]{ccc}%
\text{\textbf{Group}} & \text{\textbf{Connected?}} & \text{\textbf{Components}%
}\\
\mathsf{GL}(n;\mathbb{C}) & \text{yes} & 1\\
\mathsf{SL}\left(  n;\mathbb{C}\right)  & \text{yes} & 1\\
\mathsf{GL}(n;\mathbb{R}) & \text{no} & 2\\
\mathsf{SL}\left(  n;\mathbb{R}\right)  & \text{yes} & 1\\
\mathsf{O}(n) & \text{no} & 2\\
\mathsf{SO}(n) & \text{yes} & 1\\
\mathsf{U}(n) & \text{yes} & 1\\
\mathsf{SU}(n) & \text{yes} & 1\\
\mathsf{O}(n;1) & \text{no} & 4\\
\mathsf{SO}(n;1) & \text{no} & 2\\
\text{Heisenberg} & \text{yes} & 1\\
\mathsf{E}\left(  n\right)  & \text{no} & 2\\
\mathsf{P}(n;1) & \text{no} & 4
\end{array}
\]
Proofs of some of these results are given in Exercises \ref{so11},
\ref{connect.son}, \ref{connect.sln}, and \ref{connect.gln}. (The
connectedness of the Heisenberg group is immediate.)

\section{Simple-connectedness}

\begin{definition}
A connected matrix Lie group $G$ is said to be \textbf{simply connected} if
every loop in $G$ can be shrunk continuously to a point in $G$.

More precisely, $G$ is simply connected if given any continuous path $A(t)$,
$0\leq t\leq1$, lying in $G$ with $A(0)=A(1)$, there exists a continuous
function $A(s,t)$, $0\leq s,t\leq1$, taking values in $G$ with the following
properties: 1)$~A(s,0)=A(s,1)$ for all $s$, 2) $A(0,t)=A(t)$, and 3)
$A(1,t)=A(1,0)$ for all $t$.
\end{definition}

You should think of $A(t)$ as a loop, and $A(s,t)$ as a parameterized family
of loops which shrinks $A(t)$ to a point. Condition 1) says that for each
value of the parameter $s$, we have a loop; condition 2) says that when $s=0$
the loop is the specified loop $A(t)$; and condition 3) says that when $s=1$
our loop is a point.

It is customary to speak of simple-connectedness only for connected matrix Lie
groups, even though the definition makes sense for disconnected groups.

\begin{proposition}
\label{su2.sc}The group $\mathsf{SU}(2)$ is simply connected.
\end{proposition}

\begin{proof}
Exercise \ref{su2} shows that $\mathsf{SU}(2)$ may be thought of
(topologically) as the three-dimensional sphere $S^{3}$ sitting inside
$\mathbb{R}^{4}$. It is well-known that $S^{3}$ is simply connected.
\end{proof}

The condition of simple-connectedness is extremely important. One of our most
important theorems will be that if $G$ is simply connected, then there is a
natural one-to-one correspondence between the representations of $G$ and the
representations of its Lie algebra.

Without proof, we give the following table.
\[%
\begin{array}
[c]{cc}%
\text{\textbf{Group}} & \text{\textbf{Simply connected?}}\\
\mathsf{GL}(n;\mathbb{C}) & \text{no}\\
\mathsf{SL}\left(  n;\mathbb{C}\right)  & \text{yes}\\
\mathsf{GL}(n;\mathbb{R}) & \text{no}\\
\mathsf{SL}\left(  n;\mathbb{R}\right)  & \text{no}\\
\mathsf{SO}(n) & \text{no}\\
\mathsf{U}(n) & \text{no}\\
\mathsf{SU}(n) & \text{yes}\\
\mathsf{SO}(1;1) & \text{yes}\\
\mathsf{SO}(n;1)\text{ (}n\ge2\text{)} & \text{no}\\
\text{Heisenberg} & \text{yes}%
\end{array}
\]

\section{Homomorphisms and Isomorphisms}

\begin{definition}
\label{matrix.homomorphism}Let $G$ and $H$ be matrix Lie groups. A map $\phi$
from $G$ to $H$ is called a \textbf{Lie group homomorphism} if 1) $\phi$ is a
group homomorphism and 2) $\phi$ is continuous. If in addition, $\phi$ is
one-to-one and onto, and the inverse map $\phi^{-1}$ is continuous, then
$\phi$ is called a \textbf{Lie group isomorphism}.
\end{definition}

The condition that $\phi$ be continuous should be regarded as a technicality,
in that it is very difficult to give an example of a group homomorphism
between two matrix Lie groups which is not continuous. In fact, if
$G=\mathbb{R}$ and $H=\mathbb{C}^{\ast}$, then any group homomorphism from $G$
to $H$ which is even measurable (a very weak condition) must be continuous.
(See W. Rudin, \textit{Real and Complex Analysis,} Chap. 9, Ex. 17.)

If $G$ and $H$ are matrix Lie groups, and there exists a Lie group isomorphism
from $G$ to $H$, then $G$ and $H$ are said to be \textbf{isomorphic}, and we
write $G\cong H$. Two matrix Lie groups which are isomorphic should be thought
of as being essentially the same group. (Note that by definition, the inverse
of Lie group isomorphism is continuous, and so also a Lie group isomorphism.)

\subsection{Example: $\mathsf{SU}(2)$ and $\mathsf{SO}(3)$}

A very important topic for us will be the relationship between the groups
$\mathsf{SU}(2)$ and $\mathsf{SO}(3)$. This example is designed to show that
$\mathsf{SU}(2)$ and $\mathsf{SO}(3)$ are almost (but not quite!) isomorphic.
Specifically, there exists a Lie group homomorphism $\phi$ which maps
$\mathsf{SU}(2)$ onto $\mathsf{SO}(3)$, and which is \textit{two}-to-one. (See
Miller 7.1 and Br\"ocker, Chap. I, 6.18.)

Consider the space $V$ of all $2\times2$ complex matrices which are
self-adjoint and have trace zero. This is a three-dimensional \textit{real}
vector space with the following basis
\[%
\begin{array}
[c]{ccc}%
A_{1}=\left(
\begin{array}
[c]{cc}%
0 & 1\\
1 & 0
\end{array}
\right)  ; & A_{2}=\left(
\begin{array}
[c]{cc}%
0 & i\\
-i & 0
\end{array}
\right)  ; & A_{3}=\left(
\begin{array}
[c]{cc}%
1 & 0\\
0 & -1
\end{array}
\right)
\end{array}
\]
We may define an inner product on $V$ by the formula
\[
\left\langle A,B\right\rangle =\frac12\mathrm{trace}(AB)
\]
(Exercise: check that this is an inner product.)

Direct computation shows that $\left\{  A_{1},A_{2},A_{3}\right\}  $ is an
orthonormal basis for $V$. Having chosen an orthonormal basis for $V$, we can
identify $V$ with $\mathbb{R}^{3}$.

Now, if $U$ is an element of $\mathsf{SU}(2)$, and $A$ is an element of $V$,
then it is easy to see that $UAU^{-1}$ is in $V$. Thus for each $U\in
\mathsf{SU}(2)$, we can define a linear map $\phi_{U}$ of $V$ to itself by the
formula
\[
\phi_{U}(A)=UAU^{-1}%
\]
(This definition would work for $U\in\mathsf{U}(2)$, but we choose to restrict
our attention to $\mathsf{SU}(2)$.) Moreover, given $U\in\mathsf{SU}(2)$, and
$A,B\in V$, note that
\[
\left\langle \phi_{U}(A),\phi_{U}(B)\right\rangle =\frac12\mathrm{trace}%
(UAU^{-1}UBU^{-1})=\frac12\mathrm{trace}(AB)=\left\langle A,B\right\rangle
\]
Thus $\phi_{U}$ is an orthogonal transformation of $V\cong\mathbb{R}^{3}$,
which we can think of as an element of $\mathsf{O}(3)$.

We see, then, that the map $U\rightarrow\phi_{U}$ is a map of $\mathsf{SU}(2)
$ into $\mathsf{O}(3)$. It is very easy to check that this map is a
homomorphism (i.e., $\phi_{U_{1}U_{2}}=\phi_{U_{1}}\phi_{U_{2}}$), and that it
is continuous. Thus $U\rightarrow\phi_{U}$ is a Lie group homomorphism of
$\mathsf{SU}(2)$ into $\mathsf{O}(3)$.

Recall that every element of $\mathsf{O}(3)$ has determinant $\pm1$. Since
$\mathsf{SU}(2)$ is connected (Exercise \ref{su2}), and the map $U\rightarrow
\phi_{U}$ is continuous, $\phi_{U}$ must actually map into $\mathsf{SO}(3)$.
Thus $U\rightarrow\phi_{U}$ is a Lie group homomorphism of $\mathsf{SU}(2)$
into $\mathsf{SO}(3)$.

The map $U\rightarrow\phi_{U}$ is not one-to-one, since for any $U\in
\mathsf{SU}(2)$, $\phi_{U}=\phi_{-U}$. (Observe that if $U$ is in
$\mathsf{SU}(2)$, then so is $-U$.) It is possible to show that $\phi_{U}$ is
a two-to-one map of $\mathsf{SU}(2)$ onto $\mathsf{SO}(3)$. (See Miller.)

\section{Lie Groups\label{lie.groups}}

A Lie group is something which is simultaneously a group and a differentiable
manifold (see Definition \ref{lie.definition}). As the terminology suggests,
every matrix Lie group is a Lie group, although this requires proof (Theorem
\ref{lie.theorem}). I have decided to restrict attention to matrix Lie groups,
except in emergencies, for three reasons. First, this makes the course
accessible to students who are not familiar with the theory of differentiable
manifolds. Second, this makes the definition of the Lie algebra and of the
exponential mapping far more comprehensible. Third, all of the important
examples of Lie groups are (or can easily be represented as) matrix Lie groups.

Alas, there is a price to pay for this simplification. Certain important
topics (notably, the universal cover) are considerably complicated by
restricting to the matrix case. Nevertheless, I feel that the advantages
outweigh the disadvantages in an introductory course such as this.

\begin{definition}
\label{lie.definition}A \textbf{Lie group} is a differentiable manifold $G$
which is also a group, and such that the group product
\[
G\times G\rightarrow G
\]
and the inverse map $g\rightarrow g^{-1}$ are differentiable.
\end{definition}

For the reader who is not familiar with the notion of a differentiable
manifold, here is a brief recap. (I will consider only manifolds embedded in
some $\mathbb{R}^{n}$, which is a harmless assumption.) A subset $M$ of
$\mathbf{R}^{n}$ is called a $k$\textbf{-dimensional differentiable manifold}
if given any $m_{0}\in M$, there exists a smooth (non-linear) coordinate
system $(x^{1},\cdots x^{n})$ defined in a neighborhood $U$ of $m_{0}$ such
that
\[
M\cap U=\left\{  m\in U\left|  x^{k+1}(m)=c_{1},\cdots,x^{n}(m)=c_{n-k}%
\right.  \right\}
\]
This says that locally, after a suitable change of variables, $M$ looks like
the $k$-dimensional hyperplane in $\mathbb{R}^{n}$ obtained by setting all but
the first $k$ coordinates equal to constants.

For example, $S^{1}\subset\mathbb{R}^{2}$ is a one-dimensional differentiable
manifold because in the usual polar coordinates $(\theta,r)$, $S^{1}$ is the
set $r=1$. Of course, polar coordinates are not globally defined, because
$\theta$ is undefined at the origin, and because $\theta$ is not
``single-valued.'' But given any point $m_{0}$ in $S^{1}$, we can define polar
coordinates in a neighborhood $U$ of $m_{0}$, and then $S^{1}\cap U$ will be
the set $r=1$.

Note that while we assume that our differentiable manifolds are embedded in
some $\mathbb{R}^{n}$ (a harmless assumption), we are \textit{not} saying that
a Lie group has to be embedded in $\mathbb{R}^{n^{2}}$, or that the group
operation has to have anything to do with matrix multiplication. A Lie group
is simply a subset $G$ of some $\mathbb{R}^{n}$ which is a differentiable
manifold, together with \textit{any} map from $G\times G$ into $G$ which makes
$G$ into a group (and such that the group operations are smooth). It is
remarkable that almost (but not quite!) every Lie group is isomorphic to a
matrix Lie group.

Note also that it is far from obvious that a matrix Lie group must be a Lie
group, since our definition of a matrix Lie group $G$ does not say anything
about $G$ being a manifold. It is not too difficult to verify that all of our
examples of matrix Lie groups are Lie groups, but in fact we have the
following result which makes such verifications unnecessary:

\begin{theorem}
\label{lie.theorem}Every matrix Lie group is a Lie group.
\end{theorem}

Although I will not prove this result, I want to discuss what would be
involved. Let us consider first the group $\mathsf{GL}(n;\mathbb{R})$. The
space of all $n\times n$ real matrices can be thought of as $\mathbb{R}%
^{n^{2}}$. Since $\mathsf{GL}(n;\mathbb{R})$ is the set of all matrices $A$
with $\det A\neq0$, $\mathsf{GL}(n;\mathbb{R})$ is an open subset of
$\mathbb{R}^{n^{2}}$. (That is, given an invertible matrix $A$, there is a
neighborhood $U$ of $A$ such that every matrix $B\in U$ is also invertible.)
Thus $\mathsf{GL}(n;\mathbb{R})$ is an $n^{2}$-dimensional smooth manifold.
Furthermore, the matrix product $AB$ is clearly a smooth (even polynomial)
function of the entries of $A$ and $B$, and (in light of Kramer's rule)
$A^{-1}$ is a smooth function of the entries of $A$. Thus $\mathsf{GL}%
(n;\mathbb{R})$ is a Lie group.

Similarly, if we think of the space of $n\times n$ complex matrices as
$\mathbb{C}^{n^{2}}\cong\mathbb{R}^{2n^{2}}$, then the same argument shows
that $\mathsf{GL}(n;\mathbb{C})$ is a Lie group.

Thus, to prove that every matrix Lie group is a Lie group, it suffices to show
that a closed subgroup of a Lie group is a Lie group. This is proved in
Br\"ocker and tom Dieck, Chapter I, Theorem 3.11. The proof is not too
difficult, but it requires the exponential mapping, which we have not yet
introduced. (See Chapter 3.)

It is customary to call a map $\phi$ between two Lie groups a Lie group
homomorphism if $\phi$ is a group homomorphism and $\phi$ is \textit{smooth},
whereas we have (in Definition \ref{matrix.homomorphism}) required only that
$\phi$ be continuous. However, the following Proposition shows that our
definition is equivalent to the more standard one.

\begin{proposition}
\label{homo.smooth}Let $G$ and $H$ be Lie groups, and $\phi$ a group
homomorphism from $G$ to $H$. Then if $\phi$ is continuous it is also smooth.
\end{proposition}

Thus group homomorphisms from $G$ to $H$ come in only two varieties: the very
bad ones (discontinuous), and the very good ones (smooth). There simply aren't
any intermediate ones. (See, for example, Exercise \ref{character}.) For
proof, see Br\"ocker and tom Dieck, Chapter I, Proposition 3.12.

In light of Theorem \ref{lie.theorem}, every matrix Lie group is a (smooth)
manifold. As such, a matrix Lie group is automatically locally path connected.
It follows that a matrix Lie group is path connected if and only if it is
connected. (See Remarks following Definition \ref{connectedness}.)

\section{Exercises}

\begin{enumerate}
\item \label{not.closed}Let $a$ be an irrational real number. Show that the
set of numbers of the form $e^{2\pi ina}$, $n\in\mathbb{Z}$, is dense in
$S^{1}$. Now let $G$ be the following subgroup of $\mathsf{GL}(2;\mathbb{C}%
)$:
\[
G=\left\{  \left(
\begin{array}
[c]{cc}%
e^{it} & 0\\
0 & e^{iat}%
\end{array}
\right)  \left|  t\in\mathbb{R}\right.  \right\}
\]
Show that
\[
\overline{G}=\left\{  \left(
\begin{array}
[c]{cc}%
e^{it} & 0\\
0 & e^{is}%
\end{array}
\right)  \left|  t,s\in\mathbb{R}\right.  \right\}  ,
\]
where $\overline{G}$ denotes the closure of the set $G$ inside the space of
$2\times2$ matrices.

\textit{Note}: The group $\overline{G}$ can be thought of as the torus
$S^{1}\times S^{1}$, which in turn can be thought of as $\left[
0,2\pi\right]  \times\left[  0,2\pi\right]  $, with the ends of the intervals
identified. The set $G\subset\left[  0,2\pi\right]  \times\left[
0,2\pi\right]  $ is called an \textbf{irrational line}. Draw a picture of this
set and you should see why $G$ is dense in $\left[  0,2\pi\right]
\times\left[  0,2\pi\right]  $.

\item \label{orthogonal}\textit{Orthogonal groups}\textbf{.} Let $\left\langle
\ \right\rangle $ denote the standard inner product on $\mathbb{R}^{n}$,
$\left\langle x,y\right\rangle =\sum_{i}x_{i}y_{i}$. Show that a matrix $A$
preserves inner products if and only if the column vectors of $A$ are orthonormal.

Show that for any $n\times n$ real matrix $B$,
\[
\left\langle Bx,y\right\rangle =\left\langle x,B^{tr}y\right\rangle
\]
where $\left(  B^{tr}\right)  _{ij}=B_{ji}$. Using this fact, show that a
matrix $A$ preserves inner products if and only if $A^{tr}A=I$.

\textit{Note}: a similar analysis applies to the complex orthogonal groups
$\mathsf{O}(n;\mathbb{C})$ and $\mathsf{SO}(n;\mathbb{C})$.

\item \label{unitary}\textit{Unitary groups}\textbf{.} Let $\left\langle
\ \right\rangle $ denote the standard inner product on $\mathbb{C}^{n}$,
$\left\langle x,y\right\rangle =\sum_{i}\overline{x_{i}}y_{i}$. Following
Exercise \ref{orthogonal}, show that $A^{\ast}A=I$ if and only if
$\left\langle Ax,Ay\right\rangle =\left\langle x,y\right\rangle $ for all
$x,y\in\mathbb{C}^{n}$. ($\left(  A^{\ast}\right)  _{ij}=\overline{A_{ji}}$.)

\item \label{generalized}\textit{Generalized orthogonal groups}. Let $\left[
x,y\right]  _{n,k}$ be the symmetric bilinear form on $\mathbb{R}^{n+k}$
defined in (\ref{bilinear}). Let $g$ be the $(n+k)\times(n+k)$ diagonal matrix
with first $n$ diagonal entries equal to one, and last $k$ diagonal entries
equal to minus one:
\[
g=\left(
\begin{array}
[c]{cc}%
I_{n} & 0\\
0 & -I_{k}%
\end{array}
\right)
\]
Show that for all $x,y\in\mathbb{R}^{n+k}$,
\[
\left[  x,y\right]  _{n,k}=\left\langle x,gy\right\rangle
\]
Show that a $(n+k)\times(n+k)$ real matrix $A$ is in $\mathsf{O}(n;k)$ if and
only if $A^{tr}gA=g$. Show that $\mathsf{O}(n;k)$ and $\mathsf{SO}(n;k)$ are
subgroups of $\mathsf{GL}(n+k;\mathbb{R})$, and are matrix Lie groups.

\item \label{symplectic}\textit{Symplectic groups}. Let $B\left[  x,y\right]
$ be the skew-symmetric bilinear form on $\mathbb{R}^{2n}$ given by $B\left[
x,y\right]  =\sum_{i=1}^{n}x_{i}y_{n+i}-x_{n+i}y_{i}$. Let $J$ be the
$2n\times2n$ matrix
\[
J=\left(
\begin{array}
[c]{cc}%
0 & I\\
-I & 0
\end{array}
\right)
\]
Show that for all $x,y\in\mathbb{R}^{2n}$%
\[
B\left[  x,y\right]  =\left\langle x,Jy\right\rangle
\]
Show that a $2n\times2n$ matrix $A$ is in $\mathsf{Sp}\left(  n;\mathbb{R}%
\right)  $ if and only if $A^{tr}JA=J$. Show that $\mathsf{Sp}\left(
n;\mathbb{R}\right)  $ is a subgroup of $\mathsf{GL}(2n;\mathbb{R})$, and a
matrix Lie group.

\textit{Note}: a similar analysis applies to $\mathsf{Sp}\left(
n;\mathbb{C}\right)  $.

\item \label{so2}\textit{The groups}\emph{\ }$\mathsf{O}(2)$ \textit{and}
$\mathsf{SO}(2)$. Show that the matrix
\[
A=\left(
\begin{array}
[c]{cc}%
\cos\theta & -\sin\theta\\
\sin\theta & \cos\theta
\end{array}
\right)
\]
is in $\mathsf{SO}(2)$, and that
\[
\left(
\begin{array}
[c]{cc}%
\cos\theta & -\sin\theta\\
\sin\theta & \cos\theta
\end{array}
\right)  \left(
\begin{array}
[c]{cc}%
\cos\phi & -\sin\phi\\
\sin\phi & \cos\phi
\end{array}
\right)  =\left(
\begin{array}
[c]{cc}%
\cos(\theta+\phi) & -\sin(\theta+\phi)\\
\sin(\theta+\phi) & \cos(\theta+\phi)
\end{array}
\right)
\]
Show that every element $A$ of $\mathsf{O}(2)$ is of one of the two forms
\[
A=\left(
\begin{array}
[c]{cc}%
\cos\theta & -\sin\theta\\
\sin\theta & \cos\theta
\end{array}
\right)
\]%
\[
A=\left(
\begin{array}
[c]{cc}%
\cos\theta & \sin\theta\\
\sin\theta & -\cos\theta
\end{array}
\right)
\]
(If $A$ is of the first form, then $\det A=1$; if $A$ is of the second form,
then $\det A=-1$.)

\textit{Hint}: Recall that for $A=\left(
\begin{array}
[c]{cc}%
a & b\\
c & d
\end{array}
\right)  $ to be in $\mathsf{O}(2)$, the column vectors $\left(
\begin{array}
[c]{c}%
a\\
c
\end{array}
\right)  $ and $\left(
\begin{array}
[c]{c}%
b\\
d
\end{array}
\right)  $ must be unit vectors, and must be orthogonal.

\item \label{so11}\textit{The groups }$\mathsf{O}(1;1)$\textit{\ and
}$\mathsf{SO}(1;1)$. Show that
\[
A=\left(
\begin{array}
[c]{cc}%
\cosh t & \sinh t\\
\sinh t & \cosh t
\end{array}
\right)
\]
is in $\mathsf{SO}(1;1)$, and that
\[
\left(
\begin{array}
[c]{cc}%
\cosh t & \sinh t\\
\sinh t & \cosh t
\end{array}
\right)  \left(
\begin{array}
[c]{cc}%
\cosh s & \sinh s\\
\sinh s & \cosh s
\end{array}
\right)  =\left(
\begin{array}
[c]{cc}%
\cosh(t+s) & \sinh(t+s)\\
\sinh(t+s) & \cosh(t+s)
\end{array}
\right)
\]
Show that every element of $\mathsf{O}(1;1)$ can be written in one of the four
forms
\[
\left(
\begin{array}
[c]{cc}%
\cosh t & \sinh t\\
\sinh t & \cosh t
\end{array}
\right)
\]%
\[
\left(
\begin{array}
[c]{cc}%
-\cosh t & \sinh t\\
\sinh t & -\cosh t
\end{array}
\right)
\]%
\[
\left(
\begin{array}
[c]{cc}%
\cosh t & -\sinh t\\
\sinh t & -\cosh t
\end{array}
\right)
\]%
\[
\left(
\begin{array}
[c]{cc}%
-\cosh t & -\sinh t\\
\sinh t & \cosh t
\end{array}
\right)
\]
(Since $\cosh t$ is always positive, there is no overlap among the four cases.
Matrices of the first two forms have determinant one; matrices of the last two
forms have determinant minus one.)

\textit{Hint}: For $\left(
\begin{array}
[c]{cc}%
a & b\\
c & d
\end{array}
\right)  $ to be in $\mathsf{O}(1;1)$, we must have $a^{2}-c^{2}=1$,
$b^{2}-d^{2}=-1$, and $ab-cd=0$. The set of points $\left(  a,c\right)  $ in
the plane with $a^{2}-c^{2}=1$ (i.e., $a=\pm\sqrt{1+c^{2}}$\thinspace) is a hyperbola.

\item \label{su2}\textit{The group}\emph{\ }$\mathsf{SU}(2)$. Show that if
$\alpha,\beta$ are arbitrary complex numbers satisfying $\left|
\alpha\right|  ^{2}+\left|  \beta\right|  ^{2}=1$, then the matrix
\begin{equation}
A=\left(
\begin{array}
[c]{cc}%
\alpha & -\overline{\beta}\\
\beta & \overline{\alpha}%
\end{array}
\right)  \label{su2.form}%
\end{equation}
is in $\mathsf{SU}(2)$. Show that every $A\in\mathsf{SU}(2)$ can be expressed
in the form (\ref{su2.form}) for a unique pair $(\alpha,\beta)$ satisfying
$\left|  \alpha\right|  ^{2}+\left|  \beta\right|  ^{2}=1$. (Thus
$\mathsf{SU}(2)$ can be thought of as the three-dimensional sphere $S^{3}$
sitting inside $\mathbf{C}^{2}=\mathbb{R}^{4}$. In particular, this shows that
$\mathsf{SU}(2)$ is connected and simply connected.)

\item \label{sp1}\textit{The groups}\emph{\ }$\mathsf{Sp}\left(
1;\mathbb{R}\right)  $, $\mathsf{Sp}\left(  1;\mathbf{C}\right)  $,
\textit{and}\emph{\ }$\mathsf{Sp}\left(  1\right)  $. Show that $\mathsf{Sp}%
\left(  1;\mathbb{R}\right)  =\mathsf{SL}\left(  2;\mathbb{R}\right)  $,
$\mathsf{Sp}\left(  1;\mathbf{C}\right)  =\mathsf{SL}\left(  2;\mathbf{C}%
\right)  $, and $\mathsf{Sp}(1)=\mathsf{SU}(2)$.

\item \label{heisenberg.center}\textit{The Heisenberg group}. Determine the
center $Z(H)$ of the Heisenberg group $H$. Show that the quotient group
$H/Z(H)$ is abelian.

\item \label{connect.son}\textit{Connectedness of}\emph{\ }$\mathsf{SO}(n)$.
Show that $\mathsf{SO}(n)$ is connected, following the outline below.

For the case $n=1$, there is not much to show, since a $1\times1$ matrix with
determinant one must be $\left[  1\right]  $. Assume, then, that $n\geq2$. Let
$e_{1}$ denote the vector
\[
e_{1}=\left(
\begin{array}
[c]{c}%
1\\
0\\
\vdots\\
0
\end{array}
\right)
\]
in $\mathbb{R}^{n}$. Given any unit vector $v\in\mathbb{R}^{n}$, show that
there exists a continuous path $R(t)$ in $\mathsf{SO}(n)$ with $R(0)=I$ and
$R(1)v=e_{1}$. (Thus any unit vector can be ``continuously rotated'' to
$e_{1}$.)

Now show that any element $R$ of $\mathsf{SO}(n)$ can be connected to an
element of $\mathsf{SO}(n-1)$, and proceed by induction.

\item \label{polar}\textit{The polar decomposition of}\emph{\ }$\mathsf{SL}%
\left(  n;\mathbb{R}\right)  $. Show that every element $A$ of $\mathsf{SL}%
\left(  n;\mathbb{R}\right)  $ can be written uniquely in the form $A=RH$,
where $R$ is in $\mathsf{SO}(n)$, and $H$ is a symmetric, positive-definite
matrix with determinant one. (That is, $H^{tr}=H$, and $\left\langle
x,Hx\right\rangle \geq0$ for all $x\in\mathbb{R}^{n}$).

\textit{Hint}: If $A$ \textit{could} be written in this form, then we would
have
\[
A^{tr}A=H^{tr}R^{tr}RH=HR^{-1}RH=H^{2}%
\]
Thus $H$ would have to be the unique positive-definite symmetric square root
of $A^{tr}A$.

\textit{Note}: A similar argument gives polar decompositions for
$\mathsf{GL}(n;\mathbb{R})$, $\mathsf{SL}\left(  n;\mathbb{C}\right)  $, and
$\mathsf{GL}(n;\mathbb{C})$. For example, every element $A$ of $\mathsf{SL}%
\left(  n;\mathbb{C}\right)  $ can be written uniquely as $A=UH$, with $U$ in
$\mathsf{SU}(n)$, and $H$ a self-adjoint positive-definite matrix with
determinant one.

\item \label{connect.sln}\textit{The connectedness of}\emph{\ }$\mathsf{SL}%
\left(  n;\mathbb{R}\right)  $. Using the polar decomposition of
$\mathsf{SL}\left(  n;\mathbb{R}\right)  $ (Ex. \ref{polar}) and the
connectedness of $\mathsf{SO}(n)$ (Ex. \ref{connect.son}), show that
$\mathsf{SL}\left(  n;\mathbb{R}\right)  $ is connected.

\textit{Hint}: Recall that if $H$ is a real, symmetric matrix, then there
exists a \textit{real} orthogonal matrix $R_{1}$ such that $H=R_{1}DR_{1}%
^{-1}$, where $D$ is diagonal.

\item \label{connect.gln}\textit{The connectedness of}\emph{\ }$\mathsf{GL}%
(n;\mathbb{R})^{+}$. Show that $\mathsf{GL}(n;\mathbb{R})^{+}$ is connected.

\item  Show that the set of translations is a normal subgroup of the Euclidean
group, and also of the Poincar\'{e} group. Show that $\left(  \mathsf{E}%
(n)/\mathrm{translations}\right)  \cong\mathsf{O}(n)$.

\item \label{character}\textit{Harder. }Show that every Lie group homomorphism
$\phi$ from $\mathbb{R}$ to $S^{1}$ is of the form $\phi(x)=e^{iax}$ for some
$a\in\mathbb{R}$. In particular, every such homomorphism is smooth.
\end{enumerate}

\chapter{Lie Algebras and the Exponential Mapping}

\section{The Matrix Exponential}

The exponential of a matrix plays a crucial role in the theory of Lie groups.
The exponential enters into the definition of the Lie algebra of a matrix Lie
group (Section \ref{lie.algebra} below), and is the mechanism for passing
information from the Lie algebra to the Lie group. Since many computations are
done much more easily at the level of the Lie algebra, the exponential is indispensable.

Let $X$ be an $n\times n$ real or complex matrix. We wish to define the
exponential of $X$, $e^{X}$ or $\exp X$, by the usual power series
\begin{equation}
e^{X}=\sum_{m=0}^{\infty}\frac{X^{m}}{m!}\text{.}\label{exponential}%
\end{equation}
We will follow the convention of using letters such as $X$ and $Y$ for the
variable in the matrix exponential.

\begin{proposition}
\label{convergence}For any $n\times n$ real or complex matrix $X$, the series
(\ref{exponential}) converges. The matrix exponential $e^{X}$ is a continuous
function of $X$.
\end{proposition}

Before proving this, let us review some elementary analysis. Recall that the
norm of a vector $x$ in $\mathbb{C}^{n}$ is defined to be
\[
\left\|  x\right\|  =\sqrt{\left\langle x,x\right\rangle }=\sqrt{\sum\left|
x_{i}\right|  ^{2}}\text{.}%
\]
This norm satisfies the triangle inequality
\[
\left\|  x+y\right\|  \leq\left\|  x\right\|  +\left\|  y\right\|  \text{.}%
\]
The norm of a matrix $A$ is defined to be
\[
\left\|  A\right\|  =\sup_{x\neq0}\frac{\left\|  Ax\right\|  }{\left\|
x\right\|  }\text{.}%
\]
Equivalently, $\left\|  A\right\|  $ is the smallest number $\lambda$ such
that $\left\|  Ax\right\|  \leq\lambda\left\|  x\right\|  $ for all
$x\in\mathbf{C}^{n}$.

It is not hard to see that for any $n\times n$ matrix $A$, $\left\|
A\right\|  $ is finite. Furthermore, it is easy to see that for any matrices
$A,B$%
\begin{align}
\left\|  AB\right\|   & \leq\left\|  A\right\|  \left\|  B\right\|
\label{banach}\\
\left\|  A+B\right\|   & \leq\left\|  A\right\|  +\left\|  B\right\|  \text{.}%
\end{align}
It is also easy to see that a sequence of matrices $A_{m}$ converges to a
matrix $A$ if and only if $\left\|  A_{m}-A\right\|  \rightarrow0$. (Compare
this with Definition \ref{matrix.converge} of Chapter 2.)

A sequence of matrices $A_{m}$ is said to be a \textbf{Cauchy sequence} if
$\left\|  A_{m}-A_{l}\right\|  \rightarrow0$ as $m,l\rightarrow\infty$.
Thinking of the space of matrices as $\mathbb{R}^{n^{2}}$ or $\mathbb{C}%
^{n^{2}}$, and using a standard result from analysis, we have the following:

\begin{proposition}
\label{cauchy}If $A_{m}$ is a sequence of $n\times n$ real or complex
matrices, and $A_{m}$ is a Cauchy sequence, then there exists a unique matrix
$A$ such that $A_{m}$ converges to $A$.
\end{proposition}

\noindent That is, every Cauchy sequence converges.

Now, consider an infinite series whose terms are matrices:
\begin{equation}
A_{0}+A_{1}+A_{2}+\cdots\text{.}\label{an}%
\end{equation}
If
\[
\sum_{m=0}^{\infty}\left\|  A_{m}\right\|  <\infty
\]
then the series (\ref{an}) is said to \textbf{converge absolutely}. If a
series converges absolutely, then it is not hard to show that the partial sums
of the series form a Cauchy sequence, and hence by Proposition \ref{cauchy},
the series converges. That is, any series which converges absolutely also
converges. (The converse is not true; a series of matrices can converge
without converging absolutely.)

\begin{proof}
In light of (\ref{banach}), we see that
\[
\left\|  X^{m}\right\|  \leq\left\|  X\right\|  ^{m}\text{,}%
\]
and hence
\[
\sum_{m=0}^{\infty}\left\|  \frac{X^{m}}{m!}\right\|  \leq\sum_{m=0}^{\infty
}\frac{\left\|  X\right\|  ^{m}}{m!}=e^{\left\|  X\right\|  }<\infty\text{.}%
\]
Thus the series (\ref{exponential}) converges absolutely, and so it converges.

To show continuity, note that since $X^{m}$ is a continuous function of $X$,
the partial sums of (\ref{exponential}) are continuous. But it is easy to see
that (\ref{exponential}) converges uniformly on each set of the form $\left\{
\left\|  X\right\|  \leq R\right\}  $, and so the sum is again continuous.
\end{proof}

\begin{proposition}
\label{exp.properties}Let $X,Y$ be arbitrary $n\times n$ matrices. Then

\begin{enumerate}
\item \label{exp.zero}$e^{0}=I$.

\item \label{exp.inverse}$e^{X}$ is invertible, and $\left(  e^{X}\right)
^{-1}=e^{-X}$.

\item \label{exp.scalars}$e^{(\alpha+\beta)X}=e^{\alpha X}e^{\beta X}$ for all
real or complex numbers $\alpha,\beta$.

\item \label{exp.sum}If $XY=YX$, then $e^{X+Y}=e^{X}e^{Y}=e^{Y}e^{X}$.

\item \label{exp.adjoint}If $C$ is invertible, then $e^{CXC^{-1}}=Ce^{X}%
C^{-1}$.

\item \label{exp.norm}$\left\|  e^{X}\right\|  \leq e^{\left\|  X\right\|  }$.
\end{enumerate}
\end{proposition}

It is \textit{not} true in general that $e^{X+Y}=e^{X}e^{Y}$, although by
\ref{exp.sum}) it is true if $X$ and $Y$ commute. This is a crucial point,
which we will consider in detail later. (See the Lie product formula in
Section \ref{further} and the Baker-Campbell-Hausdorff formula in Chapter 4.)

\begin{proof}
Point \ref{exp.zero}) is obvious. Points \ref{exp.inverse}) and
\ref{exp.scalars}) are special cases of point \ref{exp.sum}). To verify point
\ref{exp.sum}), we simply multiply power series term by term. (It is left to
the reader to verify that this is legal.) Thus
\[
e^{X}e^{Y}=\left(  I+X+\frac{X^{2}}{2!}+\cdots\right)  \left(  I+Y+\frac
{Y^{2}}{2!}+\cdots\right)  \text{.}%
\]
Multiplying this out and collecting terms where the power of $X$ plus the
power of $Y$ equals $m$, we get
\begin{equation}
e^{X}e^{Y}=\sum_{m=0}^{\infty}\sum_{k=0}^{m}\frac{X^{k}}{k!}\frac{Y^{m-k}%
}{(m-k)!}=\sum_{m=0}^{\infty}\frac{1}{m!}\sum_{k=0}^{m}\frac{m!}%
{k!(m-k)!}X^{k}Y^{m-k}\text{.}\label{powers}%
\end{equation}

Now because (and \textit{only} because) $X$ and $Y$ commute,
\[
(X+Y)^{n}=\sum_{k=0}^{m}\frac{m!}{k!(m-k)!}X^{k}Y^{m-k}\text{,}%
\]
and so (\ref{powers}) becomes
\[
e^{X}e^{Y}=\sum_{m=0}^{\infty}\frac{1}{m!}(X+Y)^{m}=e^{X+Y}\text{.}%
\]

To prove \ref{exp.adjoint}), simply note that
\[
\left(  CXC^{-1}\right)  ^{m}=CX^{m}C^{-1}%
\]
and so the two sides of \ref{exp.adjoint}) are the same term by term.

Point \ref{exp.norm}) is evident from the proof of Proposition
\ref{convergence}.
\end{proof}

\begin{proposition}
\label{derivative}Let $X$ be a $n\times n$ complex matrix, and view the space
of all $n\times n$ complex matrices as $\mathbb{C}^{n^{2}}$. Then $e^{tX}$ is
a smooth curve in $\mathbb{C}^{n^{2}}$, and
\[
\frac{d}{dt}e^{tX}=Xe^{tX}=e^{tX}X\text{.}%
\]
In particular,
\[
\left.  \frac{d}{dt}\right|  _{t=0}e^{tX}=X\text{.}%
\]
\end{proposition}

\begin{proof}
Differentiate the power series for $e^{tX}$ term-by-term. (You might worry
whether this is valid, but you shouldn't. For each $i,j$, $\left(
e^{tX}\right)  _{ij}$ is given by a convergent power series in $t$, and it is
a standard theorem that you can differentiate power series term-by-term.)
\end{proof}

\section{Computing the Exponential of a Matrix\label{computing}}

\subsection{Case 1: $X$ is diagonalizable}

Suppose that $X$ is a $n\times n$ real or complex matrix, and that $X$ is
diagonalizable over $\mathbb{C}$, that is, that there exists an invertible
complex matrix $C$ such that $X=CDC^{-1}$, with
\[
D=\left(
\begin{array}
[c]{ccc}%
\lambda_{1} &  & 0\\
& \ddots & \\
0 &  & \lambda_{n}%
\end{array}
\right)  \text{.}%
\]
Observe that $e^{D}$ is the diagonal matrix with eigenvalues $e^{\lambda_{1}%
},\cdots,e^{\lambda_{n}}$, and so in light of Proposition \ref{exp.properties}%
, we have
\[
e^{X}=C\left(
\begin{array}
[c]{ccc}%
e^{\lambda_{1}} &  & 0\\
& \ddots & \\
0 &  & e^{\lambda_{n}}%
\end{array}
\right)  C^{-1}\text{.}%
\]
Thus if you can explicitly diagonalize $X$, you can explicitly compute $e^{X}%
$. Note that if $X$ is real, then although $C$ may be complex and the
$\lambda_{i}$'s may be complex, $e^{X}$ must come out to be real, since each
term in the series (\ref{exponential}) is real.

For example, take
\[
X=\left(
\begin{array}
[c]{cc}%
0 & -a\\
a & 0
\end{array}
\right)  \text{.}%
\]
Then the eigenvectors of $X$ are $\left(
\begin{array}
[c]{c}%
1\\
i
\end{array}
\right)  $ and $\left(
\begin{array}
[c]{c}%
i\\
1
\end{array}
\right)  $, with eigenvalues $-ia$ and $ia$, respectively. Thus the invertible
matrix
\[
C=\left(
\begin{array}
[c]{cc}%
1 & i\\
i & 1
\end{array}
\right)
\]
maps the basis vectors $\left(
\begin{array}
[c]{c}%
1\\
0
\end{array}
\right)  $ and $\left(
\begin{array}
[c]{c}%
0\\
1
\end{array}
\right)  $ to the eigenvectors of $X$, and so (check) $C^{-1}XC$ is a diagonal
matrix $D$. Thus $X=CDC^{-1}$:
\begin{align*}
e^{X}  & =\left(
\begin{array}
[c]{cc}%
1 & i\\
i & 1
\end{array}
\right)  \left(
\begin{array}
[c]{cc}%
e^{-ia} & 0\\
0 & e^{ia}%
\end{array}
\right)  \left(
\begin{array}
[c]{cc}%
1/2 & -i/2\\
-i/2 & 1/2
\end{array}
\right) \\
& =\left(
\begin{array}
[c]{cc}%
\cos a & -\sin a\\
\sin a & \cos a
\end{array}
\right)  \text{.}%
\end{align*}
Note that explicitly if $X$ (and hence $a$) is real, then $e^{X}$ is real.

\subsection{Case 2: $X$ is nilpotent}

An $n\times n$ matrix $X$ is said to be \textbf{nilpotent} if $X^{m}=0$ for
some positive integer $m$. Of course, if $X^{m}=0$, then $X^{l}=0$ for all
$l>m$. In this case the series (\ref{exponential}) which defines $e^{X}$
terminates after the first $m$ terms, and so can be computed explicitly.

For example, compute $e^{tX}$, where
\[
X=\left(
\begin{array}
[c]{ccc}%
0 & a & b\\
0 & 0 & c\\
0 & 0 & 0
\end{array}
\right)  \text{.}%
\]
Note that
\[
X^{2}=\left(
\begin{array}
[c]{ccc}%
0 & 0 & ac\\
0 & 0 & 0\\
0 & 0 & 0
\end{array}
\right)
\]
and that $X^{3}=0$. Thus
\[
e^{tX}=\left(
\begin{array}
[c]{ccc}%
1 & ta & tb+\dfrac12t^{2}ac\\
0 & 1 & tc\\
0 & 0 & 1
\end{array}
\right)  \text{.}%
\]

\subsection{Case 3: $X$ arbitrary}

A general matrix $X$ may be neither nilpotent nor diagonalizable. However, it
follows from the Jordan canonical form that $X$ can be written (Exercise
\ref{decomposition}) in the form $X=S+N$ with $S$ diagonalizable, $N$
nilpotent, and $SN=NS$. (See Exercise \ref{decomposition}.) Then, since $N$
and $S$ commute,
\[
e^{X}=e^{S+N}=e^{S}e^{N}%
\]
and $e^{S}$ and $e^{N}$ can be computed as above.

For example, take
\[
X=\left(
\begin{array}
[c]{cc}%
a & b\\
0 & a
\end{array}
\right)  \text{.}%
\]
Then
\[
X=\left(
\begin{array}
[c]{cc}%
a & 0\\
0 & a
\end{array}
\right)  +\left(
\begin{array}
[c]{cc}%
0 & b\\
0 & 0
\end{array}
\right)  \text{.}%
\]
The two terms clearly commute (since the first one is a multiple of the
identity), and so
\[
e^{X}=\left(
\begin{array}
[c]{cc}%
e^{a} & 0\\
0 & e^{a}%
\end{array}
\right)  \left(
\begin{array}
[c]{cc}%
1 & b\\
0 & 1
\end{array}
\right)  =\left(
\begin{array}
[c]{cc}%
e^{a} & e^{a}b\\
0 & e^{a}%
\end{array}
\right)  \text{.}%
\]

\section{The Matrix Logarithm\label{matrix.log}}

We wish to define a matrix logarithm, which should be an inverse function to
the matrix exponential. Defining a logarithm for matrices should be at least
as difficult as defining a logarithm for complex numbers, and so we cannot
hope to define the matrix logarithm for all matrices, or even for all
invertible matrices. We will content ourselves with defining the logarithm in
a neighborhood of the identity matrix.

The simplest way to define the matrix logarithm is by a power series. We
recall the situation for complex numbers:

\begin{lemma}
The function
\[
\log z=\sum_{m=1}^{\infty}(-1)^{m+1}\frac{(z-1)^{m}}{m}%
\]
is defined and analytic in a circle of radius one about $z=1$.

For all $z$ with $\left|  z-1\right|  <1$,
\[
e^{\log z}=z\text{.}%
\]
For all $u$ with $\left|  u\right|  <\log2$, $\left|  e^{u}-1\right|  <1$ and
\[
\log e^{u}=u\text{.}%
\]
\end{lemma}

\begin{proof}
The usual logarithm for real, positive numbers satisfies
\[
\frac{d}{dx}\log(1-x)=\frac{-1}{1-x}=-\left(  1+x+x^{2}+\cdots\right)
\]
for $\left|  x\right|  <1$. Integrating term-by-term and noting that $\log1=0$
gives
\[
\log(1-x)=-\left(  x+\tfrac{x^{2}}{2}+\tfrac{x^{3}}{3}+\cdots\right)  \text{.}%
\]
Taking $z=1-x$ (so that $x=1-z$), we have
\begin{align*}
\log z=-\left(  (1-z)+\tfrac{(1-z)^{2}}{2}+\tfrac{(1-z)^{3}}{3}+\cdots\right)
\\
=\sum_{m=1}^{\infty}(-1)^{m+1}\frac{(z-1)^{m}}{m}\text{.}%
\end{align*}

This series has radius of convergence one, and defines a complex analytic
function on the set $\left\{  \left|  z-1\right|  <1\right\}  $, which
coincides with the usual logarithm for real $z$ in the interval $(0,2)$. Now,
$\exp(\log z)=z$ for $z\in(0,2)$, and by analyticity this identity continues
to hold on the whole set $\left\{  \left|  z-1\right|  <1\right\}  $.

On the other hand, if $\left|  u\right|  <\log2$, then
\[
\left|  e^{u}-1\right|  =\left|  u+\tfrac{u^{2}}{2!}+\cdots\right|
\leq\left|  u\right|  +\frac{\left|  u\right|  ^{2}}{2!}+\cdots
\]
so that
\[
\left|  e^{u}-1\right|  \leq e^{\left|  u\right|  }-1<1\text{.}%
\]
Thus $\log(\exp u)$ makes sense for all such $u$. Since $\log(\exp u)=u$ for
real $u$ with $\left|  u\right|  <\log2$, it follows by analyticity that
$\log(\exp u)=u$ for all complex numbers with $\left|  u\right|  <\log2$.
\end{proof}

\begin{theorem}
\label{logarithm}The function
\begin{equation}
\log A=\sum_{m=1}^{\infty}(-1)^{m+1}\frac{(A-I)^{m}}{m}\label{log.series}%
\end{equation}
is defined and continuous on the set of all $n\times n$ complex matrices $A$
with $\left\|  A-I\right\|  <1$, and $\log A$ is real if $A$ is real.

For all $A$ with $\left\|  A-I\right\|  <1$,
\[
e^{\log A}=A\text{.}%
\]
For all $X$ with $\left\|  X\right\|  <\log2$, $\left\|  e^{X}-1\right\|  <1$
and
\[
\log e^{X}=X\text{.}%
\]
\end{theorem}

\begin{proof}
It is easy to see that the series (\ref{log.series}) converges absolutely
whenever $\left\|  A-I\right\|  <1$. The proof of continuity is essentially
the same as for the exponential. If $A$ is real, then every term in the series
(\ref{log.series}) is real, and so $\log A$ is real.

We will now show that $\exp(\log A)=A$ for all $A$ with $\left\|  A-I\right\|
<1$. We do this by considering two cases.

\textit{Case 1: }$A$\textit{\ is diagonalizable.}

Suppose that $A=CDC^{-1}$, with $D$ diagonal. Then $A-I=CDC^{-1}%
-I=C(D-I)C^{-1}$. It follows that $(A-I)^{m}$ is of the form
\[
(A-I)^{m}=C\left(
\begin{array}
[c]{ccc}%
(z_{1}-1)^{m} &  & 0\\
& \ddots & \\
0 &  & (z_{n}-1)^{m}%
\end{array}
\right)  C^{-1}\text{,}%
\]
where $z_{1},\cdots,z_{n}$ are the eigenvalues of $A$.

Now, if $\left\|  A-I\right\|  <1$, then certainly $\left|  z_{i}-1\right|
<1$ for $i=1,\cdots,n$. (Think about it.) Thus
\[
\sum_{m=1}^{\infty}(-1)^{m+1}\frac{(A-I)^{m}}{m}=C\left(
\begin{array}
[c]{ccc}%
\log z_{1} &  & 0\\
& \ddots & \\
0 &  & \log z_{n}%
\end{array}
\right)  C^{-1}%
\]
and so by the Lemma
\[
e^{\log A}=C\left(
\begin{array}
[c]{ccc}%
e^{\log z_{1}} &  & 0\\
& \ddots & \\
0 &  & e^{\log z_{n}}%
\end{array}
\right)  C^{-1}=A\text{.}%
\]

\textit{Case 2: }$A$\textit{\ is not diagonalizable.}

If $A$ is not diagonalizable, then, using the Jordan canonical form, it is not
difficult to construct a sequence $A_{m}$ of diagonalizable matrices with
$A_{m}\rightarrow A$. (See Exercise \ref{diagonal.limit}.) If $\left\|
A-I\right\|  <1$, then $\left\|  A_{m}-I\right\|  <1$ for all sufficiently
large $m$. By Case 1, $\exp(\log A_{m})=A_{m}$, and so by the continuity of
$\exp$ and $\log$, $\exp(\log A)=A$.

Thus we have shown that $\exp(\log A)=A$ for all $A$ with $\left\|
A-I\right\|  <1$. Now, the same argument as in the complex case shows that if
$\left\|  X\right\|  <\log2$, then $\left\|  e^{X}-I\right\|  <1$. But then
the same two-case argument as above shows that $\log(\exp X)=X$ for all such
$X$.
\end{proof}

\begin{proposition}
\label{log.estimate}There exists a constant $c$ such that for all $n\times n$
matrices $B$ with $\left\|  B\right\|  <\tfrac{1}{2}$%
\[
\left\|  \log(I+B)-B\right\|  \leq c\left\|  B\right\|  ^{2}\text{.}%
\]
\end{proposition}

\begin{proof}
Note that
\[
\log(I+B)-B=\sum_{m=2}^{\infty}(-1)^{m}\frac{B_{{}}^{m}}{m}=B^{2}\sum
_{m=2}^{\infty}(-1)^{m}\frac{B_{{}}^{m-2}}{m}%
\]
so that
\[
\left\|  \log(I+B)-B\right\|  \leq\left\|  B\right\|  ^{2}\sum_{m=2}^{\infty
}\frac{\left(  \tfrac{1}{2}\right)  ^{m}}{m}\text{.}%
\]
This is what we want.
\end{proof}

\begin{proposition}
\label{limit}Let $X$ be any $n\times n$ complex matrix, and let $C_{m}$ be a
sequence of matrices such that $\left\|  C_{m}\right\|  \leq\tfrac
{\mathrm{const.}}{m^{2}}$. Then
\[
\lim_{m\rightarrow\infty}\left[  I+\frac{X}{m}+C_{m}\right]  ^{m}%
=e^{X}\text{.}%
\]
\end{proposition}

\begin{proof}
The expression inside the brackets is clearly tending to $I$ as $m\rightarrow
\infty$, and so is in the domain of the logarithm for all sufficiently large
$m$. Now
\[
\log\left(  I+\frac{X}{m}+C_{m}\right)  =\frac{X}{m}+C_{m}+E_{m}%
\]
where $E_{m}$ is an error term which, by Proposition \ref{log.estimate}
satisfies $\left\|  E_{m}\right\|  \leq c\left\|  \frac{X}{m}+C_{m}\right\|
^{2}\leq\frac{\mathrm{const.}}{m^{2}}$. But then
\[
I+\frac{X}{m}+C_{m}=\exp\left(  \frac{X}{m}+C_{m}+E_{m}\right)  \text{,}%
\]
and so
\[
\left[  I+\frac{X}{m}+C_{m}\right]  ^{m}=\exp\left(  X+mC_{m}+mE_{m}\right)
\text{.}%
\]
Since both $C_{m}$ and $E_{m}$ are of order $\tfrac{1}{m^{2}}$, we obtain the
desired result by letting $m\rightarrow\infty$ and using the continuity of the exponential.
\end{proof}

\section{Further Properties of the Matrix Exponential\label{further}}

In this section we give three additional results involving the exponential of
a matrix, which will be important in our study of Lie algebras.

\begin{theorem}
[Lie Product Formula]Let $X$ and $Y$ be $n\times n$ complex matrices. Then
\[
e^{X+Y}=\lim_{m\rightarrow\infty}\left(  e^{\frac{X}{m}}e^{\frac{Y}{m}%
}\right)  ^{m}\text{.}%
\]
\end{theorem}

This theorem has a big brother, called the Trotter product formula, which
gives the same result in the case where $X$ and $Y$ are suitable unbounded
operators on an infinite-dimensional Hilbert space. The Trotter formula is
described, for example, in M. Reed and B. Simon, \textit{Methods of Modern
Mathematical Physics}, Vol. I, VIII.8.

\begin{proof}
Using the power series for the exponential and multiplying, we get
\[
e^{\frac{X}{m}}e^{\frac{Y}{m}}=I+\frac{X}{m}+\frac{Y}{m}+C_{m}\text{,}%
\]
where (check!) $\left\|  C_{m}\right\|  \leq\frac{const.}{m^{2}}$. Since
$e^{\frac{X}{m}}e^{\frac{Y}{m}}\rightarrow I$ as $m\rightarrow\infty$,
$e^{\frac{X}{m}}e^{\frac{Y}{m}}$ is in the domain of the logarithm for all
sufficiently large $m$. But
\begin{align*}
\log\left(  e^{\frac{X}{m}}e^{\frac{Y}{m}}\right)    & =\log\left(  I+\frac
{X}{m}+\frac{Y}{m}+C_{m}\right)  \\
& =\frac{X}{m}+\frac{Y}{m}+C_{m}+E_{m}%
\end{align*}
where by Proposition \ref{log.estimate} $\left\|  C_{m}\right\|  \leq
const.\left\|  \frac{X}{m}+\frac{Y}{m}+C_{m}\right\|  ^{2}\leq\frac
{const.}{m^{2}}$. Exponentiating the logarithm gives
\[
e^{\frac{X}{m}}e^{\frac{Y}{m}}=\exp\left(  \frac{X}{m}+\frac{Y}{m}+C_{m}%
+E_{m}\right)
\]
and
\[
\left(  e^{\frac{X}{m}}e^{\frac{Y}{m}}\right)  ^{m}=\exp\left(  X+Y+mC_{m}%
+mE_{m}\right)  \text{.}%
\]
Since both $C_{m}$ and $E_{m}$ are of order $\frac{1}{m^{2}}$, we have (using
the continuity of the exponential)
\[
\lim_{m\rightarrow\infty}\left(  e^{\frac{X}{m}}e^{\frac{Y}{m}}\right)
^{m}=\exp\left(  X+Y\right)
\]
which is the Lie product formula.
\end{proof}

\begin{theorem}
\label{exp.trace}Let $X$ be an $n\times n$ real or complex matrix. Then
\[
\det\left(  e^{X}\right)  =e^{\mathrm{trace}(X)}\text{.}%
\]
\end{theorem}

\begin{proof}
There are three cases, as in Section \ref{computing}.

\textit{Case 1: }$A$ \textit{is diagonalizable}. Suppose there is a complex
invertible matrix $C$ such that
\[
X=C\left(
\begin{array}
[c]{ccc}%
\lambda_{1} &  & 0\\
& \ddots & \\
0 &  & \lambda_{n}%
\end{array}
\right)  C^{-1}\text{.}%
\]
Then
\[
e^{X}=C\left(
\begin{array}
[c]{ccc}%
e^{\lambda_{1}} &  & 0\\
& \ddots & \\
0 &  & e^{\lambda_{n}}%
\end{array}
\right)  C^{-1}\text{.}%
\]
Thus $\mathrm{trace}(X)=\sum\lambda_{i}$, and $\det(e^{X})=\prod
e^{\lambda_{i}}=e^{\sum\lambda_{i}}$. (Recall that $\mathrm{trace}%
(CDC^{-1})=\mathrm{trace}(D)$.)

\textit{Case 2: }$X$\textit{\ is nilpotent}. If $X$ is nilpotent, then it
cannot have any non-zero eigenvalues (check!), and so all the roots of the
characteristic polynomial must be zero. Thus the Jordan canonical form of $X$
will be strictly upper triangular. That is, $X$ can be written as
\[
X=C\left(
\begin{array}
[c]{ccc}%
0 &  & \ast\\
& \ddots & \\
0 &  & 0
\end{array}
\right)  C^{-1}\text{.}%
\]
In that case (it is easy to see) $e^{X}$ will be upper triangular, with ones
on the diagonal:
\[
e^{X}=C\left(
\begin{array}
[c]{ccc}%
1 &  & \ast\\
& \ddots & \\
0 &  & 1
\end{array}
\right)  C^{-1}\text{.}%
\]
Thus if $X$ is nilpotent, $\mathrm{trace}(X)=0$, and $\det(e^{X})=1$.

\textit{Case 3: }$X$\textit{\ arbitrary}. As pointed out in Section
\ref{computing}, every matrix $X$ can be written as the sum of two commuting
matrices $S$ and $N$, with $S$ diagonalizable (over $\mathbb{C}$) and $N$
nilpotent. Since $S$ and $N$ commute, $e^{X}=e^{S}e^{N}$. So by the two
previous cases
\[
\det\left(  e^{X}\right)  =\det\left(  e^{S}\right)  \det\left(  e^{N}\right)
=e^{\mathrm{trace}(S)}e^{\mathrm{trace}(N)}=e^{\mathrm{trace}(X)}\text{,}%
\]
which is what we want.
\end{proof}

\begin{definition}
A function $A:\mathbb{R}\rightarrow\mathsf{GL}(n;\mathbb{C})$ is called a
\textbf{one-parameter group} if

\begin{enumerate}
\item $A$ is continuous,

\item $A(0)=I$,

\item $A(t+s)=A(t)A(s)$ for all $t,s\in\mathbb{R}$.
\end{enumerate}
\end{definition}

\begin{theorem}
[One-parameter Subgroups]\label{one.parameter}If $A$ is a one-parameter group
in $\mathsf{GL}(n;\mathbb{C})$, then there exists a unique $n\times n$ complex
matrix $X$ such that
\[
A(t)=e^{tX}\text{.}%
\]
\end{theorem}

By taking $n=1$, and noting that $\mathsf{GL}(1;\mathbb{C})\cong
\mathbb{C}^{\ast}$, this Theorem provides an alternative method of solving
Exercise \ref{character} in Chapter 2.

\begin{proof}
The uniqueness is immediate, since if there is such an $X$, then $X=\left.
\frac{d}{dt}\right|  _{t=0}A(t)$. So we need only worry about existence.

The first step is to show that $A(t)$ must be smooth. This follows from
Proposition \ref{homo.smooth} in Chapter 2 (which we did not prove), but we
give a self-contained proof. Let $f(s)$ be a smooth real-valued function
supported in a small neighborhood of zero, with $f(s)\geq0$ and $\int
f(s)ds=1$. Now look at
\begin{equation}
B(t)=\int A(t+s)f(s)\,ds\text{.}\label{convolution}%
\end{equation}
Making the change-of-variable $u=t+s$ gives
\[
B(t)=\int A(u)f(u-t)\,du\text{.}%
\]
It follows that $B(t)$ is differentiable, since derivatives in the $t$
variable go onto $f$, which is smooth.

On the other hand, if we use the identity $A(t+s)=A(t)A(s)$ in
(\ref{convolution}), we have
\[
B(t)=A(t)\int A(s)f(s)\,ds\text{.}%
\]
Now, the conditions on the function $f$, together with the continuity of $A$,
guarantee that $\int A(s)f(s)\,ds$ is close to $A(0)=I$, and hence is
invertible. Thus we may write
\begin{equation}
A(t)=B(t)\left(  \int A(s)f(s)ds\right)  ^{-1}\text{.}\label{smooth}%
\end{equation}
Since $B\left(  t\right)  $ is smooth and $\int A(s)f(s)ds$ is just a constant
matrix, this shows that $A\left(  t\right)  $ is smooth.

Now that $A(t)$ is known to be differentiable, we may define
\[
X=\left.  \tfrac{d}{dt}\right|  _{t=0}A(t)\text{.}%
\]
Our goal is to show that $A(t)=e^{tX}$. Since $A(t)$ is smooth, a standard
calculus result (extended trivially to handle matrix-valued functions) says
\[
\left\|  A(t)-(I+tX)\right\|  \leq\mathrm{const.}t^{2}\text{.}%
\]
It follows that for each fixed $t$,
\[
A\left(  \tfrac{t}{m}\right)  =I+\tfrac{t}{m}X+O\left(  \tfrac{1}{m^{2}%
}\right)  \text{.}%
\]
Then, since $A$ is a one-parameter group
\[
A(t)=\left[  A\left(  \tfrac{t}{m}\right)  \right]  ^{m}=\left[  I+\tfrac
{t}{m}X+O\left(  \tfrac{1}{m^{2}}\right)  \right]  ^{m}\text{.}%
\]
Letting $m\rightarrow\infty$ and using Proposition \ref{limit} from Section
\ref{matrix.log} shows that $A(t)=e^{tX}$.
\end{proof}

\section{The Lie Algebra of a Matrix Lie Group\label{lie.algebra}}

The Lie algebra is an indispensable tool in studying matrix Lie groups. On the
one hand, Lie algebras are simpler than matrix Lie groups, because (as we will
see) the Lie algebra is a linear space. Thus we can understand much about Lie
algebras just by doing linear algebra. On the other hand, the Lie algebra of a
matrix Lie group contains much information about that group. (See for example,
Proposition \ref{local.log} in Section \ref{exponential.mapping}, and the
Baker-Campbell-Hausdorff Formula (Chapter 4).) Thus many questions about
matrix Lie groups can be answered by considering a similar but easier problem
for the Lie algebra.

\begin{definition}
Let $G$ be a matrix Lie group. Then the \textbf{Lie algebra} of $G$, denoted
$\frak{g}$, is the set of all matrices $X$ such that $e^{tX}$ is in $G$ for
all real numbers $t$.
\end{definition}

Note that even if $G$ is a subgroup of $\mathsf{GL}(n;\mathbb{C})$ we do
\textit{not} require that $e^{tX}$ be in $G$ for all complex $t$, but only for
all \textit{real} $t$. Also, it is definitely not enough to have just $e^{X}$
in $G$. That is, it is easy to give an example of an $X$ and a $G$ such that
$e^{X}\in G$ but $e^{tX}\notin G$ for some values of $t$. Such an $X$ is not
in the Lie algebra of $G$.

It is customary to use lower case Gothic (Fraktur) characters such as
$\frak{g}$ and $\frak{h}$ to refer to Lie algebras.

\subsection{Physicists' Convention}

Physicists are accustomed to considering the map $X\rightarrow e^{iX}$ instead
of $X\rightarrow e^{X}$. Thus a physicist would think of the Lie algebra of
$G$ as the set of all matrices $X$ such that $e^{itX}\in G$ for all real $t$.
In the physics literature, the Lie algebra is frequently referred to as the
space of ``infinitesimal group elements.'' See Br\"{o}cker and tom Dieck,
Chapter I, 2.21. The physics literature does not always distinguish clearly
between a matrix Lie group and its Lie algebra.

Before examining general properties of the Lie algebra, let us compute the Lie
algebras of the matrix Lie groups introduced in the previous chapter.

\subsection{The general linear groups}

If $X$ is any $n\times n$ complex matrix, then by Proposition
\ref{exp.properties}, $e^{tX}$ is invertible. Thus the Lie algebra of
$\mathsf{GL}(n;\mathbb{C})$ is the space of all $n\times n$ complex matrices.
This Lie algebra is denoted $\mathsf{gl}(n;\mathbb{C})$.

If $X$ is any $n\times n$ real matrix, then $e^{tX}$ will be invertible and
real. On the other hand, if $e^{tX}$ is real for all real $t$, then $X=\left.
\tfrac{d}{dt}\right|  _{t=0}e^{tX}$ will also be real. Thus the Lie algebra of
$\mathsf{GL}(n;\mathbb{R})$ is the space of all $n\times n$ real matrices,
denoted $\mathsf{gl}(n;\mathbb{R})$.

Note that the preceding argument shows that if $G$ is a subgroup of
$\mathsf{GL}(n;\mathbb{R})$, then the Lie algebra of $G$ must consist entirely
of real matrices. We will use this fact when appropriate in what follows.

\subsection{The special linear groups}

Recall Theorem \ref{exp.trace}: $\det\left(  e^{X}\right)  =e^{\mathrm{trace}%
X}$. Thus if $\mathrm{trace}X=0$, then $\det\left(  e^{tX}\right)  =1$ for all
real $t$. On the other hand, if $X$ is any $n\times n$ matrix such that
$\det\left(  e^{tX}\right)  =1$ for all $t$, then $e^{(t)(\mathrm{trace}X)}=1$
for all $t$. This means that $(t)(\mathrm{trace}X)$ is an integer multiple of
$2\pi i$ for all $t$, which is only possible if $\mathrm{trace}X=0$. Thus the
Lie algebra of $\mathsf{SL}\left(  n;\mathbb{C}\right)  $ is the space of all
$n\times n$ complex matrices with trace zero, denoted $\mathsf{sl}%
(n;\mathbb{C}) $.

Similarly, the Lie algebra of $\mathsf{SL}\left(  n;\mathbb{R}\right)  $ is
the space of all $n\times n$ real matrices with trace zero, denoted
$\mathsf{sl}\left(  n;\mathbb{R}\right)  $.

\subsection{The unitary groups}

Recall that a matrix $U$ is unitary if and only if $U^{*}=U^{-1}$. Thus
$e^{tX}$ is unitary if and only if
\begin{equation}
\left(  e^{tX}\right)  ^{*}=\left(  e^{tX}\right)  ^{-1}=e^{-tX}%
\text{.}\label{unitary.condition}%
\end{equation}
But by taking adjoints term-by-term, we see that $\left(  e^{tX}\right)
^{*}=e^{tX^{*}}$, and so (\ref{unitary.condition}) becomes
\begin{equation}
e^{tX^{*}}=e^{-tX}\text{.}\label{skew}%
\end{equation}

Clearly, a sufficient condition for (\ref{skew}) to hold is that $X^{*}=-X$.
On the other hand, if (\ref{skew}) holds for all $t$, then by differentiating
at $t=0$, we see that $X^{*}=-X$ is necessary.

Thus the Lie algebra of $\mathsf{U}(n)$ is the space of all $n\times n$
complex matrices $X$ such that $X^{*}=-X$, denoted $\mathsf{u}(n)$.

By combining the two previous computations, we see that the Lie algebra of
$\mathsf{SU}(n)$ is the space of all $n\times n$ complex matrices $X$ such
that $X^{*}=-X$ and $\mathrm{trace}X=0$, denoted $\mathsf{su}(n)$.

\subsection{The orthogonal groups}

The identity component of $\mathsf{O}(n)$ is just $\mathsf{SO}(n)$. Since
(Proposition \ref{identity}) the exponential of a matrix in the Lie algebra is
automatically in the identity component, the Lie algebra of $\mathsf{O}(n) $
is the same as the Lie algebra of $\mathsf{SO}(n)$.

Now, an $n\times n$ real matrix $R$ is orthogonal if and only if
$R^{tr}=R^{-1}$. So, given an $n\times n$ real matrix $X$, $e^{tX}$ is
orthogonal if and only if $(e^{tX})^{tr}=(e^{tX})^{-1}$, or
\begin{equation}
e^{tX^{tr}}=e^{-tX}\text{.}\label{real.skew}%
\end{equation}
Clearly, a sufficient condition for this to hold is that $X^{tr}=-X$. If
(\ref{real.skew}) holds for all $t$, then by differentiating at $t=0$, we must
have $X^{tr}=-X$.

Thus the Lie algebra of $\mathsf{O}(n)$, as well as the Lie algebra of
$\mathsf{SO}(n)$, is the space of all $n\times n$ real matrices $X$ with
$X^{tr}=-X$, denoted $\mathsf{so}(n)$. Note that the condition $X^{tr}=-X$
forces the diagonal entries of $X$ to be zero, and so explicitly the trace of
$X$ is zero.

The same argument shows that the Lie algebra of $\mathsf{SO}(n;\mathbb{C})$ is
the space of $n\times n$ complex matrices satisfying $X^{tr}=-X$, denoted
$\mathsf{so}(n;\mathbb{C})$. This is not the same as $\mathsf{su}(n)$.

\subsection{The generalized orthogonal groups}

A matrix $A$ is in $\mathsf{O}(n;k)$ if and only if $A^{tr}gA=g$, where $g$ is
the $(n+k)\times(n+k)$ diagonal matrix with the first $n$ diagonal entries
equal to one, and the last $k$ diagonal entries equal to minus one. This
condition is equivalent to the condition $g^{-1}A^{tr}g=A^{-1}$, or, since
explicitly $g^{-1}=g$, $gA^{tr}g=A^{-1}$. Now, if $X$ is an $(n+k)\times(n+k)$
real matrix, then $e^{tX}$ is in $\mathsf{O}(n;k)$ if and only if
\[
ge^{tX^{tr}}g=e^{tgX^{tr}g}=e^{-tX}\text{.}%
\]
This condition holds for all real $t$ if and only if $gX^{tr}g=-X$. Thus the
Lie algebra of $\mathsf{O}(n;k)$, which is the same as the Lie algebra of
$\mathsf{SO}(n;k)$, consists of all $(n+k)\times(n+k)$ real matrices $X$ with
$gX^{tr}g=-X$. This Lie algebra is denoted $\mathsf{so}(n;k)$.

(In general, the group $\mathsf{SO}(n;k)$ will not be connected, in contrast
to the group $\mathsf{SO}(n)$. The identity component of $\mathsf{SO}(n;k)$,
which is also the identity component of $\mathsf{O}(n;k)$, is denoted
$\mathsf{SO}(n;k)_{I}$. The Lie algebra of $\mathsf{SO}(n;k)_{I}$ is the same
as the Lie algebra of $\mathsf{SO}(n;k)$.)

\subsection{The symplectic groups}

These are denoted $\mathsf{sp}\left(  n;\mathbb{R}\right)  ,$ \textsf{sp}%
$\left(  n;\mathbb{C}\right)  ,$ and $\mathsf{sp}\left(  n\right)  .$ \ The
calculation of these Lie algebras is similar to that of the generalized
orthogonal groups, and I will just record the result here. Let $J$ be the
matrix in the definition of the symplectic groups. Then $\mathsf{sp}\left(
n;\mathbb{R}\right)  $ is the space of $2n\times2n$ real matrices $X$ such
that $JX^{tr}J=X,$ \textsf{sp}$\left(  n;\mathbb{C}\right)  $ is the space of
$2n\times2n$ complex matrices satisfying the same condition, and
$\mathsf{sp}\left(  n\right)  =$\textsf{sp}$\left(  n;\mathbb{C}\right)
\cap\mathsf{u}\left(  2n\right)  .$

\subsection{The Heisenberg group}

Recall the Heisenberg group $H$ is the group of all $3\times3$ real matrices
$A$ of the form
\begin{equation}
A=\left(
\begin{array}
[c]{ccc}%
1 & a & b\\
0 & 1 & c\\
0 & 0 & 1
\end{array}
\right)  \label{heisenberg2}%
\end{equation}
Recall also that in Section \ref{computing}, Case 2, we computed the
exponential of a matrix of the form
\begin{equation}
X=\left(
\begin{array}
[c]{ccc}%
0 & \alpha & \beta\\
0 & 0 & \gamma\\
0 & 0 & 0
\end{array}
\right)  \label{upper}%
\end{equation}
and saw that $e^{X}$ was in $H$. On the other hand, if $X$ is any matrix such
that $e^{tX}$ is of the form (\ref{heisenberg2}), then all of the entries of
$X=\left.  \tfrac{d}{dt}\right|  _{t=0}e^{tX}$ which are on or below the
diagonal must be zero, so that $X$ is of form (\ref{upper}).

Thus the Lie algebra of the Heisenberg group is the space of all $3\times3$
real matrices which are strictly upper triangular.

\subsection{The Euclidean and Poincar\'{e} groups}

Recall that the Euclidean group $\mathsf{E}(n)$ is (or can be thought of as)
the group of $(n+1)\times(n+1)$ real matrices of the form
\[
\left(
\begin{array}
[c]{cccc}%
&  &  & x_{1}^{{}}\\
& R &  & \vdots\\
&  &  &  x_{n}\\
0 & \cdots & 0 & 1
\end{array}
\right)
\]
with $R\in\mathsf{O}(n)$. Now if $X$ is an $(n+1)\times(n+1)$ real matrix such
that $e^{tX}$ is in $\mathsf{E}(n)$ for all $t$, then $X=\left.  \frac{d}%
{dt}\right|  _{t=0}e^{tX}$ must be zero along the bottom row:
\begin{equation}
X=\left(
\begin{array}
[c]{cccc}%
&  &  & y_{1}^{{}}\\
& Y &  & \vdots\\
&  &  &  y_{n}\\
0 & \cdots &  & 0
\end{array}
\right) \label{euclid.al}%
\end{equation}

Our goal, then, is to determine which matrices of the form (\ref{euclid.al})
are actually in the Lie algebra of the Euclidean group. A simple computation
shows that for $n\geq1$
\[
\left(
\begin{array}
[c]{cccc}%
&  &  & y_{1}^{{}}\\
& Y &  & \vdots\\
&  &  &  y_{n}\\
0 & \cdots &  & 0
\end{array}
\right)  ^{n}=\left(
\begin{array}
[c]{cccc}%
&  &  & \\
& Y^{n} &  & Y^{n-1}y\\
&  &  & \\
0 & \cdots &  & 0
\end{array}
\right)  ,
\]
where $y$ is the column vector with entries $y_{1},\cdots,y_{n}.$ It follows
that if $X$ is as in (\ref{euclid.al}), then $e^{tX}$ is of the form
\[
e^{tX}=\left(
\begin{array}
[c]{cccc}%
&  &  & \ast\\
&  e^{tY} &  & \vdots\\
&  &  & \ast\\
0 & \cdots & 0 & 1
\end{array}
\right)  \text{.}%
\]

Now, we have already established that $e^{tY}$ is in $\mathsf{O}(n)$ for all
$t$ if and only if $Y^{tr}=-Y$. Thus we see that the Lie algebra of
$\mathsf{E}(n)$ is the space of all $(n+1)\times(n+1)$ real matrices of the
form (\ref{euclid.al}) with $Y$ satisfying $Y^{tr}=-Y$.

A similar argument shows that the Lie algebra of $\mathsf{P}(n;1)$ is the
space of all $(n+2)\times(n+2)$ real matrices of the form
\[
\left(
\begin{array}
[c]{cccc}%
&  &  & y_{1}\\
& Y &  & \vdots\\
&  &  &  y_{n+1}\\
0 & \cdots &  & 0
\end{array}
\right)
\]
with $Y\in\mathsf{so}(n;1)$.

\section{Properties of the Lie Algebra}

We will now establish various basic properties of the Lie algebra of a matrix
Lie group. The reader is invited to verify by direct calculation that these
general properties hold for the examples computed in the previous section.

\begin{proposition}
\label{identity}Let $G$ be a matrix Lie group, and $X$ an element of its Lie
algebra. Then $e^{X}$ is an element of the identity component of $G.$
\end{proposition}

\begin{proof}
By definition of the Lie algebra, $e^{tX}$ lies in $G$ for all real $t$. But
as $t$ varies from $0$ to $1$, $e^{tX}$ is a continuous path connecting the
identity to $e^{X}$.
\end{proof}

\begin{proposition}
\label{adjoint}Let $G$ be a matrix Lie group, with Lie algebra $\frak{g}$. Let
$X$ be an element of $\frak{g}$, and $A$ an element of $G$. Then $AXA^{-1}$ is
in $\frak{g}$.
\end{proposition}

\begin{proof}
This is immediate, since by Proposition \ref{exp.properties},
\[
e^{t(AXA^{-1})}=Ae^{tX}A^{-1}\text{,}%
\]
and $Ae^{tX}A^{-1}\in G$.
\end{proof}

\begin{theorem}
\label{lie.algebra.theorem}Let $G$ be a matrix Lie group, $\frak{g}$ its Lie
algebra, and $X,Y$ elements of $\frak{g}$. Then

\begin{enumerate}
\item \label{scalars}$sX\in\frak{g}$ for all real numbers $s$,

\item \label{sums}$X+Y\in\frak{g}$,

\item \label{brackets}$XY-YX\in\frak{g}$.
\end{enumerate}
\end{theorem}

If you are following the physics convention for the definition of the Lie
algebra, then condition \ref{brackets} should be replaced with the condition
$-i\left(  XY-YX\right)  \in\frak{g}$.

\begin{proof}
Point \ref{scalars} is immediate, since $e^{t(sX)}=e^{(ts)X}$, which must be
in $G$ if $X$ is in $\frak{g}$. Point \ref{sums} is easy to verify \textit{if}
$X$ and $Y$ commute, since then $e^{t(X+Y)}=e^{tX}e^{tY}$. If $X$ and $Y$ do
not commute, this argument does not work. However, the Lie product formula
says that
\[
e^{t(X+Y)}=\lim_{m\rightarrow\infty}\left(  e^{tX/m}e^{tY/m}\right)
^{m}\text{.}%
\]
Because $X$ and $Y$ are in the Lie algebra, $e^{tX/m}$ and $e^{tY/m}$ are in
$G$, as is $\left(  e^{tX/m}e^{tY/m}\right)  ^{m}$, since $G$ is a group. But
now because $G$ is a matrix Lie group, the limit of things in $G$ must be
again in $G$, provided that the limit is invertible. Since $e^{t(X+Y)}$ is
automatically invertible, we conclude that it must be in $G$. This shows that
$X+Y$ is in $\frak{g}$.

Now for point \ref{brackets}. Recall (Proposition \ref{derivative}) that
$\left.  \frac{d}{dt}\right|  _{t=0}e^{tX}=X$. It follows that $\left.
\frac{d}{dt}\right|  _{t=0}e^{tX}Y=XY$, and hence by the product rule
(Exercise \ref{product.rule})
\begin{align*}
\left.  \frac{d}{dt}\right|  _{t=0}\left(  e^{tX}Ye^{-tX}\right)
=(XY)e^{0}+(e^{0}Y)(-X)\\
=XY-YX\text{.}%
\end{align*}
But now, by Proposition \ref{adjoint}, $e^{tX}Ye^{-tX}$ is in $\frak{g}$ for
all $t$. Since we have (by points \ref{scalars} and \ref{sums}) established
that $\frak{g}$ is a real vector space, it follows that the derivative of any
smooth curve lying in $\frak{g}$ must be again in $\frak{g}$. Thus $XY-YX$ is
in $\frak{g}$.
\end{proof}

\begin{definition}
Given two $n\times n$ matrices $A$ and $B$, the \textbf{bracket} (or
\textbf{commutator}) of $A$ and $B$ is defined to be simply
\[
\left[  A,B\right]  =AB-BA\text{.}%
\]
\end{definition}

According to Theorem \ref{lie.algebra.theorem}, the Lie algebra of any matrix
Lie group is closed under brackets.

The following very important theorem tells us that a Lie group homomorphism
between two Lie groups gives rise in a natural way to a map between the
corresponding Lie algebras. In particular, this will tell us that two
isomorphic Lie groups have ``the same'' Lie algebras. (That is, the Lie
algebras are isomorphic in the sense of Section \ref{algebras}.) See Exercise
\ref{isomorphism}.

\begin{theorem}
\label{homo.theorem}Let $G$ and $H$ be matrix Lie groups, with Lie algebras
$\frak{g}$ and $\frak{h}$, respectively. Suppose that $\phi:G\rightarrow H$ be
a Lie group homomorphism. Then there exists a unique real linear map
$\widetilde{\phi}:\frak{g}\rightarrow\frak{h}$ such that
\[
\phi(e^{X})=e^{\widetilde{\phi}(X)}%
\]
for all $X\in\frak{g}$. The map $\widetilde{\phi}$ has following additional properties

\begin{enumerate}
\item $\widetilde{\phi}\left(  AXA^{-1}\right)  =\phi(A)\widetilde{\phi
}(X)\phi(A)^{-1}$, for all $X\in\frak{g}$, $A\in G$.

\item \label{homomorphism}$\widetilde{\phi}(\left[  X,Y\right]  )=\left[
\widetilde{\phi}(X),\widetilde{\phi}(Y)\right]  $, for all $X,Y\in\frak{g}$.

\item \label{differentiate}$\widetilde{\phi}(X)=\left.  \frac{d}{dt}\right|
_{t=0}\phi(e^{tX})$, for all $X\in\frak{g}$.
\end{enumerate}

If $G$, $H$, and $K$ are matrix Lie groups and $\phi:H\rightarrow K$ and
$\psi:G\rightarrow H$ are Lie group homomorphisms, then
\[
\widetilde{\phi\circ\psi}=\widetilde{\phi}\circ\widetilde{\psi}\text{.}%
\]
\end{theorem}

In practice, given a Lie group homomorphism $\phi$, the way one goes about
computing $\widetilde{\phi}$ is by using Property \ref{differentiate}. Of
course, since $\widetilde{\phi}$ is (real) linear, it suffices to compute
$\widetilde{\phi}$ on a basis for $\frak{g}$. In the language of
differentiable manifolds, Property \ref{differentiate} says that
$\widetilde{\phi}$ is the derivative (or differential) of $\phi$ at the
identity, which is the standard definition of $\widetilde{\phi}$. (See also
Exercise \ref{tangent.space}.)

A linear map with property (\ref{homomorphism}) is called a \textbf{Lie
algebra homomorphism}. (See Section \ref{algebras}.) This theorem says that
every Lie group homomorphism gives rise to a Lie algebra homomorphism. We will
see eventually that the converse is true \textit{under certain circumstances}.
Specifically, suppose that $G$ and $H$ are Lie groups, and $\widetilde{\phi
}:\frak{g}\rightarrow\frak{h}$ is a Lie algebra homomorphism. If $G$ is
\textit{connected and simply connected}, then there exists a unique Lie group
homomorphism $\phi:G\rightarrow H$ such that $\phi$ and $\widetilde{\phi}$ are
related as in Theorem \ref{homo.theorem}.

\begin{proof}
The proof is similar to the proof of Theorem \ref{lie.algebra.theorem}. Since
$\phi$ is a continuous group homomorphism, $\phi(e^{tX})$ will be a
one-parameter subgroup of $H$, for each $X\in\frak{g}$. Thus by Theorem
\ref{one.parameter}, there is a unique $Z$ such that
\begin{equation}
\phi\left(  e^{tX}\right)  =e^{tZ}\label{phi.etx}%
\end{equation}
for all $t\in\mathbb{R}$. This $Z$ must lie in $\frak{h}$ since $e^{tZ}%
=\phi\left(  e^{tX}\right)  \in H$.

We now define $\widetilde{\phi}(X)=Z$, and check in several steps that
$\widetilde{\phi}$ has the required properties.

\ 

\textit{Step 1}: $\phi(e^{X})=e^{\widetilde{\phi}(X)}$.

This follows from (\ref{phi.etx}) and our definition of $\widetilde{\phi}$, by
putting $t=1$.

\ 

\textit{Step 2}: $\widetilde{\phi}(sX)=s\widetilde{\phi}(X)$\textit{\ for all
}$s\in\mathbb{R}$.

This is immediate, since if $\phi(e^{tX})=e^{tZ}$, then $\phi(e^{tsX}%
)=e^{tsZ}$.

\ 

\textit{Step 3}: $\widetilde{\phi}(X+Y)=\widetilde{\phi}(X)+\widetilde{\phi
}(Y)$.

By Steps 1 and 2,
\[
e^{t\widetilde{\phi}(X+Y)}=e^{\widetilde{\phi}[t(X+Y)]}=\phi\left(
e^{t(X+Y)}\right)  \text{.}%
\]
By the Lie product formula, and the fact that $\phi$ is a continuous
homomorphism:
\begin{align*}
=\phi\left(  \lim_{m\rightarrow\infty}\left(  e^{tX/m}e^{tY/m}\right)
^{m}\right)  \\
=\lim_{m\rightarrow\infty}\left(  \phi\left(  e^{tX/m}\right)  \phi
(e^{tY/m})\right)  ^{m}\text{.}%
\end{align*}
But then we have
\[
e^{t\widetilde{\phi}(X+Y)}=\lim_{m\rightarrow\infty}\left(  e^{t\widetilde
{\phi}(X)/m}e^{t\widetilde{\phi}(Y)/m}\right)  ^{m}=e^{t\left(  \widetilde
{\phi}(X)+\widetilde{\phi}(Y)\right)  }\text{.}%
\]
Differentiating this result at $t=0$ gives the desired result.

\ \ 

\textit{Step 4}: $\widetilde{\phi}\left(  AXA^{-1}\right)  =\phi
(A)\widetilde{\phi}(X)\phi(A)^{-1}$.

By Steps 1 and 2,
\[
\exp t\widetilde{\phi}(AXA^{-1})=\exp\widetilde{\phi}(tAXA^{-1})=\phi\left(
\exp tAXA^{-1}\right)  \text{.}%
\]
Using a property of the exponential and Step 1, this becomes
\begin{align*}
\exp t\widetilde{\phi}(AXA^{-1})=\phi\left(  Ae^{tX}A^{-1}\right)
=\phi(A)\phi(e^{tX})\phi(A)^{-1}\\
=\phi(A)e^{t\widetilde{\phi}(X)}\phi(A)^{-1}\text{.}%
\end{align*}
Differentiating this at $t=0$ gives the desired result.

\ \ 

\textit{Step 5}: $\widetilde{\phi}(\left[  X,Y\right]  )=\left[
\widetilde{\phi}(X),\widetilde{\phi}(Y)\right]  $.

Recall from the proof of Theorem \ref{lie.algebra.theorem} that
\[
\left[  X,Y\right]  =\left.  \tfrac{d}{dt}\right|  _{t=0}e^{tX}Ye^{-tX}%
\text{.}%
\]
Hence
\[
\widetilde{\phi}\left(  \left[  X,Y\right]  \right)  =\widetilde{\phi}\left(
\left.  \tfrac{d}{dt}\right|  _{t=0}e^{tX}Ye^{-tX}\right)  =\left.  \tfrac
{d}{dt}\right|  _{t=0}\widetilde{\phi}\left(  e^{tX}Ye^{-tX}\right)
\]
where we have used the fact that a derivative commutes with a linear transformation.

But then by Step 4,
\begin{align*}
\widetilde{\phi}\left(  \left[  X,Y\right]  \right)    & =\left.  \tfrac
{d}{dt}\right|  _{t=0}\phi(e^{tX})\widetilde{\phi}(Y)\phi(e^{-tX})\\
& =\left.  \tfrac{d}{dt}\right|  _{t=0}e^{t\widetilde{\phi}(X)}\widetilde
{\phi}(Y)e^{-t\widetilde{\phi}(X)}\\
& =\left[  \widetilde{\phi}(X),\widetilde{\phi}(Y)\right]  \text{.}%
\end{align*}

\ 

\textit{Step 6}: $\widetilde{\phi}(X)=\left.  \tfrac{d}{dt}\right|  _{t=0}%
\phi(e^{tX})$.

This follows from (\ref{phi.etx}) and our definition of $\widetilde{\phi}$.

\ 

\textit{Step 7}: $\widetilde{\phi}$\textit{\ is the unique real-linear map
such that }$\phi(e^{X})=e^{\widetilde{\phi}(X)}$.

Suppose that $\psi$ is another such map. Then
\[
e^{t\psi(X)}=e^{\psi(tX)}=\phi(e^{tX})
\]
so that
\[
\psi(X)=\left.  \tfrac{d}{dt}\right|  _{t=0}\phi(e^{tX})\text{.}%
\]
Thus by Step 6, $\psi$ coincides with $\widetilde{\phi}$.

\ 

\textit{Step 8: }$\widetilde{\phi\circ\psi}=\widetilde{\phi}\circ
\widetilde{\psi}$.

For any $X\in\frak{g}$,
\[
\phi\circ\psi\left(  e^{tX}\right)  =\phi\left(  \psi\left(  e^{tX}\right)
\right)  =\phi\left(  e^{t\widetilde{\psi}(X)}\right)  =e^{t\widetilde{\phi
}(\widetilde{\psi}(X))}\text{.}%
\]
Thus $\widetilde{\phi\circ\psi}(X)=\widetilde{\phi}\circ\widetilde{\psi}(X)$.
\end{proof}

\begin{definition}
[The Adjoint Mapping]Let $G$ be a matrix Lie group, with Lie algebra
$\frak{g}$. Then for each $A\in G$, define a linear map $\mathrm{Ad}%
A:\frak{g}\rightarrow\frak{g}$ by the formula
\[
\mathrm{Ad}A(X)=AXA^{-1}\text{.}%
\]
We will let $\mathrm{Ad}$ denote the map $A\rightarrow\mathrm{Ad}A$.
\end{definition}

\begin{proposition}
Let $G$ be a matrix Lie group, with Lie algebra $\frak{g}$. Then for each
$A\in G$, $\mathrm{Ad}A$ is an invertible linear transformation of $\frak{g}$
with inverse $\mathrm{Ad}A^{-1}$, and $\mathrm{Ad}:G\rightarrow\mathsf{GL}%
(\frak{g})$ is a group homomorphism.
\end{proposition}

\begin{proof}
Easy. Note that Proposition \ref{adjoint} guarantees that $\mathrm{Ad}A(X)$ is
actually in $\frak{g}$ for all $X\in\frak{g}$.
\end{proof}

Since $\frak{g}$ is a real vector space with some dimension $k$,
$\mathsf{GL}(\frak{g})$ is essentially the same as $\mathsf{GL}(k;\mathbb{R}%
)$. Thus we will regard $\mathsf{GL}(\frak{g})$ as a matrix Lie group. It is
easy to show that $\mathrm{Ad}:G\rightarrow\mathsf{GL}(\frak{g})$ is
continuous, and so is a Lie group homomorphism. By Theorem \ref{homo.theorem},
there is an associated real linear map $\widetilde{\mathrm{Ad}}$ from the Lie
algebra of $G$ to the Lie algebra of $\mathsf{GL}(\frak{g})$, i.e., from
$\frak{g}$ to \textsf{gl}$(\frak{g})$, with the property that
\[
e^{\widetilde{\mathrm{Ad}}X}=\mathrm{Ad}\left(  e^{X}\right)  \text{.}%
\]

\begin{proposition}
\label{differentiate.Ad}Let $G$ be a matrix Lie group, let $\frak{g}$ its Lie
algebra, and let $\mathrm{Ad}:G\rightarrow\mathsf{GL}(\frak{g})$ be the Lie
group homomorphism defined above. Let $\widetilde{\mathrm{Ad}}:\frak{g}%
\rightarrow$\textsf{gl}$(\frak{g})$ be the associated Lie algebra map. Then
for all $X,Y\in\frak{g}$%
\[
\widetilde{\mathrm{Ad}}X(Y)=[X,Y]\text{.}%
\]
\end{proposition}

\begin{proof}
Recall that by Theorem \ref{homo.theorem}, $\widetilde{\mathrm{Ad}}$ can be
computed as follows:
\[
\widetilde{\mathrm{Ad}}X=\left.  \tfrac{d}{dt}\right|  _{t=0}\mathrm{Ad}%
(e^{tX})\text{.}%
\]
Thus
\begin{align*}
\widetilde{\mathrm{Ad}}X(Y)  & =\left.  \tfrac{d}{dt}\right|  _{t=0}%
\mathrm{Ad}(e^{tX})(Y)=\left.  \tfrac{d}{dt}\right|  _{t=0}e^{tX}Ye^{-tX}\\
& =[X,Y]
\end{align*}
which is what we wanted to prove. See also Exercise \ref{ad.expand}.
\end{proof}

\section{The Exponential Mapping\label{exponential.mapping}}

\begin{definition}
If $G$ is a matrix Lie group with Lie algebra $\frak{g}$, then the
\textbf{exponential mapping} for $G$ is the map
\[
\exp:\frak{g}\rightarrow G\text{.}%
\]
\end{definition}

In general the exponential mapping is neither one-to-one nor onto.
Nevertheless, it provides an crucial mechanism for passing information between
the group and the Lie algebra. The following result says that the exponential
mapping is\textit{\ locally} one-to-one and onto, a result that will be
essential later.

\begin{theorem}
\label{local.log}Let $G$ be a matrix Lie group with Lie algebra $\frak{g}$.
Then there exist a neighborhood $U$ of zero in $\frak{g}$ and a neighborhood
$V$ of $I$ in $G$ such that the exponential mapping takes $U$ homeomorphically
onto $V$.
\end{theorem}

\begin{proof}
We follow the proof of Theorem I.3.11 in Br\"{o}cker and tom Dieck. In view of
what we have proved about the matrix logarithm, we know this result for the
case of $\mathsf{GL}(n;\mathbb{C})$. To prove the general case, we consider a
matrix Lie group $G<\mathsf{GL}(n;\mathbb{C})$, with Lie algebra $\frak{g}$.

\begin{lemma}
Suppose $g_{n}$ are elements of $G$, and that $g_{n}\rightarrow I$. Let
$Y_{n}=\log g_{n}$, which is defined for all sufficiently large $n$. Suppose
$Y_{n}/\left\|  Y_{n}\right\|  \rightarrow Y\in\mathsf{gl}\left(
n;\mathbb{C}\right)  $. Then $Y\in\frak{g}$.
\end{lemma}

\begin{proof}
To show that $Y\in\frak{g}$, we must show that $\exp tY\in G$ for all
$t\in\mathbb{R}$. As $n\rightarrow\infty$, $\left(  t/\left\|  Y_{n}\right\|
\right)  Y_{n}\rightarrow tY$. Note that since $g_{n}\rightarrow I$,
$Y_{n}\rightarrow0$, and so $\left\|  Y_{n}\right\|  \rightarrow0$. Thus we
can find integers $m_{n}$ such that $\left(  m_{n}\left\|  Y_{n}\right\|
\right)  \rightarrow t$. Then $\exp\left(  m_{n}Y_{n}\right)  =\exp\left[
\left(  m_{n}\left\|  Y_{n}\right\|  \right)  \left(  Y_{n}/\left\|
Y_{n}\right\|  \right)  \right]  \rightarrow\exp\left(  tY\right)  $. But
$\exp\left(  m_{n}Y_{n}\right)  =\exp\left(  Y_{n}\right)  ^{m_{n}}=\left(
g_{n}\right)  ^{m_{n}}\in G$, and $G$ is closed, so $\exp\left(  tY\right)
\in G$.
\end{proof}

We think of $\mathsf{gl}\left(  n;\mathbb{C}\right)  $ as $\mathbb{C}^{n^{2}%
}\cong\mathbb{R}^{2n^{2}}$. Then $\frak{g}$ is a subspace of $\mathbb{R}%
^{2n^{2}}$. Let $D$ denote the orthogonal complement of $\frak{g}$ with
respect to the usual inner product on $\mathbb{R}^{2n^{2}}$. Consider the map
$\Phi:\frak{g}\oplus D\rightarrow\mathsf{GL}(n;\mathbb{C})$ given by
\[
\Phi\left(  X,Y\right)  =e^{X}e^{Y}\text{.}%
\]
Of course, we can identify $\frak{g}\oplus D$ with $\mathbb{R}^{2n^{2}}$.
Moreover, $\mathsf{GL}(n;\mathbb{C})$ is an open subset of $\mathsf{gl}\left(
n;\mathbb{C}\right)  \cong\mathbb{R}^{2n^{2}}$. Thus we can regard $\Phi$ as a
map from $\mathbb{R}^{2n^{2}}$ to itself.

Now, using the properties of the matrix exponential, we see that
\begin{align*}
\left.  \frac{d}{dt}\right|  _{t=0}\Phi\left(  tX,0\right)    & =X\\
\left.  \frac{d}{dt}\right|  _{t=0}\Phi\left(  0,tY\right)    & =Y\text{.}%
\end{align*}
This shows that the derivative of $\Phi$ at the point $0\in\mathbb{R}^{2n^{2}%
}$ is the identity. (Recall that the derivative at a point of a function from
$\mathbb{R}^{2n^{2}}$ to itself is a linear map of $\mathbb{R}^{2n^{2}}$ to
itself, in this case the identity map.) In particular, the derivative of
$\Phi$ at 0 is invertible. Thus the inverse function theorem says that $\Phi$
has a continuous local inverse, defined in a neighborhood of $I$.

Now let $U$ be any neighborhood of zero in $\frak{g}$. I want to show that
$\exp\left(  U\right)  $ contains a neighborhood of $I$ in $G$. Suppose not.
Then we can find a sequence $g_{n}\in G$ with $g_{n}\rightarrow I$ such that
no $g_{n}$ is in $\exp\left(  U\right)  $. Since $\Phi$ is locally invertible,
we can write $g_{n}$ (for large $n$) uniquely as $g_{n}=\exp\left(
X_{n}\right)  \exp\left(  Y_{n}\right)  $, with $X_{n}\in\frak{g}$ and
$Y_{n}\in D$. Since $g_{n}\rightarrow I$ and $\Phi^{-1}$ is continuous,
$X_{n}$ and $Y_{n}$ tend to zero. Thus (for large $n$), $X_{n}\in U$. So we
must have (for large $n$) $Y_{n}\neq0$, otherwise $g_{n}$ would be in
$\exp\left(  U\right)  $.

Let $\widetilde{g}_{n}=\exp\left(  Y_{n}\right)  =\exp\left(  -X_{n}\right)
g_{n}$. Note that $\widetilde{g}_{n}\in G$ and $\widetilde{g}_{n}\rightarrow
I$. Since the unit ball in $D$ is compact, we can choose a subsequence of
$\left\{  Y_{n}\right\}  $ (still called $\left\{  Y_{n}\right\}  $) so that
$Y_{n}/\left\|  Y_{n}\right\|  $ converges to some $Y\in D$, with $\left\|
Y\right\|  =1$. But then by the Lemma, $Y\in\frak{g}$! This is a
contradiction, because $D$ is the orthogonal complement of $\frak{g}$.

So for every neighborhood $U$ of zero in $\frak{g}$, $\exp\left(  U\right)  $
contains a neighborhood of the identity in $G$. If we make $U$ small enough,
then the exponential will be one-to-one on $\overline{U}$. (The existence of
the matrix logarithm implies that the exponential is one-to-one near zero.)
Let $\log$ denote the inverse map, defined on $\exp\left(  \overline
{U}\right)  $. Since $\overline{U}$ is compact, and $\exp$ is one-to-one and
continuous on $\overline{U}$, log will be continuous. (This is a standard
topological result.) So take $V$ to be a neighborhood of $I$ contained in
$\exp\left(  \overline{U}\right)  $, and let $U^{\prime}=\exp^{-1}\left(
V\right)  \cap U$. Then $U^{\prime}$ is open and the exponential takes
$U^{\prime}$ homeomorphically onto $V$.
\end{proof}

\begin{definition}
If $U$ and $V$ are as in Proposition \ref{local.log}, then the inverse map
$\exp^{-1}:V\rightarrow\frak{g}$ is called the logarithm for $G$.
\end{definition}

\begin{corollary}
\label{ex1.ex2}If $G$ is a connected matrix Lie group, then every element $A$
of $G$ can be written in the form
\begin{equation}
A=e^{X_{1}}e^{X_{2}}\cdots e^{X_{n}}\label{exp.n}%
\end{equation}
for some $X_{1},X_{2},\cdots X_{n}$ in $\frak{g}$.
\end{corollary}

\begin{proof}
Recall that for us, saying $G$ is connected means that $G$ is path-connected.
This certainly means that $G$ is connected in the usual topological sense,
namely, the only non-empty subset of $G$ that is both open and closed is $G$
itself. So let $E$ denote the set of all $A\in G$ that can be written in the
form (\ref{exp.n}). In light of the Proposition, $E$ contains a neighborhood
$V$ of the identity. In particular, $E$ is non-empty.

We first claim that $E$ is open. To see this, consider $A\in E$. Then look at
the set of matrices of the form $AB$, with $B\in V$. This will be a
neighborhood of $A$. But every such $B$ can be written as $B=e^{X}$ and $A$
can be written as $A=e^{X_{1}}e^{X_{2}}\cdots e^{X_{n}}$, so $AB=e^{X_{1}%
}e^{X_{2}}\cdots e^{X_{n}}e^{X}$.

Now we claim that $E$ is closed (in $G$). Suppose $A\in G$, and there is a
sequence $A_{n}\in E$ with $A_{n}\rightarrow A$. Then $AA_{n}^{-1}\rightarrow
I$. Thus we can choose some $n_{0}$ such that $AA_{n_{0}}^{-1}\in V$. Then
$AA_{n_{0}}^{-1}=e^{X}$ and $A=A_{n_{0}}e^{X}$. But by assumption, $A_{n_{0}%
}=e^{X_{1}}e^{X_{2}}\cdots e^{X_{n}}$, so $A=e^{X_{1}}e^{X_{2}}\cdots
e^{X_{n}}e^{X}$. Thus $A\in E$, and $E$ is closed.

Thus $E$ is both open and closed, so $E=G$.
\end{proof}

\section{Lie Algebras\label{algebras}}

\begin{definition}
A \textbf{finite-dimensional real or complex Lie algebra} is a
finite-dimensional real or complex vector space $\frak{g}$, together with a
map $\left[  \ \right]  $ from $\frak{g}\times\frak{g}$ into $\frak{g}$, with
the following properties:

\begin{enumerate}
\item $\left[  \ \right]  $ is bilinear.

\item \label{lie.skew}$\left[  X,Y\right]  =-\left[  Y,X\right]  $ for all
$X,Y\in\frak{g}$.

\item \label{jacobi}$\left[  X,\left[  Y,Z\right]  \right]  +\left[  Y,\left[
Z,X\right]  \right]  +\left[  Z,\left[  X,Y\right]  \right]  =0$ for all
$X,Y,Z\in\frak{g}$.
\end{enumerate}
\end{definition}

Condition \ref{jacobi} is called the \textbf{Jacobi identity}. Note also that
Condition \ref{lie.skew} implies that $\left[  X,X\right]  =0$ for all
$X\in\frak{g}$. The same three conditions define a Lie algebra over an
arbitrary field $\mathbf{F}$, except that if $\mathbf{F}$ has characteristic
two, then one should add the condition $\left[  X,X\right]  =0$, which doesn't
follow from skew-symmetry in characteristic two. We will deal only with
finite-dimensional Lie algebras, and will from now on interpret ``Lie
algebra'' as ``finite-dimensional Lie algebra.''

A Lie algebra is in fact an algebra in the usual sense, but the product
operation $\left[  \ \right]  $ for this algebra is neither commutative nor
associative. The Jacobi identity should be thought of as a substitute for associativity.

\begin{proposition}
The space $\mathsf{gl}(n;\mathbb{R})$ of all $n\times n$ real matrices is a
real Lie algebra with respect to the bracket operation $\left[  A,B\right]
=AB-BA$. The space $\mathsf{gl}(n;\mathbb{C})$ of all $n\times n$ complex
matrices is a complex Lie algebra with respect to the analogous bracket operation.

Let $V$ is a finite-dimensional real or complex vector space, and let
$\mathsf{gl}(V)$ denote the space of linear maps of $V$ into itself. Then
$\mathsf{gl}(V)$ becomes a real or complex Lie algebra with the bracket
operation $\left[  A,B\right]  =AB-BA$.
\end{proposition}

\begin{proof}
The only non-trivial point is the Jacobi identity. The only way to prove this
is to write everything out and see, and this is best left to the reader. Note
that each triple bracket generates four terms, for a total of twelve. Each of
the six orderings of $\left\{  X,Y,Z\right\}  $ occurs twice, once with a plus
sign and once with a minus sign.
\end{proof}

\begin{definition}
A \textbf{subalgebra} of a real or complex Lie algebra $\frak{g}$ is a
subspace $\frak{h}$ of $\frak{g}$ such that $\left[  H_{1},H_{2}\right]
\in\frak{h}$ for all $H_{1},H_{2}\in\frak{h}$. If $\frak{g}$ is a complex Lie
algebra, and $\frak{h}$ is a real subspace of $\frak{g}$ which is closed under
brackets, then $\frak{h}$ is said to be a \textbf{real subalgebra} of
$\frak{g}$.

If $\frak{g}$ and $\frak{h}$ are Lie algebras, then a linear map
$\phi:\frak{g}\rightarrow\frak{h}$ is called a \textbf{Lie algebra
homomorphism} if $\phi\left(  \left[  X,Y\right]  \right)  =\left[
\phi(X),\phi(Y)\right]  $ for all $X,Y\in\frak{g}$. If in addition $\phi$ is
one-to-one and onto, then $\phi$ is called a \textbf{Lie algebra isomorphism}.
A Lie algebra isomorphism of a Lie algebra with itself is called a \textbf{Lie
algebra automorphism}.
\end{definition}

A subalgebra of a Lie algebra is again a Lie algebra. A real subalgebra of a
complex Lie algebra is a real Lie algebra. The inverse of a Lie algebra
isomorphism is again a Lie algebra isomorphism.

\begin{proposition}
The Lie algebra $\frak{g}$ of a matrix Lie group $G$ is a real Lie algebra.
\end{proposition}

\begin{proof}
By Theorem \ref{lie.algebra.theorem}, $\frak{g}$ is a real subalgebra of
$\mathsf{gl}(n;\mathbb{C})$ complex matrices, and is thus a real Lie algebra.
\end{proof}

\begin{theorem}
[Ado]Every finite-dimensional real Lie algebra is isomorphic to a subalgebra
of $\mathsf{gl}(n;\mathbb{R})$. Every finite-dimensional complex Lie algebra
is isomorphic to a (complex) subalgebra of $\mathsf{gl}(n;\mathbb{C})$.
\end{theorem}

This remarkable theorem is proved in Varadarajan. The proof is well beyond the
scope of this course (which is after all a course on Lie \textit{groups}), and
requires a deep understanding of the structure of complex Lie algebras. The
theorem tells us that every Lie algebra is (isomorphic to) a Lie algebra of
matrices. (This is in contrast to the situation for Lie groups, where most but
not all Lie groups are matrix Lie groups.)

\begin{definition}
Let $\frak{g}$ be a Lie algebra. For $X\in\frak{g}$, define a linear map
$\mathrm{ad}X:\frak{g}\rightarrow\frak{g}$ by
\[
\mathrm{ad}X(Y)=[X,Y]\text{.}%
\]
Thus ``$\mathrm{ad}$'' (i.e., the map $X\rightarrow\mathrm{ad}X$) can be
viewed as a linear map from $\frak{g}$ into $\mathsf{gl}(\frak{g})$, where
$\mathsf{gl}(\frak{g})$ denotes the space of linear operators from $\frak{g}$
to $\frak{g}$.
\end{definition}

Since $\mathrm{ad}X(Y)$ is just $[X,Y]$, it might seem foolish to introduce
the additional ``$\mathrm{ad}$'' notation. However, thinking of $[X,Y]$ as a
linear map in $Y$ for each fixed $X$, gives a somewhat different perspective.
In any case, the ``$\mathrm{ad}$'' notation is extremely useful in some
situations. For example, instead of writing
\[
\lbrack X,[X,[X,[X,Y]]]]
\]
we can now write
\[
\left(  \mathrm{ad}X\right)  ^{4}(Y)\text{.}%
\]
This kind of notation will be essential in Section \ref{bch}.

\begin{proposition}
\label{ad.homo}If $\frak{g}$ is a Lie algebra, then
\[
\mathrm{ad}[X,Y]=\mathrm{ad}X\mathrm{ad}Y-\mathrm{ad}Y\mathrm{ad}%
X=[\mathrm{ad}X,\mathrm{ad}Y]\text{.}%
\]
That is, \textrm{ad}$:\frak{g}\rightarrow\mathsf{gl}(\frak{g})$ is a Lie
algebra homomorphism.
\end{proposition}

\begin{proof}
Observe that
\[
\mathrm{ad}[X,Y](Z)=[[X,Y],Z]
\]
whereas
\[
\lbrack\mathrm{ad}X,\mathrm{ad}Y](Z)=[X,[Y,Z]]-[Y,[X,Z]]\text{.}%
\]
So we require that
\[
\lbrack\lbrack X,Y],Z]=[X,[Y,Z]]-[Y,[X,Z]]
\]
or equivalently
\[
0=[X,[Y,Z]]+[Y,[Z,X]]+[Z,[X,Y]]
\]
which is exactly the Jacobi identity.
\end{proof}

Recall that for any $X\in\frak{g}$, and any $A\in G$, we define
\[
\mathrm{Ad}A(X)=AXA^{-1}%
\]
and that \textrm{Ad}$:G\rightarrow\mathsf{GL}(\frak{g})$ is a Lie group
homomorphism. We showed (Proposition \ref{differentiate.Ad}) that the
associated Lie algebra homomorphism $\widetilde{\mathrm{Ad}}:\frak{g}%
\rightarrow\mathsf{gl}(\frak{g})$ is given by
\[
\widetilde{\mathrm{Ad}}X(Y)=[X,Y]\text{.}%
\]
In our new notation, we may say
\[
\widetilde{\mathrm{Ad}}=\mathrm{ad}%
\]
By the defining property of $\widetilde{\mathrm{Ad}}$, we have the following
identity: For all $X\in\frak{g}$,
\begin{equation}
\mathrm{Ad}(e^{X})=e^{\mathrm{ad}X}\text{.}\label{Ad.ad}%
\end{equation}
Note that both sides of (\ref{Ad.ad}) are linear operators on the Lie algebra
$\frak{g}$. This is an important relation, which can also be verified
directly, by expanding out both sides. (See Exercise \ref{ad.expand}.)

\subsection{Structure Constants}

Let $\frak{g}$ be a finite-dimensional real or complex Lie algebra, and let
$X_{1},\cdots,X_{n}$ be a basis for $\frak{g}$ (as a vector space). Then for
each $i,j$, $[X_{i},X_{j}]$ can be written uniquely in the form
\[
\lbrack X_{i},X_{j}]=\sum_{k=1}^{n}c_{ijk}X_{k}\text{.}%
\]
The constants $c_{ijk}$ are called the \textbf{structure constants} of
$\frak{g}$ (with respect to the chosen basis). Clearly, the structure
constants determine the bracket operation on $\frak{g}$. In some of the
literature, the structure constants play an important role, although we will
not have occasion to use them in this course. (In the physics literature, the
structure constants are defined as $[X_{i},X_{j}]=\sqrt{-1}\sum_{k}%
c_{ijk}X_{k}$, reflecting the factor of $\sqrt{-1}$ difference between the
physics definition of the Lie algebra and our own.)

The structure constants satisfy the following two conditions,
\begin{align*}
c_{ijk}+c_{jik}  & =0\\
\sum_{m}(c_{ijm}c_{mkl}+c_{jkm}c_{mil}+c_{kim}c_{mjl})  & =0
\end{align*}
for all $i,j,k,l$. The first of these conditions comes from the skew-symmetry
of the bracket, and the second comes from the Jacobi identity. (The reader is
invited to verify these conditions for himself.)

\section{The Complexification of a Real Lie Algebra\label{complex}}

\begin{definition}
If $V$ is a finite-dimensional real vector space, then the
\textbf{complexification} of $V$, denoted $V_{\mathbb{C}}$, is the space of
formal linear combinations
\[
v_{1}+iv_{2}%
\]
with $v_{1},v_{2}\in V$. This becomes a real vector space in the obvious way,
and becomes a complex vector space if we define
\[
i(v_{1}+iv_{2})=-v_{2}+iv_{1}\text{.}%
\]
\end{definition}

We could more pedantically define $V_{\mathbb{C}}$ to be the space of ordered
pairs $(v_{1},v_{2})$, but this is notationally cumbersome. It is
straightforward to verify that the above definition really makes
$V_{\mathbb{C}} $ into a complex vector space. We will regard $V$ as a real
subspace of $V_{\mathbb{C}}$ in the obvious way.

\begin{proposition}
Let $\frak{g}$ be a finite-dimensional real Lie algebra, and \thinspace
$\frak{g}_{\mathbb{C}}$ its complexification (as a real vector space). Then
the bracket operation on $\frak{g}$ has a unique extension to $\frak{g}%
_{\mathbb{C}}$ which makes $\frak{g}_{\mathbb{C}}$ into a complex Lie algebra.
The complex Lie algebra $\frak{g}_{\mathbb{C}}$ is called the
\textbf{complexification} of the real Lie algebra $\frak{g}$.
\end{proposition}

\begin{proof}
The uniqueness of the extension is obvious, since if the bracket operation on
$\frak{g}_{\mathbb{C}}$ is to be bilinear, then it must be given by
\begin{equation}
\left[  X_{1}+iX_{2},Y_{1}+iY_{2}\right]  =\left(  \left[  X_{1},Y_{1}\right]
-\left[  X_{2},Y_{2}\right]  \right)  +i\left(  \left[  X_{1},Y_{2}\right]
+\left[  X_{2},Y_{1}\right]  \right)  \text{.}\label{complex.bracket}%
\end{equation}
To show existence, we must now check that (\ref{complex.bracket}) is really
bilinear and skew-symmetric, and that it satisfies the Jacobi identity. It is
clear that (\ref{complex.bracket}) is \textit{real} bilinear, and
skew-symmetric. The skew-symmetry means that if (\ref{complex.bracket}) is
complex linear in the first factor, it is also complex linear in the second
factor. Thus we need only show that
\begin{equation}
\left[  i(X_{1}+iX_{2}),Y_{1}+iY_{2}\right]  =i\left[  X_{1}+iX_{2}%
,Y_{1}+iY_{2}\right]  \text{.}\label{linearity}%
\end{equation}
Well, the left side of (\ref{linearity}) is
\[
\left[  -X_{2}+iX_{1},Y_{1}+iY_{2}\right]  =\left(  -\left[  X_{2}%
,Y_{1}\right]  -\left[  X_{1},Y_{2}\right]  \right)  +i\left(  \left[
X_{1},Y_{1}\right]  -\left[  X_{2},Y_{2}\right]  \right)
\]
whereas the right side of (\ref{linearity}) is
\begin{align*}
& i\left\{  \left(  \left[  X_{1},Y_{1}\right]  -\left[  X_{2},Y_{2}\right]
\right)  +i\left(  \left[  X_{2},Y_{1}\right]  +\left[  X_{1},Y_{2}\right]
\right)  \right\}  \\
& =\left(  -\left[  X_{2},Y_{1}\right]  -\left[  X_{1},Y_{2}\right]  \right)
+i\left(  \left[  X_{1},Y_{1}\right]  -\left[  X_{2},Y_{2}\right]  \right)
\text{,}%
\end{align*}
and indeed these are equal.

It remains to check the Jacobi identity. Of course, the Jacobi identity holds
if $X,Y,$ and $Z$ are in $\frak{g}$. But now observe that the expression on
the left side of the Jacobi identity is (complex!) linear in $X$ for fixed $Y$
and $Z$. It follows that the Jacobi identity holds if $X$ is in $\frak{g}%
_{\mathbb{C}}$, and $Y,Z$ in $\frak{g}$. The same argument then shows that we
can extend to $Y$ in $\frak{g}_{\mathbb{C}}$, and then to $Z$ in
$\frak{g}_{\mathbb{C}}$. Thus the Jacobi identity holds in $\frak{g}%
_{\mathbb{C}}$.
\end{proof}

\begin{proposition}
\label{complex.classical}The Lie algebras $\mathsf{gl}(n;\mathbb{C})$,
$\mathsf{sl}(n;\mathbb{C})$, $\mathsf{so}(n;\mathbb{C})$, and $\mathsf{sp}%
(n;\mathbb{C})$ are complex Lie algebras, as is the Lie algebra of the complex
Heisenberg group. In addition, we have the following isomorphisms of complex
Lie algebras
\[%
\begin{array}
[c]{ccc}%
\mathsf{gl}\left(  n;\mathbb{R}\right)  _{\mathbb{C}} & \cong & \mathsf{gl}%
(n;\mathbb{C})\\
\mathsf{u}(n)_{\mathbb{C}} & \cong & \mathsf{gl}(n;\mathbb{C})\\
\mathsf{sl}\left(  n;\mathbb{R}\right)  _{\mathbb{C}} & \cong & \mathsf{sl}%
(n;\mathbb{C})\\
\mathsf{so}(n)_{\mathbb{C}} & \cong & \mathsf{so}(n;\mathbb{C})\\
\mathsf{sp}(n;\mathbb{R})_{\mathbb{C}} & \cong & \mathsf{sp}(n;\mathbb{C})\\
\mathsf{sp}(n)_{\mathbb{C}} & \cong & \mathsf{sp}(n;\mathbb{C})\text{.}%
\end{array}
\]
\end{proposition}

\begin{proof}
From the computations in the previous section we see easily that the specified
Lie algebras are in fact complex subalgebras of $\mathsf{gl}(n;\mathbb{C})$,
and hence are complex Lie algebras.

Now, $\mathsf{gl}(n;\mathbb{C})$ is the space of all $n\times n$ complex
matrices, whereas $\mathsf{gl}\left(  n;\mathbb{R}\right)  $ is the space of
all $n\times n$ real matrices. Clearly, then, every $X\in\mathsf{gl}\left(
n;\mathbb{C}\right)  $ can be written uniquely in the form $X_{1}+iX_{2}$,
with $X_{1},X_{2}\in\mathsf{gl}\left(  n;\mathbb{R}\right)  $. This gives us a
complex vector space isomorphism of $\mathsf{gl}\left(  n;\mathbb{R}\right)
_{\mathbb{C}}$ with $\mathsf{gl}(n;\mathbb{C})$, and it is a triviality to
check that this is a Lie algebra isomorphism.

On the other hand, $\mathsf{u}(n)$ is the space of all $n\times n$ complex
skew-self-adjoint matrices. But if $X$ is any $n\times n$ complex matrix,
then
\begin{align*}
X  & =\frac{X-X^{\ast}}{2}+\frac{X+X^{\ast}}{2}\\
& =\frac{X-X^{\ast}}{2}+i\frac{(-iX)-(-iX)^{\ast}}{2}\text{.}%
\end{align*}
Thus $X$ can be written as a skew matrix plus $i$ times a skew matrix, and it
is easy to see that this decomposition is unique. Thus every $X$ in
$\mathsf{gl}(n;\mathbb{C})$ can be written uniquely as $X_{1}+iX_{2}$, with
$X_{1}$ and $X_{2}$ in $\mathsf{u}(n)$. It follows that $\mathsf{u}%
(n)_{\mathbb{C}}\cong\mathsf{gl}(n;\mathbb{C})$.

The verification of the remaining isomorphisms is similar, and is left as an
exercise to the reader.
\end{proof}

Note that $\mathsf{u}(n)_{\mathbb{C}}\cong\mathsf{gl}\left(  n;\mathbb{R}%
\right)  _{\mathbb{C}}$ $\cong\mathsf{gl}(n;\mathbb{C})$. However,
$\mathsf{u}(n)$ is \textit{not} isomorphic to $\mathsf{gl}\left(
n;\mathbb{R}\right)  $, except when $n=1$. The real Lie algebras
$\mathsf{u}(n)$ and $\mathsf{gl}\left(  n;\mathbb{R}\right)  $ are called
\textbf{real forms} of the complex Lie algebra $\mathsf{gl}(n;\mathbb{C})$. A
given complex Lie algebra may have several non-isomorphic real forms. See
Exercise \ref{real.forms}.

Physicists do not always clearly distinguish between a matrix Lie group and
its (real) Lie algebra, or between a real Lie algebra and its
complexification. Thus, for example, some references in the physics literature
to \textsf{SU}$(2)$ actually refer to the complexified Lie algebra,
$\mathsf{sl}(2;\mathbb{C})$.

\section{Exercises}

\begin{enumerate}
\item \label{product.rule}\textit{The product rule}\emph{.} Recall that a
matrix-valued function $A(t)$ is smooth if each $A_{ij}(t)$ is smooth. The
derivative of such a function is defined as
\[
\left(  \frac{dA}{dt}\right)  _{ij}=\frac{dA_{ij}}{dt}%
\]
or equivalently,
\[
\frac{d}{dt}A(t)=\lim_{h\rightarrow0}\frac{A(t+h)-A(t)}{h}\text{.}%
\]

Let $A(t)$ and $B(t)$ be two such functions. Prove that $A(t)B(t)$ is again
smooth, and that
\[
\frac{d}{dt}\left[  A(t)B(t)\right]  =\frac{dA}{dt}B(t)+A(t)\frac{dB}%
{dt}\text{.}%
\]

\item \label{decomposition}Using the Jordan canonical form, show that every
$n\times n$ matrix $A$ can be written as $A=S+N$, with $S$ diagonalizable
(over $\mathbb{C}$), $N$ nilpotent, and $SN=NS$. Recall that the Jordan
canonical form is block diagonal, with each block of the form
\[
\left(
\begin{array}
[c]{ccc}%
\lambda &  & \ast\\
& \ddots & \\
0 &  & \lambda
\end{array}
\right)  \text{.}%
\]

\item \label{lie.product}Let $X$ and $Y$ be $n\times n$ matrices. Show that
there exists a constant $C$ such that
\[
\left\|  e^{(X+Y)/m}-e^{X/m}e^{Y/m}\right\|  \leq\frac{C}{m^{2}}%
\]
for all integers $m\geq1$.

\item \label{diagonal.limit}Using the Jordan canonical form, show that every
$n\times n$ complex matrix $A$ is the limit of a sequence of diagonalizable matrices.

\textit{Hint}: If the characteristic polynomial of $A$ has $n$ distinct roots,
then $A$ is diagonalizable.

\item  Give an example of a matrix Lie group $G$ and a matrix $X$ such that
$e^{X}\in G$, but $X\notin\frak{g}$.

\item \label{isomorphism}Show that two isomorphic matrix Lie groups have
isomorphic Lie algebras.

\item \label{so31}\textit{The Lie algebra}\emph{\ }\textsf{so}$(3;1)$. Write
out explicitly the general form of a $4\times4$ real matrix in \textsf{so}%
$(3;1)$.

\item  Verify directly that Proposition \ref{adjoint} and Theorem
\ref{lie.algebra.theorem} hold for the Lie algebra of $\mathsf{SU}(n)$.

\item \label{cross.product}\textit{The Lie algebra}\emph{\ }$\mathsf{su}(2)$.
Show that the following matrices form a basis for the real Lie algebra
$\mathsf{su}(2)$:
\[%
\begin{array}
[c]{ccc}%
E_{1}=\tfrac{1}{2}\left(
\begin{array}
[c]{cc}%
i & 0\\
0 & -i
\end{array}
\right)   & E_{2}=\tfrac{1}{2}\left(
\begin{array}
[c]{cc}%
0 & 1\\
-1 & 0
\end{array}
\right)   & E_{3}=\tfrac{1}{2}\left(
\begin{array}
[c]{cc}%
0 & i\\
i & 0
\end{array}
\right)
\end{array}
\text{.}%
\]

Compute $[E_{1},E_{2}]$, $[E_{2},E_{3}]$, and $[E_{3},E_{1}]$. Show that there
is an invertible linear map $\phi:\mathsf{su}(2)\rightarrow\mathbb{R}^{3}$
such that $\phi(\left[  X,Y\right]  )=\phi(X)\times\phi(Y)$ for all $X,Y\in$
$\mathsf{su}(2),$ where $\times$ denotes the cross-product on $\mathbb{R}^{3}$.

\item \label{su2.iso.so3}\textit{The Lie algebras }$\mathsf{su}(2)$
\textit{and}\emph{\ }$\mathsf{so}(3)$. Show that the real Lie algebras
$\mathsf{su}(2)$ and $\mathsf{so}(3)$ are isomorphic.

\textit{Note}: Nevertheless, the corresponding \textit{groups} $\mathsf{SU}%
(2)$ and $\mathsf{SO}(3)$ are not isomorphic. (Although $\mathsf{SO}(3)$ is
isomorphic to $\mathsf{SU}(2)/\left\{  I,-I\right\}  $.)

\item \label{real.forms}\textit{The Lie algebras}\emph{\ }$\mathsf{su}%
(2)$\emph{\ \textit{and} }$\mathsf{sl}(2;\mathbb{R})$. Show that
$\mathsf{su}(2)$\emph{\ }and $\mathsf{sl}(2;\mathbb{R})$ are not isomorphic
Lie algebras, even though $\mathsf{su}(2)_{\mathbb{C}}\cong\mathsf{sl}%
(2;\mathbb{R})_{\mathbb{C}}$.

\textit{Hint}: Using Exercise \ref{cross.product}, show that $\mathsf{su}(2)$
has no two-dimensional subalgebras.

\item  Let $G$ be a matrix Lie group, and $\frak{g}$ its Lie algebra. For each
$A\in G$, show that $\mathrm{Ad}A$ is a Lie algebra automorphism of $\frak{g}$.

\item \label{ad.expand}Ad \textit{and} ad. Let $X$ and $Y$ be matrices. Show
by induction that
\[
\left(  \mathrm{ad}X\right)  ^{n}(Y)=\sum_{k=0}^{n}\binom{n}{k}X^{k}%
Y(-X)^{n-k}\text{.}%
\]
Now show by direct computation that
\[
e^{\mathrm{ad}X}(Y)=\mathrm{Ad}(e^{X})Y=e^{X}Ye^{-X}\text{.}%
\]
You may assume that it is legal to multiply power series term-by-term. (This
result was obtained indirectly in Equation \ref{Ad.ad}.)

\textit{Hint}: Recall that Pascal's Triangle gives a relationship between
things of the form $\tbinom{n+1}{k}$ and things of the form $\tbinom{n}{k}$.

\item \label{extend.lie.homo}\textit{The complexification of a real Lie
algebra}. Let $\frak{g}$ be a real Lie algebra, $\frak{g}_{\mathbb{C}}$ its
complexification, and $\frak{h}$ an arbitrary complex Lie algebra. Show that
every real Lie algebra homomorphism of $\frak{g}$ into $\frak{h}$ extends
uniquely to a complex Lie algebra homomorphism of $\frak{g}_{\mathbb{C}}$ into
$\frak{h}$. (This is the \textbf{universal property} of the complexification
of a real Lie algebra. This property can be used as an alternative definition
of the complexification.)

\item \label{exp.sl2}\textit{The exponential mapping for}\emph{\ }%
$\mathsf{SL}\left(  2;\mathbb{R}\right)  $. Show that the image of the
exponential mapping for $\mathsf{SL}\left(  2;\mathbb{R}\right)  $ consists of
precisely those matrices $A\in\mathsf{SL}\left(  2;\mathbb{R}\right)  $ such
that $\mathrm{trace}\left(  A\right)  >-2,$ together with the matrix $-I$
(which has trace $-2$). You will need to consider the possibilities for the
eigenvalues of a matrix in the Lie algebra $\mathsf{sl}\left(  2;\mathbb{R}%
\right)  $ and in the group $\mathsf{SL}\left(  2;\mathbb{R}\right)  $. In the
Lie algebra, show that the eigenvalues are of the form $\left(  \lambda
,-\lambda\right)  $ or $\left(  i\lambda,-i\lambda\right)  $ with $\lambda$
real. In the group, show that the eigenvalues are of the form $\left(
\alpha,1/a\right)  $ or $\left(  -a,-1/a\right)  $ with $a$ real and positive,
or else of the form $\left(  e^{i\theta},e^{-i\theta}\right)  ,$ with $\theta$
real. The case of a repeated eigenvalue ($\left(  0,0\right)  $ in the Lie
algebra and $\left(  1,1\right)  $ or $\left(  -1,-1\right)  $ in the group)
will have to be treated separately.

Show that the image of the exponential mapping is not dense in $\mathsf{SL}%
\left(  2;\mathbb{R}\right)  $.

\item  Using Exercise \ref{diagonal.limit}, show that the exponential mapping
for $\mathsf{GL}(n;\mathbb{C})$ maps onto a dense subset of $\mathsf{GL}%
(n;\mathbb{C})$.

\item \label{exp.heisenberg}\textit{The exponential mapping for the Heisenberg
group}. Show that the exponential mapping from the Lie algebra of the
Heisenberg group to the Heisenberg group is one-to-one and onto.

\item \label{exp.un}\textit{The exponential mapping for}\emph{\ }%
$\mathsf{U}(n)$. Show that the exponential mapping from $\mathsf{u}(n)$ to
$\mathsf{U}(n)$ is onto, but not one-to-one. (Note that this shows that
$\mathsf{U}(n)$ is connected.)

\textit{Hint}: Every unitary matrix has an orthonormal basis of eigenvectors.

\item \label{tangent.space}Let $G$ be a matrix Lie group, and $\frak{g}$ its
Lie algebra. Let $A(t)$ be a smooth curve lying in $G$, with $A(0)=I$. Let
$X=\left.  \tfrac{d}{dt}\right|  _{t=0}A(t)$. Show that $X\in\frak{g}$.

\textit{Hint}: Use Proposition \ref{limit}.

\textit{Note}: This shows that the Lie algebra $\frak{g}$ coincides with what
would be called the \textbf{tangent space at the identity} in the language of
differentiable manifolds.

\item \label{ad.diagonal}Consider the space $\mathsf{gl}(n;\mathbb{C})$ of all
$n\times n$ complex matrices. As usual, for $X\in\mathsf{gl}(n;\mathbb{C})$,
define $\mathrm{ad}X:\mathsf{gl}(n;\mathbb{C})\rightarrow\mathsf{gl}%
(n;\mathbb{C})$ by $\mathrm{ad}X(Y)=[X,Y]$. Suppose that $X$ is a
diagonalizable matrix. Show, then, that $\mathrm{ad}X$ is diagonalizable as an
operator on $\mathsf{gl}(n;\mathbb{C})$.

\textit{Hint}: Consider first the case where $X$ is actually diagonal.

\textit{Note}: The problem of diagonalizing $\mathrm{ad}X$ is an important one
that we will encounter again in Chapter 6, when we consider semisimple Lie algebras.
\end{enumerate}

\chapter{The Baker-Campbell-Hausdorff Formula}

\section{The Baker-Campbell-Hausdorff Formula for the Heisenberg
Group\label{bch}}

\ A crucial result of Chapter 5 will be the following: Let $G$ and $H$ be
matrix Lie groups, with Lie algebras $\frak{g}$ and $\frak{h}$, and suppose
that $G$ is connected and simply connected. Then if $\widetilde{\phi}%
:\frak{g}\rightarrow\frak{h}$ is a Lie algebra homomorphism, there exists a
unique Lie group homomorphism $\phi:G\rightarrow H$ such that $\phi$ and
$\widetilde{\phi}$ are related as in Theorem \ref{homo.theorem}. This result
is extremely important because it implies that if $G$ is connected and simply
connected, then there is a natural one-to-one correspondence between the
representations of $G$ and the representations of its Lie algebra $\frak{g}$
(as explained in Chapter 5). In practice, it is much easier to determine the
representations of the Lie algebra than to determine directly the
representations of the corresponding group.

This result (relating Lie algebra homomorphisms and Lie group homomorphisms)
is deep. The ``modern'' proof (e.g., Varadarajan, Theorem 2.7.5) makes use of
the Frobenius theorem, which is both hard to understand and hard to prove
(Varadarajan, Section 1.3). Our proof will instead use the
Baker-Campbell-Hausdorff formula, which is more easily stated and more easily
motivated than the Frobenius theorem, but still deep.

The idea is the following. The desired group homomorphism $\phi:G\rightarrow
H$ must satisfy
\begin{equation}
\phi\left(  e^{X}\right)  =e^{\widetilde{\phi}(X)}\text{.}\label{relating}%
\end{equation}
We would like, then, to \textit{define} $\phi$ by this relation. This approach
has two serious difficulties. First, a given element of $G$ may not be
expressible as $e^{X}$, and even if it is, the $X$ may not be unique. Second,
it is very far from clear why the $\phi$ in (\ref{relating}) (even to the
extent it is well-defined) should be a group homomorphism.

It is the second issue which the Baker-Campbell-Hausdorff formula addresses.
(The first issue will be addressed in the next chapter; it is there that the
simple connectedness of $G$ comes into play.) Specifically, (one form of) the
Baker-Campbell-Hausdorff formula says that if $X$ and $Y$ are sufficiently
small, then
\begin{equation}
\log(e^{X}e^{Y})=X+Y+\tfrac{1}{2}[X,Y]+\tfrac{1}{12}[X,[X,Y]]-\tfrac{1}%
{12}[Y,[X,Y]]+\cdots\text{.}\label{bch.series}%
\end{equation}
It is not supposed to be evident at the moment what ``$\cdots$'' refers to.
The only important point is that all of the terms in (\ref{bch.series}) are
given in terms of $X$ and $Y$, brackets of $X$ and $Y$, brackets of brackets
involving $X$ and $Y$, etc. Then because $\widetilde{\phi}$ is a Lie algebra
homomorphism,
\begin{align}
\widetilde{\phi}\left(  \log\left(  e^{X}e^{Y}\right)  \right)   &
=\widetilde{\phi}(X)+\widetilde{\phi}(Y)+\tfrac{1}{2}[\widetilde{\phi
}(X),\widetilde{\phi}(Y)]\nonumber\\
& +\tfrac{1}{12}[\widetilde{\phi}(X),[\widetilde{\phi}(X),\widetilde{\phi
}(Y)]]-\tfrac{1}{12}[\widetilde{\phi}(Y),[\widetilde{\phi}(X),\widetilde{\phi
}(Y)]]+\cdots\nonumber\\
& =\log\left(  e^{\widetilde{\phi}(X)}e^{\widetilde{\phi}(Y)}\right)
\label{switch}%
\end{align}

The relation (\ref{switch}) is extremely significant. For of course
\[
e^{X}e^{Y}=e^{\log(e^{X}e^{Y})}%
\]
and so by (\ref{relating}),
\[
\phi\left(  e^{X}e^{Y}\right)  =e^{\widetilde{\phi}(\log(e^{X}e^{Y}))}\text{.}%
\]
Thus (\ref{switch}) tells us that
\[
\phi\left(  e^{X}e^{Y}\right)  =e^{\log\left(  e^{\widetilde{\phi}%
(X)}e^{\widetilde{\phi}(Y)}\right)  }=e^{\widetilde{\phi}(X)}e^{\widetilde
{\phi}(Y)}=\phi(e^{X})\phi(e^{Y})\text{.}%
\]
Thus, the Baker-Campbell-Hausdorff formula shows that on elements of the form
$e^{X}$, with $X$ small, $\phi$ is a group homomorphism. (See Corollary
\ref{local.homo} below.)

The Baker-Campbell-Hausdorff formula shows that all the information about the
group product, at least near the identity, is ``encoded'' in the Lie algebra.
Thus if $\widetilde{\phi}$ is a Lie algebra homomorphism (which by definition
preserves the Lie algebra structure), and if we define $\phi$ near the
identity by (\ref{relating}), then we can expect $\phi$ to preserve the group
structure, i.e., to be a group homomorphism.

In this section we will look at how all of this works out in the very special
case of the Heisenberg group. In the next section we will consider the general situation.

\begin{theorem}
\label{bch.heisenberg}Suppose $X$ and $Y$ are $n\times n$ complex matrices,
and that $X$ and $Y$ commute with their commutator. That is, suppose that
\[
\left[  X,\left[  X,Y\right]  \right]  =\left[  Y,\left[  X,Y\right]  \right]
=0\text{.}%
\]
Then
\[
e^{X}e^{Y}=e^{X+Y+\frac{1}{2}\left[  X,Y\right]  }\text{.}%
\]
\end{theorem}

This is the special case of (\ref{bch.series}) in which the series terminates
after the $\left[  X,Y\right]  $ term.

\begin{proof}
Let $X$ and $Y$ be as in the statement of the theorem. We will prove that in
fact
\[
e^{tX}e^{tY}=\exp\left(  tX+tY+\frac{t^{2}}{2}\left[  X,Y\right]  \right)
\text{,}%
\]
which reduces to the desired result in the case $t=1$. Since by assumption
$\left[  X,Y\right]  $ commutes with everything in sight, the above relation
is equivalent to
\begin{equation}
e^{tX}e^{tY}e^{-\frac{t^{2}}{2}\left[  X,Y\right]  }=e^{t\left(  X+Y\right)
}\text{.}\label{atbt}%
\end{equation}

Let us call the left side of (\ref{atbt}) $A(t)$ and the right side $B\left(
t\right)  $. Our strategy will be to show that $A\left(  t\right)  $ and
$B\left(  t\right)  $ satisfy the same differential equation, with the same
initial conditions. We can see right away that
\[
\frac{dB}{dt}=B\left(  t\right)  \left(  X+Y\right)  \text{.}%
\]
On the other hand, differentiating $A\left(  t\right)  $ by means of the
product rule gives
\begin{equation}
\frac{dA}{dt}=e^{tX}Xe^{tY}e^{-\frac{t^{2}}{2}\left[  X,Y\right]  }%
+e^{tX}e^{tY}Ye^{-\frac{t^{2}}{2}\left[  X,Y\right]  }+e^{tX}e^{tY}%
e^{-\frac{t^{2}}{2}\left[  X,Y\right]  }\left(  -t\left[  X,Y\right]  \right)
\text{.}\label{part.deriv}%
\end{equation}
(You can verify that the last term on the right is correct by differentiating term-by-term.)

Now, since $X$ and $Y$ commute with $\left[  X,Y\right]  $, they also commute
with $e^{-\frac{t^{2}}{2}\left[  X,Y\right]  }$. Thus the second term on the
right in (\ref{part.deriv}) can be rewritten as
\[
e^{tX}e^{tY}e^{-\frac{t^{2}}{2}\left[  X,Y\right]  }Y\text{.}%
\]
The first term on the right in (\ref{part.deriv}) is more complicated, since
$X$ does not necessarily commute with $e^{tY}$. However,
\begin{align*}
Xe^{tY}=e^{tY}e^{-tY}Xe^{tY}\\
=e^{tY}\mathrm{Ad}\left(  e^{-tY}\right)  \left(  X\right)  \\
=e^{tY}e^{-t\mathrm{ad}Y}\left(  X\right)  \text{.}%
\end{align*}
But since $\left[  Y,\left[  Y,X\right]  \right]  =-\left[  Y,\left[
X,Y\right]  \right]  =0$,
\[
e^{-t\mathrm{ad}Y}\left(  X\right)  =X-t\left[  Y,X\right]  =X+t\left[
X,Y\right]
\]
with all higher terms being zero. Using the fact that everything commutes with
$e^{-\frac{t^{2}}{2}\left[  X,Y\right]  }$ gives
\[
e^{tX}Xe^{tY}e^{-\frac{t^{2}}{2}\left[  X,Y\right]  }=e^{tX}e^{tY}%
e^{-\frac{t^{2}}{2}\left[  X,Y\right]  }\left(  X+t\left[  X,Y\right]
\right)
\]

Making these substitutions into (\ref{part.deriv}) gives
\begin{align*}
\frac{dA}{dt}=e^{tX}e^{tY}e^{-\frac{t^{2}}{2}\left[  X,Y\right]  }\left(
X+t\left[  X,Y\right]  \right)  +e^{tX}e^{tY}e^{-\frac{t^{2}}{2}\left[
X,Y\right]  }Y+e^{tX}e^{tY}e^{-\frac{t^{2}}{2}\left[  X,Y\right]  }\left(
-t\left[  X,Y\right]  \right)  \\
=e^{tX}e^{tY}e^{-\frac{t^{2}}{2}\left[  X,Y\right]  }\left(  X+Y\right)  \\
=A\left(  t\right)  \left(  X+Y\right)  \text{.}%
\end{align*}

Thus $A\left(  t\right)  $ and $B\left(  t\right)  $ satisfy the same
differential equation. Moreover, $A\left(  0\right)  =B\left(  0\right)  =I$.
Thus by standard uniqueness results for ordinary differential equations,
$A\left(  t\right)  =B\left(  t\right)  $ for all $t$.
\end{proof}

\begin{theorem}
Let $H$ denote the Heisenberg group, and $\frak{h}$ its Lie algebra. Let $G$
be a matrix Lie group with Lie algebra $\frak{g}$, and let $\widetilde{\phi
}:\frak{h}\rightarrow\frak{g}$ be a Lie algebra homomorphism. Then there
exists a unique Lie group homomorphism $\phi:H\rightarrow G$ such that
\[
\phi\left(  e^{X}\right)  =e^{\widetilde{\phi}\left(  X\right)  }%
\]
for all $X\in\frak{h}$.
\end{theorem}

\begin{proof}
Recall that the Heisenberg group has the very special property that its
exponential mapping is one-to-one and onto. Let ``log'' denote the inverse of
this map. Define $\phi:H\rightarrow G$ by the formula
\[
\phi\left(  A\right)  =e^{\widetilde{\phi}\left(  \log A\right)  }\text{.}%
\]
We will show that $\phi$ is a Lie group homomorphism.

If $X$ and $Y$ are in the Lie algebra of the Heisenberg group ($3\times3$
strictly upper-triangular matrices), then $\left[  X,Y\right]  $ is of the
form
\[
\left(
\begin{array}
[c]{lll}%
0 & 0 & a\\
0 & 0 & 0\\
0 & 0 & 0
\end{array}
\right)  ;
\]
such a matrix commutes with both $X$ and $Y$. That is, $X$ and $Y$ commute
with their commutator. Since $\widetilde{\phi}$ is a Lie algebra homomorphism,
$\widetilde{\phi}\left(  X\right)  $ and $\widetilde{\phi}\left(  Y\right)  $
will also commute with their commutator:
\begin{align*}
\left[  \widetilde{\phi}\left(  X\right)  ,\left[  \widetilde{\phi}\left(
X\right)  ,\widetilde{\phi}\left(  Y\right)  \right]  \right]  =\widetilde
{\phi}\left(  \left[  X,\left[  X,Y\right]  \right]  \right)  =0\\
\left[  \widetilde{\phi}\left(  Y\right)  ,\left[  \widetilde{\phi}\left(
X\right)  ,\widetilde{\phi}\left(  Y\right)  \right]  \right]  =\widetilde
{\phi}\left(  \left[  Y,\left[  X,Y\right]  \right]  \right)  =0\text{.}%
\end{align*}

We want to show that $\phi$ is a homomorphism, i.e., that $\phi\left(
AB\right)  =\phi\left(  A\right)  \phi\left(  B\right)  $. Well, $A$ can be
written as $e^{X}$ for a unique $X\in\frak{h}$ and $B$ can be written as
$e^{Y}$ for a unique $Y\in\frak{h}$. Thus by Theorem \ref{bch.heisenberg}
\[
\phi\left(  AB\right)  =\phi\left(  e^{X}e^{Y}\right)  =\phi\left(
e^{X+Y+\frac{1}{2}\left[  X,Y\right]  }\right)  \text{.}%
\]
Using the definition of $\phi$ and the fact that $\widetilde{\phi}$ is a Lie
algebra homomorphism:
\[
\phi\left(  AB\right)  =\exp\left(  \widetilde{\phi}\left(  X\right)
+\widetilde{\phi}\left(  Y\right)  +\frac{1}{2}\left[  \widetilde{\phi}\left(
X\right)  ,\widetilde{\phi}\left(  Y\right)  \right]  \right)  \text{.}%
\]
Finally, using Theorem \ref{bch.heisenberg} again we have
\[
\phi\left(  AB\right)  =e^{\widetilde{\phi}\left(  X\right)  }e^{\widetilde
{\phi}\left(  Y\right)  }=\phi\left(  A\right)  \phi\left(  B\right)  \text{.}%
\]

Thus $\phi$ is a group homomorphism. It is easy to check that $\phi$ is
continuous (by checking that $\log$, exp, and $\widetilde{\phi}$ are all
continuous), and so $\phi$ is a Lie group homomorphism. Moreover, $\phi$ by
definition has the right relationship to $\widetilde{\phi}$. Furthermore,
since the exponential mapping is one-to-one and onto, there can be at most one
$\phi$ with $\phi\left(  e^{X}\right)  =e^{\widetilde{\phi}\left(  X\right)
}$. So we have uniqueness.
\end{proof}

\section{The General Baker-Campbell-Hausdorff Formula\label{bch.general}}

The importance of the Baker-Campbell-Hausdorff formula lies not in the details
of the formula, but in the fact that there is one, and the fact that it gives
$\log(e^{X}e^{Y})$ in terms of brackets of $X$ and $Y$, brackets of brackets,
etc. This tells us something very important, namely that (at least for
elements of the form $e^{X}$, $X$ small) the group product for a matrix Lie
group $G$ is \textit{completely expressible in terms of the Lie algebra}.
(This is because $\log\left(  e^{X}e^{Y}\right)  $, and hence also $e^{X}%
e^{Y}$ itself, can be computed in Lie-algebraic terms by (\ref{bch.series}).)

We will actually state and prove an integral form of the
Baker-Campbell-Hausdorff formula, rather than the series form
(\ref{bch.series}). However, the integral form is sufficient to obtain the
desired result (\ref{switch}). (See Corollary \ref{local.homo}.) The series
form of the Baker-Campbell-Hausdorff formula is stated precisely and proved in
Varadarajan, Sec. 2.15.

Consider the function
\[
g(z)=\frac{\log z}{1-\frac1z}\text{.}%
\]
This function is defined and analytic in the disk $\left\{  \left|
z-1\right|  <1\right\}  $, and thus for $z$ in this set, $g(z)$ can be
expressed as
\[
g(z)=\sum_{m=0}^{\infty}a_{m}(z-1)^{m}\text{.}%
\]
This series has radius of convergence one.

Now suppose $V$ is a finite-dimensional complex vector space. Choose an
arbitrary basis for $V$, so that $V$ can be identified with $\mathbb{C}^{n}$
and thus the norm of a linear operator on $V$ can be defined. Then for any
operator $A$ on $V$ with $\left\|  A-I\right\|  <1$, we can define
\[
g(A)=\sum_{m=0}^{\infty}a_{m}(A-1)^{m}\text{.}%
\]
We are now ready to state the integral form of the Baker-Campbell-Hausdorff formula.

\begin{theorem}
[Baker-Campbell-Hausdorff]\label{bch.theorem}For all $n\times n$ complex
matrices $X$ and $Y$ with $\left\|  X\right\|  $ and $\left\|  Y\right\|  $
sufficiently small,
\begin{equation}
\log\left(  e^{X}e^{Y}\right)  =X+\int_{0}^{1}g(e^{\mathrm{ad}X}%
e^{t\mathrm{ad}Y})(Y)\,dt\text{.}\label{bch.integral}%
\end{equation}
\end{theorem}

\begin{corollary}
\label{local.homo}Let $G$ be a matrix Lie group and $\frak{g}$ its Lie
algebra. Suppose that $\widetilde{\phi}:\frak{g}\rightarrow\mathsf{gl}%
(n;\mathbf{C})$ is a Lie algebra homomorphism. Then for all sufficiently small
$X,Y$ in $\frak{g}$, $\log\left(  e^{X}e^{Y}\right)  $ is in $\frak{g}$, and
\begin{equation}
\widetilde{\phi}\left[  \log\left(  e^{X}e^{Y}\right)  \right]  =\log\left(
e^{\widetilde{\phi}(X)}e^{\widetilde{\phi}(Y)}\right)  \text{.}\label{switch2}%
\end{equation}
\end{corollary}

Note that $e^{\mathrm{ad}X}e^{t\mathrm{ad}Y}$, and hence also
$g(e^{\mathrm{ad}X}e^{t\mathrm{ad}Y})$, is a linear operator on the space
$\mathsf{gl}(n;\mathbb{C})$ of all $n\times n$ complex matrices. In
(\ref{bch.integral}), this operator is being applied to the matrix $Y$. The
fact that $X$ and $Y$ are assumed small guarantees that $e^{\mathrm{ad}%
X}e^{t\mathrm{ad}Y}$ is close to the identity operator on $\mathsf{gl}%
(n;\mathbb{C})$ for all $0\leq t\leq1$. This ensures that $g(e^{\mathrm{ad}%
X}e^{t\mathrm{ad}Y})$ is well defined.

If $X$ and $Y$ commute, then we expect to have $\log\left(  e^{X}e^{Y}\right)
=\log(e^{X+Y})=X+Y$. Exercise \ref{bch.commute} asks you to verify that the
Baker-Campbell-Hausdorff formula indeed gives $X+Y$ in that case.

Formula (\ref{bch.integral}) is admittedly horrible-looking. However, we are
interested not in the details of the formula, but in the fact that it
expresses $\log\left(  e^{X}e^{Y}\right)  $ (and hence $e^{X}e^{Y}$) in terms
of the Lie-algebraic quantities $\mathrm{ad}X$ and $\mathrm{ad}Y$.

The goal of the Baker-Campbell-Hausdorff theorem is to compute $\log\left(
e^{X}e^{Y}\right)  $. You may well ask, ``Why don't we simply expand both
exponentials and the logarithm in power series and multiply everything out?''
Well, you can do this, and if you do it for the first several terms you will
get the same answer as B-C-H. However, there is a serious problem with this
approach, namely: How do you know that the terms in such an expansion are
expressible in terms of commutators? Consider for example the quadratic term.
It is clear that this will be a linear combination of $X^{2}$, $Y^{2}$, $XY$,
and $YX$. But to be expressible in terms of commutators it must actually be a
constant times $\left(  XY-YX\right)  $. Of course, for the quadratic term you
can just multiply it out and see, and indeed you get $\frac12\left(
XY-YX\right)  =\frac12\left[  X,Y\right]  $. But it is far from clear how to
prove that a similar result occurs for all the higher terms. See Exercise
\ref{by.series}.

\begin{proof}
We begin by proving that the corollary follows from the integral form of the
Baker-Campbell-Hausdorff formula. The proof is conceptually similar to the
reasoning in Equation (\ref{switch}). Note that if $X$ and $Y$ lie in some Lie
algebra $\frak{g}$ then $\mathrm{ad}X$ and $\mathrm{ad}Y$ will preserve
$\frak{g}$, and so also will $g(e^{\mathrm{ad}X}e^{t\mathrm{ad}Y})(Y)$. Thus
whenever formula (\ref{bch.integral}) holds, $\log\left(  e^{X}e^{Y}\right)  $
will lie in $\frak{g}$. It remains only to verify (\ref{switch2}). The idea is
that if $\widetilde{\phi}$ is Lie algebra homomorphism, then it will take a
big horrible looking expression involving `ad' and $X$ and $Y$, and turn it
into the same expression with $X$ and $Y$ replaced by $\widetilde{\phi}\left(
X\right)  $ and $\widetilde{\phi}\left(  Y\right)  $.

More precisely, since $\widetilde{\phi}$ is a Lie algebra homomorphism,
\[
\widetilde{\phi}[Y,X]=[\widetilde{\phi}(Y),\widetilde{\phi}(X)]
\]
or
\[
\widetilde{\phi}\left(  \mathrm{ad}Y\left(  X\right)  \right)  =\mathrm{ad}%
\widetilde{\phi}\left(  Y\right)  \left(  \widetilde{\phi}\left(  X\right)
\right)  \text{.}%
\]
More generally,
\[
\widetilde{\phi}\left(  \left(  \mathrm{ad}Y\right)  ^{n}\left(  X\right)
\right)  =\left(  \mathrm{ad}\widetilde{\phi}\left(  Y\right)  \right)
^{n}\left(  \widetilde{\phi}\left(  X\right)  \right)  \text{.}%
\]

This being the case,
\begin{align*}
\widetilde{\phi}\left(  e^{\mathrm{ad}Y}\left(  X\right)  \right)    &
=\sum_{m=0}^{\infty}\frac{t^{m}}{m!}\widetilde{\phi}\left(  \left(
\mathrm{ad}Y\right)  ^{n}\left(  X\right)  \right)  \\
& =\sum_{m=0}^{\infty}\frac{t^{m}}{m!}\left(  \mathrm{ad}\widetilde{\phi
}\left(  Y\right)  \right)  ^{n}\left(  \widetilde{\phi}\left(  X\right)
\right)  \\
& =e^{t\mathrm{ad}\widetilde{\phi}(Y)}\left(  \widetilde{\phi}(X)\right)
\text{.}%
\end{align*}
Similarly,
\[
\widetilde{\phi}\left(  e^{\mathrm{ad}X}e^{t\mathrm{ad}Y}(X)\right)
=e^{\mathrm{ad}\widetilde{\phi}(X)}e^{t\mathrm{ad}\widetilde{\phi}(Y)}\left(
\widetilde{\phi}(X)\right)  \text{.}%
\]
Assume now that $X$ and $Y$ are small enough that B-C-H applies to $X$ and
$Y$, and to $\widetilde{\phi}(X)$ and $\widetilde{\phi}(Y)$. Then, using the
linearity of the integral and reasoning similar to the above, we have:
\begin{align*}
\widetilde{\phi}\left(  \log\left(  e^{X}e^{Y}\right)  \right)    &
=\widetilde{\phi}(X)+\int_{0}^{1}\sum_{m=0}^{\infty}a_{m}\widetilde{\phi
}\left[  \left(  e^{\mathrm{ad}X}e^{t\mathrm{ad}Y}-I\right)  ^{n}(X)\right]
\,dt\\
& =\widetilde{\phi}(X)+\int_{0}^{1}\sum_{m=0}^{\infty}a_{m}\left(
e^{\mathrm{ad}\widetilde{\phi}(X)}e^{t\mathrm{ad}\widetilde{\phi}%
(Y)}-I\right)  ^{n}(\widetilde{\phi}(X))\,dt\\
& =\log\left(  e^{\widetilde{\phi}(X)}e^{\widetilde{\phi}(Y)}\right)  \text{.}%
\end{align*}
This is what we wanted to show.
\end{proof}

Before coming to the proof Baker-Campbell-Hausdorff formula itself, we will
obtain a result concerning derivatives of the exponential mapping. This result
is valuable in its own right, and will play a central role in our proof of the
Baker-Campbell-Hausdorff formula.

Observe that if $X$ and $Y$ commute, then
\[
e^{X+tY}=e^{X}e^{tY}%
\]
and so
\[
\left.  \tfrac d{dt}\right|  _{t=0}e^{X+tY}=e^{X}\left.  \tfrac d{dt}\right|
_{t=0}e^{tY}=e^{X}Y\text{.}%
\]
In general, $X$ and $Y$ do not commute, and
\[
\left.  \tfrac d{dt}\right|  _{t=0}e^{X+tY}\neq e^{X}Y\text{.}%
\]
This, as it turns out, is an important point. In particular, note that in the
language of multivariate calculus
\begin{equation}
\left.  \tfrac d{dt}\right|  _{t=0}e^{X+tY}=\text{ }\left\{
\begin{array}
[c]{l}%
\text{directional derivative of \textit{exp} at }X\text{,}\\
\text{in the direction of }Y
\end{array}
\right.  \text{.}\label{directional}%
\end{equation}
Thus computing the left side of (\ref{directional}) is the same as computing
all of the directional derivatives of the (matrix-valued) function
\textit{exp}. We expect the directional derivative to be a linear function of
$Y$, for each fixed $X$.

Now, the function
\[
\frac{1-e^{-z}}{z}=\frac{1-(1-z+\frac{z^{2}}{2!}-\cdots)}{z}%
\]
is an entire analytic function of $z$, even at $z=0$, and is given by the
power series
\[
\frac{1-e^{-z}}{z}=\sum_{n=1}^{\infty}(-1)^{n-1}\frac{z^{n-1}}{n!}=1-\frac
{z}{2!}+\frac{z^{2}}{3!}-\cdots\text{.}%
\]
This series (which has infinite radius of convergence), make sense when $z$ is
replaced by a linear operator $A$ on some finite-dimensional vector space.

\begin{theorem}
[Derivative of Exponential]\label{exp.derivative}Let $X$ and $Y$ be $n\times
n$ complex matrices. Then
\begin{align}
\left.  \frac{d}{dt}\right|  _{t=0}e^{X+tY}  & =e^{X}\left\{  \frac
{I-e^{-\mathrm{ad}X}}{\mathrm{ad}X}(Y)\right\}  \nonumber\\
& =e^{X}\left\{  Y-\frac{[X,Y]}{2!}+\frac{[X,[X,Y]]}{3!}-\cdots\right\}
\text{.}\label{derivative.a}%
\end{align}
More generally, if $X\left(  t\right)  $ is a smooth matrix-valued function,
then
\begin{equation}
\left.  \frac{d}{dt}\right|  _{t=0}e^{X(t)}=e^{X(0)}\left\{  \frac
{I-e^{-\mathrm{ad}X(0)}}{\mathrm{ad}X(0)}\left(  \left.  \tfrac{dX}%
{dt}\right|  _{t=0}\right)  \right\}  \text{.}\label{derivative2}%
\end{equation}
\end{theorem}

Note that the directional derivative in (\ref{derivative.a}) is indeed linear
in $Y$ for each fixed $X$. Note also that (\ref{derivative.a}) is just a
special case of (\ref{derivative2}), by taking $X(t)=X+tY$, and evaluating at
$t=0$.

Furthermore, observe that if $X$ and $Y$ commute, then only the first term in
the series (\ref{derivative.a}) survives. In that case, we obtain $\left.
\frac d{dt}\right|  _{t=0}e^{X+tY}=e^{X}Y$ as expected.

\begin{proof}
It is possible to prove this Theorem by expanding everything in a power series
and differentiating term-by-term; we will not take that approach. We will
prove only form (\ref{derivative.a}) of the derivative formula, but the form
(\ref{derivative2}) follows by the chain rule.

Let us use the Lie product formula, and let us assume for the moment that it
is legal to interchange limit and derivative. (We will consider this issue at
the end.) Then we have
\[
e^{-X}\left.  \frac{d}{dt}\right|  _{t=0}e^{X+tY}=e^{-X}\lim_{n\rightarrow
\infty}\left.  \frac{d}{dt}\right|  _{t=0}\left(  e^{X/n}e^{tY/n}\right)
^{n}\text{.}%
\]
We now apply the product rule (generalized to $n$ factors) to obtain
\begin{align*}
e^{-X}\left.  \frac{d}{dt}\right|  _{t=0}e^{X+tY}  & =e^{-X}\lim
_{n\rightarrow\infty}\sum_{k=0}^{n-1}\left[  \left(  e^{X/n}e^{tY/n}\right)
^{n-k-1}\left(  e^{X/n}e^{tY/n}Y/n\right)  \left(  e^{X/n}e^{tY/n}\right)
^{k}\right]  _{t=0}\\
& =e^{-X}\lim_{n\rightarrow\infty}\sum_{k=0}^{n-1}\left(  e^{X/n}\right)
^{n-k-1}\left(  e^{X/n}Y/n\right)  \left(  e^{X/n}\right)  ^{k}\\
& =\lim_{n\rightarrow\infty}\frac{1}{n}\sum_{k=0}^{n-1}\left(  e^{X/n}\right)
^{-k}Y\left(  e^{X/n}\right)  ^{k}\text{.}%
\end{align*}

But
\begin{align*}
\left(  e^{X/n}\right)  ^{-k}Y\left(  e^{X/n}\right)  ^{k}  & =\left[
\mathrm{Ad}\left(  e^{-X/n}\right)  \right]  ^{k}\left(  Y\right)  \\
& =\left(  e^{-\mathrm{ad}X/n}\right)  ^{k}(Y)
\end{align*}
(where we have used the relationship between \textit{Ad} and \textit{ad}). So
we have
\begin{equation}
e^{-X}\left.  \frac{d}{dt}\right|  _{t=0}e^{X+tY}=\lim_{n\rightarrow\infty
}\frac{1}{n}\sum_{k=0}^{n-1}\left(  e^{-\mathrm{ad}X/n}\right)  ^{k}%
(Y)\text{.}\label{derivative.series}%
\end{equation}

Observe now that $\sum_{k=0}^{n-1}\left(  e^{-\mathrm{ad}X/n}\right)  ^{k}$ is
a geometric series. Let us now reason for a moment at the purely formal level.
Using the usual formula for geometric series, we get
\begin{align*}
e^{-X}\left.  \frac{d}{dt}\right|  _{t=0}e^{X+tY}=\lim_{n\rightarrow\infty
}\frac{1}{n}\frac{I-\left(  e^{-\mathrm{ad}X/n}\right)  ^{n}}%
{I-e^{-\mathrm{ad}X/n}}(Y)\\
=\lim_{n\rightarrow\infty}\frac{I-e^{-\mathrm{ad}X}}{n\left[  I-\left(
I-\frac{\mathrm{ad}X}{n}+\frac{(\mathrm{ad}X)^{2}}{n^{2}2!}-\cdots\right)
\right]  }(Y)\\
=\lim_{n\rightarrow\infty}\frac{I-e^{-\mathrm{ad}X}}{\mathrm{ad}%
X-\frac{(\mathrm{ad}X)^{2}}{n2!}+\cdots}(Y)\\
=\frac{I-e^{-\mathrm{ad}X}}{\mathrm{ad}X}(Y)\text{.}%
\end{align*}
This is what we wanted to show!

Does this argument make sense at any rigorous level? In fact it does. As
usual, let us consider first the diagonalizable case. That is, assume that
$\mathrm{ad}X$ is diagonalizable as an operator on $\mathsf{gl}(n;\mathbb{C}%
)$, and assume that $Y$ is an eigenvector for $\mathrm{ad}X$. This means that
$\mathrm{ad}X(Y)=[X,Y]=\lambda Y$, for some $\lambda\in\mathbb{C}$. Now, there
are two cases, $\lambda=0$ and $\lambda\neq0$. The $\lambda=0$ case
corresponds to the case in which $X$ and $Y$ commute, and we have already
observed that the Theorem holds trivially in that case.

The interesting case, then, is the case $\lambda\neq0$. Note that $\left(
\mathrm{ad}X\right)  ^{n}(Y)=\lambda^{n}Y$, and so
\[
\left(  e^{-\mathrm{ad}X/n}\right)  ^{k}(Y)=\left(  e^{-\lambda/n}\right)
^{k}(Y)\text{.}%
\]
Thus the geometric series in (\ref{derivative.series}) becomes an ordinary
complex-valued series, with ratio $e^{-\lambda/n}$. Since $\lambda\neq0$, this
ratio will be different from one for all sufficiently large $n$. Thus we get
\[
e^{-X}\left.  \frac{d}{dt}\right|  _{t=0}e^{X+tY}=\left(  \lim_{n\rightarrow
\infty}\frac{1}{n}\frac{I-\left(  e^{-\lambda/n}\right)  ^{n}}{I-e^{-\lambda
/n}}\right)  Y\text{.}%
\]
There is now no trouble in taking the limit as we did formally above to get
\begin{align*}
e^{-X}\left.  \frac{d}{dt}\right|  _{t=0}e^{X+tY}=\frac{1-e^{-\lambda}%
}{\lambda}Y\\
=\frac{I-e^{-\mathrm{ad}X}}{\mathrm{ad}X}(Y)\text{.}%
\end{align*}

We see then that the Theorem holds in the case that $\mathrm{ad}X$ is
diagonalizable and $Y$ is an eigenvector of $\mathrm{ad}X$. If $\mathrm{ad}X$
is diagonalizable but $Y$ is not an eigenvector, then $Y$ is a linear
combination of eigenvectors and applying the above computation to each of
those eigenvectors gives the desired result.

We need, then, to consider the case where $\mathrm{ad}X$ is not
diagonalizable. But (Exercise \ref{ad.diagonal}), if $X$ is a diagonalizable
matrix, then $\mathrm{ad}X$ will be diagonalizable as an operator on
$\mathsf{gl}(n;\mathbb{C})$. Since, as we have already observed, every matrix
is the limit of diagonalizable matrices, we are essentially done. For it is
easy to see by differentiating the power series term-by-term that
$e^{-X}\left.  \tfrac{d}{dt}\right|  _{t=0}e^{X+tY}$ exists and varies
continuously with $X$. Thus once we have the Theorem for all diagonalizable
$X$ we have it for all $X$ by passing to the limit.

The only unresolved issue, then, is the interchange of limit and derivative
which we performed at the very beginning of the argument. I do not want to
spell this out in detail, but let us see what would be involved in justifying
this. A standard theorem in elementary analysis says that if $f_{n}%
(t)\rightarrow f(t)$ pointwise, and in addition $df_{n}/dt$ converges
uniformly to some function $g(t)$, then $f(t)$ is differentiable and
$df/dt=g(t)$. (E.g., Theorem 7.17 in W. Rudin's \textit{Principles of
Mathematical Analysis}.) The key requirement is that the \textit{derivatives}
converge uniformly. Uniform convergence of the $f_{n}$'s themselves is
definitely not sufficient.

In our case, $f_{n}(t)=e^{-X}\left(  e^{X/n}e^{tY/n}\right)  ^{n}$. The Lie
product formula says that this converges pointwise to $e^{-X}e^{X+tY}$. We
need, then, to show that
\[
\frac{d}{dt}e^{-X}\left(  e^{X/n}e^{tY/n}\right)  ^{n}%
\]
converges uniformly to some $g(t)$, say on the interval $-1\leq t\leq1$. This
computation is similar to what we did above, with relatively minor
modifications to account for the fact that we do not take $t=0$ and to make
sure the convergence is uniform. This part of the proof is left as an exercise
to the reader.
\end{proof}

\subsection{Proof of the Baker-Campbell-Hausdorff Formula}

We now turn to the proof of the Baker-Campbell-Hausdorff formula itself. Our
argument follows Miller, Sec. 5.1, with minor differences of convention.
(Warning: Miller's ``Ad'' is what we call ``ad.'') Define
\[
Z(t)=\log\left(  e^{X}e^{tY}\right)
\]
If $X$ and $Y$ are sufficiently small, then $Z\left(  t\right)  $ is defined
for $0\leq t\leq1$. It is left as an exercise to verify that $Z(t)$ is smooth.
Our goal is to compute $Z(1).$

By definition
\[
e^{Z(t)}=e^{X}e^{tY}%
\]
so that
\[
e^{-Z(t)}\frac d{dt}e^{Z(t)}=\left(  e^{X}e^{tY}\right)  ^{-1}e^{X}%
e^{tY}Y=Y\text{.}%
\]
On the other hand, by Theorem \ref{exp.derivative},
\[
e^{-Z(t)}\frac d{dt}e^{Z(t)}=\left\{  \frac{I-e^{-\mathrm{ad}Z(t)}%
}{\mathrm{ad}Z(t)}\right\}  \left(  \frac{dZ}{dt}\right)  \text{.}%
\]
Hence
\[
\left\{  \frac{I-e^{-\mathrm{ad}Z(t)}}{\mathrm{ad}Z(t)}\right\}  \left(
\frac{dZ}{dt}\right)  =Y\text{.}%
\]
If $X$ and $Y$ are small enough, then $Z(t)$ will also be small, so that
$\left(  I-e^{-\mathrm{ad}Z(t)}\right)  /\mathrm{ad}Z(t)$ will be close to the
identity and thus invertible. So
\begin{equation}
\frac{dZ}{dt}=\left\{  \frac{I-e^{-\mathrm{ad}Z(t)}}{\mathrm{ad}Z(t)}\right\}
^{-1}(Y)\text{.}\label{dzdt}%
\end{equation}

Recall that $e^{Z(t)}=e^{X}e^{tY}$. Applying the homomorphism `Ad' gives
\[
\mathrm{Ad}\left(  e^{Z(t)}\right)  =\mathrm{Ad}\left(  e^{X}\right)
\mathrm{Ad}\left(  e^{tY}\right)  \text{.}%
\]
By the relationship (\ref{Ad.ad}) between `Ad' and `ad,' this becomes
\[
e^{\mathrm{ad}Z(t)}=e^{\mathrm{ad}X}e^{t\mathrm{ad}Y}%
\]
or
\[
\mathrm{ad}Z(t)=\log\left(  e^{\mathrm{ad}X}e^{t\mathrm{ad}Y}\right)  \text{.}%
\]
Plugging this into (\ref{dzdt}) gives
\begin{equation}
\frac{dZ}{dt}=\left\{  \frac{I-\left(  e^{\mathrm{ad}X}e^{t\mathrm{ad}%
Y}\right)  ^{-1}}{\log\left(  e^{\mathrm{ad}X}e^{t\mathrm{ad}Y}\right)
}\right\}  ^{-1}(Y)\text{.}\label{dzdt2}%
\end{equation}

But now observe that
\[
g(z)=\left\{  \frac{1-z^{-1}}{\log z}\right\}  ^{-1}%
\]
so, formally, (\ref{dzdt2}) is the same as
\begin{equation}
\frac{dZ}{dt}=g\left(  e^{\mathrm{ad}X}e^{t\mathrm{ad}Y}\right)
(Y)\text{.}\label{dzdt3}%
\end{equation}
Reasoning as in the proof of Theorem \ref{exp.derivative} shows easily that
this formal argument is actually correct.

Now we are essentially done, for if we note that $Z(0)=X$ and integrate
(\ref{dzdt3}), we get
\[
Z(1)=X+\int_{0}^{1}g(e^{\mathrm{ad}X}e^{t\mathrm{ad}Y})(Y)\,dt
\]
which is the Baker-Campbell-Hausdorff formula.

\section{The Series Form of the Baker-Campbell-Hausdorff
Formula\label{series.section}}

Let us see how to get the first few terms of the series form of B-C-H from the
integral form. Recall the function
\begin{align*}
g\left(  z\right)   & =\frac{z\log z}{z-1}\\
& =\frac{\left[  1+\left(  z-1\right)  \right]  \left[  \left(  z-1\right)
-\frac{\left(  z-1\right)  ^{2}}2+\frac{\left(  z-1\right)  ^{3}}%
3\cdots\right]  }{\left(  z-1\right)  }\\
& =\left[  1+\left(  z-1\right)  \right]  \left[  1-\frac{z-1}2+\frac{\left(
z-1\right)  ^{2}}3\right]  \text{.}%
\end{align*}
Multiplying this out and combining terms gives
\[
g\left(  z\right)  =1+\frac12\left(  z-1\right)  -\frac16\left(  z-1\right)
^{2}+\cdots\text{.}%
\]
The closed-form expression for $g$ is
\[
g\left(  z\right)  =1+\sum_{n=1}^{\infty}\frac{\left(  -1\right)  ^{n+1}%
}{n\left(  n+1\right)  }\left(  z-1\right)  ^{n}\text{.}%
\]

Meanwhile
\begin{align*}
e^{\mathrm{ad}X}e^{t\mathrm{ad}Y}-I  & =\left(  I+\mathrm{ad}X+\frac{\left(
\mathrm{ad}X\right)  ^{2}}2+\cdots\right)  \left(  I+t\mathrm{ad}Y+\frac
{t^{2}\left(  \mathrm{ad}Y\right)  ^{2}}2+\cdots\right)  -I\\
& =\mathrm{ad}X+t\mathrm{ad}Y+t\mathrm{ad}X\mathrm{ad}Y+\frac{\left(
\mathrm{ad}X\right)  ^{2}}2+\frac{t^{2}\left(  \mathrm{ad}Y\right)  ^{2}%
}2+\cdots\text{.}%
\end{align*}
The crucial observation here is that $e^{\mathrm{ad}X}e^{t\mathrm{ad}Y}-I$ has
no zero-order term, just first-order and higher in $\mathrm{ad}X/\mathrm{ad}%
Y$. Thus $\left(  e^{\mathrm{ad}X}e^{t\mathrm{ad}Y}-I\right)  ^{n}$ will
contribute only terms of degree $n$ or higher in $\mathrm{ad}X/\mathrm{ad}Y$.

We have, then, up to degree two in $\mathrm{ad}X/\mathrm{ad}Y$%
\begin{align*}
g\left(  e^{\mathrm{ad}X}e^{t\mathrm{ad}Y}\right)   & =I+\frac{1}{2}\left[
\mathrm{ad}X+t\mathrm{ad}Y+t\mathrm{ad}X\mathrm{ad}Y+\frac{\left(
\mathrm{ad}X\right)  ^{2}}{2}+\frac{t^{2}\left(  \mathrm{ad}Y\right)  ^{2}}%
{2}+\cdots\right] \\
& -\frac{1}{6}\left[  \mathrm{ad}X+t\mathrm{ad}Y+\cdots\right]  ^{2}\\
& =I+\frac{1}{2}\mathrm{ad}X+\frac{t}{2}\mathrm{ad}Y+\frac{t}{2}%
\mathrm{ad}X\mathrm{ad}Y+\frac{\left(  \mathrm{ad}X\right)  ^{2}}{4}%
+\frac{t^{2}\left(  \mathrm{ad}Y\right)  ^{2}}{4}\\
& -\frac{1}{6}\left[  \left(  \mathrm{ad}X\right)  ^{2}+t^{2}\left(
\mathrm{ad}Y\right)  ^{2}+t\mathrm{ad}X\mathrm{ad}Y+t\mathrm{ad}%
Y\mathrm{ad}X\right] \\
& +\text{ higher-order terms.}%
\end{align*}
We now to apply $g\left(  e^{\mathrm{ad}X}e^{t\mathrm{ad}Y}\right)  $ to $Y$
and integrate. So (neglecting higher-order terms) by B-C-H, and noting that
any term with $adY$ acting first is zero:
\begin{align*}
& \log\left(  e^{X}e^{Y}\right) \\
& =X+\int_{0}^{1}\left[  Y+\frac{1}{2}\left[  X,Y\right]  +\frac{1}{4}\left[
X,\left[  X,Y\right]  \right]  -\frac{1}{6}\left[  X,\left[  X,Y\right]
\right]  -\frac{t}{6}\left[  Y,\left[  X,Y\right]  \right]  \right]  \,dt\\
& =X+Y+\frac{1}{2}\left[  X,Y\right]  +\left(  \frac{1}{4}-\frac{1}{6}\right)
\left[  X,\left[  X,Y\right]  \right]  -\frac{1}{6}\int_{0}^{1}t\,dt\left[
Y,\left[  X,Y\right]  \right]  \text{.}%
\end{align*}

Thus if we do the algebra we end up with
\begin{align*}
\log\left(  e^{X}e^{Y}\right)   & =X+Y+\frac12\left[  X,Y\right]  +\frac
1{12}\left[  X,\left[  X,Y\right]  \right]  -\frac1{12}\left[  Y,\left[
X,Y\right]  \right] \\
& +\text{ higher order terms.}%
\end{align*}
This is the expression in (\ref{bch.series}).

\section{Subgroups and Subalgebras}

Suppose that $G$ is a matrix Lie group, $H$ another matrix Lie group, and
suppose that $H\subset G$. Then certainly the Lie algebra $\frak{h}$ of $H$
will be a subalgebra of the Lie algebra $\frak{g}$ of $G$. Does this go the
other way around? That is given a Lie group $G$ with Lie algebra $\frak{g}$,
and a subalgebra $\frak{h}$ of $\frak{g}$, is there a matrix Lie group $H$
whose Lie algebra is $\frak{h}$?

In the case of the Heisenberg group, the answer is yes. This is easily seen
using the fact that the exponential mapping is one-to-one and onto, together
with the special form of the Baker-Campbell-Hausdorff formula. (See Exercise
\ref{sub.heisenberg}.)

Unfortunately, the answer in general is no. For example, let $G=\mathsf{GL}%
\left(  2;\mathbb{C}\right)  $ and let
\[
\frak{h}=\left\{  \left.  \left(
\begin{array}
[c]{cc}%
it & 0\\
0 & ita
\end{array}
\right)  \right|  t\in\mathbb{R}\right\}  \text{,}%
\]
where $a$ is irrational. If there is going to be a matrix Lie group $H$ with
Lie algebra $\frak{h}$, then $H$ would contain the set
\[
H_{0}=\left\{  \left.  \left(
\begin{array}
[c]{cc}%
e^{it} & 0\\
0 & e^{ita}%
\end{array}
\right)  \right|  t\in\mathbb{R}\right\}  \text{.}%
\]
To be a matrix Lie group, $H$ would have to be closed in $\mathsf{GL}\left(
2;\mathbb{C}\right)  $, and so it would contain the closure of $H_{0}$, which
(see ) is the set
\[
H_{1}=\left\{  \left.  \left(
\begin{array}
[c]{cc}%
e^{it} & 0\\
0 & e^{is}%
\end{array}
\right)  \right|  s,t\in\mathbb{R}\right\}  \text{.}%
\]
But then the Lie algebra of $H$ would have to contain the Lie algebra of
$H_{1}$, which is two-dimensional!

Fortunately, all is not lost. We can still get a subgroup $H$ for each
subalgebra $\frak{h}$, if we weaken the condition that $H$ be a matrix Lie
group. In the above example, the subgroup we want is $H_{0}$, despite the fact
that $H_{0}$ is not a matrix Lie group.

\begin{definition}
If $H$ is \textbf{any} subgroup of $\mathsf{GL}\left(  n;\mathbb{C}\right)  $,
define the Lie algebra $\frak{h}$ of $H$ to be the set of all matrices $X$
such that
\[
e^{tX}\in H
\]
for all real $t.$
\end{definition}

\begin{definition}
If $G$ is a matrix Lie group with Lie algebra $\frak{g}$, then $H$ is a
\textbf{connected Lie subgroup} of $G$ if

i) $H$ is a subgroup of $G$

ii) $H$ is connected

iii) the Lie algebra $\frak{h}$ of $H$ is a subspace of $\frak{g}$

iv) Every element of $H$ can be written in the form $e^{X_{1}}e^{X_{2}}\cdots
e^{X_{n}}$, with $X_{1},\cdots,X_{n}\in\frak{h}$.
\end{definition}

\begin{theorem}
If $G$ is a matrix Lie group with Lie algebra $\frak{g}$, and $H$ is a
connected Lie subgroup of $G$, then the Lie algebra $\frak{h}$ of $H$ is a
subalgebra of $\frak{g}$.
\end{theorem}

\begin{proof}
Since by definition $\frak{h}$ is a subspace of $\frak{g}$, it remains only to
show that $\frak{h}$ is closed under brackets. So assume $X,Y\in\frak{h}$.
Then $e^{tX}$ and $e^{sY}$ are in $H$, and so (since $H$ is a subgroup) is the
element
\[
e^{tX}e^{sY}e^{-tX}=\exp\left[  s\left(  e^{tX}Ye^{-tX}\right)  \right]
\text{.}%
\]
This shows that $e^{tX}Ye^{-tX}$ is in $\frak{h}$ for all $t$. But $\frak{h}$
is a subspace of $\frak{g}$, which is necessarily a closed subset of
$\frak{g}$. Thus
\[
\left[  X,Y\right]  =\left.  \frac{d}{dt}\right|  _{t=0}e^{tX}Ye^{-tX}%
=\lim_{h\rightarrow0}\frac{\left(  e^{hX}Ye^{-hX}-Y\right)  }{h}%
\]
is in $\frak{h}$. (This argument is precisely the one we used to show that the
Lie algebra of a matrix Lie group is a closed under brackets, once we had
established that it is a subspace.)
\end{proof}

We are now ready to state the main theorem of this section, which is our
second major application of the Baker-Campbell-Hausdorff formula.

\begin{theorem}
Let $G$ be a matrix Lie group with Lie algebra $\frak{g}$. Let $\frak{h}$ be a
Lie subalgebra of $\frak{g}$. Then there exists a unique connected Lie
subgroup $H$ of $G$ such that the Lie algebra of $H$ is $\frak{h}$.
\end{theorem}

Given a matrix Lie group $G$ and a subalgebra $\frak{h}$ of $\frak{g}$, the
associated connected Lie subgroup $H$ \textit{might} be a matrix Lie group.
This will happen precisely if $H$ is a closed subset of $G$. There are various
conditions under which you can prove that $H$ is closed. For example, if
$G=\mathsf{GL}\left(  n;\mathbb{C}\right)  $, and $\frak{h}$ is semisimple,
then $H$ is automatically closed, and hence a matrix Lie group. (See Helgason,
Chapter II, Exercises and Further Results, D.)

If only the Baker-Campbell-Hausdorff formula worked globally instead of only
locally the proof of this theorem would be easy. If the B-C-H formula
converged for all $X,Y$ we could just define $H$ to be the image of $\frak{h}$
under the exponential mapping. In that case B-C-H would show that this image
is a subgroup, since then we would have $e^{H_{1}}e^{H_{2}}=e^{Z},$ with
$Z=H_{1}+H_{2}+\frac{1}{2}\left[  H_{1},H_{2}\right]  +\cdots\in\frak{h}$
provided that $H_{1},H_{2}\in\frak{h}.$ Unfortunately, the B-C-H formula is
not convergent in general, and in general the image of $H$ under the
exponential mapping is not a subgroup.

\begin{proof}
Not written at this time.
\end{proof}

\section{Exercises}

\begin{enumerate}
\item \label{bch.special}The \textbf{center} of a Lie algebra $\frak{g}$ is
defined to be the set of all $X\in\frak{g}$ such that $\left[  X,Y\right]  =0$
for all $Y\in\frak{g}$. Now consider the Heisenberg group
\[
H=\left\{  \left(
\begin{array}
[c]{ccc}%
1 & a & b\\
0 & 1 & c\\
0 & 0 & 1
\end{array}
\right)  \left|  a,b,c\in\mathbb{R}\right.  \right\}
\]
with Lie algebra
\[
\frak{h=}\left\{  \left(
\begin{array}
[c]{ccc}%
0 & \alpha & \beta\\
0 & 0 & \gamma\\
0 & 0 & 0
\end{array}
\right)  \left|  \alpha,\beta,\gamma\in\mathbb{R}\right.  \right\}  \text{.}%
\]
Determine the center $Z(\frak{h)}$ of $\frak{h}$. For any $X,Y\in\frak{h}$,
show that $[X,Y]\in Z(\frak{h)}$. This implies, in particular that both $X$
and $Y$ commute with their commutator $[X,Y]$.

Show by direct computation that for any $X,Y\in\frak{h}$,
\begin{equation}
e^{X}e^{Y}=e^{X+Y+\tfrac{1}{2}[X,Y]}\text{.}\label{terminate}%
\end{equation}

\item  Let $X$ be a $n\times n$ complex matrix. Show that
\[
\frac{I-e^{-X}}{X}%
\]
is invertible if and only if $X$ has no eigenvalue of the form $\lambda=2\pi
in$, with $n$ an non-zero integer.

\textit{Hint}: When is $\left(  1-e^{-z}\right)  /z$ equal to zero?

\textit{Remark}: This exercise, combined with the formula in Theorem
\ref{exp.derivative}, gives the following result (in the language of
differentiable manifolds): The exponential mapping $\exp:\frak{g}\rightarrow
G$ is a local diffeomorphism near $X\in\frak{g}$ if and only $\mathrm{ad}X$
has no eigenvalue of the form $\lambda=2\pi in$, with $n$ a non-zero integer.

\item \label{bch.commute}Verify that the right side of the
Baker-Campbell-Hausdorff formula (\ref{bch.integral}) reduces to $X+Y$ in the
case that $X$ and $Y$ commute.

\textit{Hint}: Compute first $e^{\mathrm{ad}X}e^{t\mathrm{ad}Y}(Y)$ and
$\left(  e^{\mathrm{ad}X}e^{t\mathrm{ad}Y}-I\right)  (Y)$.

\item \label{by.series}Compute $\log\left(  e^{X}e^{Y}\right)  $ through third
order in $X/Y$ by using the power series for the exponential and the
logarithm. Show that you get the same answer as the Baker-Campbell-Hausdorff formula.

\item  Using the techniques in Section \ref{series.section}, compute the
series form of the Baker-Campbell-Hausdorff formula up through fourth-order
brackets. (We have already computed up through third-order brackets.)

\item \label{sub.heisenberg}Let $\frak{a}$ be a subalgebra of the Lie algebra
of the Heisenberg group. Show that $\exp\left(  \frak{a}\right)  $ is a
connected Lie subgroup of the Heisenberg group. Show that in fact $\exp\left(
\frak{a}\right)  $ is a matrix Lie group.

\item  Show that every connected Lie subgroup of $\mathsf{SU}\left(  2\right)
$ is closed. Show that this is not the case for $\mathsf{SU}\left(  3\right)
$.
\end{enumerate}

\chapter{Basic Representation Theory}

\section{Representations}

\begin{definition}
Let $G$ be a matrix Lie group. Then a \textbf{finite-dimensional complex
representation} of $G$ is a Lie group homomorphism
\[
\Pi:G\rightarrow\mathsf{GL}(n;\mathbb{C})
\]
($n\geq1$) or more generally a Lie group homomorphism
\[
\Pi:G\rightarrow\mathsf{GL}(V)
\]
where $V$ is a finite-dimensional complex vector space (with dim$(V)\geq1$). A
\textbf{finite-dimensional real representation} of $G$ is a Lie group
homomorphism $\Pi$ of $G$ into $\mathsf{GL}(n;\mathbb{R})$ or into
$\mathsf{GL}(V)$, where $V$ is a finite-dimensional real vector space.

If $\frak{g}$ is a real or complex Lie algebra, then a
\textbf{finite-dimensional complex representation} of $\frak{g}$ is a Lie
algebra homomorphism $\pi$ of $\frak{g}$ into $\mathsf{gl}(n;\mathbb{C})$ or
into \textsf{gl}$(V)$, where $V$ is a finite-dimensional complex vector space.
If $\frak{g}$ is a \textit{real} Lie algebra, then a
\textbf{finite-dimensional real representation} of $\frak{g}$ is a Lie algebra
homomorphism $\pi$ of $\frak{g}$ into $\mathsf{gl}(n;\mathbb{R})$ or into
$\mathsf{gl}(V)$.

If $\Pi$ or $\pi$ is a one-to-one homomorphism, then the representation is
called \textbf{faithful}.
\end{definition}

You should think of a representation as a (linear) \textbf{action} of a group
or Lie algebra on a vector space. (Since, say, to every $g\in G$ there is
associated an operator $\Pi(g)$, which acts on the vector space $V$.) In fact,
we will use terminology such as, ``Let $\Pi$ be a representation of $G$ acting
on the space $V$.'' Even if $\frak{g}$ is a real Lie algebra, we will consider
mainly complex representations of $\frak{g}$. After making a few more
definitions, we will discuss the question of why one should be interested in
studying representations.

\begin{definition}
Let $\Pi$ be a finite-dimensional real or complex representation of a matrix
Lie group $G$, acting on a space $V$. A subspace $W$ of $V$ is called
\textbf{invariant} if $\Pi(A)w\in W$ for all $w\in W$ and all $A\in G$. An
invariant subspace $W$ is called \textbf{non-trivial} if $W\neq\{0\}$ and
$W\neq V$. A representation with no non-trivial invariant subspaces is called
\textbf{irreducible}.

The terms \textbf{invariant}, \textbf{non-trivial}, and \textbf{irreducible}
are defined analogously for representations of Lie algebras.
\end{definition}

\begin{definition}
Let $G$ be a matrix Lie group, let $\Pi$ be a representation of $G$ acting on
the space $V$, and let $\Sigma$ be a representation of $G$ acting on the space
$W$. A linear map $\phi:V\rightarrow W$ is called a \textbf{morphism} (or
\textbf{intertwining map}) of representations if
\[
\phi(\Pi(A)v)=\Sigma(A)\phi(v)
\]
for all $A\in G$ and all $v\in V$. The analogous property defines morphisms of
representations of a Lie algebra.

If $\phi$ is a morphism of representations, and in addition $\phi$ is
invertible, then $\phi$ is said to be an \textbf{isomorphism} of
representations. If there exists an isomorphism between $V$ and $W$, then the
representations are said to be \textbf{isomorphic} (or \textbf{equivalent}).
\end{definition}

Two isomorphic representations should be regarded as being ``the same''
representation. A typical problem in representation theory is to determine, up
to isomorphism, all the irreducible representations of a particular group or
Lie algebra. In Section 5.4 we will determine all the finite-dimensional
complex irreducible representations of the Lie algebra $\mathsf{su}(2)$.

\begin{proposition}
\label{relating.rep}Let $G$ be a matrix Lie group with Lie algebra $\frak{g}$,
and let $\Pi$ be a (finite-dimensional real or complex) representation of $G$,
acting on the space $V$. Then there is a unique representation $\pi$ of
$\frak{g}$ acting on the same space such that
\[
\Pi(e^{X})=e^{\pi(X)}%
\]
for all $X\in\frak{g}$. The representation $\pi$ can be computed as
\[
\pi(X)=\left.  \frac{d}{dt}\right|  _{t=0}\Pi\left(  e^{tX}\right)
\]
and satisfies
\[
\pi\left(  AXA^{-1}\right)  =\Pi(A)\pi(X)\Pi(A)^{-1}%
\]
for all $X\in\frak{g}$ and all $A\in G$.
\end{proposition}

\begin{proof}
Theorem \ref{homo.theorem} in Chapter 3 states that for each Lie group
homomorphism $\phi:G\rightarrow H$ there is an associated Lie algebra
homomorphism $\widetilde{\phi}:\frak{g}\rightarrow\frak{h}$. Take
$H=\mathsf{GL}(V)$ and $\phi=\Pi$. Since the Lie algebra of $\mathsf{GL}(V)$
is $\mathsf{gl}(V)$ (since the exponential of any operator is invertible), the
associated Lie algebra homomorphism $\widetilde{\phi}=\pi$ maps from
$\frak{g}$ to $\mathsf{gl}(V)$, and so constitutes a representation of
$\frak{g}$.

The properties of $\pi$ follow from the properties of $\widetilde{\phi}$ given
in Theorem 6.
\end{proof}

\begin{proposition}
\label{rep.complex}Let $\frak{g}$ be a real Lie algebra, and $\frak{g}%
_{\mathbb{C}}$ its complexification. Then every finite-dimensional complex
representation $\pi$ of $\frak{g}$ has a unique extension to a
(complex-linear) representation of $\frak{g}_{\mathbb{C}}$, also denoted $\pi
$. The representation of $\frak{g}_{\mathbb{C}}$ satisfies
\[
\pi(X+iY)=\pi(X)+i\pi(Y)
\]
for all $X\in\frak{g}$.
\end{proposition}

\begin{proof}
This follows from Exercise \ref{extend.lie.homo} of Chapter 3.
\end{proof}

\begin{definition}
Let $G$ be a matrix Lie group, let $\mathcal{H}$ be a Hilbert space, and let
$U(\mathcal{H})$ denote the group of unitary operators on $\mathcal{H}$. Then
a homomorphism $\Pi:G\rightarrow U(\mathcal{H})$ is called a \textbf{unitary
representation} of $G$ if $\Pi$ satisfies the following continuity condition:
If $A_{n},A\in G$ and $A_{n}\rightarrow A$, then
\[
\Pi(A_{n})v\rightarrow\Pi(A)v
\]
for all $v\in\mathcal{H}$. A unitary representation with no non-trivial
\textit{closed} invariant subspaces is called \textbf{irreducible}.
\end{definition}

This continuity condition is called \textbf{strong continuity}. One could
require the even stronger condition that $\left\|  \Pi(A_{n})-\Pi(A)\right\|
\rightarrow0$, but this turns out to be too stringent a requirement. (That is,
most of the interesting representations of $G$ will not have this stronger
continuity condition.) In practice, any homomorphism of $G$ into
$U(\mathcal{H})$ you can write down explicitly will be strongly continuous.

One could try to define some analog of unitary representations for Lie
algebras, but there are serious technical difficulties associated with getting
the ``right'' definition.

\section{Why Study Representations?}

If a representation $\Pi$ is a faithful representation of a matrix Lie group
$G$, then $\left\{  \Pi(A)\left|  A\in G\right.  \right\}  $ is a group of
matrices which is isomorphic to the original group $G$. Thus $\Pi$ allows us
to \textit{represent} $G$ as a group of matrices. This is the motivation for
the term representation. (Of course, we still call $\Pi$ a representation even
if it is not faithful.)

Despite the origin of the term, the point of representation theory is
\textit{not} (at least in this course) to represent a group as a group of
matrices. After all, all of our groups are already matrix groups! While it
might seem redundant to study representations of a group which is already
represented as a group of matrices, this is precisely what we are going to do.

The reason for this is that a representation can be thought of (as we have
already noted) as an action of our group on some vector space. Such actions
(representations) arise naturally in many branches of both mathematics and
physics, and it is important to understand them.

A typical example would be a differential equation in three-dimensional space
which has rotational symmetry. If the equation has rotational symmetry, then
the space of solutions will be invariant under rotations. Thus the space of
solutions will constitute a representation of the rotation group
$\mathsf{SO}(3)$. If you know what all of the representations of
$\mathsf{SO}(3)$ are, this can help immensely in narrowing down what the space
of solutions can be. (As we will see, $\mathsf{SO}(3)$ has lots of other
representations besides the obvious one in which $\mathsf{SO}(3)$ acts on
$\mathbb{R}^{3}$.)

In fact, one of the chief applications of representation theory is to exploit
symmetry. If a system has symmetry, then the set of symmetries will form a
group, and understanding the representations of the symmetry group allows you
to use that symmetry to simplify the problem.

In addition, studying the representations of a group $G$ (or of a Lie algebra
$\frak{g}$) can give information about the group (or Lie algebra) itself. For
example, if $G$ is a \textit{finite} group, then associated to $G $ is
something called the \textbf{group algebra}. The structure of this group
algebra can be described very nicely in terms of the irreducible
representations of $G$.

In this course, we will be interested primarily in computing the
finite-dimensional irreducible complex representations of matrix Lie groups.
As we shall see, this problem can be reduced almost completely to the problem
of computing the finite-dimensional irreducible complex representations of the
associated Lie algebra. In this chapter, we will discuss the theory at an
elementary level, and will consider in detail the example of $\mathsf{SO}(3)$
and $\mathsf{SU}(2)$. In Chapter 6, we will study the representations of
$\mathsf{SU}(3)$, which is substantially more involved than that of
$\mathsf{SU}(2)$, and give an overview of the representation theory of a very
important class of Lie groups, namely, the semisimple ones.

\section{Examples of Representations\label{examples.rep}}

\subsection{The Standard Representation}

A matrix Lie group $G$ is by definition a subset of some $\mathsf{GL}%
(n;\mathbb{R})$ or $\mathsf{GL}(n;\mathbb{C})$. The inclusion map of $G$ into
$\mathsf{GL}(n)$ (i.e., $\Pi(A)=A$) is a representation of $G$, called the
\textbf{standard representation} of $G$. Thus for example the standard
representation of $\mathsf{SO}(3)$ is the one in which $\mathsf{SO}(3)$ acts
in the usual way on $\mathbb{R}^{3}$. If $G$ is a subgroup of $\mathsf{GL}%
(n;\mathbb{R})$ or $\mathsf{GL}(n;\mathbb{C})$, then its Lie algebra
$\frak{g}$ will be a subalgebra of $\mathsf{gl}(n;\mathbb{R})$ or
$\mathsf{gl}(n;\mathbb{C})$. The inclusion of $\frak{g}$ into $\mathsf{gl}%
(n;\mathbb{R})$ or $\mathsf{gl}(n;\mathbb{C})$ is a representation of
$\frak{g}$, called the \textbf{standard representation}.

\subsection{The Trivial Representation}

Consider the one-dimensional complex vector space $\mathbb{C}$. Given any
matrix Lie group $G$, we can define the \textbf{trivial representation} of $G
$, $\Pi:G\rightarrow\mathsf{GL}(1;\mathbb{C})$, by the formula
\[
\Pi(A)=I
\]
for all $A\in G$. Of course, this is an irreducible representation, since
$\mathbb{C}$ has no non-trivial subspaces, let alone non-trivial invariant
subspaces. If $\frak{g}$ is a Lie algebra, we can also define the
\textbf{trivial representation} of $\frak{g}$, $\pi:\frak{g}\rightarrow
\mathsf{gl}(1;\mathbb{C})$, by
\[
\pi(X)=0
\]
for all $X\in\frak{g}$. This is an irreducible representation.

\subsection{The Adjoint Representation}

Let $G$ be a matrix Lie group with Lie algebra $\frak{g}$. We have already
defined the adjoint mapping
\[
\mathrm{Ad}:G\rightarrow\mathsf{GL}(\frak{g)}%
\]
by the formula
\[
\mathsf{Ad}A(X)=AXA^{-1}\text{.}%
\]
Recall that \textit{Ad} is a Lie group homomorphism. Since \textit{Ad} is a
Lie group homomorphism into a group of invertible operators, we see that in
fact \textit{Ad} is a representation of $G$, acting on the space $\frak{g}$.
Thus we can now give \textit{Ad} its proper name, the \textbf{adjoint
representation} of $G$. The adjoint representation is a real representation of
$G$.

Similarly, if $\frak{g}$ is a Lie algebra, we have
\[
\mathrm{ad}:\frak{g}\rightarrow\mathsf{gl}(\frak{g})
\]
defined by the formula
\[
\mathsf{ad}X(Y)=[X,Y]\text{.}%
\]
We know that \textit{ad} is a Lie algebra homomorphism (Chapter 3, Proposition
\ref{ad.homo}), and is therefore a representation of $\frak{g}$, called the
\textbf{adjoint representation}. In the case that $\frak{g}$ is the Lie
algebra of some matrix Lie group $G$, we have already established (Chapter 3,
Proposition \ref{differentiate.Ad} and Exercise \ref{ad.expand}) that
\textit{Ad} and \textit{ad} are related as in Proposition \ref{relating.rep}.

Note that in the case of $\mathsf{SO}(3)$ the standard representation and the
adjoint representation are both three dimensional real representations. In
fact these two representations are equivalent (Exercise \ref{equiv.rep}).

\subsection{Some Representations of $\mathsf{SU}(2)$}

Consider the space $V_{m}$ of homogeneous polynomials in two complex variables
with total degree $m$ ($m\geq0$). That is, $V_{m}$ is the space of functions
of the form
\begin{equation}
f(z_{1},z_{2})=a_{0}z_{1}^{m}+a_{1}z_{1}^{m-1}z_{2}+a_{2}z_{1}^{m-2}z_{2}%
^{2}\cdots+a_{m}z_{2}^{m}\label{polynomial}%
\end{equation}
with $z_{1},z_{2}\in\mathbb{C}$ and the $a_{i}$'s arbitrary complex constants.
The space $V_{m}$ is an $(m+1)$-dimensional complex vector space.

Now by definition an element $U$ of $\mathsf{SU}(2)$ is a linear
transformation of $\mathbb{C}^{2}$. Let $z$ denote the pair $z=(z_{1},z_{2})$
in $\mathbb{C}^{2}$. Then we may define a linear transformation $\Pi_{m}(U)$
on the space $V_{m}$ by the formula
\begin{equation}
\left[  \Pi_{m}(U)f\right]  (z)=f(U^{-1}z)\text{.}\label{su2.rep}%
\end{equation}
Explicitly, if $f$ is as in (\ref{polynomial}), then
\[
\left[  \Pi_{m}(U)f\right]  (z_{1},z_{2})=\sum_{k=0}^{m}a_{k}\left(
U_{11}^{-1}z_{1}+U_{12}^{-1}z_{2}\right)  ^{m-k}\left(  U_{21}^{-1}%
z_{1}+U_{22}^{-1}z_{2}\right)  ^{k}\text{.}%
\]
By expanding out the right side of this formula we see that $\Pi_{m}(U)f$ is
again a homogeneous polynomial of degree $m$. Thus $\Pi_{m}(U)$ actually maps
$V_{m}$ into $V_{m}$.

Now, compute
\begin{align*}
\Pi_{m}\left(  U_{1}\right)  \left[  \Pi_{m}\left(  U_{2}\right)  f\right]
(z)  & =\left[  \Pi_{m}\left(  U_{2}\right)  f\right]  (U_{1}^{-1}z)=f\left(
U_{2}^{-1}U_{1}^{-1}z\right) \\
& =\Pi_{m}\left(  U_{1}U_{2}\right)  f(z)\text{.}%
\end{align*}
Thus $\Pi_{m}$ is a (finite-dimensional complex) representation of
$\mathsf{SU}(2)$. (It is very easy to do the above computation incorrectly.)
The inverse in definition (\ref{su2.rep}) is necessary in order to make
$\Pi_{m}$ a representation. It turns out that each of the representations
$\Pi_{m}$ of $\mathsf{SU}(2)$ is irreducible, and that every
finite-dimensional irreducible representation of $\mathsf{SU}(2)$ is
equivalent to one (and only one) of the $\Pi_{m}$'s. (Of course, no two of the
$\Pi_{m}$'s are equivalent, since they don't even have the same dimension.)

Let us now compute the corresponding Lie algebra representation $\pi_{m}$.
According to Proposition \ref{relating.rep}, $\pi_{m}$ can be computed as
\[
\pi_{m}(X)=\left.  \frac d{dt}\right|  _{t=0}\Pi_{m}\left(  e^{tX}\right)
\text{.}%
\]
So
\[
\left(  \pi_{m}(X)f\right)  (z)=\left.  \frac d{dt}\right|  _{t=0}f\left(
e^{-tX}z\right)  \text{.}%
\]
Now let $z(t)$ be the curve in $\mathbb{C}^{2}$ defined as $z(t)=e^{-tX}z$, so
that $z(0)=z$. Of course, $z(t)$ can be written as $z(t)=(z_{1}(t),z_{2}(t))$,
with $z_{i}(t)\in\mathbb{C}$. By the chain rule,
\[
\pi_{m}(X)f=\frac{\partial f}{\partial z_{1}}\left.  \frac{dz_{1}}{dt}\right|
_{t=0}+\frac{\partial f}{\partial z_{2}}\left.  \frac{dz_{2}}{dt}\right|
_{t=0}\text{.}%
\]
But $\left.  dz/dt\right|  _{t=0}=-Xz$, so we obtain the following formula for
$\pi_{m}(X)$
\begin{equation}
\pi_{m}(X)f=-\frac{\partial f}{\partial z_{1}}\left(  X_{11}z_{1}+X_{12}%
z_{2}\right)  -\frac{\partial f}{\partial z_{2}}\left(  X_{21}z_{1}%
+X_{22}z_{2}\right)  \text{.}\label{diff.formula}%
\end{equation}

Now, according to Proposition \ref{rep.complex}, every finite-dimensional
complex representation of the Lie algebra $\mathsf{su}(2)$ extends uniquely to
a complex-linear representation of the complexification of $\mathsf{su}(2)$.
But the complexification of $\mathsf{su}(2)$ is (isomorphic to) $\mathsf{sl}%
(2;\mathbb{C})$ (Chapter 3, Proposition \ref{complex.classical}). To see that
this is so, note that $\mathsf{sl}(2;\mathbb{C})$ is the space of all
$2\times2$ complex matrices with trace zero. But if $X$ is in $\mathsf{sl}%
(2;\mathbb{C})$, then
\[
X=\frac{X-X^{\ast}}{2}+\frac{X+X^{\ast}}{2}=\frac{X-X^{\ast}}{2}%
+i\frac{X+X^{\ast}}{2i}%
\]
where both $(X-X^{\ast})/2$ and $(X+X^{\ast})/2i$ are in $\mathsf{su}(2)$.
(Check!) It is easy to see that this decomposition is unique, so that every
$X\in\mathsf{sl}(2;\mathbb{C})$ can be written uniquely as $X=X_{1}+iY_{1}$
with $X_{1},Y_{1}\in\mathsf{su}(2)$. Thus $\mathsf{sl}(2;\mathbb{C})$ is
isomorphic as a vector space to $\mathsf{su}(2)_{\mathbf{C}}$. But this is in
fact an isomorphism of Lie algebras, since in both cases
\[
\left[  X_{1}+iY_{1},X_{2}+iY_{2}\right]  =\left[  X_{1},X_{2}\right]
-\left[  Y_{1},Y_{2}\right]  +i\left(  \left[  X_{1},Y_{2}\right]  +\left[
X_{2},Y_{1}\right]  \right)  \text{.}%
\]
(See Exercise \ref{su2.sl2}.)

So, the representation $\pi_{m}$ of $\mathsf{su}(2)$ given by
(\ref{diff.formula}) extends to a representation of $\mathsf{sl}%
(2;\mathbb{C})$, which we will also call $\pi_{m}$. I assert that in fact
formula (\ref{diff.formula}), still holds for $X\in\mathsf{sl}(2;\mathbb{C})$.
Why is this? Well, (\ref{diff.formula}) is undoubtedly (complex) linear, and
it agrees with the original $\pi_{m}$ for $X\in\mathsf{su}(2)$. But there is
only one complex linear extension of $\pi_{m}$ from $\mathsf{su}(2)$ to
$\mathsf{sl}(2;\mathbb{C})$, so this must be it!

So, for example, consider the element
\[
H=\left(
\begin{array}
[c]{cc}%
1 & 0\\
0 & -1
\end{array}
\right)
\]
in the Lie algebra $\mathsf{sl}(2;\mathbb{C})$. Applying formula
(\ref{diff.formula}) gives
\[
\left(  \pi_{m}(H)f\right)  (z)=-\frac{\partial f}{\partial z_{1}}z_{1}%
+\frac{\partial f}{\partial z_{2}}z_{2}\text{.}%
\]
Thus we see that
\begin{equation}
\pi_{m}(H)=-z_{1}\frac\partial{\partial z_{1}}+z_{2}\frac\partial{\partial
z_{2}}\text{.}\label{pi.h}%
\end{equation}
Applying $\pi_{m}(H)$ to a basis element $z_{1}^{k}z_{2}^{m-k}$ we get
\[
\pi_{m}(H)z_{1}^{k}z_{2}^{m-k}=-kz_{1}^{k}z_{2}^{m-k}+(m-k)z_{1}^{k}%
z_{2}^{m-k}=(m-2k)z_{1}^{k}z_{2}^{m-k}\text{.}%
\]
Thus $z_{1}^{k}z_{2}^{m-k}$ is an eigenvector for $\pi_{m}(H)$ with eigenvalue
$(m-2k)$. In particular, $\pi_{m}(H)$ is diagonalizable.

Let $X$ and $Y$ be the elements
\[%
\begin{array}
[c]{cc}%
X=\left(
\begin{array}
[c]{cc}%
0 & 1\\
0 & 0
\end{array}
\right)  ; & Y=\left(
\begin{array}
[c]{cc}%
0 & 0\\
1 & 0
\end{array}
\right)
\end{array}
\]
in $\mathsf{sl}(2;\mathbb{C})$. Then (\ref{diff.formula}) tells us that
\[%
\begin{array}
[c]{cc}%
\pi_{m}(X)=-z_{2}\frac{\partial}{\partial z_{1}}; & \pi_{m}(Y)=-z_{1}%
\frac{\partial}{\partial z_{2}}%
\end{array}
\]
so that
\begin{align}
\pi_{m}(X)z_{1}^{k}z_{2}^{m-k}  & =-kz_{1}^{k-1}z_{2}^{m-k+1}\nonumber\\
\pi_{m}(Y)z_{1}^{k}z_{2}^{m-k}  & =(k-m)z_{1}^{k+1}z_{2}^{m-k-1}%
\text{.}\label{poly.act}%
\end{align}

\begin{proposition}
\label{poly.irred}The representation $\pi_{m}$ is an irreducible
representation of $\mathsf{sl}(2;\mathbb{C})$.
\end{proposition}

\begin{proof}
It suffices to show that every non-zero invariant subspace of $V_{m}$ is in
fact equal to $V_{m}$. So let $W$ be such a space. Since $W$ is assumed
non-zero, there is at least one non-zero element $w$ in $W$. Then $w$ can be
written uniquely in the form
\[
w=a_{0}z_{1}^{m}+a_{1}z_{1}^{m-1}z_{2}+a_{2}z_{1}^{m-2}z_{2}^{2}\cdots
+a_{m}z_{2}^{m}%
\]
with at least one of the $a_{k}$'s non-zero. Let $k_{0}$ be the largest value
of $k$ for which $a_{k}\neq0$, and consider
\[
\pi_{m}(X)^{k_{0}}w\text{.}%
\]
Since (by (\ref{poly.act})) each application of $\pi_{m}(X)$ lowers the power
of $z_{1}$ by 1, $\pi_{m}(X)^{k_{0}}$ will kill all the terms in $w$ whose
power of $z_{1}$ is less than $k_{0}$, that is, all except the $a_{k_{0}}%
z_{1}^{k_{0}}z_{2}^{m-k_{0}}$ term. On the other hand, we compute easily that
\[
\pi_{m}(X)^{k_{0}}\left(  a_{k_{0}}z_{1}^{k_{0}}z_{2}^{m-k_{0}}\right)
=k_{0}!(-1)^{k_{0}}a_{k_{0}}z_{2}^{m}\text{.}%
\]

We see, then, that $\pi_{m}(X)^{k_{0}}w$ is a \textit{non-zero} multiple of
$z_{2}^{m}$. Since $W$ is assumed invariant, $W$ must contain this multiple of
$z_{2}^{m}$, and so also $z_{2}^{m}$ itself.

But now it follows from (\ref{poly.act}) that $\pi_{m}(Y)^{k}z_{2}^{m}$ is a
\textit{non-zero} multiple of $z_{1}^{k}z_{2}^{m-k}$. Therefore $W$ must also
contain $z_{1}^{k}z_{2}^{m-k}$ for all $0\leq k\leq m$. Since these elements
form a basis for $V_{m}$, we see that in fact $W=V_{m}$, as desired.
\end{proof}

\subsection{Two Unitary Representations of $\mathsf{SO}(3)$}

Let $\mathcal{H}=L^{2}(\mathbb{R}^{3},dx)$. For each $R\in\mathsf{SO}(3)$,
define an operator $\Pi_{1}(R)$ on $\mathcal{H}$ by the formula
\[
\left[  \Pi_{1}(R)f\right]  (x)=f\left(  R^{-1}x\right)  \text{.}%
\]
Since Lebesgue measure $dx$ is rotationally invariant, $\Pi_{1}(R)$ is a
unitary operator for each $R\in\mathsf{SO}(3)$. The calculation of the
previous subsection shows that the map $R\rightarrow\Pi_{1}(R)$ is a
homomorphism of $\mathsf{SO}(3)$ into $U(\mathcal{H})$. This map is strongly
continuous, and hence constitutes a unitary representation of $\mathsf{SO}(3)
$.

Similarly, we may consider the unit sphere $S^{2}\subset\mathbb{R}^{3}$, with
the usual surface measure $\Omega$. Of course, any $R\in\mathsf{SO}(3)$ maps
$S^{2}$ into $S^{2}$. For each $R$ we can define $\Pi_{2}(R)$ acting on
$L^{2}(S^{2},d\Omega)$ by
\[
\left[  \Pi_{2}(R)f\right]  (x)=f\left(  R^{-1}x\right)  \text{.}%
\]
Then $\Pi_{2}$ is a unitary representation of $\mathsf{SO}(3)$.

Neither of the unitary representations $\Pi_{1}$ and $\Pi_{2}$ is irreducible.
In the case of $\Pi_{2}$, $L^{2}(S^{2},d\Omega)$ has a very nice decomposition
as the orthogonal direct sum of finite-dimensional invariant subspaces. This
decomposition is the theory of ``spherical harmonics,'' which are well known
in the physics (and mathematics) literature.

\subsection{A Unitary Representation of the Reals}

Let $\mathcal{H}=L^{2}(\mathbb{R},dx)$. For each $a\in\mathbb{R}$, define
$T_{a}:\mathcal{H}\rightarrow\mathcal{H}$ by
\[
\left(  T_{a}f\right)  (x)=f(x-a)\text{.}%
\]
Clearly $T_{a}$ is a unitary operator for each $a\in\mathbb{R}$, and clearly
$T_{a}T_{b}=T_{a+b}$. The map $a\rightarrow T_{a}$ is strongly continuous, so
$T$ is a unitary representation of $\mathbb{R}$. This representation is not
irreducible. The theory of the Fourier transform allows you to determine all
the closed, invariant subspaces of $\mathcal{H}$ (W. Rudin, \textit{Real and
Complex Analysis}, Theorem 9.17).

\subsection{The Unitary Representations of the Real Heisenberg Group}

Consider the Heisenberg group
\[
H=\left\{  \left(
\begin{array}
[c]{ccc}%
1 & a & b\\
0 & 1 & c\\
0 & 0 & 1
\end{array}
\right)  \left|  a,b,c\in\mathbb{R}\right.  \right\}  \text{.}%
\]
Now consider a real, non-zero constant, which for reasons of historical
convention we will call $\hbar$ (``aitch-bar''). Now for each $\hbar
\in\mathbb{R}\backslash\{0\}$, define a unitary operator $\Pi_{\hbar}$ on
$L^{2}(\mathbb{R},dx)$ by
\begin{equation}
\Pi_{\hbar}\left(
\begin{array}
[c]{ccc}%
1 & a & b\\
0 & 1 & c\\
0 & 0 & 1
\end{array}
\right)  f=e^{-i\hbar b}e^{i\hbar cx}f(x-a)\text{.}\label{h.bar}%
\end{equation}
It is clear that the right side of (\ref{h.bar}) has the same norm as $f$, so
$\Pi_{\hbar}$ is indeed unitary.

Now compute
\begin{align*}
& \Pi_{\hbar}\left(
\begin{array}
[c]{ccc}%
1 & \widetilde{a} & \widetilde{b}\\
0 & 1 & \widetilde{c}\\
0 & 0 & 1
\end{array}
\right)  \Pi_{\hbar}\left(
\begin{array}
[c]{ccc}%
1 & a & b\\
0 & 1 & c\\
0 & 0 & 1
\end{array}
\right)  f\\
& =e^{-i\hbar\widetilde{b}}e^{i\hbar\widetilde{c}x}e^{-i\hbar b}e^{i\hbar
c(x-\widetilde{a})}f(x-\widetilde{a}-a)\\
& =e^{-i\hbar(\widetilde{b}+b+c\widetilde{a})}e^{i\hbar(\widetilde{c}%
+c)x}f\left(  x-(\widetilde{a}+a)\right)  \text{.}%
\end{align*}
This shows that the map $A\rightarrow\Pi_{\hbar}(A)$ is a homomorphism of the
Heisenberg group into $U\left(  L^{2}(\mathbb{R}\mathbf{)}\right)  $. This map
is strongly continuous, and so $\Pi_{\hbar}$ is a unitary representation of
$H$.

Note that a typical unitary operator $\Pi_{\hbar}(A)$ consists of first
translating $f$, then multiplying $f$ by the function $e^{i\hbar cx}$, and
then multiplying $f$ by the constant $e^{-i\hbar b}$. Multiplying $f$ by the
function $e^{i\hbar cx}$ has the effect of translating the Fourier transform
of $f$, or in physical language, ``translating $f$ in momentum space.'' Now,
if $U_{1}$ is an ordinary translation and $U_{2}$ is a translation of the
Fourier transform (i.e., $U_{2}=$ multiplication by some $e^{i\hbar cx}$),
then $U_{1}$ and $U_{2}$ will not commute, but $U_{1}U_{2}U_{1}^{-1}U_{2}%
^{-1}$ will be simply multiplication by a constant of absolute value one. Thus
$\left\{  \Pi_{\hbar}(A)\left|  A\in H\right.  \right\}  $ is the group of
operators on $L^{2}(\mathbb{R}\mathbf{)}$ generated by ordinary translations
and translations in Fourier space. It is this representation of the Heisenberg
group which motivates its name. (See also Exercise \ref{ccr}.)

It follows fairly easily from standard Fourier transform theory (e.g., W.
Rudin, \textit{Real and Complex Analysis}, Theorem 9.17) that for each
$\hbar\in\mathbb{R}\backslash\{0\}$ the representation $\Pi_{\hbar}$ is
irreducible. Furthermore, these are (up to equivalence) almost all of the
irreducible unitary representations of $H$. The only remaining ones are the
one-dimensional representations $\Pi_{\alpha,\beta}$%
\[
\Pi_{\alpha,\beta}\left(
\begin{array}
[c]{ccc}%
1 & a & b\\
0 & 1 & c\\
0 & 0 & 1
\end{array}
\right)  =e^{i(\alpha a+\beta c)}I
\]
with $\alpha,\beta\in\mathbb{R}$. (The $\Pi_{\alpha,\beta}$'s are the
irreducible unitary representations in which the center of $H$ acts
trivially.) The fact that $\Pi_{\hbar}$'s and the $\Pi_{\alpha,\beta}$'s are
all of the (strongly continuous) irreducible unitary representations of $H$ is
closely related to the celebrated Stone-Von Neumann theorem in mathematical
physics. See, for example, M. Reed and B. Simon, \textit{Methods of Modern
Mathematical Physics}, Vol. 3, Theorem XI.84. See also Exercise
\ref{heisenberg.p}.

\section{The Irreducible Representations of $\mathsf{su}(2)$}

In this section we will compute (up to equivalence) all the finite-dimensional
irreducible complex representations of the Lie algebra $\mathsf{su}(2)$. This
computation is important for several reasons. In the first place,
$\mathsf{su}(2)\cong\mathsf{so}(3)$, and the representations of $\mathsf{so}%
(3)$ are of physical significance. (The computation we will do here is found
in every standard textbook on quantum mechanics, under the heading ``angular
momentum.'') In the second place, the representation theory of $\mathsf{su}%
(2)$ is an illuminating example of how one uses commutation relations to
determine the representations of a Lie algebra. In the third place, in
determining the representations of general semisimple Lie algebras (Chapter
6), we will explicitly use the representation theory of $\mathsf{su}(2)$.

Now, every finite-dimensional complex representation $\pi$ of $\mathsf{su}(2)$
extends by Prop. \ref{rep.complex} to a complex-linear representation (also
called $\pi$) of the complexification of $\mathsf{su}(2)$, namely
$\mathsf{sl}(2;\mathbb{C})$.

\begin{proposition}
Let $\pi$ be a complex representation of $\mathsf{su}(2)$, extended to a
complex-linear representation of $\mathsf{sl}(2;\mathbb{C})$. Then $\pi$ is
irreducible as a representation of $\mathsf{su}(2)$ if and only if it is
irreducible as a representation of $\mathsf{sl}(2;\mathbb{C})$.
\end{proposition}

\begin{proof}
Let us make sure we are clear about what this means. Suppose that $\pi$ is a
complex representation of the (real) Lie algebra $\mathsf{su}(2)$, acting on
the complex space $V$. Then saying that $\pi$ is irreducible means that there
is no non-trivial invariant \textit{complex} subspace $W\subset V$. That is,
even though $\mathsf{su}(2)$ is a real Lie algebra, when considering complex
representations we are interested only in complex invariant subspaces.

Now, suppose that $\pi$ is irreducible as a representation of $\mathsf{su}%
(2)$. If $W$ is a (complex) subspace of $V$ which is invariant under
$\mathsf{sl}(2;\mathbb{C})$, then certainly $W$ is invariant under
$\mathsf{su}(2)\subset\mathsf{sl}(2;\mathbb{C})$. Therefore $W=\{0\}$ or
$W=V$. Thus $\pi$ is irreducible as a representation of $\mathsf{sl}%
(2;\mathbb{C})$.

On the other hand, suppose that $\pi$ is irreducible as a representation of
$\mathsf{sl}(2;\mathbb{C})$, and suppose that $W$ is a (complex) subspace of
$V$ which is invariant under $\mathsf{su}(2)$. Then $W$ will also be invariant
under $\pi(X+iY)=\pi(X)+i\pi(Y)$, for all $X,Y\in\mathsf{su}(2)$. Since every
element of $\mathsf{sl}(2;\mathbb{C})$ can be written as $X+iY$, we conclude
that in fact $W$ is invariant under $\mathsf{sl}(2;\mathbb{C})$. Thus
$W=\{0\}$ or $W=V$, so $\pi$ is irreducible as a representation of
$\mathsf{su}(2)$.
\end{proof}

We see, then that studying the irreducible representations of $\mathsf{su}(2)
$ is equivalent to studying the irreducible representations of $\mathsf{sl}%
(2;\mathbb{C})$. Passing to the complexified Lie algebra makes our
computations easier.

We will use the following basis for $\mathsf{sl}(2;\mathbb{C})$:
\[%
\begin{array}
[c]{ccc}%
H=\left(
\begin{array}
[c]{cc}%
1 & 0\\
0 & -1
\end{array}
\right)  ; & X=\left(
\begin{array}
[c]{cc}%
0 & 1\\
0 & 0
\end{array}
\right)  ; & Y=\left(
\begin{array}
[c]{cc}%
0 & 0\\
1 & 0
\end{array}
\right)
\end{array}
\]
which have the commutation relations
\[%
\begin{array}
[c]{rrr}%
\lbrack H,X] & = & 2X\\
\lbrack H,Y] & = & -2Y\\
\lbrack X,Y] & = & H
\end{array}
\text{.}%
\]
If $V$ is a (finite-dimensional complex) vector space, and $A,B,$ and $C$ are
operators on $V$ satisfying
\[%
\begin{array}
[c]{rrr}%
\lbrack A,B] & = & 2B\\
\lbrack A,C] & = & -2C\\
\lbrack B,C] & = & A
\end{array}
\]
then because of the skew-symmetry and bilinearity of brackets, the linear map
$\pi:\mathsf{sl}(2;\mathbb{C})\rightarrow\mathsf{gl}(V)$ satisfying
\[%
\begin{array}
[c]{ccc}%
\pi(H)=A; & \pi(X)=B; & \pi(Y)=C
\end{array}
\]
will be a representation of $\mathsf{sl}(2;\mathbb{C})$.

\begin{theorem}
\label{classify.rep}For each integer $m\geq0$, there is an irreducible
representation of $\mathsf{sl}(2;\mathbb{C})$ with dimension $m+1$. Any two
irreducible representations of $\mathsf{sl}(2;\mathbb{C})$ with the same
dimension are equivalent. If $\pi$ is an irreducible representation of
$\mathsf{sl}(2;\mathbb{C})$ with dimension $m+1$, then $\pi$ is equivalent to
the representation $\pi_{m}$ described in Section \ref{examples.rep}.
\end{theorem}

\begin{proof}
Let $\pi$ be an irreducible representation of $\mathsf{sl}(2;\mathbb{C})$
acting on a (finite-dimensional complex) space $V$. Our strategy is to
diagonalize the operator $\pi(H)$. Of course, \textit{a priori}, we don't know
that $\pi(H)$ is diagonalizable. However, because we are working over the
(algebraically closed) field of complex numbers, $\pi(H)$ must have at least
one eigenvector.
\end{proof}

\begin{proof}
The following lemma is the key to the entire proof.

\begin{lemma}
\label{raising.op}Let $u$ be an eigenvector of $\pi(H)$ with eigenvalue
$\alpha\in\mathbb{C}$. Then
\[
\pi(H)\pi(X)u=(\alpha+2)\pi(X)u\text{.}%
\]
Thus either $\pi(X)u=0$, or else $\pi(X)u$ is an eigenvector for $\pi(H)$ with
eigenvalue $\alpha+2$. Similarly,
\[
\pi(H)\pi(Y)u=(\alpha-2)\pi(Y)u
\]
so that either $\pi(Y)u=0$, or else $\pi(Y)u$ is an eigenvector for $\pi(H)$
with eigenvalue $\alpha-2$.
\end{lemma}

\begin{proof}
We call $\pi(X)$ the ``raising operator,'' because it has the effect of
raising the eigenvalue of $\pi(H)$ by 2, and we call $\pi(Y)$ the ``lowering
operator.'' We know that $\left[  \pi(H),\pi(X)\right]  =\pi\left(  \left[
H,X\right]  \right)  =2\pi(X)$. Thus
\[
\pi(H)\pi(X)-\pi(X)\pi(H)=2\pi(X)
\]
or
\[
\pi(H)\pi(X)=\pi(X)\pi(H)+2\pi(X)\text{.}%
\]
Thus
\begin{align*}
\pi(H)\pi(X)u=\pi(X)\pi(H)u+2\pi(X)u\\
=\pi(X)\left(  \alpha u\right)  +2\pi(X)u\\
=(\alpha+2)\pi(X)u\text{.}%
\end{align*}

Similarly, $\left[  \pi(H),\pi(Y)\right]  =-2\pi(Y)$, and so
\[
\pi(H)\pi(Y)=\pi(Y)\pi(H)-2\pi(Y)
\]
so that
\begin{align*}
\pi(H)\pi(Y)u=\pi(Y)\pi(H)u-2\pi(Y)u\\
=\pi(Y)\left(  \alpha u\right)  -2\pi(Y)u\\
=(\alpha-2)\pi(Y)u\text{.}%
\end{align*}
This is what we wanted to show.
\end{proof}

As we have observed, $\pi(H)$ must have at least one eigenvector $u$ ($u\neq
0$), with some eigenvalue $\alpha\in\mathbb{C}$. By the lemma,
\[
\pi(H)\pi(X)u=(\alpha+2)\pi(X)u
\]
and more generally
\[
\pi(H)\pi(X)^{n}u=(\alpha+2n)\pi(X)^{n}u\text{.}%
\]
This means that either $\pi(X)^{n}u=0$, or else $\pi(X)^{n}u$ is an
eigenvector for $\pi(H)$ with eigenvalue $(\alpha+2n)$.

Now, an operator on a finite-dimensional space can have only finitely many
distinct eigenvalues. Thus the $\pi(X)^{n}u$'s cannot all be different from
zero. Thus there is some $N\geq0$ such that
\[
\pi(X)^{N}u\neq0
\]
but
\[
\pi(X)^{N+1}u=0\text{.}%
\]

Define $u_{0}=\pi(X)^{N}u$ and $\lambda=\alpha+2N$. Then
\begin{align}
\pi(H)u_{0}=\lambda u_{0}\label{weight}\\
\pi(X)u_{0}=0\label{kill}%
\end{align}
Then define
\[
u_{k}=\pi(Y)^{k}u_{0}%
\]
for $k\geq0$. By the second part of the lemma, we have
\begin{equation}
\pi(H)u_{k}=\left(  \lambda-2k\right)  u_{k}\text{.}\label{lower.weight}%
\end{equation}
Since, again, $\pi(H)$ can have only finitely many eigenvalues, the $u_{k}$'s
cannot all be non-zero.

\begin{lemma}
\label{raising}With the above notation,
\begin{align*}
\pi(X)u_{k}=\left[  k\lambda-k(k-1)\right]  u_{k-1}\text{\quad}(k>0)\\
\pi\left(  X\right)  u_{0}=0.
\end{align*}
\end{lemma}

\begin{proof}
We proceed by induction on $k$. In the case $k=1$ we note that $u_{1}%
=\pi(Y)u_{0}$. Using the commutation relation $\left[  \pi(X),\pi(Y)\right]
=\pi(H)$ we have
\[
\pi(X)u_{1}=\pi(X)\pi(Y)u_{0}=\left(  \pi(Y)\pi(X)+\pi(H)\right)
u_{0}\text{.}%
\]
But $\pi(X)u_{0}=0$, so we get
\[
\pi(X)u_{1}=\lambda u_{0}%
\]
which is the lemma in the case $k=1$.

Now, by definition $u_{k+1}=\pi(Y)u_{k}$. Using (\ref{lower.weight}) and
induction we have
\begin{align*}
\pi(X)u_{k+1}=\pi(X)\pi(Y)u_{k}\\
=\left(  \pi(Y)\pi(X)+\pi(H)\right)  u_{k}\\
=\pi(Y)\left[  k\lambda-k(k-1)\right]  u_{k-1}+(\lambda-2k)u_{k}\\
=\left[  k\lambda-k(k-1)+(\lambda-2k)\right]  u_{k}\text{.}%
\end{align*}
Simplifying the last expression give the Lemma.
\end{proof}

Since $\pi(H)$ can have only finitely many eigenvalues, the $u_{k}$'s cannot
all be non-zero. There must therefore be an integer $m\geq0$ such that
\[
u_{k}=\pi(Y)^{k}u_{0}\neq0
\]
for all $k\leq m$, but
\[
u_{m+1}=\pi(Y)^{m+1}u_{0}=0\text{.}%
\]

Now if $u_{m+1}=0$, then certainly $\pi(X)u_{m+1}=0$. Then by Lemma
\ref{raising},
\[
0=\pi(X)u_{m+1}=\left[  (m+1)\lambda-m(m+1)\right]  u_{m}=(m+1)(\lambda
-m)u_{m}\text{.}%
\]
But $u_{m}\neq0$, and $m+1\neq0$ (since $m\geq0$). Thus in order to have
$(m+1)(\lambda-m)u_{m}$ equal to zero, we must have $\lambda=m$.

We have made considerable progress. Given a finite-dimensional irreducible
representation $\pi$ of $\mathsf{sl}(2;\mathbb{C})$, acting on a space $V$,
there exists an integer $m\geq0$ and non-zero vectors $u_{0},\cdots u_{m}$
such that (putting $\lambda$ equal to $m$)
\begin{align}
\pi(H)u_{k}=(m-2k)u_{k}\nonumber\\
\pi(Y)u_{k}=u_{k+1}\quad(k<m)\nonumber\\
\pi(Y)u_{m}=0\nonumber\\
\pi(X)u_{k}=\left[  km-k(k-1)\right]  u_{k-1}\text{\quad}(k>0)\nonumber\\
\pi(X)u_{0}=0\label{sl2.rep}%
\end{align}

The vectors $u_{0},\cdots u_{m}$ must be linearly independent, since they are
eigenvectors of $\pi(H)$ with distinct eigenvalues. Moreover, the
$(m+1)$-dimensional span of $u_{0},\cdots u_{m}$ is explicitly invariant under
$\pi(H)$, $\pi(X)$, and $\pi(Y)$, and hence under $\pi(Z)$ for all
$Z\in\mathsf{sl}(2;\mathbb{C})$. Since $\pi$ is irreducible, this space must
be all of $V$.

We have now shown that every irreducible representation of $\mathsf{sl}%
(2;\mathbb{C})$ is of the form (\ref{sl2.rep}). It remains to show that
everything of the form (\ref{sl2.rep}) is a representation, and that it is
irreducible. That is, if we \textit{define} $\pi(H)$, $\pi(X)$, and $\pi(Y)$
by (\ref{sl2.rep}) (where the $u_{k}$'s are basis elements for some
$(m+1)$-dimensional vector space), then we want to show that they have the
right commutation relations to form a representation of $\mathsf{sl}%
(2;\mathbb{C})$, and that this representation is irreducible.

The computation of the commutation relations of $\pi(H)$, $\pi(X)$, and
$\pi(Y)$ is straightforward, and is left as an exercise. Note that when
dealing with $\pi(Y)$, you should treat separately the vectors $u_{k}$, $k<m$,
and $u_{m}$. Irreducibility is also easy to check, by imitating the proof of
Proposition \ref{poly.irred}. (See Exercise \ref{sl2.check}.)

We have now shown that there is an irreducible representation of
$\mathsf{sl}(2;\mathbb{C})$ in each dimension $m+1$, by explicitly writing
down how $H$, $X$, and $Y$ should act (Equation \ref{sl2.rep}) in a basis. But
we have shown more than this. We also have shown that any $(m+1)$-dimensional
irreducible representation of $\mathsf{sl}(2;\mathbb{C})$ must be of the form
(\ref{sl2.rep}). It follows that any two irreducible representations of
$\mathsf{sl}(2;\mathbb{C})$ of dimension $(m+1)$ must be equivalent. For if
$\pi_{1}$ and $\pi_{2}$ are two irreducible representations of dimension
$(m+1)$, acting on spaces $V_{1}$ and $V_{2}$, then $V_{1}$ has a basis
$u_{0},\cdots u_{m}$ as in (\ref{sl2.rep}) and $V_{2}$ has a similar basis
$\widetilde{u}_{0},\cdots\widetilde{u}_{m}$. But then the map $\phi
:V_{1}\rightarrow V_{2}$ which sends $u_{k}$ to $\widetilde{u}_{k}$ will be an
isomorphism of representations. (Think about it.)

In particular, the $(m+1)$-dimensional representation $\pi_{m}$ described in
Section \ref{examples.rep} must be equivalent to (\ref{sl2.rep}).This can be
seen explicitly by introducing the following basis for $V_{m}$:
\[
u_{k}=\left[  \pi_{m}(Y)\right]  ^{k}(z_{2}^{m})=(-1)^{k}\frac{m!}%
{(m-k)!}z_{1}^{k}z_{2}^{m-k}\qquad(k\leq m)\text{.}%
\]
Then by definition $\pi_{m}(Y)u_{k}=u_{k+1}$ ($k<m$), and it is clear that
$\pi_{m}(Y)u_{m}=0$. It is easy to see that $\pi_{m}(H)u_{k}=(m-2k)u_{k}$. The
only thing left to check is the behavior of $\pi_{m}(X)$. But direct
computation shows that
\[
\pi_{m}(X)u_{k}=k(m-k+1)u_{k-1}=\left[  km-k(k-1)\right]  u_{k-1}\text{.}%
\]
as required.

This completes the proof of Theorem \ref{classify.rep}.
\end{proof}

\section{Direct Sums of Representations and Complete Reducibility}

One way of generating representations is to take some representations you know
and combine them in some fashion. We will consider two methods of generating
new representations from old ones, namely direct sums and tensor products of
representations. In this section we consider direct sums; in the next section
we look at tensor products. (There is one other standard construction of this
sort, namely the dual of a representation. See Exercise \ref{define.dual}.)

\begin{definition}
Let $G$ be a matrix Lie group, and let $\Pi_{1},\Pi_{2},\cdots\Pi_{n}$ be
representations of $G$ acting on vector spaces $V_{1},V_{2},\cdots V_{n}$.
Then the \textbf{direct sum} of $\Pi_{1},\Pi_{2},\cdots\Pi_{n}$ is a
representation $\Pi_{1}\oplus\cdots\oplus\Pi_{n}$ of $G$ acting on the space
$V_{1}\oplus\cdots\oplus V_{n}$, defined by
\[
\left[  \Pi_{1}\oplus\cdots\oplus\Pi_{n}(A)\right]  \left(  v_{1,}\cdots
v_{n}\right)  =\left(  \Pi_{1}(A)v_{1},\cdots,\Pi_{n}(A)v_{n}\right)
\]
for all $A\in G$.

Similarly, if $\frak{g}$ is a Lie algebra, and $\pi_{1},\pi_{2},\cdots\pi_{n}$
are representations of $\frak{g}$ acting on $V_{1},V_{2},\cdots V_{n}$, then
we define the \textbf{direct sum} of $\pi_{1},\pi_{2},\cdots\pi_{n}$, acting
on $V_{1}\oplus\cdots\oplus V_{n}$ by
\[
\left[  \pi_{1}\oplus\cdots\oplus\pi_{n}(X)\right]  \left(  v_{1,}\cdots
v_{n}\right)  =\left(  \pi_{1}(X)v_{1},\cdots,\pi_{n}(X)v_{n}\right)
\]
for all $X\in\frak{g}$.
\end{definition}

It is trivial to check that, say, $\Pi_{1}\oplus\cdots\oplus\Pi_{n}$ is really
a representation of $G$.

\begin{definition}
A finite-dimensional representation of a group or Lie algebra, acting on a
space $V$, is said to be \textbf{completely reducible} if the following
property is satisfied: Given an invariant subspace $W\subset V$, and a second
invariant subspace $U\subset W\subset V$, there exists a third invariant
subspace $\widetilde{U}\subset W$ such that $U\cap\widetilde{U}=\left\{
0\right\}  $ and $U+\widetilde{U}=W$.
\end{definition}

The following Proposition shows that complete reducibility is a nice property
for a representation to have.

\begin{proposition}
A finite-dimensional completely reducible representation of a group or Lie
algebra is equivalent to a direct sum of (one or more) irreducible representations.
\end{proposition}

\begin{proof}
The proof is by induction on the dimension of the space $V$. If $\dim V=1$,
then automatically the representation is irreducible, since then $V$ is has no
non-trivial subspaces, let alone non-trivial invariant subspaces. Thus $V$ is
a direct sum of irreducible representations, with just one summand, namely $V$ itself.

Suppose, then, that the Proposition holds for all representations with
dimension strictly less than $n$, and that $\dim V=n$. If $V$ is irreducible,
then again we have a direct sum with only one summand, and we are done. If $V$
is not irreducible, then there exists a non-trivial invariant subspace
$U\subset V$. Taking $W=V$ in the definition of complete reducibility, we see
that there is another invariant subspace $\widetilde{U}$ with $U\cap
\widetilde{U}=\left\{  0\right\}  $ and $U+\widetilde{U}=V$. That is, $V\cong
U\oplus\widetilde{U}$ as a vector space.

But since $U$ and $\widetilde{U}$ are invariant, they can be viewed as
representations in their own right. (That is, the \textit{action} of our group
or Lie algebra on $U$ or $\widetilde{U}$ is a representation.) It is easy to
see that in fact $V$ is isomorphic to $U\oplus\widetilde{U}$ as a
representation. Furthermore, it is easy to see that both $U$ and
$\widetilde{U}$ are completely reducible representations, since every
invariant subspace $W$ of, say, $U$ is also an invariant subspace of $V$. But
since $U$ is non-trivial (i.e., $U\neq\left\{  0\right\}  $ and $U\neq V$), we
have $\dim U<\dim V$ and $\dim\widetilde{U}<\dim V$. Thus by induction $U\cong
U_{1}\oplus\cdots U_{n}$ (as representations), with the $U_{i}$'s irreducible,
and $\widetilde{U}\cong\widetilde{U}_{1}\oplus\cdots\widetilde{U}_{m}$, with
the $\widetilde{U}_{i}$'s irreducible, so that $V\cong U_{1}\oplus\cdots
U_{n}\oplus\widetilde{U}_{1}\oplus\cdots\widetilde{U}_{m}$.
\end{proof}

Certain groups and Lie algebras have the property that \textit{every}
(finite-dimensional) representation is completely reducible. This is a very
nice property, because it implies (by the above Proposition) that every
representation is equivalent to a direct sum of irreducible representations.
(And, as it turns out, this decomposition is essentially unique.) Thus for
such groups and Lie algebras, if you know (up to equivalence) what all the
irreducible representations are, then you know (up to equivalence) what
\textit{all} the representations are.

Unfortunately, not every representation is irreducible. For example, the
standard representation of the Heisenberg group is not completely reducible.
(See Exercise \ref{heisenberg.inv}.)

\begin{proposition}
\label{unitary.reduce}Let $G$ be a matrix Lie group. Let $\Pi$ be a
finite-dimensional unitary representation of $G$, acting on a
finite-dimensional real or complex Hilbert space $V$. Then $\Pi$ is completely reducible.
\end{proposition}

\begin{proof}
So, we are assuming that our space $V$ is equipped with an inner product, and
that $\Pi(A)$ is unitary for each $A\in G$. Suppose that $W\subset V$ is
invariant, and that $U\subset W\subset V$ is also invariant. Define
\[
\widetilde{U}=U^{\perp}\cap W\text{.}%
\]
Then of course $\widetilde{U}\cap U=\left\{  0\right\}  $, and standard
Hilbert space theory implies that $\widetilde{U}+U=W$.

It remains only to show that $\widetilde{U}$ is invariant. So suppose that
$v\in U^{\perp}\cap W$. Since $W$ is assumed invariant, $\Pi(A)v$ will be in
$W$ for any $A\in G$. We need to show that $\Pi(A)v$ is perpendicular to $U$.
Well, since $\Pi(A^{-1})$ is unitary, then for any $u\in U$%
\[
\left\langle u,\Pi(A)v\right\rangle =\left\langle \Pi(A^{-1})u,\Pi(A^{-1}%
)\Pi(A)v\right\rangle =\left\langle \Pi(A^{-1})u,v\right\rangle \text{.}%
\]
But $U$ is assumed invariant, and so $\Pi(A^{-1})u\in U$. But then since $v\in
U^{\perp}$, $\left\langle \Pi(A^{-1})u,v\right\rangle =0$. This means that
\[
\left\langle u,\Pi(A)v\right\rangle =0
\]
for all $u\in U$, i.e., $\Pi(A)v\in U^{\perp}$.

Thus $\widetilde{U}$ is invariant, and we are done.
\end{proof}

\begin{proposition}
\label{finite.reduce}If $G$ is a finite group, then every finite-dimensional
real or complex representation of $G$ is completely reducible.
\end{proposition}

\begin{proof}
Suppose that $\Pi$ is a representation of $G$, acting on a space $V$. Choose
an arbitrary inner product $\left\langle \ \right\rangle $ on $V$. Then define
a new inner product $\left\langle \ \right\rangle _{G}$ on $V$ by
\[
\left\langle v_{1},v_{2}\right\rangle _{G}=\sum_{g\in G}\left\langle
\Pi(g)v_{1},\Pi(g)v_{2}\right\rangle \text{.}%
\]
It is very easy to check that indeed $\left\langle \ \right\rangle _{G}$ is an
inner product. Furthermore, if $h\in G$, then
\begin{align*}
\left\langle \Pi(h)v_{1},\Pi(h)v_{2}\right\rangle _{G}=\sum_{g\in
G}\left\langle \Pi(g)\Pi(h)v_{1},\Pi(g)\Pi(h)v_{2}\right\rangle \\
=\sum_{g\in G}\left\langle \Pi(gh)v_{1},\Pi(gh)v_{2}\right\rangle \text{.}%
\end{align*}
But as $g$ ranges over $G$, so does $gh$. Thus in fact
\[
\left\langle \Pi(h)v_{1},\Pi(h)v_{2}\right\rangle _{G}=\left\langle
v_{1},v_{2}\right\rangle _{G}\text{.}%
\]

That is, $\Pi$ is a unitary representation with respect to the inner product
$\left\langle \ \right\rangle _{G}$. Thus $\Pi$ is completely reducible by
Proposition \ref{unitary.reduce}.
\end{proof}

There is a variant of the above argument which can be used to prove the
following result:

\begin{proposition}
\label{compact.reduce}If $G$ is a compact matrix Lie group, then every
finite-dimensional real or complex representation of $G$ is completely reducible.
\end{proposition}

\begin{proof}
This proof requires the notion of Haar measure. A \textbf{left Haar measure}
on a matrix Lie group $G$ is a non-zero measure $\mu$ on the Borel $\sigma
$-algebra in $G$ with the following two properties: 1) it is locally finite,
that is, every point in $G$ has a neighborhood with finite measure, and 2) it
is left-translation invariant. Left-translation invariance means that
$\mu\left(  gE\right)  =\mu\left(  E\right)  $ for all $g\in G$ and for all
Borel sets $E\subset G$, where
\[
gE=\left\{  ge\left|  e\in E\right.  \right\}  \text{.}%
\]
It is a fact which we cannot prove here that every matrix Lie group has a left
Haar measure, and that this measure is unique up to multiplication by a
constant. (One can analogously define right Haar measure, and a similar
theorem holds for it. Left Haar measure and right Haar measure may or may not
coincide; a group for which they do is called \textbf{unimodular}.)

Now, the key fact for our purpose is that left Haar measure is finite if and
only if the group $G\,$ is compact. So if $\Pi$ is a finite-dimensional
representation of a compact group $G$ acting on a space $V$, then let
$\left\langle \ \right\rangle $ be an arbitrary inner product on $V$, and
define a new inner product $\left\langle \ \right\rangle _{G}$ on $V$ by
\[
\left\langle v_{1},v_{2}\right\rangle _{G}=\int_{G}\left\langle \Pi
(g)v_{1},\Pi(g)v_{2}\right\rangle \,d\mu\left(  g\right)  \text{,}%
\]
where $\mu$ is left Haar measure. Again, it is easy to check that
$\left\langle \ \right\rangle _{G}$ is an inner product. Furthermore, if $h\in
G$, then by the left-invariance of $\mu$
\begin{align*}
\left\langle \Pi(h)v_{1},\Pi(h)v_{2}\right\rangle _{G}  & =\int_{G}%
\left\langle \Pi(g)\Pi(h)v_{1},\Pi(g)\Pi(h)v_{2}\right\rangle \,d\mu\left(
g\right)  \\
& =\int_{G}\left\langle \Pi(gh)v_{1},\Pi(gh)v_{2}\right\rangle \,d\mu\left(
g\right)  \\
& =\left\langle v_{1},v_{2}\right\rangle _{G}\text{.}%
\end{align*}
So $\Pi$ is a unitary representation with respect to $\left\langle
\ \right\rangle _{G}$, and thus completely reducible. Note that $\left\langle
\ \right\rangle _{G}$ is well-defined only because $\mu$ is finite.
\end{proof}

\section{Tensor Products of Representations}

Let $U$ and $V$ be finite-dimensional real or complex vector spaces. We wish
to define the \textbf{tensor product} of $U$ and $V$, which is will be a new
vector space $U\otimes V$ ``built'' out of $U$ and $V$. We will discuss the
idea of this first, and then give the precise definition.

We wish to consider a formal ``product'' of an element $u$ of $U$ with an
element $v$ of $V$, denoted $u\otimes v$. The \textit{space} $U\otimes V$ is
then the space of linear combinations of such products, i.e., the space of
elements of the form
\begin{equation}
a_{1}u_{1}\otimes v_{1}+a_{2}u_{2}\otimes v_{2}+\cdots+a_{n}u_{n}\otimes
v_{n}\text{.}\label{tensors}%
\end{equation}
Of course, if ``$\otimes$'' is to be interpreted as a product, then it should
be bilinear. That is, we should have
\begin{align*}
\left(  u_{1}+au_{2}\right)  \otimes v  & =u_{1}\otimes v+au_{2}\otimes v\\
u\otimes\left(  v_{1}+av_{2}\right)   & =u\otimes v_{1}+au\otimes
v_{2}\text{.}%
\end{align*}
We do not assume that the product is commutative. (In fact, the product in the
other order, $v\otimes u$, is in a different space, namely, $V\otimes U$.)

Now, if $e_{1},e_{2},\cdots,e_{n}$ is a basis for $U$ and $f_{1},f_{2}%
,\cdots,f_{m}$ is a basis for $V$, then using bilinearity it is easy to see
that any element of the form (\ref{tensors}) can be written as a linear
combination of the elements $e_{i}\otimes f_{j\text{ }}$. In fact, it seems
reasonable to expect that $\left\{  e_{i}\otimes f_{j}\left|  0\leq i\leq
n,0\leq j\leq m\right.  \right\}  $ should be a basis for the space $U\otimes
V$. This in fact turns out to be the case.

\begin{definition}
If $U$ and $V$ are finite-dimensional real or complex vector spaces, then a
\textbf{tensor product} of $U$ with $V$ is a vector space $W$, together with a
bilinear map $\phi:U\times V\rightarrow W$ with the following property: If
$\psi$ is any bilinear map of $U\times V$ into a vector space $X$, then there
exists a unique linear map $\widetilde{\psi}$ of $W$ into $X$ such that the
following diagram commutes:
\[%
\begin{array}
[c]{ccc}%
U\times V & \overset{\phi}{\rightarrow} & W\\
\psi\searrow &  & \swarrow\widetilde{\psi}\\
& X &
\end{array}
\text{.}%
\]
\end{definition}

Note that the \textit{bilinear} map $\psi$ from $U\times V$ into $X$ turns
into the \textit{linear} map $\widetilde{\psi}$ of $W$ into $X$. This is one
of the points of tensor products: bilinear maps on $U\times V$ turn into
linear maps on $W$.

\begin{theorem}
\label{tensor.exist}If $U$ and $V$ are any finite-dimensional real or complex
vector spaces, then a tensor product $(W,\phi)$ exists. Furthermore,
$(W,\phi)$ is unique up to canonical isomorphism. That is, if $(W_{1},\phi
_{1})$ and $(W_{2},\phi_{2})$ are two tensor products, then there exists a
unique vector space isomorphism $\Phi:W_{1}\rightarrow W_{2}$ such that the
following diagram commutes
\[%
\begin{array}
[c]{ccc}%
U\times V & \overset{\phi_{1}}{\rightarrow} & W_{1}\\
\phi_{2}\searrow &  & \swarrow\Phi\\
&  W_{2} &
\end{array}
\text{.}%
\]

Suppose that $(W,\phi)$ is a tensor product, and that $e_{1},e_{2}%
,\cdots,e_{n}$ is a basis for $U$ and $f_{1},f_{2},\cdots,f_{m}$ is a basis
for $V$. Then $\left\{  \phi(e_{i},f_{j})\left|  0\leq i\leq n,0\leq j\leq
m\right.  \right\}  $ is a basis for $W$.
\end{theorem}

\begin{proof}
Exercise \ref{universal}.
\end{proof}

\begin{notation}
Since the tensor product of $U$ and $V$ is essentially unique, we will let
$U\otimes V$ denote an arbitrary tensor product space, and we will write
$u\otimes v$ instead of $\phi(u,v)$. In this notation, the Theorem says that
$\left\{  e_{i}\otimes f_{j}\left|  0\leq i\leq n,0\leq j\leq m\right.
\right\}  $ is a basis for $U\otimes V$, as expected. Note in particular that
\[
\dim\left(  U\otimes V\right)  =\left(  \dim U\right)  \left(  \dim V\right)
\]
(not $\dim U+\dim V$).
\end{notation}

The defining property of $U\otimes V$ is called the \textbf{universal
property} of tensor products. While it may seem that we are taking a simple
idea and making it confusing, in fact there is a point to this universal
property. Suppose we want to define a linear map $T$ from $U\otimes V$ into
some other space. The most sensible way to define this is to define $T$ on
elements of the form $u\otimes v$. (You might try defining it on a basis, but
this forces you to worry about whether things depend on the choice of basis.)
Now, every element of $U\otimes V$ is a linear combination of things of the
form $u\otimes v$. However, this representation is far from unique. (Since,
say, if $u=u_{1}+u_{2}$, then you can rewrite $u\otimes v$ as $u_{1}\otimes
v+u_{2}\otimes v$.)

Thus if you try to define $T$ by what it does to elements of the form
$u\otimes v$, you have to worry about whether $T$ is well-defined. This is
where the universal property comes in. Suppose that $\psi(u,v)$ is some
bilinear expression in $u,v$. Then the universal property says precisely that
there is a unique linear map $T$ ($=\widetilde{\psi}$) such that
\[
T(u\otimes v)=\psi(u,v)\text{.}%
\]
(Think about it and make sure that you see that this is really what the
universal property says.)

The conclusion is this: You can define a linear map $T$ on $U\otimes V$ by
defining it on elements of the form $u\otimes v$, and this will be
well-defined, \textit{provided} that $T(u\otimes v)$ is bilinear in $(u,v)$.
The following Proposition shows how to make use of this idea.

\begin{proposition}
Let $U$ and $V$ be finite-dimensional real or complex vector spaces. Let
$A:U\rightarrow U$ and $B:V\rightarrow V$ be linear operators. Then there
exists a unique linear operator from $U\otimes V$ to $U\otimes V$, denoted
$A\otimes B$, such that
\[
A\otimes B(u\otimes v)=\left(  Au\right)  \otimes\left(  Bv\right)
\]
for all $u\in U$, $v\in V$.

If $A_{1},A_{2}$ are linear operators on $U$ and $B_{1},B_{2}$ are linear
operators on $V$, then
\[
\left(  A_{1}\otimes B_{1}\right)  \left(  A_{2}\otimes B_{2}\right)  =\left(
A_{1}A_{2}\right)  \otimes\left(  B_{1}B_{2}\right)  \text{.}%
\]
\end{proposition}

\begin{proof}
Define a map $\psi$ from $U\times V$ into $U\otimes V$ by
\[
\psi(u,v)=\left(  Au\right)  \otimes\left(  Bv\right)  \text{.}%
\]
Since $A$ and $B$ are linear, and since $\otimes$ is bilinear, $\psi$ will be
a bilinear map of $U\times V$ into $U\otimes V$. But then the universal
property says that there is an associated linear map $\widetilde{\psi
}:U\otimes V\rightarrow U\otimes V$ such that
\[
\widetilde{\psi}(u\otimes v)=\psi(u,v)=\left(  Au\right)  \otimes\left(
Bv\right)  \text{.}%
\]
Then $\widetilde{\psi}$ is the desired map $A\otimes B$.

Now, if $A_{1},A_{2}$ are operators on $U$ and $B_{1},B_{2}$ are operators on
$V$, then compute that
\begin{align*}
\left(  A_{1}\otimes B_{1}\right)  \left(  A_{2}\otimes B_{2}\right)
(u\otimes v)=\left(  A_{1}\otimes B_{1}\right)  \left(  A_{2}u\otimes
B_{2}v\right)  \\
=A_{1}A_{2}u\otimes B_{1}B_{2}v\text{.}%
\end{align*}
This shows that $\left(  A_{1}\otimes B_{1}\right)  \left(  A_{2}\otimes
B_{2}\right)  =\left(  A_{1}A_{2}\right)  \otimes\left(  B_{1}B_{2}\right)  $
are equal on elements of the form $u\otimes v$. Since every element of
$U\otimes V$ can be written as a linear combination of things of the form
$u\otimes v$ (in fact of $e_{i}\otimes f_{j}$), $\left(  A_{1}\otimes
B_{1}\right)  \left(  A_{2}\otimes B_{2}\right)  $ and $\left(  A_{1}%
A_{2}\right)  \otimes\left(  B_{1}B_{2}\right)  $ must be equal on the whole space.
\end{proof}

We are now ready to define tensor products of representations. There are two
different approaches to this, both of which are important. The first approach
starts with a representation of a group $G$ acting on a space $V$ and a
representation of another group $H$ acting on a space $U,$ and produces a
representation of the product group $G\times H$ acting on the space $U\otimes
V$. The second approach starts with two different representations of the same
group $G$, acting on spaces $U$ and $V$, and produces a representation of $G$
acting on $U\otimes V$. Both of these approaches can be adapted to apply to
Lie algebras.

\begin{definition}
\label{tensor.2}Let $G$ and $H$ be matrix Lie groups. Let $\Pi_{1}$ be a
representation of $G$ acting on a space $U$ and let $\Pi_{2}$ be a
representation of $H$ acting on a space $V$. The the \textbf{tensor product}
of $\Pi_{1}$ and $\Pi_{2}$ is a representation $\Pi_{1}\otimes\Pi_{2}$ of
$G\times H$ acting on $U\otimes V$ defined by
\[
\Pi_{1}\otimes\Pi_{2}(A,B)=\Pi_{1}(A)\otimes\Pi_{2}(B)
\]
for all $A\in G$ and $B\in H.$
\end{definition}

Using the above Proposition, it is very easy to check that indeed $\Pi
_{1}\otimes\Pi_{2}$ is a representation of $G\times H$.

Now, if $G$ and $H$ are matrix Lie groups, that is, $G$ is a closed subgroup
of $\mathsf{GL}(n;\mathbb{C})$ and $H$ is a closed subgroup of $\mathsf{GL}%
(m;\mathbb{C})$, then $G\times H$ can be regarded in an obvious way as a
closed subgroup of $\mathsf{GL}(n+m;\mathbb{C})$. Thus the direct product of
matrix Lie groups can be regarded as a matrix Lie group. It is easy to check
that the Lie algebra of $G\times H$ is isomorphic to the direct sum of the Lie
algebra of $G$ and the Lie algebra of $H$. See Exercise \ref{direct.product}.

In light of Proposition \ref{relating.rep}, the representation $\Pi_{1}%
\otimes\Pi_{2}$ of $G\times H$ gives rise to a representation of the Lie
algebra of $G\times H$, namely $\frak{g}\oplus\frak{h}$. The following
Proposition shows that this representation of $\frak{g}\oplus\frak{h}$ is not
what you might expect at first.

\begin{proposition}
Let $G$ and $H$ be matrix Lie groups, let $\Pi_{1}$, $\Pi_{2}$ be
representations of $G,H$ respectively, and consider the representation
$\Pi_{1}\otimes\Pi_{2}$ of $G\times H$. Let $\pi_{1}\otimes\pi_{2}$ denote the
associated representation of the Lie algebra of $G\times H$, namely
$\frak{g}\oplus\frak{h}$. Then for all $X\in\frak{g}$ and $Y\in\frak{h}$%
\[
\pi_{1}\otimes\pi_{2}(X,Y)=\pi_{1}(X)\otimes I+I\otimes\pi_{2}(Y)\text{.}%
\]
\end{proposition}

\begin{proof}
Suppose that $u(t)$ is a smooth curve in $U$ and $v(t)$ is a smooth curve in
$V$. Then we verify the product rule in the usual way:
\begin{align*}
\lim_{h\rightarrow0}\frac{u(t+h)\otimes v(t+h)-u(t)\otimes v(t)}{h}\\
=\lim_{h\rightarrow0}\frac{u(t+h)\otimes v(t+h)-u(t+h)\otimes v(t)}{h}%
+\frac{u(t+h)\otimes v(t)-u(t)\otimes v(t)}{h}\\
=\lim_{h\rightarrow0}\left[  u(t+h)\otimes\frac{\left(  v(t+h)-v\left(
t\right)  \right)  }{h}\right]  +\lim_{h\rightarrow0}\left[  \frac{\left(
u(t+h)-u\left(  t\right)  \right)  }{h}\otimes v(t)\right]  \text{.}%
\end{align*}
Thus
\[
\frac{d}{dt}\left(  u(t)\otimes v(t)\right)  =\frac{du}{dt}\otimes
v(t)+u(t)\otimes\frac{dv}{dt}\text{.}%
\]

This being the case, we can compute $\pi_{1}\otimes\pi_{2}(X,Y)$:
\begin{align*}
\pi_{1}\otimes\pi_{2}(X,Y)(u\otimes v)=\left.  \frac{d}{dt}\right|  _{t=0}%
\Pi_{1}\otimes\Pi_{2}(e^{tX},e^{tY})(u\otimes v)\\
=\left.  \frac{d}{dt}\right|  _{t=0}\Pi_{1}(e^{tX})u\otimes\Pi_{2}(e^{tY})v\\
=\left(  \left.  \frac{d}{dt}\right|  _{t=0}\Pi_{1}(e^{tX})u\right)  \otimes
v+u\otimes\left(  \left.  \frac{d}{dt}\right|  _{t=0}\Pi_{2}(e^{tY})v\right)
\text{.}%
\end{align*}
This shows that $\pi_{1}\otimes\pi_{2}(X,Y)=\pi_{1}(X)\otimes I+I\otimes
\pi_{2}(Y)$ on elements of the form $u\otimes v$, and therefore on the whole
space $U\otimes V$.
\end{proof}

\begin{definition}
\label{tensor.algebra.2}Let $\frak{g}$ and $\frak{h}$ be Lie algebras, and let
$\pi_{1}$ and $\pi_{2}$ be representations of $\frak{g}$ and $\frak{h}$,
acting on spaces $U$ and $V$. Then the \textbf{tensor product} of $\pi_{1}$
and $\pi_{2}$, denoted $\pi_{1}\otimes\pi_{2}$, is a representation of
$\frak{g}\oplus\frak{h}$ acting on $U\otimes V$, given by
\[
\pi_{1}\otimes\pi_{2}(X,Y)=\pi_{1}(X)\otimes I+I\otimes\pi_{2}(Y)
\]
for all $X\in\frak{g}$ and $Y\in\frak{h}.$
\end{definition}

It is easy to check that this indeed defines a representation of
$\frak{g}\oplus\frak{h}$. Note that if we defined $\pi_{1}\otimes\pi
_{2}(X,Y)=\pi_{1}(X)\otimes\pi_{2}(Y)$, this would \textit{not} be a
representation of $\frak{g}\oplus\frak{h}$, for this is not even a linear map.
(E.g., we would then have $\pi_{1}\otimes\pi_{2}(2X,2Y)=4\pi_{1}\otimes\pi
_{2}(X,Y)$!) Note also that the above definition applies even if $\pi_{1}$ and
$\pi_{2}$ do not come from a representation of any matrix Lie group.

\begin{definition}
Let $G$ be a matrix Lie group, and let $\Pi_{1}$ and $\Pi_{2}$ be
representations of $G$, acting on spaces $V_{1}$ and $V_{2}$. Then the
\textbf{tensor product} of $\Pi_{1}$ and $\Pi_{2}$ is a representation of $G$
acting on $V_{1}\otimes V_{2}$ defined by
\[
\Pi_{1}\otimes\Pi_{2}(A)=\Pi_{1}(A)\otimes\Pi_{2}(A)
\]
for all $A\in G.$
\end{definition}

\begin{proposition}
With the above notation, the associated representation of the Lie algebra
$\frak{g}$ satisfies
\[
\pi_{1}\otimes\pi_{2}(X)=\pi_{1}(X)\otimes I+I\otimes\pi_{2}(X)
\]
for all $X\in\frak{g}.$
\end{proposition}

\begin{proof}
Using the product rule,
\begin{align*}
\pi_{1}\otimes\pi_{2}(X)\left(  u\otimes v\right)    & =\left.  \frac{d}%
{dt}\right|  _{t=0}\Pi_{1}\left(  e^{tX}\right)  u\otimes\Pi_{2}\left(
e^{tX}\right)  v\\
& =\pi_{1}\left(  X\right)  u\otimes v+v\otimes\pi_{2}\left(  X\right)
u\text{.}%
\end{align*}
This is what we wanted to show.
\end{proof}

\begin{definition}
\label{algebra.tensor}If $\frak{g}$ is a Lie algebra, and $\pi_{1}$ and
$\pi_{2}$ are representations of $\frak{g}$ acting on spaces $V_{1}$ and
$V_{2}$, then the \textbf{tensor product} of $\pi_{1}$ and $\pi_{2}$ is a
representation of $\frak{g}$ acting on the space $V_{1}\otimes V_{2}$ defined
by
\[
\pi_{1}\otimes\pi_{2}(X)=\pi_{1}(X)\otimes I+I\otimes\pi_{2}(X)
\]
for all $X\in\frak{g}.$
\end{definition}

It is easy to check that $\Pi_{1}\otimes\Pi_{2}$ and $\pi_{1}\otimes\pi_{2}$
are actually representations of $G$ and $\frak{g}$, respectively. There is
some ambiguity in the notation, say, $\Pi_{1}\otimes\Pi_{2}$. For even if
$\Pi_{1}$ and $\Pi_{2}$ are both representations of the same group $G$, we
could still regard $\Pi_{1}\otimes\Pi_{2}$ as a representation of $G\times G$,
by taking $H=G$ in definition \ref{tensor.2}. We will rely on context to make
clear whether we are thinking of $\Pi_{1}\otimes\Pi_{2}$ as a representation
of $G\times G$ or as representation of $G$.

Suppose $\Pi_{1}$ and $\Pi_{2}$ are \textit{irreducible} representations of a
group $G$. If we regard $\Pi_{1}\otimes\Pi_{2}$ as a representation of $G$, it
may no longer be irreducible. If it is not irreducible, one can attempt to
decompose it as a direct sum of irreducible representations. This process is
called \textbf{Clebsch-Gordan} theory. In the case of $\mathsf{SU}(2)$, this
theory is relatively simple. (In the physics literature, the problem of
analyzing tensor products of representations of $\mathsf{SU}(2)$ is called
``addition of angular momentum.'') See Exercise \ref{c.g}.

\section{Schur's Lemma}

Let $\Pi$ and $\Sigma$ be representations of a matrix Lie group $G$, acting on
spaces $V$ and $W$. Recall that a \textbf{morphism} of representations is a
linear map $\phi:V\rightarrow W$ with the property that
\[
\phi\left(  \Pi(A)v\right)  =\Sigma(A)\left(  \phi(v)\right)
\]
for all $v\in V$ and all $A\in G$. Schur's Lemma is an extremely important
result which tells us about morphisms of irreducible representations. Part of
Schur's Lemma applies to both real and complex representations, but part of it
applies only to complex representations.

It is desirable to be able to state Schur's lemma simultaneously for groups
and Lie algebras. In order to do so, we need to indulge in a common abuse of
notation. If, say, $\Pi$ is a representation of $G$ acting on a space $V$, we
will refer to $V$ as the representation, without explicit reference to $\Pi$.

\begin{theorem}
[Schur's Lemma]%
\begin{enumerate}
\item \label{schur1}Let $V$ and $W$ be \textit{irreducible} real or complex
representations of a group or Lie algebra, and let $\phi:V\rightarrow W$ be a
morphism. Then either $\phi=0$ or $\phi$ is an isomorphism.

\item \label{schur2}Let $V$ be an irreducible \textit{complex} representation
of a group or Lie algebra, and let $\phi:V\rightarrow V$ be a morphism of $V$
with itself. Then $\phi=\lambda I$, for some $\lambda\in\mathbb{C}$.

\item \label{schur3}Let $V$ and $W$ be irreducible complex representations of
a group or Lie algebra, and let $\phi_{1},\phi_{2}:V\rightarrow W$ be non-zero
morphisms. Then $\phi_{1}=\lambda\phi_{2}$, for some $\lambda\in\mathbb{C}$.
\end{enumerate}
\end{theorem}

\begin{corollary}
\label{schur.center}Let $\Pi$ be an irreducible complex representation of a
matrix Lie group $G$. If $A$ is in the center of $G$, then $\Pi(A)=\lambda I$.
Similarly, if $\pi$ is an irreducible complex representation of a Lie algebra
$\frak{g}$, and if $X$ is in the center of $\frak{g}$ (i.e., $[X,Y]=0$ for all
$Y\in\frak{g}$), then $\pi(X)=\lambda I$.
\end{corollary}

\begin{proof}
We prove the group case; the proof of the Lie algebra case is the same. If $A$
is in the center of $G$, then for all $B\in G$,
\[
\Pi(A)\Pi(B)=\Pi(AB)=\Pi(BA)=\Pi(B)\Pi(A)\text{.}%
\]
But this says exactly that $\Pi(A)$ is a morphism of $\Pi$ with itself. So by
Point 2 of Schur's lemma, $\Pi(A)$ is a multiple of the identity.
\end{proof}

\begin{corollary}
\label{schur.commute}An irreducible complex representation of a
\textit{commutative} group or Lie algebra is one-dimensional.
\end{corollary}

\begin{proof}
Again, we prove only the group case. If $G$ is commutative, then the center of
$G$ is all of $G$, so by the previous corollary $\Pi(A)$ is a multiple of the
identity for each $A\in G$. But this means that \textit{every} subspace of $V$
is invariant! Thus the only way that $V$ can fail to have a non-trivial
invariant subspace is for it not to have any non-trivial subspaces. This means
that $V$ must be one-dimensional. (Recall that we do not allow $V$ to be zero-dimensional.)
\end{proof}

\begin{proof}
As usual, we will prove just the group case; the proof of the Lie algebra case
requires only the obvious notational changes.

\textit{Proof of \ref{schur1}}. Saying that $\phi$ is a morphism means
$\phi(\Pi(A)v)=\Sigma(A)\left(  \phi(v)\right)  $ for all $v\in V$ and all
$A\in G$. Now suppose that $v\in\ker(\phi)$. Then
\[
\phi(\Pi(A)v)=\Sigma(A)\phi(v)=0\text{.}%
\]
This shows that $\ker\phi$ is an invariant subspace of $V$. Since $V$ is
irreducible, we must have $\ker\phi=0$ or $\ker\phi=V$. Thus $\phi$ is either
one-to-one or zero.

Suppose $\phi$ is one-to-one. Then the image of $\phi$ is a non-zero subspace
of $W$. On the other hand, the image of $\phi$ is invariant, for if $w\in W$
is of the form $\phi(v)$ for some $v\in V$, then
\[
\Sigma(A)w=\Sigma(A)\phi(v)=\phi(\Pi(A)v)\text{.}%
\]
Since $W$ is irreducible and image$(V)$ is non-zero and invariant, we must
have image$(V)=W$. Thus $\phi$ is either zero or one-to-one and onto.

\textit{Proof of \ref{schur2}}. Suppose now that $V$ is an irreducible
\textit{complex} representation, and that $\phi:V\rightarrow V$ is a morphism
of $V$ to itself. This means that $\phi\Pi(A)=\Pi(A)\phi$ for all $A\in G$,
i.e., that $\phi$ commutes with all of the $\Pi(A)$'s. Now, since we are over
an algebraically complete field, $\phi$ must have at least one eigenvalue
$\lambda\in\mathbb{C}$. Let $U$ denote the eigenspace for $\phi$ associated to
the eigenvalue $\lambda$, and let $u\in U$. Then for each $A\in G$%
\[
\phi\left(  \Pi(A)u\right)  =\Pi(A)\phi(v)=\lambda\Pi(A)u\text{.}%
\]
Thus applying $\Pi(A)$ to an eigenvector of $\phi$ with eigenvalue $\lambda$
yields another eigenvector of $\phi$ with eigenvalue $\lambda$. That is, $U$
is invariant.

Since $\lambda$ is an eigenvalue, $U\neq0$, and so we must have $U=V$. But
this means that $\phi(v)=\lambda v$ for all $v\in V$, i.e., that $\phi=\lambda
I$.

\textit{Proof of \ref{schur3}}. If $\phi_{2}\neq0$, then by (1) $\phi_{2}$ is
an isomorphism. Now look at $\phi_{1}\circ\phi_{2}^{-1}$. As is easily
checked, the composition of two morphisms is a morphism, so $\phi_{1}\circ
\phi_{2}^{-1}$ is a morphism of $W$ with itself. Thus by (2), $\phi_{1}%
\circ\phi_{2}^{-1}=\lambda I$, whence $\phi_{1}=\lambda\phi_{2}$.
\end{proof}

\section{Group Versus Lie Algebra Representations\label{group.algebra}}

We know from Chapter 3 (Theorem \ref{homo.theorem}) that every Lie group
homomorphism gives rise to a Lie algebra homomorphism. In particular, this
shows (Proposition \ref{relating.rep}) that every representation of a matrix
Lie group gives rise to a representation of the associated Lie algebra. The
goal of this section is to investigate the reverse process. That is, given a
representation of the Lie algebra, under what circumstances is there an
associated representation of the Lie group?

The climax of this section is Theorem \ref{exp.rep}, which states that if $G$
is a \textit{connected} and \textit{simply connected} matrix Lie group with
Lie algebra $\frak{g}$, and if $\pi$ is a representation of $\frak{g}$, then
there is a unique representation $\Pi$ of $G$ such that $\Pi$ and $\pi$ are
related as in Proposition \ref{relating.rep}. Our proof of this theorem will
make use of the Baker-Campbell-Hausdorff formula from Chapter 4. Before
turning to this general theorem, we will examine two special cases, namely
$\mathsf{SO}(3)$ and $\mathsf{SU}(2)$, for which we can work things out by
hand. See Br\"{o}cker and tom Dieck, Chapter II, Section 5.

We have shown (Theorem \ref{classify.rep}) that every irreducible complex
representation of $\mathsf{su}(2)$ is equivalent to one of the representations
$\pi_{m}$ described in Section \ref{examples.rep}. (Recall that the
irreducible complex representations of $\mathsf{su}(2)$ are in one-to-one
correspondence with the irreducible representations of $\mathsf{sl}%
(2;\mathbb{C})$.) Each of the representations $\pi_{m}$ of $\mathsf{su}(2)$
was constructed from the corresponding representation $\Pi_{m}$ of the group
$\mathsf{SU}(2)$. Thus we see, by brute force computation, that every
irreducible complex representation of $\mathsf{su}(2)$ actually comes from a
representation of the group $\mathsf{SU}(2)$! This is consistent with the fact
that $\mathsf{SU}(2)$ is simply connected (Chapter 2, Prop. \ref{su2.sc}).

Let us now consider the situation for $\mathsf{SO}(3)$. (Which is not simply
connected.) We know from Exercise \ref{su2.iso.so3} of Chapter 3 that the Lie
algebras $\mathsf{su}(2)$ and $\mathsf{so}(3)$ are isomorphic. In particular,
if we take the basis
\[%
\begin{array}
[c]{ccc}%
E_{1}=\tfrac12\left(
\begin{array}
[c]{cc}%
i & 0\\
0 & -i
\end{array}
\right)  & E_{2}=\tfrac12\left(
\begin{array}
[c]{cc}%
0 & 1\\
-1 & 0
\end{array}
\right)  & E_{3}=\tfrac12\left(
\begin{array}
[c]{cc}%
0 & i\\
i & 0
\end{array}
\right)
\end{array}
\]
for $\mathsf{su}(2)$ and the basis
\[%
\begin{array}
[c]{ccc}%
F_{1}=\left(
\begin{array}
[c]{ccc}%
0 & 0 & 0\\
0 & 0 & -1\\
0 & 1 & 0
\end{array}
\right)  & F_{2}=\left(
\begin{array}
[c]{ccc}%
0 & 0 & 1\\
0 & 0 & 0\\
-1 & 0 & 0
\end{array}
\right)  & F_{3}=\left(
\begin{array}
[c]{ccc}%
0 & -1 & 0\\
1 & 0 & 0\\
0 & 0 & 0
\end{array}
\right)
\end{array}
\]
then direct computation shows that $\left[  E_{1},E_{2}\right]  =E_{3}$,
$\left[  E_{2},E_{3}\right]  =E_{1}$, $\left[  E_{3},E_{1}\right]  =E_{2}$,
and similarly with the $E $'s replaced by the $F$'s. Thus the map
$\phi:\mathsf{so}(3)\rightarrow\mathsf{su}(2)$ which takes $F_{i}$ to $E_{i}$
will be a Lie algebra isomorphism.

Since $\mathsf{su}(2)$ and $\mathsf{so}(3)$ are isomorphic Lie algebras, they
must have ``the same'' representations. Specifically, if $\pi$ is a
representation of $\mathsf{su}(2)$, then $\pi\circ\phi$ will be a
representation of $\mathsf{so}(3)$, and every representation of $\mathsf{so}%
(3)$ is of this form. In particular, the irreducible representations of
$\mathsf{so}(3)$ are precisely of the form $\sigma_{m}=\pi_{m}\circ\phi$. We
wish to determine, for a particular $m$, whether there is a representation
$\Sigma_{m} $ of the group $\mathsf{SO}(3)$ such that $\sigma_{m}$ and
$\Sigma_{m}$ are related as in Proposition \ref{relating.rep}.

\begin{proposition}
\label{so3.odd}Let $\sigma_{m}=\pi_{m}\circ\phi$ be the irreducible complex
representations of the Lie algebra $\mathsf{so}(3)$ ($m\geq0$). If $m$ is
even, then there is a representation $\Sigma_{m}$ of the group $\mathsf{SO}%
(3)$ such that $\sigma_{m}$ and $\Sigma_{m}$ are related as in Proposition
\ref{relating.rep}. If $m$ is odd, then there is no such representation of
$\mathsf{SO}(3)$.
\end{proposition}

Note that the condition that $m$ be even is equivalent to the condition that
$\dim V_{m}=m+1$ be odd. Thus it is the odd-dimensional representations of the
Lie algebra $\mathsf{so}(3)$ which come from group representations.

In the physics literature, the representations of $\mathsf{su}(2)/\mathsf{so}%
(3)$ are labeled by the parameter $l=m/2$. In terms of this notation, a
representation of $\mathsf{so}(3)$ comes from a representation of
$\mathsf{SO}(3)$ if and only if $l$ is an integer. The representations with
$l$ an integer are called ``integer spin''; the others are called
``half-integer spin.''

\subsubsection{Proof}

\begin{proof}
\textit{Case 1: m odd}. In this case, we want to prove that there is no
representation $\Sigma_{m}$ such that $\sigma_{m}$ and $\Sigma_{m}$ are
related as in Proposition \ref{relating.rep}. (We have already considered the
case $m=1$ in Exercise \ref{spin.half}.) Suppose, to the contrary, that there
is such a $\Sigma_{m}$. Then Proposition \ref{relating.rep} says that
\[
\Sigma_{m}(e^{X})=e^{\sigma_{m}(X)}%
\]
for all $X\in\mathsf{so}(3)$. In particular, take $X=2\pi F_{1}$. Then,
computing as in Chapter 3, Section \ref{computing} we see that
\[
e^{2\pi F_{1}}=\left(
\begin{array}
[c]{ccc}%
1 & 0 & 0\\
0 & \cos2\pi & -\sin2\pi\\
0 & \sin2\pi & \cos2\pi
\end{array}
\right)  =I\text{.}%
\]
Thus on the one hand $\Sigma_{m}\left(  e^{2\pi F_{1}}\right)  =\Sigma
_{m}(I)=I$, while on the other hand $\Sigma_{m}\left(  e^{2\pi F_{1}}\right)
=e^{2\pi\sigma_{m}(F_{1})}$.

Let us compute $e^{2\pi\sigma_{m}(F_{1})}$. By definition, $\sigma_{m}%
(F_{1})=\pi_{m}(\phi(F_{1}))=\pi_{m}(E_{1})$. But, $E_{1}=\frac{i}{2}H$, where
as usual
\[
H=\left(
\begin{array}
[c]{cc}%
1 & 0\\
0 & -1
\end{array}
\right)  \text{.}%
\]
We know that there is a basis $u_{0},u_{1},\cdots,u_{m}$ for $V_{m}$ such that
$u_{k}$ is an eigenvector for $\pi_{m}(H)$ with eigenvalue $m-2k$. This means
that $u_{k}$ is also an eigenvector for $\sigma_{m}(F_{1})=\frac{i}{2}\pi
_{m}(H)$, with eigenvalue $\frac{i}{2}(m-2k)$. Thus in the basis $\left\{
u_{k}\right\}  $ we have
\[
\sigma_{m}(F_{1})=\left(
\begin{array}
[c]{cccc}%
\frac{i}{2}m &  &  & \\
& \frac{i}{2}(m-2) &  & \\
&  & \ddots & \\
&  &  & \frac{i}{2}(-m)
\end{array}
\right)  \text{.}%
\]

But we are assuming the $m$ is odd! This means that $m-2k$ is an odd integer.
Thus $e^{2\pi\frac{i}{2}(m-2k)}=-1$, and in the basis $\left\{  u_{k}\right\}
$%
\[
e^{2\pi\sigma_{m}(F_{1})}=\left(
\begin{array}
[c]{cccc}%
e^{2\pi\frac{i}{2}m} &  &  & \\
& e^{2\pi\frac{i}{2}(m-2)} &  & \\
&  & \ddots & \\
&  &  &  e^{2\pi\frac{i}{2}(-m)}%
\end{array}
\right)  =-I\text{.}%
\]
Thus on the one hand, $\Sigma_{m}\left(  e^{2\pi F_{1}}\right)  =\Sigma
_{m}(I)=I$, while on the other hand $\Sigma_{m}\left(  e^{2\pi F_{1}}\right)
=e^{2\pi\sigma_{m}(F_{1})}=-I$. This is a contradiction, so there can be no
such group representation $\Sigma_{m}$.

\textit{Case 2: m is even}. We will use the following:

\begin{lemma}
\label{su2.so3}There exists a Lie group homomorphism $\Phi:\mathsf{SU}%
(2)\rightarrow\mathsf{SO}(3)$ such that

1) $\Phi$ maps $\mathsf{SU}(2)$ onto $\mathsf{SO}(3),$

2) $\ker\Phi=\{I,-I\},$ and

3) the associated Lie algebra homomorphism $\widetilde{\Phi}:\mathsf{su}%
(2)\rightarrow\mathsf{so}(3)$ is an isomorphism which takes $E_{i}$ to $F_{i}%
$. That is, $\widetilde{\Phi}=\phi^{-1}$.
\end{lemma}

\begin{proof}
Exercise \ref{prove.su2.so3}.
\end{proof}

Now consider the representations $\Pi_{m}$ of $\mathsf{SU}(2)$. I claim that
if $m$ is even, then $\Pi_{m}(-I)=I$. To see this, note that
\[
e^{2\pi E_{1}}=\exp\left(
\begin{array}
[c]{cc}%
\pi i & 0\\
0 & -\pi i
\end{array}
\right)  =-I\text{.}%
\]
Thus $\Pi_{m}(-I)=\Pi_{m}(e^{2\pi E_{1}})=e^{\pi_{m}(2\pi E_{1})}$. But as in
Case 1,
\[
e^{\pi_{m}(2\pi E_{1})}=\left(
\begin{array}
[c]{cccc}%
e^{2\pi\frac{i}{2}m} &  &  & \\
& e^{2\pi\frac{i}{2}(m-2)} &  & \\
&  & \ddots & \\
&  &  &  e^{2\pi\frac{i}{2}(-m)}%
\end{array}
\right)  \text{.}%
\]
Only, this time, $m$ is even, and so $\frac{i}{2}(m-2k)$ is an integer, so
that $\Pi_{m}(-I)=e^{\pi_{m}(2\pi E_{1})}=I$.

Since $\Pi_{m}(-I)=I$, $\Pi_{m}(-U)=\Pi_{m}(U)$ for all $U\in\mathsf{SU}(2)$.
According to Lemma \ref{su2.so3}, for each $R\in\mathsf{SO}(3)$, there is a
unique pair of elements $\left\{  U,-U\right\}  $ such that $\Phi
(U)=\Phi(-U)=R$. Since $\Pi_{m}(U)=\Pi_{m}(-U)$, it makes sense to define
\[
\Sigma_{m}(R)=\Pi_{m}(U)\text{.}%
\]
It is easy to see that $\Sigma_{m}$ is a Lie group homomorphism (hence, a
representation). By construction, we have
\begin{equation}
\Pi_{m}=\Sigma_{m}\circ\Phi\text{.}\label{compose}%
\end{equation}

Now, if $\widetilde{\Sigma}_{m}$ denotes the Lie algebra representation
associated to $\Sigma_{m}$, then it follows from (\ref{compose}) that
\[
\pi_{m}=\widetilde{\Sigma}_{m}\circ\widetilde{\Phi}\text{.}%
\]
But the Lie algebra homomorphism $\widetilde{\Phi}$ takes $E_{i}$ to $F_{i}$,
that is, $\widetilde{\Phi}=\phi^{-1}$. So $\pi_{m}=\widetilde{\Sigma}_{m}%
\circ\phi^{-1}$, or $\widetilde{\Sigma}_{m}=\pi_{m}\circ\phi$. Thus
$\widetilde{\Sigma}_{m}=\sigma_{m}$, which is what we want to show.
\end{proof}

It is now time to state the main theorem.

\begin{theorem}
\label{exp.rep}

\begin{enumerate}
\item  Let $G,H$ be a matrix Lie groups, let $\phi_{1},\phi_{2}:G\rightarrow
H$ be Lie group homomorphisms, and let $\widetilde{\phi}_{1},\widetilde{\phi
}_{2}:\frak{g}\rightarrow\frak{h}$ be the associated Lie algebra
homomorphisms. If $G$ is \textit{connected} and $\widetilde{\phi}%
_{1}=\widetilde{\phi}_{2}$, then $\phi_{1}=\phi_{2}$.

\item  Let $G,H$ be a matrix Lie groups with Lie algebras $\frak{g}$ and
$\frak{h}$. Let $\widetilde{\phi}:\frak{g\rightarrow h}$ be a Lie algebra
homomorphism. If $G$ is \textit{connected} and \textit{simply}
\textit{connected}, then there exists a unique Lie group homomorphism
$\phi:G\rightarrow H$ such that $\phi$ and $\widetilde{\phi}$ are related as
in Theorem \ref{homo.theorem} of Chapter 3.
\end{enumerate}
\end{theorem}

This has the following corollaries.

\begin{corollary}
Suppose $G$ and $H$ are connected, simply connected matrix Lie groups with Lie
algebras $\frak{g}$ and $\frak{h}$. If $\frak{g}\cong\frak{h}$ then $G\cong H$.
\end{corollary}

\begin{proof}
Let $\widetilde{\phi}:\frak{g}\rightarrow\frak{h}$ be a Lie algebra
isomorphism. By Theorem \ref{exp.rep}, there exists an associated Lie group
homomorphism $\phi:G\rightarrow H$. Since $\widetilde{\phi}^{-1}%
:\frak{h}\rightarrow\frak{g}$ is also a Lie algebra homomorphism, there is a
corresponding Lie group homomorphism $\psi:H\rightarrow G$. We want to show
that $\phi$ and $\psi$ are inverses of each other.

Well, $\widetilde{\phi\circ\psi}=\widetilde{\phi}\circ\widetilde{\psi
}=I_{\frak{h}}$, so by the Point 1 of the Theorem, $\phi\circ\psi=I_{H}$.
Similarly, $\psi\circ\phi=I_{G}$.
\end{proof}

\begin{corollary}
\begin{enumerate}
\item \label{sc.reps}Let $G$ be a \textit{connected} matrix Lie group, let
$\Pi_{1}$ and $\Pi_{2}$ be representations of $G$, and let $\pi_{1}$ and
$\pi_{2}$ be the associated Lie algebra representations. If $\pi_{1}$ and
$\pi_{2}$ are equivalent, then $\Pi_{1}$ and $\Pi_{2}$ are equivalent.

\item  Let $G$ be \textit{connected }and \textit{simply connected}. If $\pi$
is a representation of $\frak{g}$, then there exists a representation $\Pi$ of
$G$, acting on the same space, such that $\Pi$ and $\pi$ are related as in
Proposition \ref{relating.rep}.
\end{enumerate}
\end{corollary}

\begin{proof}
For (1), let $\Pi_{1}$ act on $V$ and $\Pi_{2}$ on $W$. We assume that the
associated Lie algebra representations are equivalent, i.e., that there exists
an invertible linear map $\phi:V\rightarrow W$ such that
\[
\phi\left(  \pi_{1}(X)v\right)  =\pi_{2}(X)\phi(v)
\]
for all $X\in\frak{g}$ and all $v\in V$. This is the same as saying that
$\phi\pi_{1}(X)=\pi_{2}(X)\phi$, or equivalently that $\phi\pi_{1}(X)\phi
^{-1}=\pi_{2}(X)$ (for all $X\in\frak{g}$).

Now define a map $\Sigma_{2}:G\rightarrow\mathsf{GL}(W)$ by the formula
\[
\Sigma_{2}(A)=\phi\Pi_{1}(A)\phi^{-1}\text{.}%
\]
It is trivial to check that $\Sigma_{2}$ is a homomorphism. Furthermore,
differentiation shows that the associated Lie algebra homomorphism is
\[
\sigma_{2}(X)=\phi\pi_{1}(X)\phi^{-1}=\pi_{2}(X)
\]
for all $X$. Then by (1) in the Theorem, we must also have $\Sigma_{2}=\Pi
_{2}$, i.e.,
\[
\phi\Pi_{1}(A)\phi^{-1}=\Pi_{2}(A)
\]
for all $A\in G$. But this shows that $\Pi_{1}$ and $\Pi_{2}$ are equivalent.

Point (2) of the Corollary follows immediately from Point (2) of the Theorem,
by taking $H=\mathsf{GL}(V)$. $\square$
\end{proof}

\ We now proceed with the proof of Theorem \ref{exp.rep}.

\begin{proof}
\textit{Step 1: Verify Point (1) of the Theorem.}

Since $G$ is connected, Corollary \ref{ex1.ex2} of Chapter 3 tells us that
every element $A$ of $G$ is a finite product of the form $A=\exp X_{1}\exp
X_{2}\cdots\exp X_{n}$, with $X_{i}\in\frak{g}$. But then if $\widetilde{\phi
}_{1}=\widetilde{\phi}_{2}$, we have
\[
\phi_{1}\left(  e^{X_{1}}\cdots e^{X_{n}}\right)  =e^{\widetilde{\phi}%
_{1}(X_{1})}\cdots e^{\widetilde{\phi}_{1}(X_{n})}=e^{\widetilde{\phi}%
_{2}(X_{1})}\cdots e^{\widetilde{\phi}_{2}(X_{n})}=\phi_{2}\left(  e^{X_{1}%
}\cdots e^{X_{n}}\right)  \text{.}%
\]

So we now need only prove Point (2).

\ 

\textit{Step 2: Define }$\phi$ \textit{in a neighborhood of the identity.}

Proposition \ref{local.log} of Chapter 3 says that the exponential mapping for
$G$ has a local inverse which maps a neighborhood $V$ of the identity into the
Lie algebra $\frak{g}$. On this neighborhood $V$ we can \textit{define}
$\phi:V\rightarrow H$ by
\[
\phi(A)=\exp\left\{  \widetilde{\phi}(\log A)\right\}  \text{.}%
\]
That is
\[
\phi=\exp\circ\widetilde{\phi}\circ\log\text{.}%
\]
(Note that if there is to be a homomorphism $\phi$ as in Theorem
\ref{homo.theorem} of Chapter 3, then on $V$, $\phi$ must be $\exp
\circ\widetilde{\phi}\circ\log$.)

It follows from Corollary \ref{local.homo} to the Baker-Campbell-Hausdorff
formula that this $\phi$ is a ``local homomorphism.'' That is, if $A$ and $B$
are in $V$, and \textit{if} $AB$ happens to be in $V$ as well, then
$\phi(AB)=\phi(A)\phi(B)$. (See the discussion at the beginning of Chapter 4.)

\ 

\textit{Step 3: Define }$\phi$ \textit{along a path.}

Recall that when we say $G$ is connected, we really mean that $G$ is
path-connected. Thus for any $A\in G$, there exists a path $A(t)\in G$ with
$A(0)=I$ and $A(1)=A$. A compactness argument shows that there exists numbers
$0=t_{0}<t_{1}<t_{2}\cdots<t_{n}=1$ such that
\begin{equation}
A(s)A(t_{i})^{-1}\in V\label{ratio}%
\end{equation}
for all $s$ between $t_{i}$ and $t_{i+1}$.

In particular, for $i=0$, we have $A(s)\in V$ for $0\leq s\leq t_{1}$. Thus we
can define $\phi\left(  A(s)\right)  $ by Step 2 for $s\in\lbrack0,t_{1}]$.
Now, for $s\in\lbrack t_{1},t_{2}]$ we have by (\ref{ratio}) $A(s)A(t_{1}%
)^{-1}\in V$. Moving the $A(t_{1})$ to the other side, this means that for
$s\in\lbrack t_{1},t_{2}]$ we can write
\[
A(s)=\left[  A(s)A(t_{1})^{-1}\right]  A(t_{1})\text{.}%
\]
with $A(s)A(t_{1})^{-1}\in V$. If $\phi$ is to be a homomorphism, we must
have
\begin{equation}
\phi\left(  A(s)\right)  =\phi\left(  \left[  A(s)A(t_{1})^{-1}\right]
A(t_{1})\right)  =\phi\left(  A(s)A(t_{1})^{-1}\right)  \phi\left(
A(t_{1})\right)  \text{.}\label{more.rat}%
\end{equation}
But $\phi\left(  A(t_{1})\right)  $ has already been defined, and we can
define $\phi\left(  A(s)A(t_{1})^{-1}\right)  $ by Step 2. In this way we can
use (\ref{more.rat}) to define $\phi\left(  A(s)\right)  $ for $s\in\lbrack
t_{1},t_{2}]$.

Proceeding on in the same way, we can define $\phi\left(  A(s)\right)  $
successively on each interval $[t_{i},t_{i+1}]$ until eventually we have
defined $\phi\left(  A(s)\right)  $ on the whole time interval $[0,1]$. This
in particular serves to define $\phi\left(  A(1)\right)  =\phi(A)$.

\ 

\textit{Step 4: Prove independence of path.}

In Step 3, we ``defined'' $\phi(A)$ by defining $\phi$ along a path joining
the identity to $A$. For this to make sense as a definition of $\phi(A)$ we
have to prove that the answer is independent of the choice of path, and also,
for a particular path, independent of the choice of partition $(t_{0}%
,t_{1},\cdots t_{n})$.

To establish independence of partition, we first show that passing from a
particular partition to a refinement of that partition doesn't change the
answer. (A refinement of a partition is one which contains all the points of
the original partition, plus some other ones.) This is proved by means of the
Baker-Campbell-Hausdorff formula. For example, suppose we insert an extra
partition point $s$ between $t_{0}$ and $t_{1}$. Under the old partition we
have
\begin{equation}
\phi\left(  A(t_{1})\right)  =\exp\circ\widetilde{\phi}\circ\log\left(
A(t_{1})\right)  \text{.}\label{old}%
\end{equation}
Under the new partition we write
\[
A(t_{1})=\left[  A(t_{1})A(s)^{-1}\right]  A(s)
\]
so that
\begin{equation}
\phi\left(  A(t_{1})\right)  =\exp\circ\widetilde{\phi}\circ\log\left(
A(t_{1})A(s)^{-1}\right)  \exp\circ\widetilde{\phi}\circ\log\left(
A(s)\right)  \text{.}\label{new}%
\end{equation}

But (as noted in Step 2), Corollary \ref{local.homo} of the
Baker-Campbell-Hausdorff formula (Chapter 4, Section \ref{bch.general})
implies that for $A$ and $B$ sufficiently near the identity
\[
\exp\circ\widetilde{\phi}\circ\log(AB)=\left[  \exp\circ\widetilde{\phi}%
\circ\log(A)\right]  \left[  \exp\circ\widetilde{\phi}\circ\log(B)\right]
\text{.}%
\]
Thus the right sides of (\ref{old}) and (\ref{new}) are equal. Once we know
that passing to a refinement doesn't change the answer, we have independence
of partition. For any two partitions of $\left[  0,1\right]  $ have a common
refinement, namely, the union of the two.

Once we know independence of partition, we need to prove independence of path.
It is at this point that we use the fact that $G$ is simply connected. In
particular, because of simple connectedness, any two paths $A_{1}(t)$ and
$A_{2}(t)$ joining the identity to $A$ will be homotopic with endpoints fixed.
(This is a standard topological fact.) Using this, we want to prove that Step
3 gives the same answer for $A_{1}$ and $A_{2}$.

Our strategy is to deform $A_{1}$ into $A_{2}$ in a series of steps, where
during each step we only change the path in a small time interval
$(t,t+\epsilon)$, keeping everything fixed on $[0,t]$ and on $[t+\epsilon,1]$.
Since we have independence of partition, we can take $t$ and $t+\epsilon$ to
be partition points. Since the time interval is small, we can assume there are
no partition points between $t$ and $t+\epsilon$. Then we have
\[
\phi\left(  A(t+\epsilon)\right)  =\phi\left(  A(t+\epsilon)A(t)^{-1}\right)
\phi\left(  A(t)\right)
\]
where $\phi\left(  A(t+\epsilon)A(t)^{-1}\right)  $ is defined as in Step 2.

But notice that our value for $\phi\left(  A(t+\epsilon)\right)  $ depends
only on $A\left(  t\right)  $ and $A\left(  t+\epsilon\right)  $, not on how
we get from $A\left(  t\right)  $ to $A\left(  t+\epsilon\right)  $! Thus the
value $\phi\left(  A(t+\epsilon)\right)  $ doesn't change as we deform the
path. But if $\phi\left(  A(t+\epsilon)\right)  $ doesn't change as we deform
the path, neither does $\phi\left(  A(1)\right)  $, since the path isn't
changing on $[t+\epsilon,1]$.

Since $A_{1}$ and $A_{2}$ are homotopic with endpoints fixed, it is possible
(by a standard topological argument) to deform $A_{1}$ into $A_{2}$ in a
series of small steps as above.

\ 

\textit{Step 5: Prove that }$\phi$\textit{\ is a homomorphism, and is properly
related to }$\widetilde{\phi}$.

Now that we have independence of path (and partition), we can give a simpler
description of how to compute $\phi$. Given any group element $A$, $A$ can be
written in the form
\[
A=C_{n}C_{n-1}\cdots C_{1}%
\]
with each $C_{i}$ in $V$. (This follows from the (path-)connectedness of $G$.)
We can then choose a path $A(t)$ which starts at the identity, then goes to
$C_{1}$, then to $C_{2}C_{1}$, and so on to $C_{n}C_{n-1}\cdots C_{1}=A$. We
can choose a partition so that $A(t_{i})=C_{i}C_{i-1}\cdots C_{1}$. By the way
we have defined things
\[
\phi(A)=\phi\left(  A(1)A(t_{n-1})^{-1}\right)  \phi\left(  A(t_{n-1}%
)A(t_{n-2})^{-1}\right)  \cdots\phi\left(  A(t_{1})A(0)\right)  \text{.}%
\]
But
\[
A(t_{i})A(t_{i-1})^{-1}=\left(  C_{i}C_{i-1}\cdots C_{1}\right)  \left(
C_{i-1}\cdots C_{1}\right)  ^{-1}=C_{i}%
\]
so
\[
\phi(A)=\phi(C_{n})\phi(C_{n-1})\cdots\phi(C_{1})\text{.}%
\]

Now suppose that $A$ and $B$ are two elements of $G$ and we wish to compute
$\phi(AB)$. Well, write
\begin{align*}
A=C_{n}C_{n-1}\cdots C_{1}\\
B=D_{n}D_{n-1}\cdots D_{1}\text{.}%
\end{align*}
Then
\begin{align*}
\phi\left(  AB\right)  =\phi\left(  C_{n}C_{n-1}\cdots C_{1}D_{n}D_{n-1}\cdots
D_{1}\right)  \\
=\left[  \phi(C_{n})\cdots\phi(C_{1})\right]  \left[  \phi(D_{n})\cdots
\phi(D_{1})\right]  \\
=\phi(A)\phi(B)\text{.}%
\end{align*}

We see then that $\phi$ is a homomorphism. It remains only to verify that
$\phi$ has the proper relationship to $\widetilde{\phi}$. But since $\phi$ is
defined near the identity to be $\phi=\exp\circ\widetilde{\phi}\circ\log$, we
see that
\[
\left.  \frac{d}{dt}\right|  _{t=0}\phi\left(  e^{tX}\right)  =\left.
\frac{d}{dt}\right|  _{t=0}e^{t\widetilde{\phi}(X)}=\widetilde{\phi
}(X)\text{.}%
\]
Thus $\widetilde{\phi}$ is the Lie algebra homomorphism associated to the Lie
group homomorphism $\phi$.

This completes the proof of Theorem \ref{exp.rep}.
\end{proof}

\section{Covering Groups}

It is at this point that we pay the price for our decision to consider only
\textit{matrix} Lie groups. For the universal covering group of a matrix Lie
group (defined below) is always a Lie group, but \textit{not} always a matrix
Lie group. For example, the universal covering group of $\mathsf{SL}\left(
n;\mathbb{R}\right)  $ ($n\ge2$) is a Lie group, but not a matrix Lie group.
(See Exercise \ref{not.matrix}.)

The notion of a universal cover allows us to determine, in the case of a
non-simply connected group, which representations of the Lie algebra
correspond to representations of the group. See Theorem \ref{non.simple} below.

\begin{definition}
Let $G$ be a connected matrix Lie group. A \textbf{universal covering group}
of $G$ (or just \textbf{universal cover}) is a connected, simply connected Lie
group $\widetilde{G}$, together with a Lie group homomorphism $\phi
:\widetilde{G}\rightarrow G$ (called the \textbf{projection map}) with the
following properties:

\begin{enumerate}
\item $\phi$ maps $\widetilde{G}$ onto $G$.

\item  There is a neighborhood $U$ of $I$ in $\widetilde{G}$ which maps
homeomorphically under $\phi$ onto a neighborhood $V$ of $I$ in $G$.
\end{enumerate}
\end{definition}

\begin{proposition}
If $G$ is any connected matrix Lie group, then a universal covering group
$\widetilde{G}$ of $G$ exists and is unique up to canonical isomorphism.
\end{proposition}

We will not prove this theorem, but the idea of proof is as follows. We assume
that $G$ is a matrix Lie group, hence a Lie group (that is, a manifold). As a
manifold, $G$ has a topological universal cover $\widetilde{G}$ which is a
connected, simply connected manifold. The universal cover comes with a
``projection map'' $\phi:\widetilde{G}\rightarrow G$ which is a local
homeomorphism. Now, since $G$ is not only a manifold but also a group,
$\widetilde{G}$ also becomes a group, and the projection map $\phi$ becomes a homomorphism.

\begin{proposition}
\label{cover.iso}Let $G$ be a connected matrix Lie group, $\widetilde{G}$ its
universal cover, and $\phi$ the projection map from $\widetilde{G}$ to $G$.
Suppose that $\widetilde{G}$ is a matrix Lie group with Lie algebra
$\widetilde{\frak{g}}$. Then the associated Lie algebra map
\[
\widetilde{\phi}:\widetilde{\frak{g}}\rightarrow\frak{g}%
\]
is an isomorphism.
\end{proposition}

In light of this Proposition, we often say that $G$ and $\widetilde{G}$ have
the same Lie algebra.

The above Proposition is true even if $\widetilde{G}$ is not a matrix Lie
group. But to make sense out of the Proposition in that case, we need the
definition of the Lie algebra of a general Lie group, which we have not defined.

\begin{proof}
Exercise \ref{prove.cover.iso}.
\end{proof}

\subsection{Examples}

The universal cover of $S^{1}$ is $\mathbb{R}$, and the projection map is the
map $x\rightarrow e^{ix}$. The universal cover of $\mathsf{SO}(3)$ is
$\mathsf{SU}(2)$, and the projection map is the homomorphism described in
Lemma \ref{su2.so3}.

More generally, we can consider $\mathsf{SO}(n)$ for $n\geq3$. As it turns
out, for $n\geq3$ the universal cover of $\mathsf{SO}(n)$ is a double cover.
(That is, the projection map $\phi$ is two-to-one.) The universal cover of
$\mathsf{SO}(n)$ is called $\mathsf{Spin}(n)$, and may be constructed as a
certain group of invertible elements in the \textbf{Clifford algebra} over
$\mathbb{R}^{n}$. See Br\"ocker and tom Dieck, Chapter I, Section 6,
especially Propositions I.6.17 and I.6.19. In particular, $\mathsf{Spin}(n)$
\textit{is} a matrix Lie group.

The case $n=4$ is quite special. It turns out that the universal cover of
$\mathsf{SO}(4)$ (i.e., $\mathsf{Spin}(4)$) is isomorphic to $\mathsf{SU}%
(2)\times\mathsf{SU}(2)$. This is best seen by regarding $\mathbb{R}^{4}$ as
the quaternion algebra.

\begin{theorem}
\label{cover.homo}Let $G$ be a matrix Lie group, and suppose that
$\widetilde{G}$ is also a matrix Lie group. Identify the Lie algebra of
$\widetilde{G}$ with the Lie algebra $\frak{g}$ of $G$ as in Proposition
\ref{cover.iso}. Suppose that $H$ is a matrix Lie group with Lie algebra
$\frak{h}$, and that $\widetilde{\phi}:\frak{g}\rightarrow\frak{h}$ is a
homomorphism. Then there exists a unique Lie group homomorphism $\phi
:\widetilde{G}\rightarrow H$ such that $\phi$ and $\widetilde{\phi}$ are
related as in Theorem \ref{homo.theorem} of Chapter 3.
\end{theorem}

\begin{proof}
$\widetilde{G}$ is simply connected.
\end{proof}

\begin{corollary}
Let $G$ and $\widetilde{G}$ be as in Theorem \ref{cover.homo}, and let $\pi$
be a representation of $\frak{g}$. Then there exists a unique representation
$\widetilde{\Pi}$ of $\widetilde{G}$ such that
\[
\pi(X)=\left.  \frac{d}{dt}\right|  _{t=0}\widetilde{\Pi}\left(
e^{tX}\right)
\]
for all $X\in\frak{g}$.
\end{corollary}

\begin{theorem}
\label{non.simple}Let $G$ and $\widetilde{G}$ be as in Theorem
\ref{cover.homo}, and let $\phi:\widetilde{G}\rightarrow G$. Now let $\pi$ be
a representation of $\frak{g}$, and $\widetilde{\Pi}$ the associated
representation of $\widetilde{G}$, as in the Corollary. Then there exists a
representation $\Pi$ of $G$ corresponding to $\pi$ if and only if
\[
\ker\widetilde{\Pi}\supset\ker\phi\text{.}%
\]
\end{theorem}

\begin{proof}
Exercise \ref{prove.non.simple}.
\end{proof}

\section{Exercises}

\begin{enumerate}
\item  Let $G$ be a matrix Lie group, and $\frak{g}$ its Lie algebra. Let
$\Pi_{1}$ and $\Pi_{2}$ be representations of $G$, and let $\pi_{1}$ and
$\pi_{2}$ be the associated representations of $\frak{g}$ (Proposition
\ref{relating.rep}). Show that if $\Pi_{1}$ and $\Pi_{2}$ are equivalent
representations of $G$, then $\pi_{1}$ and $\pi_{2}$ are equivalent
representations of $\frak{g}$. Show that if $G$ is connected, and if $\pi_{1}$
and $\pi_{2}$ are equivalent representations of $\frak{g}$, then $\Pi_{1}$ and
$\Pi_{2}$ are equivalent representations of $G$.

\textit{Hint}: Use Corollary \ref{ex1.ex2} of Chapter 3.

\item  Let $G$ be a connected matrix Lie group with Lie algebra $\frak{g}$.
Let $\Pi$ be a representation of $G$ acting on a space $V$, and let $\pi$ be
the associated Lie algebra representation. Show that a subspace $W\subset V$
is invariant for $\Pi$ if and only if it is invariant for $\pi$. Show that
$\Pi$ is irreducible if and only if $\pi$ is irreducible.

\item  Suppose that $\Pi$ is a finite-dimensional unitary representation of a
matrix Lie group $G$. (That is, $V$ is a finite-dimensional Hilbert space, and
$\Pi$ is a continuous homomorphism of $G$ into $U(V)$.) Let $\pi$ be the
associated representation of the Lie algebra $\frak{g}$. Show that for each
$X\in\frak{g}$, $\pi(X)^{\ast}=-\pi(X)$.

\item \label{equiv.rep}Show explicitly that the adjoint representation and the
standard representation are equivalent representations of the Lie algebra
$\mathsf{so}(3)$. Show that the adjoint and standard representations of the
group $\mathsf{SO}(3)$ are equivalent.

\item \label{su2.sl2}Consider the elements $E_{1}$, $E_{2}$, and $E_{3}$ in
$\mathsf{su}(2)$ defined in Exercise \ref{cross.product} of Chapter 3. These
elements form a basis for the real vector space $\mathsf{su}(2)$. Show
directly that $E_{1}$, $E_{2}$, and $E_{3}$ form a basis for the complex
vector space $\mathsf{sl}(2;\mathbb{C})$.

\item \label{sl2.check}Define a vector space with basis $u_{0},u_{1}\cdots
u_{m}$. Now define operators $\pi(H)$, $\pi(X)$, and $\pi(Y)$ by formula
(\ref{sl2.rep}). Verify by direct computation that the operators defined by
(\ref{sl2.rep}) satisfy the commutation relations $\left[  \pi(H),\pi
(X)\right]  =2\pi(X)$, $\left[  \pi(H),\pi(Y)\right]  =-2\pi(Y)$, and $\left[
\pi(X),\pi(Y)\right]  =\pi(H)$. (Thus $\pi(H)$, $\pi(X)$, and $\pi(Y)$ define
a representation of $\mathsf{sl}(2;\mathbb{C})$.) Show that this
representation is irreducible.

\textit{Hint}: It suffices to show, for example, that $\left[  \pi
(H),\pi(X)\right]  =2\pi(X)$ on each basis element. When dealing with $\pi
(Y)$, don't forget to treat separately the case of $u_{k}$, $k<m$, and the
case of $u_{m}$.

\item \label{spin.half}We can define a two-dimensional representation of
$\mathsf{so}(3)$ as follows:
\[
\pi\left(
\begin{array}
[c]{ccc}%
0 & 0 & 0\\
0 & 0 & 1\\
0 & -1 & 0
\end{array}
\right)  =\frac{1}{2}\left(
\begin{array}
[c]{cc}%
i & 0\\
0 & -i
\end{array}
\right)  ;
\]%
\[
\pi\left(
\begin{array}
[c]{ccc}%
0 & 0 & 1\\
0 & 0 & 0\\
-1 & 0 & 0
\end{array}
\right)  =\frac{1}{2}\left(
\begin{array}
[c]{cc}%
0 & 1\\
-1 & 0
\end{array}
\right)  ;
\]%
\[
\pi\left(
\begin{array}
[c]{ccc}%
0 & 1 & 0\\
-1 & 0 & 0\\
0 & 0 & 0
\end{array}
\right)  =\frac{1}{2}\left(
\begin{array}
[c]{cc}%
0 & i\\
i & 0
\end{array}
\right)  \text{.}%
\]
(You may assume that this actually gives a representation.) Show that there is
no group representation $\Pi$ of $\mathsf{SO}(3)$ such that $\Pi$ and $\pi$
are related as in Proposition \ref{relating.rep}.

\textit{Hint}: If $X\in\mathsf{so}(3)$ is such that $e^{X}=I$, and $\Pi$ is
any representation of $\mathsf{SO}(3)$, then $\Pi(e^{X})=\Pi(I)=I$.

\textit{Remark}: In the physics literature, this non-representation of
$\mathsf{SO}(3)$ is called ``spin $\frac{1}{2}$.''

\item \label{heisenberg.inv}Consider the standard representation of the
Heisenberg group, acting on $\mathbb{C}^{3}$. Determine all subspaces of
$\mathbb{C}^{3}$ which are invariant under the action of the Heisenberg group.
Is this representation completely reducible?

\item  Give an example of a representation of the commutative group
$\mathbb{R}$ which is not completely reducible.

\item \label{ccr}Consider the unitary representations $\Pi_{\hbar}$ of the
real Heisenberg group. Assume that there is some sort of associated
representation $\pi_{\hbar}$ of the Lie algebra, which should be given by
\[
\pi_{\hbar}(X)f=\left.  \frac{d}{dt}\right|  _{t=0}\Pi_{\hbar}\left(
e^{tX}\right)  f
\]
(We have not proved any theorem of this sort for infinite-dimensional unitary representations.)

Computing in a purely formal manner (that is, ignoring all technical issues)
compute
\[%
\begin{array}
[c]{ccc}%
\pi_{\hbar}\left(
\begin{array}
[c]{ccc}%
0 & 1 & 0\\
0 & 0 & 0\\
0 & 0 & 0
\end{array}
\right)  ; & \pi_{\hbar}\left(
\begin{array}
[c]{ccc}%
0 & 0 & 0\\
0 & 0 & 1\\
0 & 0 & 0
\end{array}
\right)  ; & \pi_{\hbar}\left(
\begin{array}
[c]{ccc}%
0 & 0 & 1\\
0 & 0 & 0\\
0 & 0 & 0
\end{array}
\right)
\end{array}
\text{.}%
\]
Verify (still formally) that these operators have the right commutation
relations to generate a representation of the Lie algebra of the real
Heisenberg group. (That is, verify that on this basis, $\pi_{\hbar}%
[X,Y]=[\pi_{\hbar}(X),\pi_{\hbar}(Y)]$.)

Why is this computation not rigorous?

\item \label{heisenberg.p}Consider the Heisenberg group over the field
$\mathbb{Z}_{p}$ of integers mod $p$, with $p$ prime, namely
\[
H_{p}=\left\{  \left(
\begin{array}
[c]{lll}%
1 & a & b\\
0 & 1 & c\\
0 & 0 & 1
\end{array}
\right)  \left|  a,b,c\in\mathbb{Z}_{p}\right.  \right\}  \text{.}%
\]
This is a subgroup of the group $\mathsf{GL}\left(  3;\mathbb{Z}_{p}\right)
$, and has $p^{3}$ elements.

Let $V_{p}$ denote the space of complex-valued functions on $\mathbb{Z}_{p}$,
which is a $p$-dimensional complex vector space. For each non-zero
$n\in\mathbb{Z}_{p}$, define a representation of $H_{p}$ by the formula
\[
\left(  \Pi_{n}f\right)  \left(  x\right)  =e^{-i2\pi nb/p}e^{i2\pi
ncx/p}f\left(  x-a\right)  \bigskip\ x\in\mathbb{Z}_{p}\text{.}%
\]
(These representations are analogous to the unitary representations of the
real Heisenberg group, with the quantity $2\pi n/p$ playing the role of
$\hbar$.)

a) Show that for each $n$, $\Pi_{n}$ is actually a representation of $H_{p}$,
and that it is irreducible.

b) Determine (up to equivalence) all the one-dimensional representations of
$H_{p}$.

c) Show that every irreducible representation of $H_{p}$ is either
one-dimensional or equivalent to one of the $\Pi_{n}$'s.

\item \label{universal}Prove Theorem \ref{tensor.exist}.

\textit{Hints}: For existence, choose bases $\left\{  e_{i}\right\}  $ and
$\left\{  f_{j}\right\}  $ for $U$ and $V$. Then define a space $W$ which has
as a basis $\left\{  w_{ij}\left|  0\leq i\leq n,0\leq j\leq m\right.
\right\}  $. Define $\phi(e_{i},f_{j})=w_{ij}$ and extend by bilinearity. For
uniqueness, use the universal property.

\item \label{direct.product}Let $\frak{g}$ and $\frak{h}$ be Lie algebras, and
consider the vector space $\frak{g}\oplus\frak{h}$. Show that the following
operation makes $\frak{g}\oplus\frak{h}$ into a Lie algebra
\[
\left[  (X_{1},Y_{1}),(X_{2},Y_{2})\right]  =\left(  [X_{1},X_{2}%
],[Y_{1},Y_{2}]\right)  \text{.}%
\]

Now let $G$ and $H$ be matrix Lie groups, with Lie algebras $\frak{g}$ and
$\frak{h}$. Show that $G\times H$ can be regarded as a matrix Lie group in an
obvious way, and that the Lie algebra of $G\times H$ is isomorphic to
$\frak{g}\oplus\frak{h}$.

\item \label{define.dual}Suppose that $\pi$ is a representation of a Lie
algebra $\frak{g}$ acting on a finite-dimensional vector space $V$. Let
$V^{\ast}$ denote as usual the dual space of $V$, that is, the space of linear
functionals on $V$. If $A$ is a linear operator on $V$, let $A^{tr}$ denote
the dual or transpose operator on $V^{\ast\text{ }}$,
\[
\left(  A^{tr}\phi\right)  \left(  v\right)  =\phi\left(  Av\right)
\]
for $\phi\in V^{\ast}$, $v\in V$. Define a representation $\pi^{\ast}$ of
$\frak{g}$ on $V^{\ast}$ by the formula
\[
\pi^{\ast}\left(  X\right)  =-\pi\left(  X^{tr}\right)  \text{.}%
\]

a) Show that $\pi^{\ast}$ is really a representation of $\frak{g}$.

b) Show that $\left(  \pi^{\ast}\right)  ^{\ast}$ is isomorphic to $\pi$.

c) Show that $\pi^{\ast}$ is irreducible if and only if $\pi$ is.

d) What is the analogous construction of the dual representation for
representations of groups?

\item \label{c.g}Recall the spaces $V_{m}$ introduced in Section
\ref{examples.rep}, viewed as representations of the Lie algebra
$\mathsf{sl}(2;\mathbb{C})$. In particular, consider the space $V_{1}$ (which
has dimension 2).

a) Regard $V_{1}\otimes V_{1}$ as a representation of $\mathsf{sl}%
(2;\mathbb{C})$, as in Definition \ref{algebra.tensor}. Show that this
representation is not irreducible.

b) Now view $V_{1}\otimes V_{1}$ as a representation of $\mathsf{sl}%
(2;\mathbb{C})\oplus\mathsf{sl}(2;\mathbb{C})$, as in Definition
\ref{tensor.algebra.2}. Show that this representation is irreducible.

c) More generally, show that $V_{m}\otimes V_{n}$ is irreducible as a
representation of $\mathsf{sl}(2;\mathbb{C})\oplus\mathsf{sl}(2;\mathbb{C})$,
but reducible (except if one of $n$ or $m$ is zero) as a representation of
$\mathsf{sl}(2;\mathbb{C})$.

\item  Show explicitly that $\exp:\mathsf{so}(3)\rightarrow\mathsf{SO}(3)$ is onto.

\textit{Hint}: Using the fact that $\mathsf{SO}(3)\subset\mathsf{SU}(3)$, show
that the eigenvalues of $R\in\mathsf{SO}(3)$ must be of one of the three
following forms: $(1,1,1)$, $(1,-1,-1)$, or $(1,e^{i\theta},e^{-i\theta})$. In
particular, $R$ must have an eigenvalue equal to one. Now show that in a
suitable orthonormal basis, $R$ is of the form
\[
R=\left(
\begin{array}
[c]{ccc}%
1 & 0 & 0\\
0 & \cos\theta & \sin\theta\\
0 & -\sin\theta & \cos\theta
\end{array}
\right)  \text{.}%
\]

\item \label{prove.su2.so3}\textit{Proof of Lemma \ref{su2.so3}}.

Let $\left\{  E_{1},E_{2},E_{3}\right\}  $ be the usual basis for
$\mathsf{su}(2)$, and $\left\{  F_{1},F_{2},F_{3}\right\}  $ be the basis for
$\mathsf{so}(3)$ introduced in Section \ref{group.algebra}. Identify
$\mathsf{su}(2)$ with $\mathbb{R}^{3}$ by identifying the basis $\left\{
E_{1},E_{2},E_{3}\right\}  $ with the standard basis for $\mathbb{R}^{3}$.
Consider $\mathrm{ad}E_{1}$, $\mathrm{ad}E_{2}$, and $\mathrm{ad}E_{3}$ as
operators on $\mathsf{su}(2)$, hence on $\mathbb{R}^{3}$. Show that
$\mathrm{ad}E_{i}=F_{i}$, for $i=1,2,3$. In particular, \textit{ad} is a Lie
algebra isomorphism of $\mathsf{su}(2)$ onto $\mathsf{so}(3)$.

Now consider $\mathrm{Ad}:\mathsf{SU}(2)\rightarrow\mathsf{GL}\left(
\mathsf{SU}(2)\right)  =\mathsf{GL}\left(  3;\mathbb{R}\right)  $. Show that
the image of \textit{Ad} is precisely $\mathsf{SO}(3)$. Show that the kernel
of \textit{Ad} is $\{I,-I\}$.

Show that $\mathrm{Ad}:\mathsf{SU}(2)\rightarrow\mathsf{SO}(3)$ is the
homomorphism $\Phi$ required by Lemma \ref{su2.so3}.

\item \label{prove.cover.iso}\textit{Proof of Proposition \ref{cover.iso}.}

Suppose that $G$ and $\widetilde{G}$ are matrix Lie groups. Suppose that
$\phi:\widetilde{G}\rightarrow G$ is a Lie group homomorphism such that $\phi$
maps some neighborhood $U$ of $I$ in $\widetilde{G}$ homeomorphically onto a
neighborhood $V$ of $I$ in $G$. Prove that the associated Lie algebra map
$\widetilde{\phi}:\widetilde{\frak{g}}\rightarrow\frak{g}$ is an isomorphism.

\textit{Hints}: Suppose that $\widetilde{\phi}$ were not one-to-one. Show,
then, that there exists a sequence of points $A_{n}$ in $\widetilde{G}$ with
$A_{n}\neq I$, $A_{n}\rightarrow I$ and $\phi(A_{n})=I$, giving a contradiction.

To show that $\widetilde{\phi}$ is onto, use Step 1 of the proof of Theorem
\ref{exp.rep} to show that on a sufficiently small neighborhood of zero in
$\widetilde{\frak{g}}$,
\[
\widetilde{\phi}=\log\circ\phi\circ\exp\text{.}%
\]
Use this to show that the image of $\widetilde{\phi}$ contains a neighborhood
of zero in $\frak{g}$. Now use linearity to show that the image of
$\widetilde{\phi}$ is all of $\frak{g}$.

\item \label{prove.non.simple}\textit{Proof of Theorem \ref{non.simple}.}

First suppose that $\ker\widetilde{\Pi}\supset\ker\phi$. Then construct $\Pi$
as in the proof of Proposition \ref{so3.odd}.

Now suppose that there is a representation $\Pi$ of $G$ for which the
associated Lie algebra representation is $\pi$. We want to show, then, that
$\ker\widetilde{\Pi}\supset\ker\phi$. Well, define a new representation
$\Sigma$ of $\widetilde{G}$ by
\[
\Sigma=\Pi\circ\phi\text{.}%
\]
Show that the associated Lie algebra homomorphism $\sigma$ is equal to $\pi$,
so that, by Point (1) of Theorem \ref{exp.rep}, $\widetilde{\Pi}=\Sigma$. What
can you say about the kernel of $\Sigma$?

\item \label{not.matrix}Fix an integer $n\geq2$.

a) Show that every (finite-dimensional complex) representation of the Lie
algebra $\mathsf{sl}\left(  n;\mathbb{R}\right)  $ gives rise to a
representation of the group $\mathsf{SL}\left(  n;\mathbb{R}\right)  $, even
though $\mathsf{SL}\left(  n;\mathbb{R}\right)  $ is not simply connected.
(You may use the fact that $\mathsf{SL}\left(  n;\mathbb{C}\right)  $ is
simply connected.)

b) Show that the universal cover of $\mathsf{SL}\left(  n;\mathbb{R}\right)  $
is not isomorphic to any matrix Lie group. (You may use the fact that
$\mathsf{SL}\left(  n;\mathbb{R}\right)  $ is not simply connected.)

\item  Let $G$ be a matrix Lie group with Lie algebra $\frak{g}$, let
$\frak{h}$ be a subalgebra of $\frak{g}$, and let $H$ be the unique connected
Lie subgroup of $G$ with Lie algebra $\frak{h}$. Suppose that there exists a
compact simply connected matrix Lie group $K$ such that the Lie algebra of $K$
is isomorphic to $\frak{h}$. Show that $H$ is closed. Is $H$ necessarily
isomorphic to $K$?
\end{enumerate}

\chapter{The Representations of $\mathsf{SU}(3)$, and Beyond}

\section{Preliminaries}

There is a theory of the representations of semisimple groups/Lie algebras
which includes as a special case the representation theory of $\mathsf{SU}(3)
$. However, I feel that it is worthwhile to examine the case of $\mathsf{SU}%
(3)$ separately. I feel this way partly because $\mathsf{SU}(3)$ is an
important group in physics, but chiefly because the general semisimple theory
is difficult to digest. Considering a non-trivial example makes it much
clearer what is going on. In fact, all of the elements of the general theory
are present already in the case of $\mathsf{SU}(3)$, so we do not lose too
much by considering at first just this case.

The main result of this chapter is Theorem \ref{classify.sl3}, which states
that an irreducible finite-dimensional representation of $\mathsf{SU}(3)$ can
be classified in terms of its ``highest weight.'' This is analogous to
labeling the irreducible representations $V_{m}$ of $\mathsf{SU}%
(2)/\mathsf{sl}(2;\mathbb{C})$ by the highest eigenvalue of $\pi_{m}(H)$. (The
highest eigenvalue of $\pi_{m}(H)$ in $V_{m}$ is precisely $m$.) We will then
discuss, without proofs, what the corresponding results are for general
semisimple Lie algebras.

The group $\mathsf{SU}(3)$ is connected and simply connected (Br\"ocker and
tom Dieck), so by Corollary \ref{sc.reps} of Chapter 5, the finite-dimensional
representations of $\mathsf{SU}(3)$ are in one-to-one correspondence with the
finite-dimensional representations of the Lie algebra $\mathsf{su}(3)$.
Meanwhile, the complex representations of $\mathsf{su}(3)$ are in one-to-one
correspondence with the complex-linear representations of the complexified Lie
algebra $\mathsf{su}(3)_{\mathbb{C}}$. But $\mathsf{su}(3)_{\mathbb{C}}%
\cong\mathsf{sl}\left(  3;\mathbb{C}\right)  $, as is easily verified.
Moreover, since $\mathsf{SU}(3)$ is connected, it follows that a subspace
$W\subset V$ is invariant under the action of $\mathsf{SU}(3)$ if and only if
it is invariant under the action of $\mathsf{sl}\left(  3;\mathbb{C}\right)
$. Thus we have the following:

\begin{proposition}
There is a one-to-one correspondence between the finite-dimensional complex
representations $\Pi$ of $\mathsf{SU}(3)$ and the finite-dimensional
complex-linear representations $\pi$ of $\mathsf{sl}\left(  3;\mathbb{C}%
\right)  $. This correspondence is determined by the property that
\[
\Pi\left(  e^{X}\right)  =e^{\pi(X)}%
\]
for all $X\in\mathsf{su}(3)\subset\mathsf{sl}\left(  3;\mathbb{C}\right)  $.

The representation $\Pi$ is irreducible if and only the representation $\pi$
is irreducible. Moreover, a subspace $W\subset V$ is invariant for $\Pi$ if
and only if it is invariant for $\pi$.
\end{proposition}

Since $\mathsf{SU}(3)$ is compact, Proposition \ref{compact.reduce} of Chapter
5 tells us that all the finite-dimensional representations of $\mathsf{SU}(3)$
are completely reducible. The above proposition then implies that all the
finite-dimensional representations of $\mathsf{sl}\left(  3;\mathbb{C}\right)
$ are completely reducible.

Moreover, we can apply the same reasoning to the group $\mathsf{SU}(2)$, its
Lie algebra $\mathsf{su}(2)$, and its complexified Lie algebra $\mathsf{sl}%
(2;\mathbb{C})$. Since $\mathsf{SU}(2)$ is simply connected, there is a
one-to-one correspondence between the complex representations of
$\mathsf{SU}(2)$ and the representations of the complexified Lie algebra
$\mathsf{sl}(2;\mathbb{C})$. Since $\mathsf{SU}(2)$ is compact, all of the
representations of $\mathsf{SU}(2)$--and therefore also of $\mathsf{sl}%
(2;\mathbb{C})$--are completely reducible. Thus we have established the following.

\begin{proposition}
\label{reduce.sl3}Every finite-dimensional (complex-linear) representation of
$\mathsf{sl}(2;\mathbb{C})$ or $\mathsf{sl}\left(  3;\mathbb{C}\right)  $ is
completely reducible. In particular, every finite-dimensional representation
of $\mathsf{sl}(2;\mathbb{C})$ or $\mathsf{sl}\left(  3;\mathbb{C}\right)  $
decomposes as a direct sum of irreducible invariant subspaces.
\end{proposition}

We will use the following basis for $\mathsf{sl}\left(  3;\mathbb{C}\right)
$:
\[%
\begin{array}
[c]{ccc}%
H_{1}=\left(
\begin{array}
[c]{ccc}%
1 & 0 & 0\\
0 & -1 & 0\\
0 & 0 & 0
\end{array}
\right)  & H_{2}=\left(
\begin{array}
[c]{ccc}%
0 & 0 & 0\\
0 & 1 & 0\\
0 & 0 & -1
\end{array}
\right)  & \\
&  & \\
X_{1}=\left(
\begin{array}
[c]{ccc}%
0 & 1 & 0\\
0 & 0 & 0\\
0 & 0 & 0
\end{array}
\right)  & X_{2}=\left(
\begin{array}
[c]{ccc}%
0 & 0 & 0\\
0 & 0 & 1\\
0 & 0 & 0
\end{array}
\right)  & X_{3}=\left(
\begin{array}
[c]{ccc}%
0 & 0 & 1\\
0 & 0 & 0\\
0 & 0 & 0
\end{array}
\right) \\
&  & \\
Y_{1}=\left(
\begin{array}
[c]{ccc}%
0 & 0 & 0\\
1 & 0 & 0\\
0 & 0 & 0
\end{array}
\right)  & Y_{2}=\left(
\begin{array}
[c]{ccc}%
0 & 0 & 0\\
0 & 0 & 0\\
0 & 1 & 0
\end{array}
\right)  & Y_{3}=\left(
\begin{array}
[c]{ccc}%
0 & 0 & 0\\
0 & 0 & 0\\
1 & 0 & 0
\end{array}
\right)  \text{.}%
\end{array}
\]

Note that the span of $\left\{  H_{1},X_{1},Y_{1}\right\}  $ is a subalgebra
of $\mathsf{sl}\left(  3;\mathbb{C}\right)  $ which is isomorphic to
$\mathsf{sl}(2;\mathbb{C})$, by ignoring the third row and the third column.
Similarly, the span of $\left\{  H_{2},X_{2},Y_{2}\right\}  $ is a subalgebra
isomorphic to $\mathsf{sl}(2;\mathbb{C})$, by ignoring the first row and first
column. Thus we have the following commutation relations%

\[%
\begin{array}
[c]{rrrrrrr}%
\left[  H_{1},X_{1}\right]  & = & 2X_{1} &  & \left[  H_{2},X_{2}\right]  &
= & 2X_{2}^{}\\
\left[  H_{1},Y_{1}\right]  & = & -2Y_{1} &  & \left[  H_{2},Y_{2}\right]  &
= & -2Y_{2}\\
\left[  X_{1},Y_{1}\right]  & = & H_{1} &  & \left[  X_{2},Y_{2}\right]  & = &
H_{2}\text{.}%
\end{array}
\]

We now list all of the commutation relations among the basis elements which
involve at least one of $H_{1}$ and $H_{2}$. (This includes some repetitions
of the commutation relations above.)
\begin{equation}%
\begin{array}
[c]{rrrrrrr}%
\left[  H_{1},H_{2}\right]  & = & 0 &  &  &  & \\
&  &  &  &  &  & \\
\left[  H_{1},X_{1}\right]  & = & 2X_{1} &  & \left[  H_{1},Y_{1}\right]  &
= & -2Y_{1}\\
\left[  H_{2},X_{1}\right]  & = & -X_{1} &  & \left[  H_{2},Y_{1}\right]  &
= & Y_{1}\\
&  &  &  &  &  & \\
\left[  H_{1},X_{2}\right]  & = & -X_{2} &  & \left[  H_{1},Y_{2}\right]  &
= & Y_{2}\\
\left[  H_{2},X_{2}\right]  & = & 2X_{2} &  & \left[  H_{2},Y_{2}\right]  &
= & -2Y_{2}\\
&  &  &  &  &  & \\
\left[  H_{1},X_{3}\right]  & = & X_{3} &  & \left[  H_{1},Y_{3}\right]  & = &
-Y_{3}\\
\left[  H_{2},X_{3}\right]  & = & X_{3} &  & \left[  H_{2},Y_{3}\right]  & = &
-Y_{3}%
\end{array}
\label{h.rel}%
\end{equation}

We now list all of the remaining commutation relations.
\[%
\begin{array}
[c]{ccccccc}%
\left[  X_{1},Y_{1}\right]  & = & H_{1} &  &  &  & \\
\left[  X_{2},Y_{2}\right]  & = & H_{2} &  &  &  & \\
\left[  X_{3},Y_{3}\right]  & = & H_{1}+H_{2} &  & \qquad & \ \ \quad &
\end{array}
\]%
\[%
\begin{array}
[c]{ccccccc}%
\left[  X_{1},X_{2}\right]  & = & X_{3} &  & \left[  Y_{1},Y_{2}\right]  & = &
-Y_{3}\\
\left[  X_{1},Y_{2}\right]  & = & 0 &  & \left[  X_{2},Y_{1}\right]  & = & 0\\
&  &  &  &  &  & \\
\left[  X_{1},X_{3}\right]  & = & 0 &  & \left[  Y_{1},Y_{3}\right]  & = & 0\\
\left[  X_{2},X_{3}\right]  & = & 0 &  & \left[  Y_{2},Y_{3}\right]  & = & 0\\
&  &  &  &  &  & \\
\left[  X_{2},Y_{3}\right]  & = & Y_{1} &  & \left[  X_{3},Y_{2}\right]  & = &
X_{1}\\
\left[  X_{1},Y_{3}\right]  & = & -Y_{2} &  & \left[  X_{3},Y_{1}\right]  &
= & -X_{2}%
\end{array}
\]

Note that there is a kind of symmetry between the $X_{i}$'s and the $Y_{i}$'s.
If a relation in the first column involves an $X_{i}$ and/or a $Y_{j}$, the
corresponding relation in the second column will involve a $Y_{i}$ and/or an
$X_{j}$. (E.g., we have the relation $\left[  H_{1},X_{2}\right]  =-X_{2}$ in
the first column, and the relation $\left[  H_{2},Y_{2}\right]  =Y_{2}$ in the
second column.) See Exercise \ref{relations.tr}.

All of the analysis we will do for the representations of $\mathsf{sl}\left(
3;\mathbb{C}\right)  $ will be in terms of the above basis. From now on, all
representations of $\mathsf{sl}\left(  3;\mathbb{C}\right)  $ will be assumed
to be finite-dimensional and complex-linear.

\section{Weights and Roots}

Our basic strategy in classifying the representations of $\mathsf{sl}\left(
3;\mathbb{C}\right)  $ is to simultaneously diagonalize $\pi(H_{1})$ and
$\pi(H_{2})$. Since $H_{1}$ and $H_{2}$ commute, $\pi(H_{1})$ and $\pi(H_{2})$
will also commute, and so there is at least a chance that $\pi(H_{1})$ and
$\pi(H_{2})$ can be simultaneously diagonalized.

\begin{definition}
If $\left(  \pi,V\right)  $ is a representation of $\mathsf{sl}\left(
3;\mathbb{C}\right)  $, then an ordered pair $\mu=\left(  \mu_{1},\mu
_{2}\right)  \in\mathbb{C}^{2}$ is called a \textbf{weight} for $\pi$ if there
exists $v\neq0$ in $V$ such that
\begin{align}
\pi(H_{1})v=\mu_{1}v\nonumber\\
\pi(H_{2})v=\mu_{2}v\text{.}\label{weight.def}%
\end{align}
The vector $v$ is called a \textbf{weight vector} corresponding to the weight
$\mu$. If $\mu=\left(  \mu_{1},\mu_{2}\right)  $ is a weight, then the space
of all vectors $v$ satisfying (\ref{weight.def}) is the \textbf{weight space}
corresponding to the weight $\mu$.
\end{definition}

Thus a weight is simply a pair of simultaneous eigenvalues for $\pi(H_{1})$
and $\pi(H_{2})$.

\begin{proposition}
\label{weight.exist}Every representation of $\mathsf{sl}\left(  3;\mathbb{C}%
\right)  $ has at least one weight.
\end{proposition}

\begin{proof}
Since we are working over the complex numbers, $\pi(H_{1})$ has at least one
eigenvalue $\mu_{1}$. Let $W\subset V$ be the eigenspace for $\pi(H_{1})$ with
eigenvalue $\mu_{1}$. I assert that $W$ is invariant under $\pi(H_{2})$. To
see this consider $w\in W$, and compute
\begin{align*}
\pi(H_{1})\left(  \pi(H_{2})w\right)    & =\pi(H_{2})\pi(H_{1})w\\
& =\pi(H_{2})\left(  \mu_{1}w\right)  =\mu_{1}\pi(H_{2})w\text{.}%
\end{align*}
This shows that $\pi(H_{2})w$ is either zero or an eigenvector for $\pi
(H_{1})$ with eigenvalue $\mu_{1}$; thus $W$ is invariant.

Thus $\pi(H_{2})$ can be viewed as an operator on $W$. Again, since we are
over $\mathbb{C}$, the restriction of $\pi(H_{2})$ to $W$ must have at least
one eigenvector $w$ with eigenvalue $\mu_{2}$. But then $w$ is a simultaneous
eigenvector for $\pi(H_{1})$ and $\pi(H_{2})$ with eigenvalues $\mu_{1}$ and
$\mu_{2}$.
\end{proof}

Now, every representation $\pi$ of $\mathsf{sl}\left(  3;\mathbb{C}\right)  $
can be viewed, by restriction, as a representation of the subalgebra $\left\{
H_{1},X_{1},Y_{1}\right\}  \cong\mathsf{sl}(2;\mathbb{C})$. Note that, even if
$\pi$ is irreducible as a representation of $\mathsf{sl}\left(  3;\mathbb{C}%
\right)  $, there is no reason to expect that it will still be irreducible as
a representation of the subalgebra $\left\{  H_{1},X_{1},Y_{1}\right\}  $.
Nevertheless, $\pi$ restricted to $\left\{  H_{1},X_{1},Y_{1}\right\}  $ must
be \textit{some} finite-dimensional representation of $\mathsf{sl}%
(2;\mathbb{C})$. The same reasoning applies to the restriction of $\pi$ to the
subalgebra $\left\{  H_{2},X_{2},Y_{2}\right\}  $, which is also isomorphic to
$\mathsf{sl}(2;\mathbb{C})$.

\begin{proposition}
\label{sl2.integer}Let $\left(  \pi,V\right)  $ be any finite-dimensional
complex-linear representation of $\mathsf{sl}(2;\mathbb{C})=\left\{
H,X,Y\right\}  $. Then all the eigenvalues of $\pi(H)$ are integers.
\end{proposition}

\begin{proof}
By Proposition \ref{reduce.sl3}, $V$ decomposes as a direct sum of irreducible
invariant subspaces $V_{i}$. Each $V_{i}$ must be one of the irreducible
representations of $\mathsf{sl}(2;\mathbb{C})$, which we have classified. In
particular, in each $V_{i}$, $\pi(H)$ can be diagonalized, and the eigenvalues
of $\pi(H)$ are integers. Thus $\pi(H)$ can be diagonalized on the whole space
$V$, and all of the eigenvalues are integers.
\end{proof}

\begin{corollary}
If $\pi$ is a representation of $\mathsf{sl}\left(  3;\mathbb{C}\right)  $,
then all of the weights of $\pi$ are of the form
\[
\mu=(m_{1},m_{2})
\]
with $m_{1}$ and $m_{2}$ integers.
\end{corollary}

\begin{proof}
Apply Proposition \ref{sl2.integer} to the restriction of $\pi$ to $\left\{
H_{1},X_{1},Y_{1}\right\}  $, and to the restriction of $\pi$ to $\left\{
H_{2},X_{2},Y_{2}\right\}  $.
\end{proof}

Our strategy now is to begin with one simultaneous eigenvector for $\pi(H_{1})
$ and $\pi(H_{2})$, and then to apply $\pi(X_{i})$ or $\pi(Y_{i})$, and see
what the effect is. The following definition is relevant in this context. (See
Lemma \ref{weight.root} below.)

\begin{definition}
An ordered pair $\alpha=\left(  \alpha_{1},\alpha_{2}\right)  \in
\mathbb{C}^{2}$ is called a \textbf{root} if

\begin{enumerate}
\item $\alpha_{1}$ and $\alpha_{2}$ are not both zero, and

\item  there exists $Z\in\mathsf{sl}\left(  3;\mathbb{C}\right)  $ such that
\begin{align*}
\left[  H_{1},Z\right]  =\alpha_{1}Z\\
\left[  H_{2},Z\right]  =\alpha_{2}Z\text{.}%
\end{align*}
\end{enumerate}

The element $Z$ is called a \textbf{root vector} corresponding to the root
$\alpha$.
\end{definition}

That is, a root is a non-zero weight for the adjoint representation. The
commutation relations (\ref{h.rel}) tell us what the roots for $\mathsf{sl}%
\left(  3;\mathbb{C}\right)  $ are. There are six roots.
\begin{equation}%
\begin{array}
[c]{cc}%
\mathbf{\alpha} & \mathbf{Z}\\
\left(  2,-1\right)  & X_{1}\\
(-1,2) & X_{2}\\
(1,1) & X_{3}\\
(-2,1) & Y_{1}\\
(1,-2) & Y_{2}\\
(-1,-1) & Y_{3}%
\end{array}
\label{the.roots}%
\end{equation}

It is convenient to single out the two roots corresponding to $X_{1}$ and
$X_{2}$ and give them special names:
\begin{align}
\alpha^{(1)}  & =\left(  2,-1\right) \nonumber\\
\alpha^{(2)}  & =(-1,2)\text{.}\label{simple.roots}%
\end{align}
The roots $\alpha^{(1)}$ and $\alpha^{(2)}$ are called the \textbf{simple
roots}. They have the property that all of the roots can be expressed as
linear combinations of $\alpha^{(1)}$ and $\alpha^{(2)}$ with \textit{integer}
coefficients, and these coefficients are either all greater than or equal to
zero or all less than or equal to zero. This is verified by direct
computation:
\[%
\begin{array}
[c]{ccc}%
(2,-1) & = & \alpha^{(1)}\\
(-1,2) & = & \alpha^{(2)}\\
(1,1) & = & \alpha^{(1)}+\alpha^{(2)}\\
(-2,1) & = & -\alpha^{(1)}\\
(1,-2) & = & -\alpha^{(2)}\\
(-1,-1) & = & -\alpha^{(1)}-\alpha^{(2)}\text{.}%
\end{array}
\]

The significance of the roots for the representation theory of $\mathsf{sl}%
\left(  3;\mathbb{C}\right)  $ is contained in the following Lemma. Although
its proof is very easy, this Lemma plays a crucial role in the classification
of the representations of $\mathsf{sl}\left(  3;\mathbb{C}\right)  $. Note
that this Lemma is the analog of Lemma \ref{raising.op} of Chapter 5, which
was the key to the classification of the representations of $\mathsf{sl}%
(2;\mathbb{C})$.

\begin{lemma}
\label{weight.root}Let $\alpha=(\alpha_{1},\alpha_{2})$ be a root, and
$Z_{\alpha}\neq0$ a corresponding root vector in $\mathsf{sl}\left(
3;\mathbf{C}\right)  $. Let $\pi$ be a representation of $\mathsf{sl}\left(
3;\mathbb{C}\right)  $, $\mu=(m_{1},m_{2})$ a weight for $\pi$, and $v\neq0$ a
corresponding weight vector. Then
\begin{align*}
\pi(H_{1})\pi(Z_{\alpha})v=(m_{1}+\alpha_{1})\pi(Z_{\alpha})v\\
\pi(H_{2})\pi(Z_{\alpha})v=(m_{2}+\alpha_{2})\pi(Z_{\alpha})v\text{.}%
\end{align*}
Thus either $\pi(Z_{\alpha})v=0$ or else $\pi(Z_{\alpha})v$ is a new weight
vector with weight
\[
\mu+\alpha=(m_{1}+\alpha_{1},m_{2}+\alpha_{2})\text{.}%
\]
\end{lemma}

\begin{proof}
The definition of a root tells us that we have the commutation relation
$\left[  H_{1}Z_{\alpha}\right]  =\alpha_{1}Z_{\alpha}$. Thus
\begin{align*}
\pi(H_{1})\pi(Z_{\alpha})v=\left(  \pi(Z_{\alpha})\pi(H_{1})+\alpha_{1}%
\pi(Z_{a})\right)  v\\
=\pi(Z_{\alpha})(m_{1}v)+\alpha_{1}\pi(Z_{\alpha})v\\
=(m_{1}+\alpha_{1})\pi(Z_{\alpha})v\text{.}%
\end{align*}
A similar argument allows us to compute $\pi(H_{2})\pi(Z_{\alpha})v$.
\end{proof}

\section{Highest Weights and the Classification Theorem}

We see then that if we have a representation with a weight $\mu=(m_{1},m_{2}%
)$, then by applying the root vectors $X_{1},X_{2},X_{3},Y_{1},Y_{2},Y_{3}$ we
can get some new weights of the form $\mu+\alpha$, where $\alpha$ is the root.
Of course, some of the weight vectors may simply give zero. In fact, since our
representation is finite-dimensional, there can be only finitely many weights,
so we must get zero quite often. By analogy to the classification of the
representations of $\mathsf{sl}(2;\mathbf{C})$, we would like to single out in
each representation a ``highest'' weight, and then work from there. The
following definition gives the ``right'' notion of highest.

\begin{definition}
Let $\alpha^{(1)}=(2,-1)$ and $\alpha^{(2)}=(-1,2)$ be the roots introduced in
(\ref{simple.roots}). Let $\mu_{1}$ and $\mu_{2}$ be two weights. Then
$\mu_{1}$ is \textbf{higher} than $\mu_{2}$ (or equivalently, $\mu_{2}$ is
\textbf{lower} than $\mu_{1}$) if $\mu_{1}-\mu_{2}$ can be written in the
form
\[
\mu_{1}-\mu_{2}=a\alpha^{(1)}+b\alpha^{(2)}%
\]
with $a\geq0$ and $b\geq0$. This relationship is written as $\mu_{1}\succeq
\mu_{2}$ or $\mu_{2}\preceq\mu_{1}$.

If $\pi$ is a representation of $\mathsf{sl}\left(  3;\mathbb{C}\right)  $,
then a weight $\mu_{0}$ for $\pi$ is said to be a \textbf{highest weight} if
for all weights $\mu$ of $\pi$, $\mu\preceq\mu_{0}$.
\end{definition}

Note that the relation of ``higher'' is only a \textit{partial} ordering. That
is, one can easily have $\mu_{1}$ and $\mu_{2}$ such that $\mu_{1}$ is neither
higher nor lower than $\mu_{2}$. For example, $\alpha^{(1)}-\alpha^{(2)}$ is
neither higher nor lower than $0$. This in particular means that a finite set
of weights need not have a highest element. (E.g., the set $\left\{
0,\alpha^{(1)}-\alpha^{(2)}\right\}  $ has no highest element.)

We are now ready to state the main theorem regarding the irreducible
representations of $\mathsf{sl}\left(  3;\mathbb{C}\right)  $.

\begin{theorem}
\begin{enumerate}
\item \label{classify.sl3}Every irreducible representation $\pi$ of
$\mathsf{sl}\left(  3;\mathbb{C}\right)  $ is the direct sum of its weight
spaces. That is, $\pi(H_{1})$ and $\pi(H_{2})$ are simultaneously diagonalizable.

\item  Every irreducible representation of $\mathsf{sl}\left(  3;\mathbb{C}%
\right)  $ has a unique highest weight $\mu_{0}$, and two equivalent
irreducible representations have the same highest weight.

\item  Two irreducible representations of $\mathsf{sl}\left(  3;\mathbb{C}%
\right)  $ with the same highest weight are equivalent.

\item  If $\pi$ is an irreducible representation of $\mathsf{sl}\left(
3;\mathbb{C}\right)  $, then the highest weight $\mu_{0}$ of $\pi$ is of the
form
\[
\mu_{0}=(m_{1},m_{2})
\]
with $m_{1}$ and $m_{2}$ non-negative integers.

\item  Conversely, if $m_{1}$ and $m_{2}$ are non-negative integers, then
there exists a unique irreducible representation $\pi$ of $\mathsf{sl}\left(
3;\mathbb{C}\right)  $ with highest weight $\mu_{0}=(m_{1},m_{2})$.
\end{enumerate}
\end{theorem}

Note the parallels between this result and the classification of the
irreducible representations of $\mathsf{sl}(2;\mathbb{C})$: In each
irreducible representation of $\mathsf{sl}(2;\mathbb{C})$, $\pi(H)$ is
diagonalizable, and there is a largest eigenvalue of $\pi(H)$. Two irreducible
representations of $\mathsf{sl}(2;\mathbb{C})$ with the same largest
eigenvalue are equivalent. The highest eigenvalue is always a non-negative
integer, and conversely, for every non-negative integer $m$, there is an
irreducible representation with highest eigenvalue $m$.

However, note that in the classification of the representations of
$\mathsf{sl}\left(  3;\mathbb{C}\right)  $ the notion of ``highest'' does not
mean what we might have thought it should mean. For example, the weight
$(1,1)$ is higher than the weights $(-1,2)$ and $(2,-1)$. (In fact, $(1,1)$ is
the highest weight for the adjoint representation, which is irreducible.)

It is possible to obtain much more information about the irreducible
representations besides the highest weight. For example, we have the following
formula for the dimension of the representation with highest weight
$(m_{1},m_{2})$.

\begin{theorem}
The dimension of the irreducible representation with highest weight
$(m_{1},m_{2})$ is
\[
\frac{1}{2}(m_{1}+1)(m_{2}+1)(m_{1}+m_{2}+2)\text{.}%
\]
\end{theorem}

We will not prove this formula. It is a consequence of the ``Weyl character
formula.'' See Humphreys, Section 24.3. Humphreys refers to $\mathsf{sl}%
\left(  3;\mathbb{C}\right)  $ as $A_{2}$.

\section{Proof of the Classification Theorem}

It will take us some time to prove Theorem \ref{classify.sl3}. The proof will
consist of a series of Propositions.

\begin{proposition}
\label{sl3.weights}In every irreducible representation $\left(  \pi,V\right)
$ of $\mathsf{sl}\left(  3;\mathbb{C}\right)  $, $\pi(H_{1})$ and $\pi(H_{2})$
can be simultaneously diagonalized. That is, $V$ is the direct sum of its
weight spaces.
\end{proposition}

\begin{proof}
Let $W$ be the direct sum of the weight spaces in $V$. Equivalently, $W$ is
the space of all vectors $w\in V$ such that $w$ can be written as a linear
combination of simultaneous eigenvectors for $\pi(H_{1})$ and $\pi(H_{2})$.
Since (Proposition \ref{weight.exist}) $\pi$ always has at least one weight,
$W\neq\left\{  0\right\}  $.

On the other hand, Lemma \ref{weight.root} tells us that if $Z_{\alpha}$ is a
root vector corresponding to the root $\alpha$, then $\pi(Z_{\alpha})$ maps
the weight space corresponding to $\mu$ into the weight space corresponding to
$\mu+\alpha$. Thus $W$ is invariant under the action of all of the root
vectors, namely, under the action $X_{1},X_{2},X_{3},Y_{1},Y_{2},$ and $Y_{3}%
$. Since $W$ is certainly invariant under the action of $H_{1}$ and $H_{2}$,
$W$ is invariant. Thus by irreducibility, $W=V$.
\end{proof}

\begin{definition}
A representation $\left(  \pi,V\right)  $ of $\mathsf{sl}\left(
3;\mathbb{C}\right)  $ is said to be a \textbf{highest weight cyclic
representation with weight }$\mu_{0}=(m_{1},m_{2})$ if there exists $v\neq0$
in $V$ such that

\begin{enumerate}
\item $v$ is a weight vector with weight $\mu_{0}$.

\item $\pi(X_{1})v=\pi(X_{2})v=0$.

\item  The smallest invariant subspace of $V$ containing $v$ is all of $V$.
\end{enumerate}

\noindent The vector $v$ is called a \textbf{cyclic vector} for $\pi$.
\end{definition}

\begin{proposition}
\label{highest.highest}Let $\left(  \pi,V\right)  $ be a highest weight cyclic
representation of $\mathsf{sl}\left(  3;\mathbb{C}\right)  $ with weight
$\mu_{0}$. Then

\begin{enumerate}
\item $\pi$ has highest weight $\mu_{0}$.

\item  The weight space corresponding to the highest weight $\mu_{0}$ is one-dimensional.
\end{enumerate}
\end{proposition}

\subsubsection{Proof}

\begin{proof}
Let $v$ be as in the definition. Consider the subspace $W$ of $V$ spanned by
elements of the form
\begin{equation}
w=\pi(Y_{i_{1}})\pi(Y_{i_{2}})\cdots\pi(Y_{i_{n}})v\label{lower.cyc}%
\end{equation}
with each $i_{l}=1,2$, and $n\geq0$. (If $n=0$, it is understood that $w\ $in
(\ref{lower.cyc}) is equal to $v$.) I assert that $W$ is invariant. To see
this, it suffices to check that $W$ is invariant under each of the basis elements.

By definition, $W$ is invariant under $\pi(Y_{1})$ and $\pi(Y_{2})$. It is
thus also invariant under $\pi(Y_{3})=-\left[  \pi(Y_{1}),\pi(Y_{2})\right]  $.

Now, Lemma \ref{weight.root} tells us that applying a root vector $Z_{\alpha
}\in\mathsf{sl}\left(  3;\mathbb{C}\right)  $ to a weight vector $v$ with
weight $\mu$ gives either zero, or else a new weight vector with weight
$\mu+\alpha$. Now, by assumption, $v$ is a weight vector with weight $\mu_{0}%
$. Furthermore, $Y_{1}$ and $Y_{2}$ are root vectors with roots $-\alpha
^{(1)}=(-2,1)$ and $-\alpha^{(2)}=(1,-2)$, respectively. (See Equation
(\ref{the.roots}).) Thus each application of $\pi(Y_{1})$ or $\pi(Y_{2})$
subtracts $\alpha^{(1)}$ or $\alpha^{(2)}$ from the weight. In particular,
each non-zero element of the form (\ref{lower.cyc}) is a simultaneous
eigenvector for $\pi(H_{1})$ and $\pi(H_{2})$. Thus $W$ is invariant under
$\pi(H_{1})$ and $\pi(H_{2})$.

To show that $W$ is invariant under $\pi(X_{1})$ and $\pi(X_{2})$, we argue by
induction on $n$. For $n=0$, we have $\pi(X_{1})v=\pi(X_{2})v=0\in W$. Now
consider applying $\pi(X_{1})$ or $\pi(X_{2})$ to a vector of the form
(\ref{lower.cyc}). Recall the commutation relations involving an $X_{1}$ or
$X_{2}$ and a $Y_{1}$ or $Y_{2}$:
\[%
\begin{array}
[c]{ccccccc}%
\left[  X_{1},Y_{1}\right]   & = & H_{1} &  & \left[  X_{1},Y_{2}\right]   &
= & 0\\
\left[  X_{2},Y_{1}\right]   & = & 0 &  & \left[  X_{2},Y_{2}\right]   & = &
H_{2}\text{.}%
\end{array}
\]
Thus (for $i$ and $j$ equal to 1 or 2) $\pi(X_{i})\pi(Y_{j})=\pi(Y_{j}%
)\pi(X_{i})+\pi(H_{ij})$, where $H_{ij}$ is either $H_{1}$ or $H_{2}$ or zero.
Hence (for $i$ equal to 1 or 2)
\begin{align*}
\pi(X_{i})\pi(Y_{i_{1}})\pi(Y_{i_{2}})\cdots\pi(Y_{i_{n}})v\\
=\pi(Y_{i_{1}})\pi(X_{i})\pi(Y_{i_{2}})\cdots\pi(Y_{i_{n}})v+\pi(H_{ij}%
)\pi(Y_{i_{2}})\cdots\pi(Y_{i_{n}})v\text{.}%
\end{align*}
But $\pi(X_{i})\pi(Y_{i_{2}})\cdots\pi(Y_{i_{n}})v$ is in $W$ by induction,
and $\pi(H_{ij})\pi(Y_{i_{2}})\cdots\pi(Y_{i_{n}})v$ is in $W$ since $W$ is
invariant under $\pi(H_{1})$ and $\pi(H_{2})$.

Finally, $W$ is invariant under $\pi(X_{3})$ since $\pi(X_{3})=\left[
\pi(X_{1}),\pi(X_{2})\right]  $. Thus $W$ is invariant. Since by definition
$W$ contains $v$, we must have $W=V$.

Since $Y_{1}$ is a root vector with root $-\alpha^{(1)}$ and $Y_{2}$ is a root
vector with root $-\alpha^{(2)}$, Lemma \ref{weight.root} tells us that each
element of the form (\ref{lower.cyc}) is either zero or a weight vector with
weight $\mu_{0}-\alpha^{(i_{1})}-\cdots-\alpha^{(i_{n})}$. Thus $V=W$ is
spanned by $v$ together with weight vectors with weights lower than $\mu_{0}$.
Thus $\mu_{0}$ is the highest weight for $V$.

Furthermore,every element of $W$ can be written as a multiple of $v$ plus a
linear combination of weight vectors with weights lower than $\mu_{0}$. Thus
the weight space corresponding to $\mu_{0}$ is spanned by $v$; that is, the
weight space corresponding to $\mu_{0}$ is one-dimensional.
\end{proof}

\begin{proposition}
Every irreducible representation of $\mathsf{sl}\left(  3;\mathbb{C}\right)  $
is a highest weight cyclic representation, with a unique highest weight
$\mu_{0}$.
\end{proposition}

\begin{proof}
Uniqueness is immediate, since by the previous Proposition, $\mu_{0}$ is the
highest weight, and two distinct weights cannot both be highest.

We have already shown that every irreducible representation is the direct sum
of its weight spaces. Since the representation is finite-dimensional, there
can be only finitely many weights. It follows that there must exist a weight
$\mu_{0}$ such that there is no weight $\mu\neq\mu_{0}$ with $\mu\succeq
\mu_{0}$. This says that there is no weight higher than $\mu_{0}$ (which is
\textit{not} the same as saying the $\mu_{0}$ is highest). But if there is no
weight higher than $\mu_{0}$, then for any non-zero weight vector $v$ with
weight $\mu_{0}$, we must have
\[
\pi(X_{1})v=\pi(X_{2})v=0\text{.}%
\]
(For otherwise, say, $\pi(X_{1})v$ will be a weight vector with weight
$\mu_{0}+\alpha^{(1)}\succ\mu_{0}$.)

Since $\pi$ is assumed irreducible, the smallest invariant subspace containing
$v$ must be the whole space; therefore the representation is highest weight
cyclic. $\square$
\end{proof}

\begin{proposition}
\label{highest.irred}Every highest weight cyclic representation of
$\mathsf{sl}\left(  3;\mathbb{C}\right)  $ is irreducible.
\end{proposition}

\begin{proof}
Let $\left(  \pi,V\right)  $ be a highest weight cyclic representation with
highest weight $\mu_{0}$ and cyclic vector $v$. By complete reducibility
(Proposition \ref{reduce.sl3}), $V$ decomposes as a direct sum of irreducible
representations
\begin{equation}
V\cong\bigoplus_{i}V_{i}\text{.}\label{v.sum}%
\end{equation}

By Proposition \ref{sl3.weights}, each of the $V_{i}$'s is the direct sum of
its weight spaces. Thus since the weight $\mu_{0}$ occurs in $V$, it must
occur in some $V_{i}$. On the other hand, Proposition \ref{highest.highest}
says that the weight space corresponding to $\mu_{0}$ is one-dimensional, that
is, $v$ is (up to a constant) the \textit{only} vector in $V$ with weight
$\mu_{0}$. Thus $V_{i}$ must contain $v$. But then that $V_{i}$ is an
invariant subspace containing $v$, so $V_{i}=V$. Thus there is only one term
in the sum (\ref{v.sum}), and $V$ is irreducible.
\end{proof}

\begin{proposition}
Two irreducible representations of $\mathsf{sl}\left(  3;\mathbb{C}\right)  $
with the same highest weight are equivalent.
\end{proposition}

\begin{proof}
We now know that a representation is irreducible if and only if it is highest
weight cyclic. Suppose that $\left(  \pi,V\right)  $ and $\left(
\sigma,W\right)  $ are two such representations with the same highest weight
$\mu_{0}$. Let $v$ and $w$ be the cyclic vectors for $V$ and $W$,
respectively. Now consider the representation $V\oplus W$, and let $U$ be
smallest invariant subspace of $V\oplus W$ which contains the vector $(v,w)$.

By definition, $U$ is a highest weight cyclic representation, therefore
irreducible by Proposition. \ref{highest.irred}. Consider the two
``projection'' maps $P_{1}:V\oplus W\rightarrow V$, $P_{1}(v,w)=v$ and
$P_{2}:V\oplus W\rightarrow W$, $P_{1}(v,w)=w$. It is easy to check that
$P_{1}$ and $P_{2}$ are morphisms of representations. Therefore the
restrictions of $P_{1}$ and $P_{2}$ to $U\subset V\oplus W$ will also be morphisms.

Clearly neither $\left.  P_{1}\right|  _{U}$ nor $\left.  P_{2}\right|  _{U}$
is the zero map (since both are non-zero on $\left(  v,w\right)  $). Moreover,
$U$, $V$, and $W$ are all irreducible. Therefore, by Schur's Lemma, $\left.
P_{1}\right|  _{U}$ is an isomorphism of $U$ with $V$, and $\left.
P_{2}\right|  _{U}$ is an isomorphism of $U$ with $W$. Thus $V\cong U\cong W$.
\end{proof}

\begin{proposition}
If $\pi$ is an irreducible representation of $\mathsf{sl}\left(
3;\mathbb{C}\right)  $, then the highest weight of $\pi$ is of the form
\[
\mu=(m_{1},m_{2})
\]
with $m_{1}$ and $m_{2}$ non-negative integers.
\end{proposition}

\begin{proof}
We already know that \textit{all} of the weights of $\pi$ are of the form
$(m_{1},m_{2})$, with $m_{1}$ and $m_{2}$ integers. We must show that if
$\mu_{0}=(m_{1},m_{2})$ is the highest weight, then $m_{1}$ and $m_{2}$ are
both non-negative. For this, we again use what we know about the
representations of $\mathsf{sl}(2;\mathbb{C})$. The following result can be
obtained from the proof of the classification of the irreducible
representations of $\mathsf{sl}(2;\mathbb{C})$.

Let $\left(  \pi,V\right)  $ be any finite-dimensional representation of
$\mathsf{sl}(2;\mathbb{C})$. Let $v$ be an eigenvector for $\pi(H)$ with
eigenvalue $\lambda$. If $\pi(X)v=0$, then $\lambda$ is a non-negative integer.

Now, if $\pi$ is an irreducible representation of $\mathsf{sl}\left(
3;\mathbf{C}\right)  $ with highest weight $\mu_{0}=(m_{1},m_{2})$, and if
$v\neq0$ is a weight vector with weight $\mu_{0}$, then we must have
$\pi(X_{1})v=\pi(X_{2})v=0$. (Otherwise, $\mu_{0}$ wouldn't be highest.) Thus
applying the above result to the restrictions of $\pi$ to $\left\{
H_{1},X_{1},Y_{1}\right\}  $ and to $\left\{  H_{2},X_{2},Y_{2}\right\}  $
shows that $m_{1}$ and $m_{2}$ must be non-negative.
\end{proof}

\begin{proposition}
If $m_{1}$ and $m_{2}$ are non-negative integers, then there exists an
irreducible representation of $\mathsf{sl}\left(  3;\mathbb{C}\right)  $ with
highest weight $\mu=(m_{1},m_{2})$.
\end{proposition}

\begin{proof}
Note that the trivial representation is an irreducible representation with
highest weight $\left(  0,0\right)  $. So we need only construct
representations with at least one of $m_{1}$ and $m_{2}$ positive.

First, we construct two irreducible representations with highest weights
$\left(  1,0\right)  $ and $\left(  0,1\right)  $. (These are the so-called
\textbf{fundamental representations}.) The standard representation of
$\mathsf{sl}\left(  3;\mathbb{C}\right)  $ is an irreducible representation
with highest weight $\left(  1,0\right)  $, as is easily checked. To construct
an irreducible representation with weight $\left(  0,1\right)  $ we modify the
standard representation. Specifically, we define
\begin{equation}
\pi(Z)=-Z^{tr}\label{three.bar}%
\end{equation}
for all $Z\in\mathsf{sl}\left(  3;\mathbb{C}\right)  $. Using the fact that
$\left(  AB\right)  ^{tr}=B^{tr}A^{tr}$, it is easy to check that
\[
-\left[  Z_{1},Z_{2}\right]  ^{tr}=\left[  -Z_{1}^{tr},-Z_{2}^{tr}\right]
\]
so that $\pi$ is really a representation. (This is isomorphic to the dual of
the standard representation, as defined in Exercise \ref{define.dual} of
Chapter 5.) It is easy to see that $\pi$ is an irreducible representation with
highest weight $\left(  0,1\right)  $.

Let $\left(  \pi_{1},V_{1}\right)  $ denote $\mathbb{C}^{3}$ acted on by the
standard representation, and let $v_{1}$ denote a weight vector corresponding
to the highest weight $\left(  1,0\right)  $. (So, $v_{1}=(1,0,0)$.) Let
$\left(  \pi_{2},V_{2}\right)  $ denote $\mathbb{C}^{3}$ acted on by the
representation (\ref{three.bar}), and let $v_{2}$ denote a weight vector for
the highest weight $\left(  0,1\right)  $. (So, $v_{2}=(0,0,1)$.) Now consider
the representation
\[
V_{1}\otimes V_{1}\cdots\otimes V_{1}\otimes V_{2}\otimes V_{2}\cdots V_{2}%
\]
where $V_{1}$ occurs $m_{1}$ times, and $V_{2}$ occurs $m_{2}$ times. Note
that the action of $\mathsf{sl}\left(  3;\mathbb{C}\right)  $ on this space
is
\begin{align}
Z\rightarrow\left(  \pi_{1}(Z)\otimes I\cdots\otimes I\right)  \nonumber\\
+\left(  I\otimes\pi_{1}(Z)\otimes I\cdots\otimes I\right)  +\cdots+\left(
I\otimes\cdots I\otimes\pi_{2}(Z)\right)  \text{.}\label{tensor.n.m}%
\end{align}
Let $\pi_{m_{1},m_{2}}$ denote this representation.

Consider the vector
\[
v_{m_{1},m_{2}}=v_{1}\otimes v_{1}\cdots\otimes v_{1}\otimes v_{2}\otimes
v_{2}\cdots\otimes v_{2}\text{.}%
\]
Then applying (\ref{tensor.n.m}) shows that
\begin{align}
\pi_{m_{1},m_{2}}(H_{1})v_{m_{1},m_{2}}=m_{1}v_{m_{1},m_{2}}\nonumber\\
\pi_{m_{1},m_{2}}(H_{2})v_{m_{1},m_{2}}=m_{2}v_{m_{1},m_{2}}\nonumber\\
\pi_{m_{1},m_{2}}(X_{1})v_{m_{1},m_{2}}=0\nonumber\\
\pi_{m_{1},m_{2}}(X_{2})v_{m_{1},m_{2}}=0\text{.}\label{cyclic.nm}%
\end{align}

Now, the representation $\pi_{m_{1},m_{2}}$ is \textit{not} irreducible
(unless $(m_{1},m_{2})=\left(  1,0\right)  $ or $\left(  0,1\right)  $).
However, if we let $W$ denote the smallest invariant subspace containing the
vector $v_{m_{1},m_{2}}$, then in light of (\ref{cyclic.nm}), $W$ will be
highest weight cyclic with highest weight $(m_{1},m_{2})$. Therefore by
Proposition \ref{highest.irred}, $W$ is irreducible with highest weight
$(m_{1},m_{2})$.

Thus $W$ is the representation we want.
\end{proof}

We have now completed the proof of Theorem \ref{classify.sl3}.

\section{An Example: Highest Weight $\left(  1,1\right)  $}

To obtain the irreducible representation with highest weight $\left(
1,1\right)  $ we are supposed to take the tensor product of the irreducible
representations with highest weights $\left(  1,0\right)  $ and $\left(
0,1\right)  $, and then extract a certain invariant subspace. Let us establish
some notation for the representations $\left(  1,0\right)  $ and $\left(
0,1\right)  $. In the standard representation, the weight vectors for
\[%
\begin{array}
[c]{cc}%
H_{1}=\left(
\begin{array}
[c]{ccc}%
1 & 0 & 0\\
0 & -1 & 0\\
0 & 0 & 0
\end{array}
\right)  ; & H_{2}=\left(
\begin{array}
[c]{ccc}%
0 & 0 & 0\\
0 & 1 & 0\\
0 & 0 & -1
\end{array}
\right)  ;
\end{array}
\]
are the standard basis elements for $\mathbb{C}^{3}$, namely, $e_{1}$, $e_{2}%
$, and $e_{3}$. The corresponding weights are $\left(  1,0\right)  $, $\left(
-1,1\right)  $, and $\left(  0,-1\right)  $. The highest weight is $\left(
1,0\right)  $.

Recall that
\[%
\begin{array}
[c]{cc}%
Y_{1}=\left(
\begin{array}
[c]{ccc}%
0 & 0 & 0\\
1 & 0 & 0\\
0 & 0 & 0
\end{array}
\right)  ; & Y_{2}=\left(
\begin{array}
[c]{ccc}%
0 & 0 & 0\\
0 & 0 & 0\\
0 & 1 & 0
\end{array}
\right)  \text{.}%
\end{array}
\]
Thus
\begin{equation}%
\begin{array}
[c]{ccccccc}%
Y_{1}(e_{1}) & = & e_{2} &  & Y_{2}(e_{1}) & = & 0\\
Y_{1}(e_{2}) & = & 0 &  & Y_{2}(e_{2}) & = & e_{3}\\
Y_{1}(e_{3}) & = & 0 &  & Y_{2}(e_{3}) & = & 0\text{.}%
\end{array}
\label{weight1.0}%
\end{equation}

Now, the representation with highest weight $\left(  0,1\right)  $ is the
representation $\pi(Z)=-Z^{tr}$, for $Z\in\mathsf{sl}\left(  3;\mathbb{C}%
\right)  $. Let us define
\[
\overline{Z}=-Z^{tr}%
\]
for all $Z\in\mathsf{sl}\left(  3;\mathbb{C}\right)  $. Thus $\pi
(Z)=\overline{Z}$. Note that
\[%
\begin{array}
[c]{cc}%
\overline{H_{1}}=\left(
\begin{array}
[c]{ccc}%
-1 & 0 & 0\\
0 & 1 & 0\\
0 & 0 & 0
\end{array}
\right)  ; & \overline{H_{2}}=\left(
\begin{array}
[c]{ccc}%
0 & 0 & 0\\
0 & -1 & 0\\
0 & 0 & 1
\end{array}
\right)  \text{.}%
\end{array}
\]
The weight vectors are again $e_{1}$, $e_{2}$, and $e_{3}$, with weights
$\left(  -1,0\right)  $, $\left(  1,-1\right)  $, and $\left(  0,1\right)  $.
The highest weight is $\left(  0,1\right)  $.

Define new basis elements
\[%
\begin{array}
[c]{ccc}%
f_{1} & = & e_{3}\\
f_{2} & = & -e_{2}\\
f_{3} & = & e_{1}\text{.}%
\end{array}
\]
Then since
\[%
\begin{array}
[c]{cc}%
\overline{Y_{1}}=\left(
\begin{array}
[c]{ccc}%
0 & -1 & 0\\
0 & 0 & 0\\
0 & 0 & 0
\end{array}
\right)  ; & \overline{Y_{2}}=\left(
\begin{array}
[c]{ccc}%
0 & 0 & 0\\
0 & 0 & -1\\
0 & 0 & 0
\end{array}
\right)  ;
\end{array}
\]
we have
\begin{equation}%
\begin{array}
[c]{ccccccc}%
\overline{Y_{1}}(f_{1}) & = & 0 &  & \overline{Y_{2}}(f_{1}) & = & f_{2}\\
\overline{Y_{1}}(f_{2}) & = & f_{3} &  & \overline{Y_{2}}(f_{2}) & = & 0\\
\overline{Y_{1}}(f_{3}) & = & 0 &  & \overline{Y_{2}}(f_{3}) & = & 0\text{.}%
\end{array}
\label{weight0.1}%
\end{equation}
Note that the highest weight vector is $f_{1}=e_{3}$.

So, to obtain an irreducible representation with highest weight $\left(
1,1\right)  $ we are supposed to take the tensor product of the
representations with highest weights $\left(  1,0\right)  $ and $\left(
0,1\right)  $, and then take the smallest invariant subspace containing the
vector $e_{1}\otimes f_{1}$. In light of the proof of Proposition
\ref{highest.highest}, this smallest invariant subspace is obtained by
starting with $e_{1}\otimes f_{1}$ and applying all possible combinations of
$Y_{1}$ and $Y_{2}$.

Recall that if $\pi_{1}$ and $\pi_{2}$ are two representations of the Lie
algebra $\mathsf{sl}\left(  3;\mathbb{C}\right)  $, then
\begin{align*}
\left(  \pi_{1}\otimes\pi_{2}\right)  (Y_{1})  & =\pi_{1}(Y_{1})\otimes
I+I\otimes\pi_{2}(Y_{1})\\
\left(  \pi_{1}\otimes\pi_{2}\right)  (Y_{2})  & =\pi_{1}(Y_{2})\otimes
I+I\otimes\pi_{2}(Y_{2})\text{.}%
\end{align*}
In our case we want $\pi_{1}(Y_{i})=Y_{i}$ and $\pi_{2}(Y_{i})=\overline
{Y_{i}}$. Thus
\begin{align*}
\left(  \pi_{1}\otimes\pi_{2}\right)  (Y_{1})  & =Y_{1}\otimes I+I\otimes
\overline{Y_{1}}\\
\left(  \pi_{1}\otimes\pi_{2}\right)  (Y_{2})  & =Y_{2}\otimes I+I\otimes
\overline{Y_{2}}\text{.}%
\end{align*}
The actions of $Y_{i}$ and $\overline{Y_{i}}$ are described in
(\ref{weight1.0}) and (\ref{weight0.1}).

Note that $\pi_{1}\otimes\pi_{2}$ is \textit{not} an irreducible
representation. The representation $\pi_{1}\otimes\pi_{2}$ has dimension 9,
whereas the smallest invariant subspace containing $e_{1}\otimes f_{1}$ has,
as it turns out, dimension 8.

So, it remains only to begin with $e_{1}\otimes f_{1}$, apply $Y_{1}$ and
$Y_{2}$ repeatedly until we get zero, and then figure out what dependence
relations exist among the vectors we get. These computations are done on a
supplementary page. Note that the weight $\left(  0,0\right)  $ has
multiplicity two. This is because, starting with $e_{1}\otimes f_{1}$,
applying $Y_{1}$ and then $Y_{2}$ gives something different than applying
$Y_{2}$ and then $Y_{1}$.

\section{The Weyl Group}

The set of weights of an arbitrary irreducible representation of
$\mathsf{sl}\left(  3;\mathbb{C}\right)  $ has a certain symmetry associated
to it. This symmetry is in terms of something called the ``Weyl group.'' (My
treatment of the Weyl group follows Br\"ocker and tom Dieck, Chap. IV, 1.3.)
We consider the following subgroup of $\mathsf{SU}(3)$:
\[
W=\left\{
\begin{array}
[c]{ccc}%
w_{0}=\left(
\begin{array}
[c]{ccc}%
1 & 0 & 0\\
0 & 1 & 0\\
0 & 0 & 1
\end{array}
\right)  ; & w_{1}=\left(
\begin{array}
[c]{ccc}%
0 & 0 & 1\\
1 & 0 & 0\\
0 & 1 & 0
\end{array}
\right)  ; & w_{2}=\left(
\begin{array}
[c]{ccc}%
0 & 1 & 0\\
0 & 0 & 1\\
1 & 0 & 0
\end{array}
\right) \\
w_{3}=-\left(
\begin{array}
[c]{ccc}%
0 & 1 & 0\\
1 & 0 & 0\\
0 & 0 & 1
\end{array}
\right)  ; & w_{4}=-\left(
\begin{array}
[c]{ccc}%
0 & 0 & 1\\
0 & 1 & 0\\
1 & 0 & 0
\end{array}
\right)  ; & w_{5}=-\left(
\begin{array}
[c]{ccc}%
1 & 0 & 0\\
0 & 0 & 1\\
0 & 1 & 0
\end{array}
\right)
\end{array}
\right\}  \text{.}%
\]
These are simply the matrices which permute the standard basis elements of
$\mathbb{C}^{3}$, with an adjustment of overall sign when necessary to make
the determinant equal one.

Now, for any $A\in\mathsf{SU}(3)$, we have the associated map $\mathrm{Ad}%
A:\mathsf{su}(3)\rightarrow\mathsf{su}(3)$, where
\[
\mathrm{Ad}A(X)=AXA^{-1}\text{.}%
\]
Now, since each element of $\mathsf{sl}\left(  3;\mathbb{C}\right)  $ is of
the form $Z=X+iY$ with $X,Y\in\mathsf{su}(3)$, it follows that $\mathsf{sl}%
\left(  3;\mathbb{C}\right)  $ is invariant under the map $Z\rightarrow
AZA^{-1}$. That is, we can think of $\mathrm{Ad}A$ as a map of $\mathsf{sl}%
\left(  3;\mathbb{C}\right)  $ to itself.

The reason for selecting the above group is the following: If $w\in W$, then
$\mathrm{Ad}w(H_{1})$ and $\mathrm{Ad}w(H_{2})$ are linear combinations of
$H_{1}$ and $H_{2}$. That is, each $\mathrm{Ad}w$ preserves the space spanned
by $H_{1}$ and $H_{2} $. (There are other elements of $\mathsf{SU}(3)$ with
this property, notably, the diagonal elements. However, these actually commute
with $H_{1}$ and $H_{2}$. Thus the adjoint action of these elements on the
span of $H_{1}$ and $H_{2}$ is trivial and therefore uninteresting. See
Exercise \ref{weyl.quotient}.)

Now, for each $w\in W$ and each irreducible representation $\pi$ of
$\mathsf{sl}\left(  3;\mathbb{C}\right)  $, let's define a new representation
$\pi_{w}$ by the formula
\[
\pi_{w}(X)=\pi\left(  \mathrm{Ad}w^{-1}(X)\right)  =\pi(w^{-1}Xw)\text{.}%
\]
Since $\mathrm{Ad}w^{-1}$ is a Lie algebra automorphism, $\pi_{w}$ will in
fact be a representation of $\mathsf{sl}\left(  3;\mathbb{C}\right)  $.

Recall that since $\mathsf{SU}(3)$ is simply connected, then for each
representation $\pi$ of $\mathsf{sl}\left(  3;\mathbb{C}\right)  $ there is an
associated representation $\Pi$ of $\mathsf{SU}(3)$ (acting on the same space)
such that
\[
\Pi\left(  e^{X}\right)  =e^{\pi(X)}%
\]
for all $X\in\mathsf{su}(3)\subset\mathsf{sl}\left(  3;\mathbb{C}\right)  $.
The representation $\Pi$ has the property that
\begin{equation}
\pi(AXA^{-1})=\Pi(A)\pi(X)\Pi(A)^{-1}\label{ad.su3}%
\end{equation}
for all $X\in\mathsf{su}(3)$. Again since every element of $\mathsf{sl}\left(
3;\mathbb{C}\right)  $ is of the form $X+iY$ with $X,Y\in\mathsf{su}(3)$, it
follows that (\ref{ad.su3}) holds also for $X\in\mathsf{sl}\left(
3;\mathbb{C}\right)  $.

In particular, taking $A=w^{-1}\in W$ we have
\begin{equation}
\pi_{w}(X)=\pi(w^{-1}Xw)=\Pi(w)^{-1}\pi(X)\Pi(w)\label{ad.weyl}%
\end{equation}
for all $X\in\mathsf{sl}\left(  3;\mathbb{C}\right)  $.

\begin{proposition}
For each representation $\pi$ of $\mathsf{sl}\left(  3;\mathbb{C}\right)  $
and for each $w\in W$, the representation $\pi_{w}$ is equivalent to the
representation $\pi$.
\end{proposition}

\begin{proof}
We need a map $\phi:V\rightarrow V$ with the property that
\[
\phi\left(  \pi_{w}(X)v\right)  =\pi(X)\phi(v)
\]
for all $v\in V$. This is the same as saying that $\phi\pi_{w}(X)=\pi(X)\phi$,
or equivalently that $\pi_{w}(X)=\phi^{-1}\pi(X)\phi$. But in light of
(\ref{ad.weyl}), we can take $\phi=\Pi(w)$.
\end{proof}

Although $\pi$ and $\pi_{w}$ are equivalent, they are not equal. That is, in
general $\pi(X)\neq\pi_{w}(X)$. You should think of $\pi$ and $\pi_{w}$ as
differing by a change of basis on $V$, where the change-of-basis matrix is
$\Pi(w)$. Two representations that differ just by a change of basis are
automatically equivalent.

\begin{corollary}
\label{weyl.weights}Let $\pi$ be a representation of $\mathsf{sl}\left(
3;\mathbb{C}\right)  $ and $w\in W$. Then a pair $\mu=(m_{1},m_{2})$ is a
weight for $\pi$ if and only if it is a weight for $\pi_{w}$. The multiplicity
of $\mu$ as a weight of $\pi$ is the same as the multiplicity of $\mu$ as a
weight for $\pi_{w}$.
\end{corollary}

\begin{proof}
Equivalent representations must have the same weights and the same multiplicities.
\end{proof}

Let us now compute explicitly the action of $\mathrm{Ad}w^{-1}$ on the span of
$H_{1}$ and $H_{2}$, for each $w\in W$. This is a straightforward
computation.
\begin{equation}%
\begin{array}
[c]{ccccccc}%
w_{0}^{-1}H_{1}w_{0} & = & H_{1} &  & w_{3}^{-1}H_{1}w_{3} & = & -H_{1}\\
w_{0}^{-1}H_{2}w_{0} & = & H_{2} &  & w_{3}^{-1}H_{2}w_{3} & = & H_{1}+H_{2}\\
&  &  &  &  &  & \\
w_{1}^{-1}H_{1}w_{1} & = & -H_{1}-H_{2} &  & w_{4}^{-1}H_{1}w_{4} & = &
-H_{2}\\
w_{1}^{-1}H_{2}w_{1} & = & H_{1} &  & w_{4}^{-1}H_{2}w_{4} & = & -H_{1}\\
&  &  &  &  &  & \\
w_{2}^{-1}H_{1}w_{2} & = & H_{2} &  & w_{5}^{-1}H_{1}w_{5} & = & H_{1}+H_{2}\\
w_{2}^{-1}H_{2}w_{2} & = & -H_{1}-H_{2} &  & w_{5}^{-1}H_{2}w_{5} & = &
-H_{2}\text{.}%
\end{array}
\label{weyl.action}%
\end{equation}

We can now see the significance of the Weyl group. Let $\pi$ be a
representation of $\mathsf{sl}\left(  3;\mathbb{C}\right)  $, $\mu
=(m_{1},m_{2})$ a weight, and $v\neq0$ a weight vector with weight $\mu$.
Then, for example,
\begin{align*}
\pi_{w_{1}}(H_{1})v  & =\pi(w_{1}^{-1}H_{1}w_{1})v=\pi(-H_{1}-H_{2}%
)v=(-m_{1}-m_{2})v\\
\pi_{w_{1}}(H_{2})v  & =\pi(w_{1}^{-1}H_{2}w_{1})v=\pi(H_{1})v=m_{1}v\text{.}%
\end{align*}
Thus $v$ is a weight vector for $\pi_{w}$ with weight $(-m_{1}-m_{2},m_{1})$.
But by Corollary \ref{weyl.weights}, the weights of $\pi$ and of $\pi_{w}$ are
the same!

\begin{quote}
\textbf{Conclusion}: If $\mu=(m_{1},m_{2})$ is a weight for $\pi$, so is
$(-m_{1}-m_{2},m_{1})$. The multiplicities of $(m_{1},m_{2})$ and
$(-m_{1}-m_{2},m_{1})$ are the same.
\end{quote}

Of course, a similar argument applies to each of the other elements of the
Weyl group. Specifically, if $\mu$ is a weight for some representation $\pi$,
and $w$ is an element of $W$, then there will be some new weight which must
also be a weight of $\pi$. We will denote this new weight $w\cdot\mu$. For
example, if $\mu=(m_{1},m_{2})$, then $w_{1}\cdot\mu=(-m_{1}-m_{2},m_{1})$.
(We define $w\cdot\mu$ so that if $v$ is a weight vector for $\pi$ with weight
$\mu$, then $v$ will be a weight for $\pi_{w}$ with weight $w\cdot\mu$.) From
(\ref{weyl.action}) we can read off what $w\cdot\mu$ is for each $w$.
\begin{equation}%
\begin{array}
[c]{ccccccc}%
w_{0}\cdot(m_{1},m_{2}) & = & (m_{1},m_{2}) &  & w_{3}\cdot(m_{1},m_{2}) & = &
(-m_{1},m_{1}+m_{2})\\
w_{1}\cdot(m_{1},m_{2}) & = & (-m_{1}-m_{2},m_{1}) &  & w_{4}\cdot(m_{1}%
,m_{2}) & = & (-m_{2},-m_{1})\\
w_{2}\cdot(m_{1},m_{2}) & = & (m_{2},-m_{1}-m_{2}) &  & w_{5}\cdot(m_{1}%
,m_{2}) & = & (m_{1}+m_{2},-m_{2})
\end{array}
\label{w.act}%
\end{equation}
It is straightforward to check that
\begin{equation}
w_{i}\cdot(w_{j}\cdot\mu)=(w_{i}w_{j})\cdot\mu\text{.}\label{w.group}%
\end{equation}

We have now proved the following.

\begin{theorem}
If $\mu=(m_{1},m_{2})$ is a weight and $w$ is an element of the Weyl group,
let $w\cdot\mu$ be defined by (\ref{w.act}). If $\pi$ is a finite-dimensional
representation of $\mathsf{sl}\left(  3;\mathbb{C}\right)  $, then $\mu$ is a
weight for $\pi$ if and only if $w\cdot\mu$ is a weight for $\pi$. The
multiplicity of $\mu$ is the same as the multiplicity of $w\cdot\mu$.
\end{theorem}

If we think of the weights $\mu=(m_{1},m_{2})$ as sitting inside
$\mathbb{R}^{2}$, then we can think of (\ref{w.act}) as a finite group of
linear transformations of $\mathbb{R}^{2}$. (The fact that this is a
\textit{group} of transformations follows form (\ref{w.group}).) Since this is
a \textit{finite} group of transformations, it is possible to choose an inner
product on $\mathbb{R}^{2}$ such that the action of $W$ is orthogonal. (As in
the proof of Proposition \ref{finite.reduce} in Chapter 5.) In fact, there is
(up to a constant) exactly one such inner product. In this inner product, the
action (\ref{w.act}) of the Weyl group is generated by a $120^{\circ}$
rotation and a reflection about the $y$-axis. Equivalently, the Weyl group is
the symmetry group of an equilateral triangle centered at the origin with one
vertex on the $y$-axis.

\section{Complex Semisimple Lie Algebras}

This section gives a brief synopsis of the structure theory and representation
theory of complex semisimple Lie algebras. The moral of the story is that all
such Lie algebras look and feel a lot like $\mathsf{sl}\left(  3;\mathbb{C}%
\right)  $. This section will not contain any (non-trivial) proofs.

If $\frak{g}$ is a Lie algebra, a subspace $I\subset\frak{g}$ is said to be an
\textbf{ideal} if $\left[  X,Y\right]  \in I$ for all $X\in\frak{g}$ and all
$Y\in I$. A Lie algebra $\frak{g}$ is a said to be \textbf{simple} if
$\dim\frak{g}\geq2$ and $\frak{g}$ has no ideals other than $\left\{
0\right\}  $ and $\frak{g}$. A Lie algebra $\frak{g}$ is said to be
\textbf{semisimple} if $\frak{g}$ can be written as the direct sum of simple
Lie algebras.

In this section we consider semisimple Lie algebras over the complex numbers.
Examples of complex semisimple Lie algebras include $\mathsf{sl}\left(
n;\mathbb{C}\right)  $, $\mathsf{so}(n;\mathbb{C})$ ($n\geq3$), and
$\mathsf{sp}(n;\mathbb{C})$. All of these are actually simple, except for
$\mathsf{so}(4;\mathbb{C})$ which is isomorphic to $\mathsf{sl}(2;\mathbb{C}%
)\oplus\mathsf{sl}(2;\mathbb{C})$.

\begin{definition}
Let $\frak{g}$ be a complex semisimple Lie algebra. A subspace $\frak{h}$ of
$\frak{g}$ is said to be a \textbf{Cartan subalgebra} if

\begin{enumerate}
\item $\frak{h}$ is abelian. That is, $\left[  H_{1},H_{2}\right]  =0$ for all
$H_{1},H_{2}\in\frak{h}$.

\item $\frak{h}$ is maximal abelian. That is, if $X\in\frak{g}$ satisfies
$\left[  H,X\right]  =0$ for all $H\in\frak{h}$, then $X\in\frak{h}$.

\item  For all $H\in\frak{h}$, $\mathrm{ad}H:\frak{g}\rightarrow\frak{g}$ is diagonalizable.
\end{enumerate}
\end{definition}

Since all the $H$'s commute, so do the $\mathrm{ad}H$'s. (I.e., $\left[
\mathrm{ad}H_{1},\mathrm{ad}H_{2}\right]  =\mathrm{ad}\left[  H_{1}%
,H_{2}\right]  =0$.) By assumption, each $\mathrm{ad}H$ is diagonalizable, and
they commute, therefore the $\mathrm{ad}H$'s are simultaneously
diagonalizable. (Using a standard linear algebra fact.) Let $\frak{h}^{*}$
denote the dual of $\frak{h}$, namely, the space of linear functionals on
$\frak{h}$.

\begin{definition}
If $\frak{g}$ is a complex semisimple Lie algebra and $\frak{h}$ a Cartan
subalgebra, then an element $\alpha$ of $\frak{h}^{\ast}$ is said to be a
\textbf{root} (for $\frak{g}$ with respect to $\frak{h}$) if $\alpha$ is
non-zero and there exists $Z\neq0$ in $\frak{g}$ such that
\begin{equation}
\left[  H,Z\right]  =\alpha(H)Z\label{root.vector}%
\end{equation}
for all $H\in\frak{h}$. (Thus a root is a non-zero set of simultaneous
eigenvalues for the $\mathrm{ad}H$'s.)

The vector $Z$ is called a \textbf{root vector} corresponding to the root
$\alpha$, and the space of all $Z\in\frak{g}$ satisfying (\ref{root.vector})
is the \textbf{root space} corresponding to $\alpha$. This space is denoted
$\frak{g}^{\alpha}$.

The set of all roots will be denoted $\Delta$.
\end{definition}

Note that if $\frak{g}=\mathsf{sl}\left(  3;\mathbb{C}\right)  $, then one
Cartan subalgebra is the space spanned by $H_{1}$ and $H_{2}$. The roots (with
respect to this Cartan subalgebra) have been calculated in (\ref{the.roots}).

\begin{theorem}
If $\frak{g}$ is a complex semisimple Lie algebra, then a Cartan subalgebra
$\frak{h}$ exists. If $\frak{h}_{1}$ and $\frak{h}_{2}$ are two Cartan
subalgebras, then there is an automorphism of $\frak{g}$ which takes
$\frak{h}_{1}$ to $\frak{h}_{2}$. In particular, any two Cartan subalgebras
have the same dimension.
\end{theorem}

From now on, $\frak{g}$ will denote a complex semisimple Lie algebra, and
$\frak{h}$ a fixed Cartan subalgebra in $\frak{g}$.

\begin{definition}
The \textbf{rank} of a complex semisimple Lie algebra is the dimension of a
Cartan subalgebra.
\end{definition}

For example, the rank of $\mathsf{sl}\left(  n;\mathbb{C}\right)  $ is $n-1$.
One Cartan subalgebra in $\mathsf{sl}\left(  n;\mathbb{C}\right)  $ is the
space of diagonal matrices with trace zero. (Note that in the case $n=3$ the
space of diagonal matrices with trace zero is precisely the span of $H_{1}$
and $H_{2}$.) Both $\mathsf{so}(2n;\mathbb{C})$ and $\mathsf{so}%
(2n+1;\mathbb{C})$ have rank $n$.

\begin{definition}
Let $\left(  \pi,V\right)  $ be a finite-dimensional, complex-linear
representation of $\frak{g}$. Then $\mu\in\frak{h}^{\ast}$ is called a
\textbf{weight} for $\pi$ if there exists $v\neq0$ in $V$ such that
\[
\pi(H)v=\mu(H)v
\]
for all $H\in\frak{h}$. The vector $v$ is called a \textbf{weight vector} for
the weight $\mu$.
\end{definition}

Note that the roots are precisely the non-zero weights for the adjoint representation.

\begin{lemma}
Let $\alpha$ be a root and $Z$ a corresponding root vector. Let $\mu$ be a
weight for a representation $\pi$ and $v$ a corresponding weight vector. Then
either $\pi(Z)v=0$ or else $\pi(Z)v$ is a weight vector with weight
$\mu+\alpha$.
\end{lemma}

\begin{proof}
Same as for $\mathsf{sl}\left(  3;\mathbf{C}\right)  $.
\end{proof}

\begin{definition}
A set of roots $\left\{  \alpha_{1},\cdots\alpha_{l}\right\}  $ is called a
\textbf{simple system} (or \textbf{basis}) if

\begin{enumerate}
\item $\left\{  \alpha_{1},\cdots\alpha_{l}\right\}  \,$ is a vector space
basis for $\frak{h}^{\ast}$.

\item  Every root $\alpha\in\Delta$ can be written in the form
\[
\alpha=n_{1}\alpha_{1}+n_{2}\alpha_{2}+\cdots+n_{l}\alpha_{l}%
\]
with each $n_{i}$ an integer, and such that the $n_{i}$'s are either all
non-negative or all non-positive.
\end{enumerate}

A root $\alpha$ is said to be \textbf{positive} (with respect to the given
simple system) if the $n_{i}$'s are non-negative; otherwise $\alpha$ is
\textbf{negative}.
\end{definition}

If $\frak{g}=\mathsf{sl}\left(  3;\mathbb{C}\right)  $ and $\frak{h=}\left\{
H_{1},H_{2}\right\}  $, then one simple system of roots is $\left\{
\alpha^{(1)},\alpha^{(2)}\right\}  =\left\{  \left(  2,-1\right)  ,\left(
-1,2\right)  \right\}  $ (with the corresponding root vectors being $X_{1}$
and $X_{2}$). The positive roots are $\left\{  \left(  2,-1\right)  ,\left(
-1,2\right)  ,\left(  1,1\right)  \right\}  $. The negative roots are
$\left\{  \left(  -2,1\right)  ,\left(  1,-2\right)  ,\left(  -1,-1\right)
\right\}  $.

\begin{definition}
Let $\left\{  \alpha_{1},\cdots\alpha_{l}\right\}  $ be a simple system of
roots and let $\mu_{1}$ and $\mu_{2}$ be two weights. Then $\mu_{1}$ is
\textbf{higher} than $\mu_{2}$ (or $\mu_{2}$ is \textbf{lower} than $\mu_{1}$)
if $\mu_{1}-\mu_{2}$ can be written as
\[
\mu_{1}-\mu_{2}=a_{1}\alpha_{1}+a_{2}\alpha_{2}+\cdots+a_{l}\alpha_{l}%
\]
with $a_{i}\geq0.$ This relation is denoted $\mu_{1}\succeq\mu_{2}$ or
$\mu_{2}\preceq\mu_{1}$.

A weight $\mu_{0}$ for a representation $\pi$ is \textbf{highest} if all the
weights $\mu$ of $\pi$ satisfy $\mu\preceq\mu_{0}$.
\end{definition}

The following deep theorem captures much of the structure theory of semisimple
Lie algebras.

\begin{theorem}
\label{roots.exist}Let $\frak{g}$ be a complex semisimple Lie algebra,
$\frak{h}$ a Cartan subalgebra, and $\Delta$ the set of roots. Then

\begin{enumerate}
\item  For each root $\alpha\in\Delta$, the corresponding root space
$\frak{g}^{\alpha}$ is one-dimensional.

\item  If $\alpha$ is a root, then so is $-\alpha$.

\item  A simple system of roots $\left\{  \alpha_{1},\cdots\alpha_{l}\right\}
$ exists.
\end{enumerate}
\end{theorem}

We now need to identify the correct set of weights to be highest weights of
irreducible representations.

\begin{theorem}
\label{sl2s}Let $\left\{  \alpha_{1},\cdots\alpha_{l}\right\}  $ denote a
simple system of roots, $X_{i}$ an element of the root space $\frak{g}%
^{\alpha_{i}}$ and $Y_{i}$ an element of the root space $\frak{g}^{-\alpha
_{i}}$. Define
\[
H_{i}=\left[  X_{i},Y_{i}\right]  \text{.}%
\]
Then it is possible to choose $X_{i}$ and $Y_{i}$ such that

\begin{enumerate}
\item  Each $H_{i}$ is non-zero and contained in $\frak{h}$.

\item  The span of $\left\{  H_{i},X_{i},Y_{i}\right\}  $ is a subalgebra of
$\frak{g}$ isomorphic (in the obvious way) to $\mathsf{sl}(2;\mathbb{C})$.

\item  The set $\left\{  H_{1},\cdots H_{l}\right\}  $ is a basis for
$\frak{h}$.
\end{enumerate}
\end{theorem}

Note that (in most cases) the set of all $H_{i}$'s, $X_{i}$'s, and $Y_{i}$'s
($i=1,2,\cdots l$) do \textit{not} span $\frak{g}$. In the case $\frak{g}%
=\mathsf{sl}\left(  3;\mathbb{C}\right)  $, $l=2$, and the span of
$H_{1},X_{1},Y_{1},H_{2},X_{2},Y_{2}$ represents only six of the eight
dimensions of $\mathsf{sl}\left(  3;\mathbb{C}\right)  $. Nevertheless the
subalgebras $\left\{  H_{i},X_{i},Y_{i}\right\}  $ play an important role.

We are now ready to state the main theorem.

\begin{theorem}
\label{classify.ss.reps}Let $\frak{g}$ be a complex semisimple Lie algebra,
$\frak{h}$ a Cartan subalgebra, and $\left\{  \alpha_{1},\cdots\alpha
_{l}\right\}  $ a simple system of roots. Let $\left\{  H_{1},\cdots
H_{l}\right\}  $ be as in Theorem \ref{sl2s}. Then

\begin{enumerate}
\item  In each irreducible representation $\pi$ of $\frak{g}$, the $\pi(H)$'s
are simultaneously diagonalizable.

\item  Each irreducible representation of $\frak{g}$ has a unique highest weight.

\item  Two irreducible representations of $\frak{g}$ with the same highest
weight are equivalent.

\item  If $\mu_{0}$ is the highest weight of an irreducible representation of
$\frak{g}$, then for $i=1,2,\cdots l$, $\mu_{0}(H_{i})$ is a non-negative integer.

\item  Conversely, if $\mu_{0}\in\frak{h}^{\ast}$ is such that $\mu_{0}%
(H_{i})$ is a non-negative integer for all $i=1,2,\cdots l$, then there is an
irreducible representation of $\frak{g}$ with highest weight $\mu_{0}$.
\end{enumerate}
\end{theorem}

The weights $\mu_{0}$ as in 4) and 5) are called \textbf{dominant integral
weights}.

\section{Exercises}

\begin{enumerate}
\item \label{relations.tr}Show that for any pair of $n\times n$ matrices $X$
and $Y$,
\[
\left[  X^{tr},Y^{tr}\right]  =-\left[  X,Y\right]  ^{tr}\text{.}%
\]
Using this fact and the fact that $X_{i}^{tr}=Y_{i}$ for $i=1,2,3$, explain
the symmetry between $X$'s and $Y$'s in the commutation relations for
$\mathsf{sl}\left(  3;\mathbb{C}\right)  $. For example, show that the
relation $\left[  Y_{1},Y_{2}\right]  =-Y_{3}$ can be obtained from the
relation $\left[  X_{1},X_{2}\right]  =X_{3}$ by taking transposes. Show that
the relation $\left[  H_{1},Y_{2}\right]  =Y_{2}$ follows from the relation
$\left[  H_{1},X_{2}\right]  =-X_{2}$.

\item  Recall the definition of the dual $\pi^{\ast}$ of a representation
$\pi$ from Exercise \ref{define.dual} of Chapter 5. Consider this for the case
of representations of $\mathsf{sl}\left(  3;\mathbb{C}\right)  $.

a) Show that the weights of $\pi^{\ast}$ are the negatives of the weights of
$\pi$.

b) Show that if $\pi$ is the irreducible representation of $\mathsf{sl}\left(
3;\mathbb{C}\right)  $ with highest weight $\left(  m_{1},m_{2}\right)  $ then
$\pi^{\ast}$ is the irreducible representation with highest weight $\left(
m_{2},m_{1}\right)  $.

\textit{Hint}: If you identify $V$ and $V^{\ast}$ by choosing a basis for $V$,
then $A^{tr}$ is just the usual matrix transpose.

\item \label{weyl.quotient}Let $\frak{h}$ denote the subspace of
$\mathsf{sl}\left(  3;\mathbb{C}\right)  $ spanned by $H_{1}$ and $H_{2}$. Let
$G$ denote the group of all matrices $A\in\mathsf{SU}(3)$ such that
$\mathrm{Ad}A$ preserves $\frak{h}$. Now let $G_{0}$ denote the group of all
matrices $A\in\mathsf{SU}(3)$ such that $\mathrm{Ad}A$ is the identity on
$\frak{h}$, i.e., such that $\mathrm{Ad}A(H_{1})=H_{1}$ and $\mathrm{Ad}%
A(H_{2})=H_{2}$. Show that $G_{0}$ is a normal subgroup of $G$. Compute $G$
and $G_{0}$. Show that $G/G_{0}$ is isomorphic to the Weyl group $W$.

\item  a) Verify Theorems \ref{roots.exist} and \ref{sl2s} explicitly for the
case $\frak{g}=\mathsf{sl}\left(  n;\mathbb{C}\right)  $.

b) Consider the task of trying to prove Theorem \ref{classify.ss.reps} for the
case of $\mathsf{sl}\left(  n;\mathbb{C}\right)  $. Now that you have done
(a), what part of the proof goes through the same way as for $\mathsf{sl}%
\left(  3;\mathbb{C}\right)  $? At what points in the proof of the
corresponding theorem for $\mathsf{sl}\left(  3;\mathbb{C}\right)  $ did we
use special properties of $\mathsf{sl}\left(  3;\mathbb{C}\right)  $?

\textit{Hint}: Most of it is the same, but there is one critical point which
we do something which does not generalize to $\mathsf{sl}\left(
n;\mathbb{C}\right)  $.
\end{enumerate}

\chapter{Cumulative exercises}

\begin{enumerate}
\item  Let $G$ be a connected matrix Lie group, and let $\mathrm{Ad}%
:G\rightarrow\mathsf{GL}(\frak{g})$ be the adjoint representation of $G$. Show
that
\[
\ker(\mathrm{Ad})=Z(G)
\]
where $Z(G)$ denotes the center of $G$. If $G=\mathsf{O}(2)$, compute
$\ker(\mathrm{Ad})$ and $Z(G)$ and show that they are not equal.

\textit{Hint}: You should use the fact that if $G$ is connected, then every
$A\in G$ can be written in the form $A=e^{X_{1}}e^{X_{2}}\cdots e^{X_{n}}$,
with $X_{i}\in\frak{g}$.

\item  Let $G$ be a finite, commutative group. Show that the number of
equivalence classes of irreducible complex representations of $G$ is equal to
the number of elements in $G$.

\textit{Hint}: Use the fact that every finite, commutative group is a product
of cyclic groups.

\item  a) Show that if $R\in\mathsf{O}(2)$, and $\det R=-1$, then $R$ has two
real, orthogonal eigenvectors with eigenvalues $1$ and $-1$.

b) Let $R$ be in $\mathsf{O}(n)$. Show that there exists a subspace $W$ of
$\mathbb{R}^{n}$ which is invariant under both $R$ and $R^{-1}$, and such that
$\dim W=1$ or $2$. Show that $W^{\perp}$ (the orthogonal complement of $W$) is
also invariant under $R$ and $R^{-1}$. Show that the restrictions of $R$ and
$R^{-1}$ to $W$ and to $W^{\perp}$ are orthogonal. (That is, show that these
restrictions preserve inner products.)

c) Let $R$ be in $\mathsf{O}(n)$. Show that $\mathbb{R}^{n}$ can be written as
the orthogonal direct sum of subspaces $W_{i}$ such that

\begin{enumerate}
\item 1) Each $W_{i}$ is invariant under $R$ and $R^{-1}$,

\item 2) Each $W_{i}$ has dimension $1$ or $2$, and

\item 3) If $\dim W_{i}=2$, then the restriction of $R$ to $W_{i}$ has
determinant one.
\end{enumerate}

d) Show that the exponential mapping for $\mathsf{SO}(n)$ is onto. Make sure
you use the fact that the elements of $\mathsf{SO}(n)$ have determinant one.

\textit{Note}: This provides an alternative proof that the group
$\mathsf{SO}(n)$ is connected.

\item  Determine, up to equivalence, all of the finite-dimensional,
irreducible (complex-linear) representations of the Lie algebra $\mathsf{sl}%
(2;\mathbb{C})\oplus\mathsf{sl}(2;\mathbb{C})$. Can your answer be expressed
in terms of a sort of ``highest weight''?

\textit{Hint}: Imitate the proof of the classification of the irreducible
representations of $\mathsf{sl}(2;\mathbb{C})$.

\item  Consider the irreducible representation $\left(  \pi,V\right)  $ of
$\mathsf{sl}\left(  3;\mathbb{C}\right)  $ with highest weight $\left(
0,2\right)  $. Following the procedure in Chapter 6, Section 5, determine

1) The dimension of $V$.

2) All of the weights of $\pi$.

3) The multiplicity of each of the weights. (That is, the dimension of the
corresponding weight spaces.)
\end{enumerate}

\chapter{Bibliography}

\begin{enumerate}
\item \textbf{Theodor Br\"{o}cker and Tammo tom Dieck},
\textit{Representations of Compact Lie Groups}. Springer-Verlag, 1985.

A good reference for basic facts on compact groups and their representations,
including characters and orthogonality relations. Analyzes representations
from a more analytic and less algebraic viewpoint than other authors.

\item \textbf{William Fulton and Joe Harris}, \textit{Representation theory. A
First Course.} Graduate Texts in Mathematics, 129. Readings in Mathematics,
Springer-Verlag, 1991. Has lots of examples. Written from an algebraic point
of view.

\item \textbf{Sigurdur Helgason}, \textit{Differential Geometry, Lie Groups,
and Symmetric Spaces}. Academic Press, 1978.

A good reference for a lot of things. Includes structure theory of semisimple groups.

\item \textbf{James E. Humphreys}, \textit{Introduction to Lie Algebras and
Representation Theory}. Springer-Verlag, 1972.

A standard reference for the Lie algebra side of things (no Lie groups).

\item \textbf{N. Jacobson}, \textit{Lie Algebras}. Interscience Tracts No. 10,
John Wiley and Sons, 1962.

Another good reference for Lie algebras.

\item \textbf{Anthony W. Knapp}, \textit{Lie groups: beyond an introduction.}
Birkhauser, 1996. Good complement to Helgason on such matters as structure
theory of Lie groups. As title suggests, not the place to start, but a good reference.

\item \textbf{W. Miller}, \textit{Symmetry Groups and Their Applications}.
Academic Press.

Oriented toward applications to physics. Includes theory of finite groups.

\item \textbf{Jean-Pierre Serre}, \textit{Complex Semisimple Lie Algebras}.
Springer-Verlag, 1987.

A very concise summary of structure theory and representation theory of
semisimple Lie algebras.

\item \textbf{Jean-Pierre Serre}, \textit{Linear Representations of Finite
Groups}. Springer-Verlag.

An introduction to both complex and modular representations of finite groups.

\item \textbf{Barry Simon}, \textit{Representations of finite and compact Lie
groups}, American Mathematical Society, 1996. Covers much of the same material
as Br\"{o}cker and tom Dieck, but from a more analytical perspective.

\item \textbf{Frank W. Warner}, \textit{Foundations of Differentiable
Manifolds and Lie Groups}. Springer-Verlag, 1983.

Key word in the title is \textit{foundations}. Gives a modern treatment of
differentiable manifolds, and then proves some important, non-trivial theorems
about Lie groups, including the relationship between subgroups and
subalgebras, and the relationship between representations of the Lie algebra
and of the Lie group.

\item \textbf{V.S. Varadarajan}, \textit{Lie Groups, Lie Algebras, and Their
Representations}. Springer-Verlag, 1974.

A comprehensive treatment of both Lie groups and Lie algebras.
\end{enumerate}
\end{document}